\documentclass[
aps, prd, reprint, superscriptaddress, nofootinbib
]{revtex4-2}

\usepackage{graphicx}
\usepackage[usenames,dvipsnames]{xcolor}
\usepackage{xspace}
\usepackage{amsmath,amsfonts,amssymb,amsthm,bm}
\usepackage[utf8]{inputenc}
\usepackage[normalem]{ulem}
\xdefinecolor{mylinkcolor}{rgb}{0,0,0.5}
\usepackage[
	breaklinks=true,
	colorlinks=true, filecolor=mylinkcolor, citecolor=mylinkcolor,
	linkcolor=mylinkcolor, urlcolor=mylinkcolor, menucolor=mylinkcolor,
]{hyperref}
\usepackage[depth=subsection]{bookmark}

\allowdisplaybreaks

\def\mr{\mathrm}
\def\di{\mr d}
\def\cross{\times}
\DeclareMathOperator{\Order}{\mathcal{O}}
\def\mF{\mathcal{F}}

\def\lhat{\bm{l}}
\def\lNhat{\bm{l}_{\rm N}}
\def\dotlNhat{\dot{\bm{l}}_{\rm N}}
\def\lamNhat{\bm{\lambda}_{\rm N}}
\def\Lscal{\tilde{L}}

\newcommand{\AEI}{\affiliation{Max Planck Institute for Gravitational Physics (Albert Einstein Institute), Am M\"uhlenberg 1, Potsdam 14476, Germany}}
\newcommand{\Maryland}{\affiliation{Department of Physics, University of Maryland, College Park, MD 20742, USA}}
\newcommand{\PI}{\affiliation{Perimeter Institute for Theoretical Physics, 31 Caroline Street North, Waterloo, ON N2L 2Y5, Canada}}

\begin{document}
\title{Theoretical groundwork supporting the precessing-spin two-body dynamics of the effective-one-body waveform models \texttt{SEOBNRv5}}

\author{Mohammed Khalil}\email{mkhalil@perimeterinstitute.ca}\PI\AEI\Maryland
\author{Alessandra Buonanno}\AEI\Maryland
\author{H\'ector Estell\'es}\AEI
\author{\\Deyan P. Mihaylov}\AEI
\author{Serguei Ossokine}\AEI
\author{Lorenzo Pompili}\AEI
\author{Antoni Ramos-Buades}\AEI

\begin{abstract}
Waveform models are essential for gravitational-wave (GW) detection and parameter estimation of coalescing compact-object binaries. 
More accurate models are required for the increasing sensitivity of current and future GW detectors.
The effective-one-body (EOB) formalism combines the post-Newtonian (PN) and small mass-ratio approximations with numerical-relativity results, and produces highly accurate inspiral-merger-ringdown waveforms.
In this paper, we derive the analytical precessing-spin two-body dynamics for the \texttt{SEOBNRv5} waveform model, which has been developed for the upcoming LIGO-Virgo-KAGRA observing run.  
We obtain an EOB Hamiltonian that reduces to the exact Kerr Hamiltonian in the test-mass limit. It includes the full 4PN precessing-spin information, and is valid for generic compact objects (i.e., for black holes or neutron stars). 
We also build an efficient and accurate EOB Hamiltonian that includes partial precessional effects, notably orbit-averaged in-plane spin effects for circular orbits, and derive 4PN-expanded precessing-spin equations of motion, consistent with such an EOB Hamiltonian. 
The results were used to build the computationally-efficient precessing-spin multipolar \texttt{SEOBNRv5PHM} waveform model.
\end{abstract}

\date{\today}

\maketitle

\section{Introduction}
Since 2015, gravitational-wave (GW) observations~\cite{LIGOScientific:2018mvr,LIGOScientific:2020ibl,LIGOScientific:2021djp} by the LIGO and Virgo detectors~\cite{LIGOScientific:2014pky,VIRGO:2014yos} have significantly improved our understanding of binary black holes (BHs) and neutron stars (NSs), and their astrophysical formation channels~\cite{LIGOScientific:2018jsj,LIGOScientific:2020kqk,LIGOScientific:2021psn}.
Making these detections and inferring the properties of GW sources
requires accurate waveform models, and the accuracy requirements for
these models will increase significantly~\cite{Purrer:2019jcp} with
upgrades to current GW detectors~\cite{KAGRA:2013rdx}, and with future
detectors in space and on the ground, such as LISA~\cite{LISA:2017pwj}, the Einstein
Telescope~\cite{Punturo:2010zz} and Cosmic
Explorer~\cite{Reitze:2019iox,Evans:2021gyd}.

Numerical relativity (NR) simulations~\cite{Pretorius:2005gq,Campanelli:2005dd,Baker:2005vv} provide very accurate waveforms, but they are computationally expensive, which makes it important to develop waveform models that combine analytical approximation methods with NR results to produce longer waveforms and to cover the entire parameter space of binary systems.
The most commonly used approaches for GW sources of ground-based detectors are post-Newtonian (PN), NR surrogate, phenomenological, and effective-one-body (EOB) waveform models.

The PN approximation is a small-velocity and weak-field expansion
(see, e.g., Refs.~\cite{Futamase:2007zz,Blanchet:2013haa,Schafer:2018kuf,Levi:2015msa,Porto:2016pyg,Levi:2018nxp,Isoyama:2020lls}
for reviews), and PN-based Taylor-expanded waveform
models~\cite{Poisson:1995ef,Damour:1997ub,Droz:1999qx,Damour:2000gg,Damour:2000zb,Buonanno:2002ft,Buonanno:2002fy,Damour:2002kr,Arun:2004hn,Gopakumar:2007vh,Boyle:2007ft,Buonanno:2009zt,Ajith:2011ec,Klein:2013qda,Chatziioannou:2013dza,Chatziioannou:2017tdw,Mishra:2016whh,Isoyama:2017tbp,Moore:2016qxz,Moore:2018kvz,Moore:2019xkm}
produce fast-to-evaluate waveforms, but are only accurate for the early
inspiral.  NR surrogate
models~\cite{Blackman:2015pia,Blackman:2017dfb,Blackman:2017pcm,Varma:2018mmi,Varma:2019csw,Williams:2019vub,Rifat:2019ltp,Islam:2021mha,Islam:2022laz}
interpolate NR waveforms, which is possible with the recent increase
in the number of NR
catalogs~\cite{Gonzalez:2008bi,Buchman:2012dw,Chu:2009md,Mroue:2013xna,Boyle:2019kee,Healy:2017psd,Healy:2022wdn,Jani:2016wkt,Foucart:2018lhe};
hence, they are very accurate, but they are limited to regions of the parameter space for which NR simulations exist.  
Phenomenological models~\cite{Pan:2007nw,Ajith:2007qp,Ajith:2009bn,Santamaria:2010yb,Hannam:2013oca,Husa:2015iqa,Khan:2015jqa,London:2017bcn,Khan:2018fmp,Khan:2019kot,Dietrich:2019kaq,Pratten:2020fqn,Pratten:2020ceb,Garcia-Quiros:2020qpx,Estelles:2020osj,Estelles:2020twz,Estelles:2021gvs,Hamilton:2021pkf}
combine PN and EOB waveforms for the inspiral with fits to NR
results for the late inspiral and merger-ringdown parts of the waveform.

The EOB formalism~\cite{Buonanno:1998gg,Buonanno:2000ef,Damour:2000we,Damour:2001tu,Buonanno:2005xu} combines information from several analytical approximation methods with NR results. It maps the dynamics of a compact binary to that of a test mass or test spin in a deformed Schwarzschild or Kerr background, with the deformation parameter being the symmetric mass ratio, implying that it contains the exact strong-field test-body limit.
EOB waveform models have been constructed for nonspinning \cite{Damour:2000we,Buonanno:2007pf,Damour:2007yf,Damour:2008gu,Buonanno:2009qa,
Pan:2011gk,Damour:2012ky,Damour:2015isa,Nagar:2019wds}, spinning~\cite{Damour:2001tu,Buonanno:2005xu,Damour:2007vq,
Damour:2008qf,Pan:2009wj,Damour:2008te,Barausse:2009xi,Barausse:2011ys,Nagar:2011fx,Damour:2014sva,Balmelli:2015zsa,Khalil:2020mmr,Taracchini:2012ig,Taracchini:2013rva,Bohe:2016gbl,Cotesta:2018fcv,Pan:2013rra,Babak:2016tgq,Ossokine:2020kjp,Nagar:2018plt,Nagar:2018zoe,Akcay:2020qrj,Gamba:2021ydi}, and eccentric binaries~\cite{Bini:2012ji,Hinderer:2017jcs,Nagar:2021gss,Khalil:2021txt,Ramos-Buades:2021adz,Albanesi:2022xge}.
In addition, tidal effects~\cite{Bernuzzi:2014owa,Akcay:2018yyh,Steinhoff:2021dsn,Matas:2020wab,Gonzalez:2022prs} and information from the post-Minkowskian ~\cite{Damour:2016gwp,Damour:2017zjx,Antonelli:2019ytb,Damgaard:2021rnk,Khalil:2022ylj,Damour:2022ybd} and small mass-ratio approximations~\cite{Damour:2009sm,Yunes:2009ef,Yunes:2010zj,Barausse:2011dq,Akcay:2012ea,Antonelli:2019fmq,Nagar:2022fep} have been incorporated in EOB models.
To reduce the computational cost of EOB waveforms, surrogate or reduced-order frequency-domain models have be developed in Refs.~\cite{Field:2013cfa,Purrer:2014fza,Purrer:2015tud,Lackey:2016krb,Lackey:2018zvw,Cotesta:2020qhw,Gadre:2022sed,Tissino:2022thn,Khan:2020fso,Thomas:2022rmc}.

EOB waveform models consist of three main components: 
(i) the Hamiltonian, which describes the conservative binary dynamics, and from which one obtains the equations of motion (EOMs);
(ii) the inspiral-merger-ringdown waveform, which resums PN information for the inspiral and includes functional fits to NR results for the plunge and merger-ringdown parts;
and (iii) the radiation reaction (RR) force, which is computed from the inspiral waveform modes and is added to the EOMs to account for the energy and angular momentum losses due to the emitted GWs.

Two main families of EOB waveform models exist: \texttt{SEOBNR} (e.g., see Refs.~\cite{Bohe:2016gbl,Cotesta:2018fcv,Ossokine:2020kjp}) and \texttt{TEOBResumS} (e.g., see Refs.~\cite{Nagar:2018zoe,Nagar:2020pcj,Gamba:2021ydi}).
In this paper, we derive the analytical precessing-spin two-body dynamics of the \texttt{SEOBNRv5} waveform model\footnote{
\texttt{SEOBNRv5} is publicly available through the python package \texttt{pySEOBNR} \href{https://git.ligo.org/waveforms/software/pyseobnr}{\texttt{git.ligo.org/waveforms/software/pyseobnr}}. Stable versions of \texttt{pySEOBNR} are published through the Python Package Index (PyPI), and can be installed via ~\texttt{pip install pyseobnr}.
}, which has been developed for the upcoming LIGO-Virgo-KAGRA observing run (O4)~\cite{LVKO4}. This waveform model has been built in \texttt{Python} language and is publicly available. Details of the model are provided in Ref.~\cite{Mihaylovv5} for the software (\texttt{pySEOBNR}), in Ref.~\cite{Pompiliv5} for the aligned-spin model (\texttt{SEOBNRv5HM}), in Ref.~\cite{RamosBuadesv5} for the precessing-spin model (\texttt{SEOBNRv5PHM}), and in Ref.~\cite{VandeMeentv5} for the inclusion of second-order gravitational self-force results in the nonspinning dissipative sector of the \texttt{SEOBNRv5} dynamics.

This paper is organized as follows. In Sec.~\ref{sec:Hamiltonian}, we derive a Hamiltonian, $H_{\rm EOB}^{\rm prec}$, that includes the full 4PN precessing-spin information, and reduces in the test-mass limit to the exact Hamiltonian of a nonspinning point particle in Kerr background. The Hamiltonian is valid for generic orbits (inclined, circular or eccentric), and for generic compact objects (BHs or NSs), since we include the spin-multipole constants, which account for the tidal deformability of the compact object due to its spin. 
As realized in Refs.~\cite{Pan:2013rra,Babak:2016tgq,Ossokine:2020kjp,Knowles:2018hqq,Devine:2016ovp}, solving the EOMs when using the full precessing-spin EOB Hamiltonian can be computationally expensive. Therefore, to develop a more efficient model, we first build a simpler EOB Hamiltonian that includes partial precessional effects, $H_{\rm EOB}^{\rm pprec}$. Notably, we incorporate in such Hamiltonian in-plane spin effects only for circular orbits, and average them over an orbit, while neglecting fourth order spin terms. 

Then, building  
on previous studies~\cite{SpinTaylorNotes,Estelles:2020twz,Akcay:2020qrj,Gamba:2021ydi}, which employed an aligned-spin orbital dynamics 
in the co-precessing frame~\cite{Apostolatos:1994mx,Buonanno:2002fy,Boyle:2011gg,Schmidt:2012rh} and PN-expanded precessing-spin equations, we derive in Sec.~\ref{sec:PNEOMs} PN-expanded, orbit-averaged, precessing-spin equations for quasi-circular orbits, and couple them consistently with $H_{\rm EOB}^{\rm pprec}$, which is not restricted to aligned spins.
We include in the PN-expanded EOMs, the spin-orbit (SO) and spin-spin (SS) couplings to next-to-next-to-leading order (NNLO), which generalizes some results in the literature~\cite{SpinTaylorNotes,Akcay:2020qrj,Chatziioannou:2013dza,Bohe:2013cla} to higher PN orders for the SS coupling. Furthermore, even for the SO contributions, our results for the EOMs employ a different gauge and spin-supplementary condition (SSC), to be consistent with the Hamiltonian, which leads to some differences compared to previous results in the literature.

We summarize our results in Sec.~\ref{sec:conclusions}, and include a few Appendices with more details about some aspects of the calculations.
We provide our results as \texttt{Mathematica} files in the Supplemental Material~\cite{ancmaterial}.

\subsection*{Notation}
We use geometric units in which $c = G =1$.

We consider a binary with masses $m_1$ and $m_2$, with $m_1 \geq m_2$, and  define the total mass $M$, reduced mass $\mu$, symmetric mass ratio $\nu$, anti-symmetric mass ratio $\delta$, and relative masses $X_{\mr i}$ as follows:
\begin{equation}
\begin{gathered}
M\equiv m_1 + m_2, \qquad \mu \equiv \frac{m_1m_2}{M}, \qquad \nu \equiv \frac{\mu}{M}, \\ 
\delta \equiv\frac{m_1 - m_2}{M},  \qquad
X_{\mr i} \equiv \frac{m_{\mr i}}{M},
\end{gathered}
\end{equation}
where $\mr i = 1,2$.

We denote the spin vector of each body by $\bm{S}_{\mr i}$, and define the dimensionless spins $\bm{\chi}_{\mr i}$ as
\begin{gather}
\bm{\chi}_{\mr i} \equiv \frac{\bm{a}_{\mr i}}{m_{\mr i}} \equiv \frac{\bm{S}_{\mr i}}{m_{\mr i}^2}, 
\end{gather}
along with the intermediate definition for $\bm{a}_{\mr i}$. 
The spin magnitudes $\chi_{\mr i}$ vary between -1 and 1, with positive spins being in the direction of the orbital angular momentum. 
We define the following combinations of $\bm{a}_{\mr i}$:
\begin{equation}
\bm{a}_\pm \equiv \bm{a}_1 \pm \bm{a}_2.
\end{equation}

The spin quadrupole, octupole, and hexadecapole constants are denoted $C_{\rm iES^2}$, $C_{\rm iBS^3}$, and $C_{\rm iES^4}$, respectively. 
These constants equal one for BHs, but are greater than one for NSs. We define 
\begin{equation} 
\begin{aligned}
\widetilde{C}_{\rm iES^2} &\equiv C_{\rm iES^2} - 1, \\
\widetilde{C}_{\rm iBS^3} &\equiv C_{\rm iBS^3} - 1, \\
\widetilde{C}_{\rm iES^4} &\equiv C_{\rm iES^4} - 1, 
\end{aligned}
\end{equation}
such that expressions for BHs can be easily recovered by setting $\widetilde{C}_\text{...}\to 0$.
To simplify some expressions, we define the following combinations of spins and multipole constants:
\begin{equation}
\label{multipoleCoefs}
\begin{aligned}
C_\pm^{a^2} &\equiv \widetilde{C}_{\rm 1ES^2}\, a_1^2 \pm \widetilde{C}_{\rm 2ES^2} a_2^2, \\
C_\pm^{n\cdot a^2} &\equiv \widetilde{C}_{\rm 1ES^2} (\bm{n}\cdot\bm{a}_1)^2 \pm \widetilde{C}_{\rm 2ES^2} (\bm{n}\cdot\bm{a}_2)^2, \\
C_\pm^{a^3} &\equiv \widetilde{C}_{\rm 1BS^3} a_1^3 \pm \widetilde{C}_{\rm 2BS^3} a_2^3, \\
C_\pm^{a^4} &\equiv \widetilde{C}_{\rm 1ES^4} a_1^4 \pm \widetilde{C}_{\rm 2ES^4} a_2^4,
\end{aligned}
\end{equation}
which are zero for BHs (see below the definition of $\bm{n}$).

In the binary's center-of-mass, we denote the relative position and momentum vectors, $\bm{r}$ and $\bm{p}$, with
\begin{equation}
\bm p^2 = p_r^2 + \frac{L^2}{r^2}, \quad
p_r= \bm{n}\cdot\bm{p}, \quad
\bm{L}=\bm{r}\cross\bm{p},
\end{equation}
where $\bm{n}\equiv\bm{r}/r$, $\bm{L}$ is the orbital angular momentum with magnitude $L$, and $\bm{J} = \bm{L} + \bm{S}_1 + \bm{S}_2$ is the total angular momentum.
For precessing spins, we use the spherical-coordinates phase-space variables $\{r,\theta,\phi,p_r,p_\theta,p_\phi\}$, where $\theta$ is the polar angle, $\phi$ is the azimuthal angle, and $p_\phi$ and $p_\theta$ are their conjugate momenta.
For equatorial orbits (aligned spins), the angular momentum reduces to $L = p_\phi$.

For precessing spins, we use two orthonormal frames: $\{\lhat,\bm{n},\bm{\lambda}\}$ and $\{\lNhat,\bm{n},\lamNhat\}$.
In both frames, $\bm{n}$ is the unit vector in the direction of $\bm{r}$.
The vector $\lhat$ is the direction of $\bm{L}$, while $\lNhat$ is the direction of $\bm{L}_{\rm N}\equiv \mu \bm{r}\cross \bm{\mr v}$, where $\bm{\mr v}\equiv \dot{\bm{r}}$ is the velocity with $\dot{~}\equiv \di/\di t$ being the time derivative.
The other unit vectors are defined by $\bm{\lambda} \equiv \lhat \cross \bm{n}$ and $\lamNhat \equiv \lNhat \cross \bm{n}$.

The orbital angular frequency is denoted $\Omega$, and we define the velocity parameter $v \equiv (M\Omega)^{1/3}$.
We also often use $u \equiv M/r$ instead of $r$.

\section{Hamiltonian}
\label{sec:Hamiltonian}
In the EOB formalism, the Hamiltonian $H_\text{EOB}$, describing the conservative binary dynamics, is related to an effective Hamiltonian $H_\text{eff}$, describing the dynamics of a test body in a deformed BH background, with $\nu$ being the deformation, via the energy map~\cite{Buonanno:1998gg}
\begin{equation}
\label{EOBmap}
H_\text{EOB} = M \sqrt{1 + 2 \nu \left(\frac{H_\text{eff}}{\mu} - 1\right)}\,.
\end{equation}
For nonspinning binaries, in the $\nu\to0$ limit, $H_\text{eff}$ reduces to the Hamiltonian of a (nonspinning) test mass in a Schwarzschild background.
The nonspinning EOB Hamiltonian was first derived in Refs.~\cite{Buonanno:1998gg,Buonanno:2000ef} with 2PN information, and then extended to 3PN~\cite{Damour:2000we} and 4PN~\cite{Damour:2015isa}, with partial information at 5PN~\cite{Bini:2019nra,Bini:2020wpo,Blumlein:2021txe} and 6PN~\cite{Bini:2020nsb,Bini:2020hmy}.

For spinning binaries, one can follow two strategies: either map the spinning binary dynamics into that of a {\it test mass} or a {\it test spin} in a deformed Kerr background.
Indeed, the first spinning EOB Hamiltonian~\cite{Damour:2001tu} was constructed based on the Hamiltonian for the geodesic motion of a test mass in Kerr spacetime, while including leading-order (LO) SO and SS effects.
This was later extended to the next-to-leading order (NLO)~\cite{Damour:2008qf} and NNLO~\cite{Nagar:2011fx} SO levels, in addition to the NLO SS level for aligned~\cite{Balmelli:2013zna,Balmelli:2015lva,Damour:2014sva} and precessing spins~\cite{Balmelli:2015zsa}, then to NNLO SS for aligned spins and circular orbits~\cite{Nagar:2018plt}, which was used to build the (publicly available) \texttt{TEOBResumS} waveform model~\cite{Akcay:2020qrj,Gamba:2021ydi}. The complete 4PN conservative dynamics for precessing spins and generic orbits was incorporated in EOB Hamiltonians in Ref.~\cite{Khalil:2020mmr}, and the 4.5PN SO dynamics in Refs.~\cite{Antonelli:2020aeb,Antonelli:2020ybz}.

The second strategy, which maps the spinning binary dynamics into that of a test spin, was first developed in Ref.~\cite{Barausse:2009aa} (to pole-dipole order) with NLO SO and LO SS corrections~\cite{Barausse:2009xi}, and then extended to NNLO SO in Ref.~\cite{Barausse:2011ys}. 
Such a Hamiltonian is applicable for generic (precessing) spins, and reproduces (resums) spin-orbit couplings at all PN orders in the test-body limit, which makes it more complicated than Hamiltonians based on the dynamics of a test mass.
The test-spin dynamics was augmented to quadrupolar order in Ref.~\cite{Vines:2016unv}, and the EOB Hamiltonian was extended to 4PN order in Ref.~\cite{Khalil:2020mmr}.

The first \texttt{SEOBNR} waveform model developed for aligned-spin BHs~\cite{Pan:2009wj} used an effective Hamiltonian for a test mass in a deformed Kerr spacetime. Subsequently, the \texttt{SEOBNRv1}~\cite{Taracchini:2012ig}, \texttt{SEOBNRv2}~\cite{Taracchini:2013rva}, \texttt{SEOBNRv3}~\cite{Pan:2013rra,Babak:2016tgq}, and \texttt{SEOBNRv4}~\cite{Bohe:2016gbl,Cotesta:2018fcv,Ossokine:2020kjp} models, publicly available in the LIGO Algorithm Library~\cite{lalsuite}, employed an effective Hamiltonian for a test spin in a deformed Kerr background. Here, to build the \texttt{SEOBNRv5} model~\cite{Pompiliv5,RamosBuadesv5}, we take the effective Hamiltonian to be a deformation of the test-mass Kerr Hamiltonian. The masses of the background BH and test mass are identified to be $M = m_1 + m_2$ and $\mu = m_1 m_2 / M$, respectively, while the Kerr spin $\bm{a}$ is mapped to be
\begin{equation}
\label{spinmap}
\bm{a}  = \bm{a}_1 + \bm{a}_2 \equiv \bm{a}_+.
\end{equation}
An advantage of this map, besides its simplicity, is that the Kerr Hamiltonian reproduces all even-in-spin leading PN orders for binary BHs~\cite{Vines:2016qwa}. 

To include PN information in the EOB Hamiltonian, we write an ansatz for the coefficients of $H_\text{eff}$ and solve for the unknowns such that $H_\text{EOB}$ is related to a PN-expanded Hamiltonian $H_\text{PN}$ in another gauge by a canonical transformation.
To obtain that transformation, we write an ansatz for a generating function $\mathcal{G}$, perform the transformation using Poisson brackets, such that
\begin{equation}
H_\text{EOB} = H_\text{PN} + \lbrace \mathcal{G}, H_\text{PN} \rbrace
+ \frac{1}{2!}\lbrace \mathcal{G},\lbrace \mathcal{G}, H_\text{PN} \rbrace\rbrace  + \dots,
\end{equation}
where each bracket introduces a factor of $1/c^2$,
and finally match the right and left hand sides of the above equation to solve for the unknown coefficients in $H_\text{eff}$ and the generating function (see, e.g., Refs.~\cite{Balmelli:2015zsa,Khalil:2020mmr} for more details).

We include in the Hamiltonian all 4PN information for precessing spins, in addition to most of the 5PN nonspinning contributions~\cite{Bini:2019nra,Bini:2020wpo,Blumlein:2021txe}.
Many studies have contributed to deriving the 4PN conservative dynamics, for nonspinning binaries~\cite{Ohta:1974pq,Damour:1985mt,Blanchet:2000ub,Blanchet:2000nv,Jaranowski:1997ky,Jaranowski:1999qd,Itoh:2003fy,Damour:2014jta,Damour:2015isa,Damour:2016abl,Bernard:2017ktp,Marchand:2017pir,Foffa:2019rdf,Foffa:2019yfl,Blumlein:2020pog}, at SO level~\cite{Tulczyjew:1959,Tagoshi:2000zg,Porto:2005ac,Faye:2006gx,Damour:2007nc,Steinhoff:2008zr,Hartung:2011te,Perrodin:2010dy,Porto:2010tr,Hartung:2013dza,Marsat:2012fn, Bohe:2012mr,Levi:2015uxa,Levi:2020kvb}, SS~\cite{Barker:1975ae,Barker:OConnell:1979,DEath:1975wqz,Hergt:2008jn,Porto:2006bt,Porto:2008jj,Porto:2008tb,Steinhoff:2007mb,Steinhoff:2009hb,Hergt:2010pa,Hartung:2011ea,Levi:2011eq,Levi:2015ixa}, and at higher orders in the spins~\cite{Levi:2014gsa,Levi:2016ofk,Vines:2016qwa}.
We start from the 4PN precessing-spin Hamiltonian in the gauge of Ref.~\cite{Levi:2016ofk}, then perform a canonical transformation to EOB coordinates.
Thus, our EOB Hamiltonian includes the NNLO SO and SS information, as well as the LO cubic- and quartic-in-spin contributions.

In the following subsections, we begin by reviewing the Kerr Hamiltonian, and by building on it, we construct the effective Hamiltonian, which we first present for nonspinning binaries, then for aligned and precessing spins. 
We end this section by describing the differences between our EOB Hamiltonian and others in the literature.
In Sec.~\eqref{sec:simpHam}, we obtain a more computationally efficient precessing-spin Hamiltonian, albeit with partial precession effects. More specifically, when PN expanded, 
such a simplified Hamiltonian reduces to the (PN-expanded) precessing-spin Hamiltonian at cubic-in-spin order, with orbit-averaged in-plane--spin effects for circular orbits. For convenience, 
Table~\ref{tab:HeffSummary} summarizes the Hamiltonians used in this paper.

\begin{table}[th]
\caption{Summary of the Hamiltonians and their relations to each other.
For each of the effective Hamiltonians, the corresponding EOB Hamiltonian is obtained via the energy map in Eq.~\eqref{EOBmap}.}
\label{tab:HeffSummary}
\begin{ruledtabular}
\def\arraystretch{1.5}
\begin{tabular}{p{0.13\linewidth}l p{0.75\linewidth}}
symbol & Eq. & description \\
\hline
$H^\text{Kerr}$ & \eqref{KerrHam} & Kerr Hamiltonian for a (nonspinning) test mass in a \emph{generic} orbit \\
$H^\text{Kerr\,eq}$ & \eqref{KerreqHam} & Kerr Hamiltonian for a (nonspinning) test mass in an \emph{equatorial} orbit \\
$H^\text{Schw}$ & \eqref{HSchw} & Schwarzschild Hamiltonian for a test mass \\
$H_\text{eff}^\text{noS}$ & \eqref{Heffpm} & effective Hamiltonian for \emph{nonspinning} binaries; it reduces to $H^\text{Schw}$ when $\nu\to 0$ \\
$H_\text{eff}^\text{align}$ & \eqref{HeffAnzAlign} & effective Hamiltonian for \emph{aligned-spin} binaries; it reduces to $H_\text{eff}^\text{noS}$ in the zero-spin limit and to $H^\text{Kerr\,eq}$ when $\nu\to 0$ \\
$H_\text{eff}^\text{prec}$ & \eqref{HeffAnzPrec} & effective Hamiltonian for \textit{precessing-spin} binaries with full precessional (prec) effects; it reduces to $H_\text{eff}^\text{align}$ for aligned spins and to $H^\text{Kerr}$ when $\nu\to 0$ \\
$H_\text{eff}^\text{pprec}$ & \eqref{HeffAnzSimp} & effective Hamiltonian for precessing-spin binaries with \textit{partial precessional} (pprec) effects; it reduces to $H_\text{eff}^\text{align}$ for aligned spins and, when PN expanded, agrees with $H_\text{eff}^\text{prec}$ to $\Order(S^3)$ (included), with orbit-averaged in-plane--spin effects for circular orbits ($p_r = 0$) \\
\end{tabular}
\end{ruledtabular}
\end{table}

\subsection{Kerr Hamiltonian}
In Boyer-Lindquist coordinates $(t, r, \theta, \phi)$, the (inverse) Kerr metric $g_\text{Kerr}^{\mu\nu}$ can be expressed by the line element (see, e.g., Refs.~\cite{Visser:2007fj,Vines:2016unv})
\begin{align}
\label{KerrMetric}
\di s^2 &=
g_\text{Kerr}^{\mu\nu} \partial_\mu \partial_\nu
\nonumber\\
&=- \frac{\Lambda}{\Delta \Sigma} \partial_t^2+\frac{\Delta}{\Sigma}\partial_r^2+\frac{1}{\Sigma}\partial_\theta^2 \nonumber\\
&\quad +\frac{\Sigma-2Mr}{\Sigma\Delta\sin^2\theta}\partial_\phi^2-\frac{4Mra}{\Sigma\Delta}\partial_t\partial_\phi,
\end{align}
where $M$ is the mass of the BH, $a$ is its spin, and
\begin{equation}
\begin{gathered}
\Sigma \equiv r^2 + a^2 \cos^2\theta, \qquad 
\Delta \equiv r^2 - 2 M r + a^2 , \\
\Lambda \equiv (r^2+a^2)^2-a^2 \Delta \sin^2\theta .
\end{gathered}
\end{equation}
The Kerr Hamiltonian for a nonspinning test mass $H^\text{Kerr}$ can be obtained by solving the mass-shell constraint $g_\text{Kerr}^{\mu\nu}p_\mu p_\nu = -\mu^2$ for $H^\text{Kerr}$, where $\mu$ is the mass of the test mass and $p_\mu=(-H^\text{Kerr},p_r,p_\theta,p_\phi)$.

Instead of using components to express the Kerr Hamiltonian, we transform to a 3-vector notation, following Ref.~\cite{Balmelli:2015zsa}, by treating the Boyer-Lindquist coordinates as spherical coordinates, with $\bm{r} = r(\sin\theta\cos\phi,\sin\theta\sin\phi,\cos\theta)$ and $\bm{a} = (0,0,a)$, in addition to writing the momentum components in terms of the momentum vector $\bm{p}$ using
\begin{equation}
\begin{gathered}
p_r = \bm{n}\cdot\bm{p}, \qquad p_\phi = L_z = (\bm{r} \times \bm{p})_z, \\
\frac{p_\theta^2}{r^2} = \bm{p}^2 - p_r^2 - \frac{p_\phi^2}{r^2 \sin^2 \theta}.
\end{gathered}
\end{equation}

The Kerr Hamiltonian can then be written as~\cite{Balmelli:2015zsa,Khalil:2020mmr}
\begin{align}
\label{KerrHam}
H^\text{Kerr} &=  \frac{2 M r}{\Lambda} \bm{L} \cdot \bm{a} \,+
\bigg[ 
A^\text{Kerr}
\Big(\mu^2 + B_{np}^\text{Kerr} (\bm{n}\cdot\bm{p})^2  \nonumber\\
&\qquad
+ B_p^\text{Kerr}\bm{p}^2 
+ B_{npa}^\text{Kerr} (\bm{n} \cross \bm{p} \cdot \bm{a})^2
\Big)
\bigg]^{1/2}.
\end{align}
The first term in Eq.~\eqref{KerrHam} only contains odd-in-spin contributions, while the square root is the even-in-spin part, with
\begin{equation}
\label{potsKerr}
\begin{aligned}
A^\text{Kerr} &= \frac{\Delta \Sigma}{\Lambda},  \quad
&B_{np}^\text{Kerr} &=  \frac{r^2}{\Sigma} \left[\frac{\Delta}{r^2} - 1\right],  \\
B_{p}^\text{Kerr} &= \frac{r^2}{\Sigma} , \quad
&B_{npa}^\text{Kerr}  &= -\frac{r^2}{\Sigma \Lambda} (\Sigma + 2 M r) ,
\end{aligned}
\end{equation}
and
\begin{equation}
\begin{gathered}
\Sigma  = r^2 + (\bm{n} \cdot \bm{a})^2 , \qquad
\Delta  = r^2 - 2 M r + a^2 , \\
\Lambda  = \left(r^2 + a^2\right)^2 - \Delta a^2 + \Delta (\bm{n} \cdot \bm{a})^2,
\end{gathered}
\end{equation}
where we used $a\cos\theta=\bm{n}\cdot\bm{a}$, $a^2\sin^2\theta=a^2-(\bm{n}\cdot\bm{a})^2$, and $a^2p_\phi^2/r^2=(\bm{n}\cross\bm{p}\cdot\bm{a})^2$.

For equatorial orbits, the Kerr Hamiltonian reduces to
\begin{align}
\label{KerreqHam}
H^\text{Kerr\,eq} &= \frac{2M p_\phi a}{r^3+a^2 (r+2M)} + \bigg[A^\text{Kerr\,eq} \bigg(
\mu^2 + p^2  \nonumber\\
&\qquad
+ B_{np}^\text{Kerr\,eq} p_r^2 + B_{npa}^\text{Kerr\,eq} \frac{p_\phi^2 a^2}{r^2} 
\bigg)\bigg]^{1/2},
\end{align}
where
\begin{equation}
\label{potsKerreq}
\begin{aligned}
A^\text{Kerr\,eq} &= \frac{1-2 M/r+a^2/r^2}{1+ (1+2 M/r)\, a^2/r^2}, \\
B_{np}^\text{Kerr\,eq} &= \frac{a^2}{r^2} - \frac{2M}{r}, \\
B_{npa}^\text{Kerr\,eq} &= -\frac{1+2M/r}{r^2+a^2 (1+2M/r)}.
\end{aligned}
\end{equation}

In the zero-spin limit, we obtain the Schwarzschild Hamiltonian
\begin{equation}
\label{HSchw}
H^\text{Schw} = \sqrt{\left(1 - \frac{2M}{r}\right)
\left[\mu^2 + \left(1 - \frac{2M}{r}\right) p_r^2 + \frac{p_\phi^2}{r^2}\right]}\,.
\end{equation}

\subsection{Effective Hamiltonian for nonspinning binaries}
The effective Hamiltonian for nonspinning (noS) binaries can be expressed as
\begin{equation}
\label{Heffpm}
H_\text{eff}^\text{noS} = \sqrt{A_\text{noS}\left[\mu^2 + A_\text{noS} \bar{D}_\text{noS}\, p_r^2 + \frac{p_\phi^2}{r^2} + Q_\text{noS}\right]}\,,
\end{equation}
where $Q_\text{noS}(r,p_r)$ is at least quartic in $p_r$.
In the test-mass limit, we have
\begin{equation}
\begin{gathered}
A_\text{noS}(r) \overset{\nu = 0}{\longrightarrow} 1 - \frac{2M}{r}, \qquad
\bar{D}_\text{noS}(r) \overset{\nu = 0}{\longrightarrow} 1, \\
Q_\text{noS}(r,p_r) \overset{\nu = 0}{\longrightarrow} 0,
\end{gathered}
\end{equation} 
and the effective Hamiltonian reduces to Eq.~\eqref{HSchw}.
For the potentials $A_\text{noS}$, $\bar{D}_\text{noS}$, and $Q_\text{noS}$, we use the results of Ref.~\cite{Bini:2020wpo} (see Table IV there), which are missing two quadratic-in-$\nu$ coefficients in $A_\text{noS}$ and $\bar{D}_\text{noS}$ at 5PN. 

The 5PN Taylor-expanded potential $A_\text{noS}$ is given by
\begin{align}
\label{ApmTay}
A_\text{noS}^\text{Tay} &= 1 - 2 u + 2\nu u^3 + \nu \left(\frac{94}{3}-\frac{41 \pi ^2}{32}\right) u^4 \nonumber\\
&\quad
+ \bigg[\nu  \left(\frac{2275 \pi ^2}{512}-\frac{4237}{60}+\frac{128 \gamma_E }{5}+\frac{256 \ln 2}{5}\right) \nonumber\\
&\qquad
+ \left(\frac{41 \pi ^2}{32}-\frac{221}{6}\right) \nu ^2 +\frac{64}{5} \nu \ln u\bigg] u^5\nonumber\\
&\quad
 + \left[\nu a_6 - \nu\left(\frac{144 \nu}{5}+\frac{7004}{105}\right) \ln u\right] u^6, 
\end{align}
where $u\equiv M/r$, $\gamma_E\simeq 0.5772$ is the Euler gamma constant, and we replaced the coefficient of $u^6$ in $A_\text{noS}$, except for the log part, by the parameter $a_6$, which is calibrated to quasi-circular NR simulations. 
Note that we pull out a factor of $\nu$ from $a_6$ compared to its definition in Ref.~\cite{Bini:2020wpo}.
Then, we perform a (1,5) Pad\'e resummation of $A_\text{noS}^\text{Tay}(u)$, while treating $\ln u$ as a constant, i.e., we use
\begin{equation}
\label{ApmPade}
A_\text{noS} = P^1_5[A_\text{noS}^\text{Tay}(u)].
\end{equation}
The Pad\'e resummation of $A_\text{noS}$ was first introduced in Ref.~\cite{Damour:2000we} at 3PN order to ensure the presence of an innermost stable circular orbit (ISCO) in the EOB dynamics for any mass ratio. 
It was then adopted in the initial nonspinning and spinning \texttt{EOBNR} models (e.g., see 
Refs.~\cite{Buonanno:2007pf,Pan:2011gk,Pan:2011gk,Pan:2009wj}), and in all \texttt{TEOBResumS} models 
(e.g., see Refs.~\cite{Damour:2007yf,Damour:2008qf,Nagar:2018zoe,Nagar:2020pcj,Gamba:2021ydi}).

The 5PN potential $\bar{D}_\text{noS}$ reads
\begin{widetext}
\begin{align}
\bar{D}_\text{noS}^\text{Tay} &= 1 + 6 \nu u^2 + \left(52 \nu -6 \nu ^2\right)u^3 + \bigg[\nu  \left(-\frac{533}{45}-\frac{23761 \pi ^2}{1536}+\frac{1184 \gamma_E}{15}-\frac{6496 \ln 2}{15}+\frac{2916 \ln 3}{5}\right) \nonumber\\
&\qquad
+\left(\frac{123 \pi ^2}{16}-260\right) \nu ^2+\frac{592 \nu }{15} \ln u\bigg] u^4 
+ \left(-\frac{3392 \nu ^2}{15}-\frac{1420 \nu }{7}\right)  u^5 \ln u \nonumber\\
&\quad
+ \bigg[
\nu \left(\frac{294464}{175}-\frac{2840 \gamma_E}{7}-\frac{63707 \pi ^2}{512}+\frac{120648 \ln 2}{35}-\frac{19683 \ln 3}{7}\right)
+ \left(\frac{1069}{3}-\frac{205 \pi ^2}{16}\right) \nu^3 \nonumber\\
&\qquad
+ \left(d_5^{\nu^2}-\frac{6784 \gamma_E}{15}+\frac{67736}{105}+\frac{58320 \ln 3}{7}-\frac{326656 \ln 2}{21}\right) \nu^2
\bigg] u^5, 
\end{align}
where we set the remaining unknown coefficient $d_5^{\nu^2}$ to zero, but it can be determined in the future from PN calculations, or replaced by a calibration parameter to NR results for eccentric orbits.
To improve agreement with NR, we perform a (2,3) Pad\'e resummation of $\bar{D}_\text{noS}^\text{Tay}(u)$, i.e.,
\begin{equation}
\label{DbpmPade}
\bar{D}_\text{noS} = P^2_3[\bar{D}_\text{noS}^\text{Tay}(u)].
\end{equation}
The 5.5PN contributions to $A_\text{noS}$ and $\bar{D}_\text{noS}$ are known~\cite{Damour:2015isa,Bini:2020wpo}; however, since we Pad\'e resum these potentials, we find it more convenient to stop at 5PN.

For $Q_\text{noS}$, we use the full 5.5PN expansion, which is also expanded in eccentricity to $\Order(p_r^8)$, and it reads~\cite{Bini:2020wpo,Bini:2020hmy}
\begin{align}
\label{QpmTay}
Q_\text{noS} &= \frac{p_r^4}{\mu^2} \bigg\lbrace 2 (4 - 3 \nu) \nu u^2 
+u^3 \left[10 \nu ^3-83 \nu ^2+\nu  \left(-\frac{5308}{15}+\frac{496256 \ln 2}{45}-\frac{33048 \ln 3}{5}\right)\right]
+ u^4 \bigg[\left(640-\frac{615 \pi ^2}{32}\right) \nu ^3  \nonumber\\
&\quad\qquad
+\nu ^2 \left(\frac{31633 \pi ^2}{512}-\frac{1184 \gamma }{5}+\frac{150683}{105}+\frac{33693536 \ln 2}{105}-\frac{6396489 \ln 3}{70}-\frac{9765625 \ln 5}{126}\right) \nonumber\\
&\quad\qquad
+\nu  \left(\frac{1295219}{350}-\frac{93031 \pi ^2}{1536}+\frac{10856 \gamma_E }{105}-\frac{40979464}{315} \ln 2+\frac{14203593 \ln 3}{280}+\frac{9765625 \ln 5}{504}\right) \nonumber\\
&\quad\qquad
+\nu \left(\frac{5428}{105}-\frac{592 \nu}{5}\right) \ln u \bigg] 
+ \frac{88703 \pi  \nu  u^{9/2}}{1890} \bigg\rbrace 
\nonumber\\
&\quad 
+ \frac{p_r^6}{\mu^4} \bigg\lbrace
u^2 \left[6 \nu ^3-\frac{27 \nu ^2}{5}+\nu  \left(-\frac{827}{3}-\frac{2358912}{25}  \ln 2+\frac{1399437 \ln 3}{50}+\frac{390625 \ln 5}{18}\right)\right] \nonumber\\
&\quad\qquad
+ u^3 \bigg[-14 \nu ^4+116 \nu ^3
+\nu ^2 \left(\frac{159089}{75}-\frac{4998308864 \ln 2}{1575}+\frac{26171875 \ln 5}{18}-\frac{45409167 \ln 3}{350}\right) \nonumber\\
&\qquad\qquad
+\nu  \left(\frac{2613083}{1050}+\frac{6875745536 \ln 2}{4725}-\frac{23132628 \ln 3}{175}-\frac{101687500 \ln 5}{189}\right)\bigg]
-\frac{2723471 \pi  \nu  u^{7/2}}{756000} \bigg\rbrace 
\nonumber\\
&\quad
+ \frac{p_r^8}{\mu^6} \bigg\lbrace
u \nu  \left(-\frac{35772}{175}+\frac{21668992 \ln 2}{45}+\frac{6591861 \ln 3}{350}-\frac{27734375 \ln 5}{126}\right)
+ u^2 \bigg[-6 \nu ^4+\frac{24 \nu ^3}{7} \nonumber\\
&\qquad\qquad
+\nu ^2 \left(\frac{870976}{525}+\frac{703189497728 \ln 2}{33075}+\frac{332067403089 \ln 3}{39200}-\frac{13841287201 \ln 7}{4320}-\frac{468490234375 \ln 5}{42336}\right) \nonumber\\
&\qquad\qquad
+\nu  \left(\frac{5790381}{2450}-\frac{16175693888 \ln 2}{1575}+\frac{875090984375 \ln 5}{169344}+\frac{13841287201 \ln 7}{17280}-\frac{393786545409 \ln 3}{156800}\right)\bigg] \nonumber\\
&\quad\qquad
+\frac{5994461 \pi  \nu  u^{5/2}}{12700800}
\bigg\rbrace.
\end{align}
\end{widetext}

\subsection{Effective Hamiltonian for aligned spins}
\label{sec:HamAlign}
\def\SOfootnote{
The SO part of EOB Hamiltonians is often expressed in terms of $S\equiv S_1+S_2$ and $S_*\equiv S_1 m_2/m_1+ S_2 m_1/m_2$, i.e., $H_\text{SO} \propto \left(g_S S + g_{S_*} S_*\right) p_\phi/r^3$.
The relation between the gyro-gravitomagnetic factors in this case and our definition in Eq.~\eqref{HeffAnzAlign} is that
\begin{equation}
g_{a_+} = \frac{1}{2} \left(g_S + g_{S_*}\right), \qquad
g_{a_-} = \frac{1}{2} \left(g_S - g_{S_*}\right).
\end{equation}}

For aligned spins, the effective Hamiltonian reduces to the equatorial Kerr Hamiltonian~\eqref{KerreqHam} in the test-mass limit. 
To include PN information for arbitrary mass ratios, we use the following ansatz:
\begin{align}
\label{HeffAnzAlign}
H_\text{eff}^\text{align} &= \frac{
M p_\phi (g_{a_+} a_+ + g_{a_-} \delta a_-) + \text{SO}_\text{calib}
+ G^\text{align}_{a^3}
}{r^3+a_+^2 (r+2M)}  \nonumber\\
& \quad +
\Bigg[A^\text{align} \Bigg(
\mu^2 + \frac{p_\phi^2}{r^2} + (1+ B^\text{align}_{np}) p_r^2  \nonumber \\
&\qquad\qquad 
+ B_{npa}^\text{Kerr\,eq}\, \frac{p_\phi^2 a_+^2}{r^2} + Q^\text{align}  
\Bigg)\Bigg]^{1/2},
\end{align}
where the gyro-gravitomagnetic factors\footnote{\SOfootnote} $g_{a_+}$ and $g_{a_-}$ include the SO corrections, $\text{SO}_\text{calib}$ is a calibration term to NR results, and $G^\text{align}_{a^3}$ contains cubic-in-spin corrections.
The nonspinning and SS contributions are included in $A^\text{align}$, $B_{np}^\text{align}$ and $Q^\text{align}$, while the quartic-in-spin corrections are added in $A^\text{align}$.
The potential $B_{npa}^\text{Kerr\,eq}$ is kept the same as in the Kerr Hamiltonian for equatorial orbits.

In some papers~\cite{Damour:2008qf,Nagar:2011fx}, the gyro-gravitomagnetic factors in the SO part of the Hamiltonian were chosen to be in a gauge such that they are functions of $1/r$ and $p_r^2$ only, but other papers~\cite{Barausse:2009xi,Barausse:2011ys} made different choices. For \texttt{SEOBNRv5}, we find better results when using a gauge in which $g_{a_+}$ and $g_{a_-}$ depend on $1/r$ and $L^2/r^2$, but not on $p_r^2$, such that
\begin{subequations}
\label{gyros35PN}
\begin{align}
g_{a_+}^\text{3.5PN} &= \frac{7}{4} 
+ \left[\Lscal^2 u^2 \left(-\frac{45 \nu }{32}-\frac{15}{32}\right)
+u\left(\frac{23 \nu }{32}-\frac{3}{32}\right)\right] \nonumber\\
&\quad
+ \bigg[\Lscal^4u^4 \left(\frac{345 \nu ^2}{256}+\frac{75 \nu }{128}+\frac{105}{256}\right)\nonumber\\
&\qquad
+\Lscal^2u^3 \left(-\frac{1591 \nu ^2}{768}-\frac{267 \nu }{128}+\frac{59}{256}\right)\nonumber\\
&\qquad
+u^2\left(\frac{109 \nu ^2}{192}-\frac{177 \nu }{32}-\frac{5}{64}\right)\bigg], 
\end{align}
\begin{align}
g_{a_-}^\text{3.5PN} &= \frac{1}{4} 
+ \left[\Lscal^2u^2 \left(\frac{15}{32}-\frac{9 \nu }{32}\right)
+u\left(\frac{11 \nu }{32}+\frac{3}{32}\right)\right] \nonumber\\
&\quad
+\bigg[\Lscal^4u^4 \left(\frac{75 \nu ^2}{256}-\frac{45 \nu }{128}-\frac{105}{256}\right)\nonumber\\
&\qquad
+\Lscal^2u^3 \left(-\frac{613 \nu ^2}{768}-\frac{35 \nu }{128}-\frac{59}{256}\right)\nonumber\\
&\qquad
+u^2\left(\frac{103 \nu ^2}{192}-\frac{\nu }{32}+\frac{5}{64}\right)\bigg],
\end{align}
\end{subequations}
where the square brackets collect different PN orders, and we defined $\tilde{L}\equiv L/(M\mu)$.
These PN expressions were obtained by canonically transforming the 3.5PN results of, e.g., Ref.~\cite{Levi:2015uxa}.

The 4.5PN SO coupling was derived in Refs.~\cite{Antonelli:2020aeb,Antonelli:2020ybz,Mandal:2022nty,Kim:2022pou}, and can be included in the effective Hamiltonian.
However, we found that using a calibration term at 5.5PN had a small effect on the dynamics, and thus only included the 3.5PN information with a 4.5PN calibration term of the form
\begin{equation}
\text{SO}_\text{calib} = \nu d_\text{SO} \frac{M^4}{r^3} p_\phi a_+.
\end{equation}

For completeness, we write the 4.5PN part in terms of $L^2/r^2$ instead of $p_r^2$, which we obtained by canonically transforming Eq.~(5.6) of Ref.~\cite{Antonelli:2020ybz}, leading to
\begin{subequations}
\begin{align}
&g_{a_+}^\text{4.5PN} = g_{a_+}^\text{3.5PN} \nonumber\\
&\quad
+ \bigg\lbrace
\Lscal^6u^6 \left(-\frac{5425 \nu ^3}{4096}-\frac{1785 \nu ^2}{2048}-\frac{1715 \nu }{4096}-\frac{1575}{4096}\right) \nonumber\\
&\qquad
+\Lscal^4u^5 \left(\frac{75187 \nu ^3}{20480}+\frac{37603 \nu ^2}{10240}+\frac{3717 \nu }{4096}-\frac{1023}{4096}\right) \nonumber\\
&\qquad
+\Lscal^2u^4 \left(\frac{209}{1024}-\frac{15093 \nu ^3}{5120}+\frac{80189 \nu ^2}{7680}-\frac{13059 \nu }{1024}\right)\nonumber\\
&\qquad
+u^3 \bigg[\frac{1079 \nu ^3}{2048}-\frac{24131 \nu ^2}{3072}+\left(\frac{487 \pi ^2}{384}-\frac{525331}{18432}\right)\nu \nonumber\\
&\qquad\qquad
-\frac{175}{2048}\bigg]
\bigg\rbrace,
\end{align}
\begin{align}
&g_{a_-}^\text{4.5PN} = g_{a_+}^\text{3.5PN} \nonumber\\
&\quad
+ \bigg\lbrace
\Lscal^6u^6 \left(-\frac{1225 \nu ^3}{4096}+\frac{525 \nu ^2}{2048}+\frac{1785 \nu }{4096}+\frac{1575}{4096}\right)\nonumber\\
&\qquad
+\Lscal^4u^5 \left(\frac{26491 \nu ^3}{20480}+\frac{4801 \nu ^2}{10240}-\frac{4843 \nu }{20480}+\frac{1023}{4096}\right)\nonumber\\
&\qquad
+\Lscal^2u^4 \left(-\frac{9549 \nu ^3}{5120}-\frac{1777 \nu ^2}{7680}-\frac{28883 \nu }{5120}-\frac{209}{1024}\right)\nonumber\\
&\qquad
+u^3 \bigg[\frac{1823 \nu ^3}{2048}-\frac{1025 \nu ^2}{3072}-\left(\frac{5 \pi ^2}{384}  +\frac{50215 }{18432}\right)\nu \nonumber\\
&\qquad\qquad
+\frac{175}{2048}\bigg]
\bigg\rbrace.
\end{align}
\end{subequations}

For the cubic-in-spin term $G^\text{align}_{a^3}$ in Eq.~\eqref{HeffAnzAlign}, we obtain
\begin{align}
G^\text{align}_{a^3} &= \frac{Mp_\phi}{r^2} \bigg[
-\frac{a_+^3}{4}+\frac{\delta}{4} a_- a_+^2
+ C_+^{a^3}-\frac{3}{2} a_- C_-^{a^2} \nonumber\\
&\quad\qquad
+ \frac{3}{8}C_+^{a^2} \left(\delta a_- +3 a_+\right) \bigg],
\end{align}
where only the first two terms contribute for BHs. 
The coefficients $C_\pm^{...}$ are defined in Eq.~\eqref{multipoleCoefs}.

As mentioned, we include SS and S$^4$ PN information in the effective Hamiltonian~\eqref{HeffAnzAlign} through the following ansatz (cf. Eqs.~\eqref{potsKerreq}):
\begin{equation}
\label{ABQanzAligned}
\begin{aligned}
A^\text{align} &= \frac{a_+^2/r^2+A_\text{noS}+A_\text{SS}^\text{align} + A_{\mr S^4}^\text{align}}{1+(1+2M/r)a_+^2/r^2}, \\
B_{np}^\text{align} &= -1 + \frac{a_+^2}{r^2} + A_\text{noS} \bar{D}_\text{noS} + B_{np,\text{SS}}^\text{align}, \\
Q^\text{align} &= Q_\text{noS} + Q_\text{SS}^\text{align},
\end{aligned}
\end{equation}
where the nonspinning contributions $A_\text{noS}$, $\bar{D}_\text{noS}$ and $Q_\text{noS}$ are given by Eqs.~\eqref{ApmPade}, \eqref{DbpmPade} and \eqref{QpmTay}, respectively.

For the SS contributions, we obtain
\begin{widetext}
\begin{subequations}
\label{SSalign}
\begin{align}
A_\text{SS}^\text{align} &= -\frac{MC_+^{a^2}}{r^3}
+ \frac{M^2}{r^4}\left[
\frac{9 a_+^2}{8}-\frac{5}{4} \delta a_- a_+ +a_-^2 \left(\frac{\nu }{2}+\frac{1}{8}\right)
-\delta  C_-^{a^2}-C_+^{a^2}\right] 
+\frac{M^3}{r^5}\bigg[
a_+^2 \left(-\frac{175 \nu }{64}-\frac{225}{64}\right)\nonumber\\
&\qquad
+\delta a_- a_+ \left(\frac{117}{32}-\frac{39 \nu }{16}\right)
+a_-^2 \left(\frac{21 \nu ^2}{16}-\frac{81 \nu }{64}-\frac{9}{64}\right)
-\frac{51}{28} \delta  C_-^{a^2} + \left(\frac{207 \nu }{28}-\frac{51}{28}\right) C_+^{a^2}\bigg], \\
%%%%
B_{np,\text{SS}}^\text{align} &= \frac{M}{r^3} \left[a_+^2 \left(3 \nu +\frac{45}{16}\right)-\frac{21}{8}\delta a_- a_+ +a_-^2 \left(\frac{3 \nu }{4}-\frac{3}{16}\right)+(3 \nu -3) C_+^{a^2}\right]
+ \frac{M^2}{r^4} \bigg[
a_+^2 \left(-\frac{1171 \nu }{64}-\frac{861}{64}\right) \nonumber\\
&\qquad
+\delta a_- a_+ \left(\frac{13 \nu }{16}+\frac{449}{32}\right)
+a_-^2 \left(\frac{\nu ^2}{16}+\frac{115 \nu }{64}-\frac{37}{64}\right)
+ \left(6 \nu -\frac{19}{4}\right)\delta C_-^{a^2}
+\left(\frac{111 \nu }{4}-\frac{3}{4}\right) C_+^{a^2}
\bigg], \\
%%%%
Q_\text{SS}^\text{align} &= \frac{Mp_r^4}{\mu^2r^3} \bigg[
a_+^2 \left(\frac{25}{32}-5 \nu ^2+\frac{165 \nu }{32}\right)
+\delta a_- a_+ \left(\frac{45 \nu }{8}-\frac{5}{16}\right)
+a_-^2 \left(-\frac{15 \nu ^2}{8}+\frac{75 \nu }{32}-\frac{15}{32}\right) \nonumber\\
&\qquad
+\left(\frac{35 \nu }{4}-5 \nu ^2\right) C_+^{a^2}
\bigg].
\end{align}
\end{subequations}
\end{widetext}

The quartic-in-spin contribution in $A$ is given by
\begin{align}
A_{\mr S^4}^\text{align} &= \frac{M}{r^5} \bigg[
-\frac{3}{4} C_+^{a^4}-\frac{3}{2} a_+ C_+^{a^3}+\frac{3}{2} a_- C_-^{a^3}\nonumber\\
&\qquad
+\left(-\frac{9}{8} a_-^2-\frac{5 a_+^2}{8}\right) C_+^{a^2}-\frac{9}{8} (C_+^{a^2})^2\nonumber\\
&\qquad
+\frac{9}{8} (C_-^{a^2})^2+\frac{9}{4} a_- a_+ C_-^{a^2}
\bigg],
\end{align}
which vanishes for BHs since the Kerr Hamiltonian with the mapping~\eqref{spinmap} reproduces it~\cite{Vines:2016qwa}.\\

\subsection{Effective Hamiltonian for precessing spins}
\label{sec:HamPrec}
For precessing spins, we derive an effective Hamiltonian that reduces to the Kerr Hamiltonian in Eq.~\eqref{KerrHam}, and includes higher PN information through the following ansatz:
\begin{align}
\label{HeffAnzPrec}
H_\text{eff}^\text{prec} &= \frac{Mr}{\Lambda} \Big[
\bm{L}\cdot \left(g_{a_+} \bm{a}_+ + g_{a_-} \delta \bm{a}_-\right) + \text{SO}_\text{calib} + G^\text{prec}_{a^3} \Big] \nonumber\\
&\quad
+ \Big[ 
A^\text{prec} \Big(\mu^2 + B_p^\text{prec} \bm{p}^2  + B_{np}^\text{prec} (\bm{n}\cdot\bm{p})^2 \nonumber\\
&\quad\qquad
 + B_{npa}^\text{Kerr}\, (\bm{n} \cross \bm{p} \cdot \bm{a}_+)^2 + Q^\text{prec} 
\Big)
\Big]^{1/2}.
\end{align}
Similarly to the aligned-spin case, $g_{a_+}$ and $g_{a_-}$ include the SO corrections, $\text{SO}_\text{calib}$ is an NR calibration term, and $G^\text{prec}_{a^3}$ contains S$^3$ corrections.
The nonspinning and SS contributions are included in $A^\text{prec}$, $B_p^\text{prec}$, $B_{np}^\text{prec}$ and $Q^\text{prec}$, while the S$^4$ corrections are added in $A^\text{prec}$.
The potential $B_{npa}^\text{Kerr}$ is the same as in the Kerr Hamiltonian.

The gyro-gravitomagnetic factors $g_{a_+}$ and $g_{a_-}$ are given by Eq.~\eqref{gyros35PN}; the same as in the aligned-spin case, since they are independent of spin.
The calibration term is also similar, except for adding a dot product
\begin{equation}
\label{SOcalibPrec}
\text{SO}_\text{calib} = \nu d_\text{SO} \frac{M^4}{r^3} \bm{L}\cdot\bm{a}_+.
\end{equation}

For the cubic-in-spin term $G^\text{prec}_{a^3}$, we obtain
\begin{widetext}
\begin{align}
\label{Ga3Prec}
G^\text{prec}_{a^3} &= \bm{L}\cdot \bm{a}_+ \Bigg\lbrace
\frac{L^2}{\mu^2r^3} \left[
\frac{\delta}{2}(\bm{n}\cdot\bm{a}_-)(\bm{n}\cdot\bm{a}_+)
-\frac{\left(\bm{n}\cdot\bm{a}_+\right)^2}{4}
\right]
+ \frac{p_r^2}{\mu^2r} \left[
\frac{5 }{4}\left(\bm{n}\cdot\bm{a}_+\right)^2
-\delta\frac{3}{2}(\bm{n}\cdot\bm{a}_-) (\bm{n}\cdot\bm{a}_+)
\right] \nonumber\\
&\qquad
+\frac{M}{r^2} \Bigg[
-\frac{a_+^2}{4} + \left(\bm{n}\cdot\bm{a}_+\right)^2
+\delta\frac{5}{24} (\bm{a}_+\cdot\bm{a}_-)
-\delta\frac{5}{3} (\bm{n}\cdot\bm{a}_-) (\bm{n}\cdot\bm{a}_+)\nonumber\\
&\qquad\quad
+\frac{1}{8} a_1^2 \left(3\delta \widetilde{C}_{\rm 1ES^2} +4 \widetilde{C}_{\rm 1BS^3}\right)
+\frac{1}{8} a_2^2 \left(4 \widetilde{C}_{\rm 2BS^3}-3 \widetilde{C}_{\rm 2ES^2} \delta\right)
-\frac{3}{8} (\bm{a}_1\cdot\bm{a}_2) \left(\widetilde{C}_{\rm 1ES^2} (\delta -3)-\widetilde{C}_{\rm 2ES^2} (\delta +3)\right) \nonumber\\
&\qquad\quad
+ \left(\bm{n}\cdot\bm{a}_1\right)^2 \left(-\frac{3}{8} \widetilde{C}_{\rm 1ES^2} (2 \delta +3)-\frac{5 \widetilde{C}_{\rm 1BS^3}}{2}\right)
+\frac{3}{4} (\bm{n}\cdot\bm{a}_1) (\bm{n}\cdot\bm{a}_2) \left(\widetilde{C}_{\rm 1ES^2} (2 \delta -7)-\widetilde{C}_{\rm 2ES^2} (2 \delta +7)\right) \nonumber\\
&\qquad\quad
+\frac{1}{8} \left(\bm{n}\cdot\bm{a}_2\right)^2 \left(\widetilde{C}_{\rm 2ES^2} (6 \delta -9)-20 \widetilde{C}_{\rm 2BS^3}\right)
\Bigg]
\Bigg\rbrace \nonumber\\
&\quad
+ \bm{L}\cdot \bm{a}_- \Bigg\lbrace
\frac{p_r^2\delta\left(\bm{n}\cdot\bm{a}_+\right)^2}{4\mu^2 r}
-\frac{L^2 \delta\left(\bm{n}\cdot\bm{a}_+\right)^2}{4\mu^2 r^3}
+\frac{M}{r^2}\Bigg[
\frac{\delta a_+^2}{24} + \frac{2}{3}\delta \left(\bm{n}\cdot\bm{a}_+\right)^2 \nonumber\\
&\qquad\quad
+\frac{1}{8} a_1^2 \left(4 \widetilde{C}_{\rm 1BS^3}-3 \widetilde{C}_{\rm 1ES^2}\right)
+\frac{1}{8} a_2^2 \left(3 \widetilde{C}_{\rm 2ES^2}-4 \widetilde{C}_{\rm 2BS^3}\right)
-\frac{3}{8} (\bm{a}_1\cdot\bm{a}_2) \left(\widetilde{C}_{\rm 1ES^2} (\delta -3)+\widetilde{C}_{\rm 2ES^2} (\delta +3)\right)\nonumber\\
&\qquad\quad
+\left(\bm{n}\cdot\bm{a}_1\right)^2 \left(-\frac{3}{8} \widetilde{C}_{\rm 1ES^2} (\delta -6)-\frac{5 \widetilde{C}_{\rm 1BS^3}}{2}\right)
+\frac{3}{4} (\bm{n}\cdot\bm{a}_1) (\bm{n}\cdot\bm{a}_2) \left(\widetilde{C}_{\rm 1ES^2} (2 \delta -7)+\widetilde{C}_{\rm 2ES^2} (2 \delta +7)\right) \nonumber\\
&\qquad\quad
+\left(\bm{n}\cdot\bm{a}_2\right)^2 \left(\frac{5 \widetilde{C}_{\rm 2BS^3}}{2}-\frac{3}{8} \widetilde{C}_{\rm 2ES^2} (\delta +6)\right)
\Bigg]
\Bigg\rbrace.
\end{align}

We include SS and S$^4$ PN information in the effective Hamiltonian~\eqref{HeffAnzPrec} through the following ansatz for the potentials  (cf. Eq.~\eqref{potsKerr}):
\begin{equation}
\label{ABQanzPrec}
\begin{aligned}
A^\text{prec} &= \frac{\left[a_+^2/r^2+A_\text{noS}+A_\text{SS}^\text{prec} + A_{\mr S^4}^\text{prec}\right] \left[1 + (\bm{n}\cdot\bm{a}_+)^2/r^2 + {A}_\text{SS}^\text{in\,plane} + {A}_{\mr S^4}^\text{in\,plane} \right]}{1 + a_+^2/r^2 + 2 M a_+^2/r^3 + (\bm{n}\cdot\bm{a}_+)^2/r^2 - 2 M(\bm{n}\cdot\bm{a}_+)^2/r^3 + a_+^2 (\bm{n}\cdot\bm{a}_+)^2/r^4}, \\
%%%
B^\text{prec}_{p} &= \frac{1}{1 + (\bm{n}\cdot\bm{a}_+)^2/r^2 + {B}_{p,\text{SS}}^\text{in\,plane}}, \\ 
%%%
B_{np}^\text{prec} &= \frac{-1 + a_+^2/r^2 + A_\text{noS} \bar{D}_\text{noS} + B_{np,\text{SS}}^\text{prec} + {B}_{np,\text{SS}}^\text{in\,plane}}{1 + (\bm{n}\cdot\bm{a}_+)^2/r^2},\\
%%%
Q^\text{prec} &= Q_\text{noS} + Q_\text{SS}^\text{prec} + {Q}_\text{SS}^\text{in\,plane},
\end{aligned}
\end{equation}
where the nonspinning contributions $A_\text{noS}$, $\bar{D}_\text{noS}$ and $Q_\text{noS}$ are given by Eqs.~\eqref{ApmPade}, \eqref{DbpmPade} and \eqref{QpmTay}, respectively, while $A_\text{SS}^\text{prec}$, $B_{np,\text{SS}}^\text{prec}$, and $Q_\text{SS}^\text{prec}$ are the same as in Eqs.~\eqref{SSalign} except for replacing $a_+a_-$ by $\bm{a}_+\cdot\bm{a}_-$.
The other terms ${A}_\text{SS}^\text{in\,plane}$, ${B}_{p,\text{SS}}^\text{in\,plane}$, ${B}_{np,\text{SS}}^\text{in\,plane}$, ${Q}_\text{SS}^\text{in\,plane}$ and $A_{\mr S^4}^\text{in\,plane}$ only contain in-plane spin components, which vanish in the aligned-spin case.

The spin-spin contributions read
\begin{subequations}
\label{SSsolnPrec}
\begin{align}
A_\text{SS}^\text{prec} &=  -\frac{MC_+^{a^2}}{r^3}
+ \frac{M^2}{r^4}\left[
\frac{9 a_+^2}{8}-\frac{5}{4} \delta \bm{a}_- \cdot \bm{a}_+ +a_-^2 \left(\frac{\nu }{2}+\frac{1}{8}\right)
-\delta  C_-^{a^2}-C_+^{a^2}\right] 
+\frac{M^3}{r^5}\bigg[
a_+^2 \left(-\frac{175 \nu }{64}-\frac{225}{64}\right)\nonumber\\
&\qquad
+\delta \bm{a}_- \cdot\bm{a}_+ \left(\frac{117}{32}-\frac{39 \nu }{16}\right)
+a_-^2 \left(\frac{21 \nu ^2}{16}-\frac{81 \nu }{64}-\frac{9}{64}\right)
-\frac{51}{28} \delta  C_-^{a^2} + \left(\frac{207 \nu }{28}-\frac{51}{28}\right) C_+^{a^2}\bigg], \\
%%%
{A}_\text{SS}^\text{in\,plane} &= \frac{3 MC_+^{n\cdot a^2}}{r^3}
+ \frac{M^2}{r^4} \bigg[\frac{33}{8} \delta  (\bm{n}\cdot\bm{a}_-) (\bm{n}\cdot\bm{a}_+) +\left(-\frac{\nu }{2}-\frac{3}{8}\right) \left(\bm{n}\cdot\bm{a}_-\right)^2+\left(\frac{7 \nu }{4}-\frac{15}{4}\right) \left(\bm{n}\cdot\bm{a}_+\right)^2 \nonumber\\
&\qquad
+3 \delta  C_-^{n\cdot a^2}+(3 \nu +6) C_+^{n\cdot a^2}\bigg]
+ \frac{M^3}{r^5} \bigg[\delta  (17 \nu +8) (\bm{n}\cdot\bm{a}_-) (\bm{n}\cdot\bm{a}_+) +\left(-\frac{41 \nu ^2}{8}+\frac{551 \nu }{32}-\frac{219}{64}\right) \left(\bm{n}\cdot\bm{a}_-\right)^2\nonumber\\
&\qquad
+\!\left(\frac{1771 \nu }{96}-\frac{11 \nu ^2}{8}-\frac{293}{64}\right) \left(\bm{n}\cdot\bm{a}_+\right)^2 
+\delta  \left(\frac{81 \nu }{16}+\frac{1245}{224}\right) C_-^{n\cdot a^2}\!
+\!\left(\frac{3555}{224}-\frac{13 \nu ^2}{16}+\frac{515 \nu }{56}\right) C_+^{n\cdot a^2}\bigg]
, \\
%%%
{B}_{p,\text{SS}}^\text{in\,plane} &= \frac{M}{r^3} \left[-\frac{3}{4} \delta  (\bm{n}\cdot\bm{a}_-) (\bm{n}\cdot\bm{a}_+) +\left(\frac{3 \nu }{4}-\frac{3}{16}\right) \left(\bm{n}\cdot\bm{a}_-\right)^2+\left(\frac{7 \nu }{4}+\frac{15}{16}\right) \left(\bm{n}\cdot\bm{a}_+\right)^2+(3 \nu -3) C_+^{n\cdot a^2}\right] \nonumber\\
&\quad
+ \frac{M^2}{r^4} \bigg[
\delta  \left(\frac{49 \nu }{4}+\frac{43}{8}\right) \bm{n}\cdot\bm{a}_- \bm{n}\cdot\bm{a}_++\!\left(\frac{545 \nu }{32}-\frac{19 \nu ^2}{8}-\frac{219}{64}\right) \left(\bm{n}\cdot\bm{a}_-\right)^2
+\!\left(\frac{805 \nu }{96}-\frac{11 \nu ^2}{8}-\frac{125}{64}\right) \left(\bm{n}\cdot\bm{a}_+\right)^2 \nonumber\\
&\qquad
+\delta  \left(\frac{81 \nu }{16}-\frac{189}{32}\right) C_-^{n\cdot a^2}+\left(-\frac{13 \nu ^2}{16}+\frac{203 \nu }{8}-\frac{51}{32}\right) C_+^{n\cdot a^2}
\bigg], \\
%%%
{B}_{np,\text{SS}}^\text{prec} &= \frac{M}{r^3} \left[a_+^2 \left(3 \nu +\frac{45}{16}\right)-\frac{21}{8}\delta \bm{a}_- \cdot\bm{a}_+ +a_-^2 \left(\frac{3 \nu }{4}-\frac{3}{16}\right)+(3 \nu -3) C_+^{a^2}\right]
+ \frac{M^2}{r^4} \bigg[
a_+^2 \left(-\frac{1171 \nu }{64}-\frac{861}{64}\right) \nonumber\\
&\qquad
+\delta \bm{a}_- \cdot\bm{a}_+ \left(\frac{13 \nu }{16}+\frac{449}{32}\right)
+a_-^2 \left(\frac{\nu ^2}{16}+\frac{115 \nu }{64}-\frac{37}{64}\right)
+ \left(6 \nu -\frac{19}{4}\right)\delta C_-^{a^2}
+\left(\frac{111 \nu }{4}-\frac{3}{4}\right) C_+^{a^2}
\bigg], \\
%%%
{B}_{np,\text{SS}}^\text{in\,plane} &= \frac{M}{r^3} \left[\frac{45}{8} \delta  (\bm{n}\cdot\bm{a}_-) (\bm{n}\cdot\bm{a}_+) +\left(-\frac{15 \nu }{4}-\frac{45}{8}\right) \left(\bm{n}\cdot\bm{a}_+\right)^2\right]
+ \frac{M^2}{r^4} \bigg[
\delta  \left(\frac{129 \nu }{4}-\frac{17}{8}\right) (\bm{n}\cdot\bm{a}_-) (\bm{n}\cdot\bm{a}_+) \nonumber\\
&\qquad
+\left(-\frac{33 \nu ^2}{4}+\frac{981 \nu }{16}-\frac{165}{16}\right) \left(\bm{n}\cdot\bm{a}_-\right)^2
+\left(-\frac{11 \nu ^2}{2}+\frac{1901 \nu }{48}+\frac{199}{16}\right) \left(\bm{n}\cdot\bm{a}_+\right)^2
+\delta  \left(\frac{9 \nu }{4}-\frac{75}{8}\right) C_-^{n\cdot a^2} \nonumber\\
&\qquad
+\left(-\frac{13 \nu ^2}{4}+\frac{37 \nu }{4}+\frac{39}{8}\right) C_+^{n\cdot a^2}
\bigg], \\
%%%
Q_\text{SS}^\text{prec} &= \frac{Mp_r^4}{\mu^2r^3} \bigg[
a_+^2 \left(-5 \nu ^2+\frac{165 \nu }{32}+\frac{25}{32}\right)
+\delta \bm{a}_- \cdot\bm{a}_+ \left(\frac{45 \nu }{8}-\frac{5}{16}\right)
+a_-^2 \left(-\frac{15 \nu ^2}{8}+\frac{75 \nu }{32}-\frac{15}{32}\right)\nonumber\\
&\qquad
+\left(\frac{35 \nu }{4}-5 \nu ^2\right) C_+^{a^2}
\bigg], \\
%%%
\label{QSSprec}
{Q}_\text{SS}^\text{in\,plane} &= \frac{Mp_r^4}{\mu^2r^3} \bigg[\delta  \left(\frac{35}{16}-\frac{273 \nu }{8}\right) (\bm{n}\cdot\bm{a}_-) (\bm{n}\cdot\bm{a}_+) +\left(\frac{105 \nu ^2}{8}-\frac{525 \nu }{32}+\frac{105}{32}\right) \left(\bm{n}\cdot\bm{a}_-\right)^2\nonumber\\
&\qquad
+\left(\frac{119 \nu ^2}{4}-\frac{2849 \nu }{96}-\frac{175}{32}\right) \left(\bm{n}\cdot\bm{a}_+\right)^2
+\left(35 \nu ^2-\frac{245 \nu }{4}\right) C_+^{n\cdot a^2}\bigg]\nonumber\\
&\quad
+ \frac{Mp_r^3}{\mu^2 r^3} \bigg[
\delta  \left(\frac{69 \nu }{8}-\frac{5}{8}\right) (\bm{p}\cdot\bm{a}_-) (\bm{n}\cdot\bm{a}_+ )
+\left(-\frac{59 \nu ^2}{4}+\frac{341 \nu }{24}+\frac{25}{8}\right) (\bm{p}\cdot\bm{a}_+) (\bm{n}\cdot\bm{a}_+ ) \nonumber\\
&\qquad
+  \delta  \left(\frac{69 \nu }{8}-\frac{5}{8}\right) (\bm{p}\cdot\bm{a}_+) (\bm{n}\cdot\bm{a}_-) 
+\left(-\frac{15 \nu ^2}{2}+\frac{75 \nu }{8}-\frac{15}{8}\right) (\bm{p}\cdot\bm{a}_-)(\bm{n}\cdot\bm{a}_-) \nonumber\\
&\qquad
+ \left(35 \nu -20 \nu ^2\right)\left(\widetilde{C}_{\rm 1ES^2} (\bm{n}\cdot\bm{a}_1) (\bm{p}\cdot\bm{a}_1) + \widetilde{C}_{\rm 2ES^2} (\bm{n}\cdot\bm{a}_2) (\bm{p}\cdot\bm{a}_2)\right)
\bigg],
\end{align}
\end{subequations}
while the quartic-in-spin contributions in $A^\text{prec}$ are given by
\begin{subequations}
\begin{align}
\label{AS4prec}
A_{\mr S^4}^\text{prec} &= \frac{M}{r^5} \bigg\lbrace
-\frac{1}{4} a_1^2 a_2^2 \left[\widetilde{C}_{\rm 1ES^2} (3 \widetilde{C}_{\rm 2ES^2}+2)+2 \widetilde{C}_{\rm 2ES^2}\right]
-\frac{3}{2} (\bm{a}_1\cdot\bm{a}_2)^2 \left[\widetilde{C}_{\rm 1ES^2} \widetilde{C}_{\rm 2ES^2}+\widetilde{C}_{\rm 1ES^2}+\widetilde{C}_{\rm 2ES^2}\right] \nonumber\\
&\qquad
+a_1^2 (\bm{a}_1\cdot\bm{a}_2) (\widetilde{C}_{\rm 1ES^2}-3 \widetilde{C}_{\rm 1BS^3})+\frac{1}{4} a_1^4 (2 \widetilde{C}_{\rm 1ES^2}-3 \widetilde{C}_{\rm 1ES^4}) + 1 \leftrightarrow 2
\bigg\rbrace,\\
{A}_{\mr S^4}^\text{in\,plane} &= \frac{M}{r^5} \bigg\lbrace
\left(\bm{n}\cdot\bm{a}_1\right){}^4 \left(\frac{21 \widetilde{C}_{\rm 1ES^2}}{2}-\frac{35 \widetilde{C}_{\rm 1ES^4}}{4}\right)
+ (\bm{n}\cdot\bm{a}_2) \left(\bm{n}\cdot\bm{a}_1\right)^3 (21 \widetilde{C}_{\rm 1ES^2}-35 \widetilde{C}_{\rm 1BS^3}) \nonumber\\
&\qquad
+ \left(\bm{n}\cdot\bm{a}_1\right)^2\Bigg[
\left(\bm{n}\cdot\bm{a}_2\right)^2 \left(-\frac{105}{4}  \widetilde{C}_{\rm 1ES^2} \widetilde{C}_{\rm 2ES^2}-21 \widetilde{C}_{\rm 1ES^2}-21 \widetilde{C}_{\rm 2ES^2}\right)
+(\bm{a}_1\cdot\bm{a}_2) (15 \widetilde{C}_{\rm 1BS^3}-12 \widetilde{C}_{\rm 1ES^2}) \nonumber\\
&\qquad\quad
+a_1^2 \left(\frac{15 \widetilde{C}_{\rm 1ES^4}}{2}-9 \widetilde{C}_{\rm 1ES^2}\right)+\frac{15}{4} a_2^2 \widetilde{C}_{\rm 1ES^2} \widetilde{C}_{\rm 2ES^2}+\frac{3 a_2^2 \widetilde{C}_{\rm 1ES^2}}{2}+3 a_2^2 \widetilde{C}_{\rm 2ES^2}
\Bigg] \nonumber\\
&\qquad
+ (\bm{n}\cdot\bm{a}_1) a_1^2 (\bm{n}\cdot\bm{a}_2) (15 \widetilde{C}_{\rm 1BS^3}-6 \widetilde{C}_{\rm 1ES^2})
+ a_1^2 \left(\bm{n}\cdot\bm{a}_2\right)^2 \left(\frac{15 \widetilde{C}_{\rm 1ES^2} \widetilde{C}_{\rm 2ES^2}}{4}+3 \widetilde{C}_{\rm 1ES^2}+\frac{3 \widetilde{C}_{\rm 2ES^2}}{2}\right) \nonumber\\
&\qquad
+ (\bm{n}\cdot\bm{a}_1)  (\bm{n}\cdot\bm{a}_2) (\bm{a}_1\cdot\bm{a}_2) \left(15 \widetilde{C}_{\rm 1ES^2} \widetilde{C}_{\rm 2ES^2}+\frac{27 \widetilde{C}_{\rm 1ES^2}}{2}+\frac{27 \widetilde{C}_{\rm 2ES^2}}{2}\right)
+ 1 \leftrightarrow 2
\bigg\rbrace,
\end{align}
\end{subequations}
which are zero for BHs.
\end{widetext}

\subsection{Hamiltonian in tortoise coordinates}
EOB waveform models often use the tortoise-coordinate $p_{r_*}$ instead of $p_r$, since it improves stability of the EOMs near the event horizon~\cite{Damour:2007xr,Pan:2009wj}.
In the nonspinning case, the tortoise-coordinate $r_*$ is defined by
\begin{equation}
\frac{dr_*}{dr} = \frac{1}{\xi(r)}, \qquad
\xi(r) \equiv A_\text{noS}(r)\sqrt{\bar{D}_\text{noS}(r)}\,,
\end{equation}
and the conjugate momentum $p_{r_*}$ is given by
\begin{equation}
\label{prstar}
p_{r_*} =  p_r \xi(r).
\end{equation}

The nonspinning effective Hamiltonian in Eq.~\eqref{Heffpm} can be written in terms of $p_{r_*}$ as
\begin{equation}
H_\text{eff}^\text{noS} = \sqrt{p_{r_*}^2 + A_\text{noS}(r) \left[\mu^2 + p_\phi^2/r^2 + Q_\text{noS}(r,p_{r_*})\right]},
\end{equation}
where we obtain $Q_\text{noS}(r,p_{r_*})$ from Eq.~\eqref{QpmTay} by converting $p_r$ to $p_{r_*}$ using Eq.~\eqref{prstar}, then PN expand to 5.5PN.

For both aligned and precessing spins, a convenient choice for $\xi(r)$ is
\begin{equation}
\xi(r) = \frac{\sqrt{\bar{D}_\text{noS}}\left(A_\text{noS} + a_+^2/r^2\right)}{1 + a_+^2/r^2},
\end{equation}
which is similar to what was used for $\xi$ in \texttt{SEOBNRv4}~\cite{Pan:2009wj,Taracchini:2012ig} except for the different resummation and PN orders in $A_\text{noS}$ and $\bar{D}_\text{noS}$.
In the $\nu\to 0$ limit, $\xi$ reduces to the Kerr value $(dr/dr_*)_\text{Kerr} = (r^2-2Mr+a_+^2)/(r^2+a_+^2)$.

The PN expansion of $\xi(r)$ is given by
\begin{equation}
\label{xiExpansion}
\xi(r) \simeq 1 - \frac{2M}{r} + \frac{3\nu M^2}{r^2} + \frac{2 Ma^2}{r^3} + \dots,
\end{equation}
which equals one at LO, while the spin contribution enters at 3PN.
Hence, we can directly replace $p_r$ by $p_{r_*}$ in the 3.5PN S$^3$ and 4PN SS contributions in the Hamiltonian.

\subsection{Comparison with other models}

\begin{table*}[ht]
\caption{Summary of the main differences of the \texttt{SEOBNRv5} Hamiltonian derived here, which builds on the results of Refs.~\cite{Balmelli:2015zsa,Khalil:2020mmr}, compared to that of \texttt{SEOBNRv4} and \texttt{TEOBResumS}.}
\label{tab:compHam}
\begin{ruledtabular}
\def\arraystretch{1.4}
\begin{tabular}{p{0.15\linewidth}| p{0.27\linewidth} |p{0.27\linewidth} | p{0.27\linewidth}}
 & \texttt{SEOBNRv5} & \texttt{SEOBNRv4}~\cite{Barausse:2009xi,Barausse:2011ys,Bohe:2016gbl,Ossokine:2020kjp} & \texttt{TEOBResumS}~\cite{Damour:2014sva,Nagar:2018zoe,Nagar:2018plt} \\
\hline
nonspinning part & 4PN with partial 5PN in $A_\text{noS}$ and $\bar{D}_\text{noS}$, 5.5PN in $Q_\text{noS}$  & 4PN in $A_\text{noS}$, 3PN in $\bar{D}_\text{noS}$ and $Q_\text{noS}$ &  4PN with partial 5PN in $A_\text{noS}$, 3PN in $\bar{D}_\text{noS}$ and $Q_\text{noS}$ \\ 
%%%%
$A_\text{noS}$ resummation  & (1,5) Pad\'e & horizon factorization and log resummation & (1,5) Pad\'e \\
%%%%
$\bar{D}_\text{noS}$ resummation  & (2,3) Pad\'e & log & Taylor expanded ($D_\text{noS} \equiv 1/\bar{D}_\text{noS}$ is inverse-Taylor resummed) \\
%%%%
Hamiltonian in the $\nu\to 0$ limit & reduces to Kerr Hamiltonian for a \emph{test mass} in a generic orbit & reduces to Kerr Hamiltonian for a \emph{test spin}, to linear order in spin, in a generic orbit & the $A$ potential reduces to Kerr, but not the full Hamiltonian \\
%%%%
spin-orbit part & 3.5PN, in $(r,L^2)$ gauge, Taylor expanded & 3.5PN, added in the spin map & 3.5PN, in $(r,p_r^2)$ gauge, inverse-Taylor resummed \\
higher-order spin information & NNLO SS (4PN), LO S$^3$ (3.5PN), LO S$^4$ (4PN) & LO SS (2PN) & NNLO SS (4PN) for circular orbits \\
%%%%
precessing-spin Hamiltonian  & yes & yes & no \\
%%%%
spin-multipole constants included & yes & no & yes (in the SS contributions for circular orbits) \\
\end{tabular}
\end{ruledtabular}
\end{table*}
In this Section, we obtained an EOB Hamiltonian that reduces in the $\nu\to 0$ limit to the Kerr Hamiltonian for a nonspinning test mass in a generic orbit.
The nonspinning part of the Hamiltonian contains 4PN and partial 5PN results, which are Pad\'e resummed.
The Hamiltonian also includes the full 4PN precessing-spin contributions, which are the same PN information included in the Hamiltonians derived in Ref.~\cite{Khalil:2020mmr}, which extended the results of Ref.~\cite{Balmelli:2015zsa} to higher orders; however, we use different resummations/factorizations from those employed in the above references.

Table~\ref{tab:compHam} summarizes the main features of the \texttt{SEOBNRv5} Hamiltonian, and compares it to two other waveform models: \texttt{SEOBNRv4}~\cite{Barausse:2009xi,Barausse:2011ys,Bohe:2016gbl,Ossokine:2020kjp} and \texttt{TEOBResumS}~\cite{Damour:2014sva,Nagar:2018zoe,Nagar:2018plt}.

\section{Computationally efficient precessing-spin dynamics}
\label{sec:PNEOMs}

In Sec.~\ref{sec:HamPrec}, we derived an effective Hamiltonian for precessing spins that reduces to the Kerr Hamiltonian for generic orbits.
The EOMs from that Hamiltonian read
\begin{equation}
\label{EOMsPrecess}
\begin{gathered}
\dot{\bm{r}} = \frac{\partial H_\text{EOB}^\text{prec}}{\partial \bm{p}}, \qquad 
\dot{\bm{p}} = -\frac{\partial H_\text{EOB}^\text{prec}}{\partial \bm{r}} + \bm{\mF}, \\
\dot{\bm{S}}_{\mr i} =  \frac{\partial H_\text{EOB}^\text{prec}}{\partial \bm{S}_{\mr i}} \cross \bm{S}_{\mr i} + \dot{\bm{S}}_{\mr i}^\text{RR},
\end{gathered}
\end{equation}
where $\bm{\mF}$ is the RR force, and $\dot{\bm{S}}_{\mr i}^\text{RR}$ is the RR contribution to the spin-evolution equations, which starts at $\Order(v^{11}S^2)$~\cite{Will:2005sn,Maia:2017yok} and is thus neglected to the order we consider here.

These equations are computationally expensive to evolve numerically. Therefore, we simplify them such that we can solve the two-body dynamics 
more efficiently without losing much accuracy when describing the precessional effects.
It was shown in Refs.~\cite{Apostolatos:1994mx,Buonanno:2002fy,Boyle:2011gg,Schmidt:2012rh} that precessing-spin waveforms can be built starting from aligned-spin waveforms in the co-precessing frame, in which the $z$-axis remains perpendicular to the instantaneous orbital plane, and then applying a suitable rotation to the inertial frame. The precessing-spin \texttt{SEOBNRv3} and \texttt{SEOBNRv4} models employed the full EOB precessing-spin Hamiltonian~\cite{Barausse:2009xi,Barausse:2011ys} to evolve the dynamics in the co-precessing frame. To build the precessing-spin \texttt{TEOBResumS} model and speed-up the computational time, Refs.~\cite{Akcay:2020qrj,Gamba:2021ydi} used an aligned-spin EOB Hamiltonian when evolving the EOMs in the co-precessing frame. Also, the \texttt{IMRPhenomT} model~\cite{Estelles:2020twz} is built using a purely aligned-spin dynamics in the co-precessing frame. 

Here, to improve the accuracy in describing precession effects, we find it important to incorporate at least partial precessing-spin information in the Hamiltonian used in the co-precessing frame, as studies in Ref.~\cite{RamosBuadesv5} have demonstrated. 
To do that, we first obtain a precessing-spin Hamiltonian simpler than the full one derived in Sec.~\ref{sec:HamPrec}, such that it reduces to $H_\text{eff}^\text{align}$ for aligned spins, and only includes the in-plane spin components for circular orbits ($p_r=0$). 
Then, we orbit average the in-plane spin components in the Hamiltonian, and use it to evolve the EOMs for the dynamical variables $r,p_r,\phi$ and $p_\phi$, while the evolution equations for the spin and angular momentum vectors are computed in a PN-expanded, orbit-averaged form for quasi-circular orbits. 
The procedure to obtain the PN-expanded EOMs, and the appropriate dynamical variables, is similar to what was used in Refs.~\cite{SpinTaylorNotes,Estelles:2020twz,Akcay:2020qrj,Gamba:2021ydi}, but we include higher PN orders in the EOMs, and derive them from the \texttt{SEOBNRv5} EOB Hamiltonian, employing a different gauge and SSC. Other differences in the waveform model from previous work are described in Ref.~\cite{RamosBuadesv5}.

In the following subsections, we present the Hamiltonian, then derive the PN-expanded EOMs for precessing spins, in EOB coordinates, up to NNLO SO (3.5PN) and NNLO SS (4PN).

\subsection{Hamiltonian with partial precessing-spin dynamics}
\label{sec:simpHam}
The precessing-spin Hamiltonian presented in Sec.~\ref{sec:HamPrec} reduces to the exact Kerr Hamiltonian in Eq.~\eqref{KerrHam} for generic orbits.
Here, we consider a simpler Hamiltonian that starts with an ansatz similar to the aligned-spin Hamiltonian in Eqs.~\eqref{HeffAnzAlign} and \eqref{ABQanzAligned}, then complement it with precessing-spin corrections for circular orbits only (i.e., we do not include in-plane spin terms proportional to $p_r$). 
Thus, this Hamiltonian reduces to $H_\text{eff}^\text{align}$ from Sec.~\ref{sec:HamAlign} for aligned spins, and to $H^\text{Kerr\,eq}$ for $\nu \rightarrow 0$, but does not reduce to the full precessing-spin Hamiltonian $H_\text{eff}^\text{prec}$ and the Kerr Hamiltonian for generic orbits.

We use the following ansatz for the partial precessing-spin (pprec) effective Hamiltonian (cf. Eqs.~\eqref{HeffAnzAlign} and \eqref{HeffAnzPrec}) 
\begin{align}
\label{HeffAnzSimp}
H_\text{eff}^\text{pprec} &= \frac{
M \bm{L}\cdot (g_{a_+} \bm{a}_+ + g_{a_-} \delta \bm{a}_-) + \text{SO}_\text{calib}
+ \langle {G}^\text{pprec}_{a^3}\rangle }{r^3+a_+^2 (r+2 M)} \nonumber\\
& +
\bigg[A^\text{pprec} \bigg(
\mu^2 + B_p^\text{pprec} \frac{\bm{L}^2}{r^2} + (1 + B_{np}^\text{pprec}) (\bm{n} \cdot \bm{p})^2 \nonumber \\
&\qquad + B_{npa}^\text{Kerr\,eq}\, \frac{(\bm{L}\cdot\bm{a}_+)^2}{r^2} + Q^\text{pprec}  
\bigg)\bigg]^{1/2},
\end{align}
where the gyro-gravitomagnetic factors and SO calibration term are the same as in Eqs.~\eqref{gyros35PN} and Eq.~\eqref{SOcalibPrec}, with the same value of $d_\text{SO}$ as the aligned-spin model.

The SS corrections are added such that (cf. Eqs.~\eqref{ABQanzAligned} and \eqref{ABQanzPrec})
\begin{align}
\label{AnzSimp}
A^\text{pprec} &= \frac{a_+^2/r^2+A_\text{noS}+ A_\text{SS}^\text{prec} + A_{\mr S^4}^\text{prec} + \langle \tilde{A}_\text{SS}^\text{in\,plane}\rangle }{1+ (1+2M/r)a_+^2/r^2}, \nonumber\\
B_p^\text{pprec} &= 1 + \langle \tilde{B}_{p,\text{SS}}^\text{in\,plane}\rangle , \nonumber \\
B_{np}^\text{pprec} &= -1 + a_+^2/r^2 + A_\text{noS} \bar{D}_\text{noS} +  B_{np,\text{SS}}^\text{prec} , \nonumber\\
Q^\text{pprec} &= Q_\text{noS} + Q_\text{SS}^\text{prec},
\end{align}
where the nonspinning contributions are given by Eqs.~\eqref{ApmTay}--\eqref{QpmTay}, the SS corrections $A_\text{SS}^\text{prec}$, $B_{np,\text{SS}}^\text{prec}$ and $Q_\text{SS}^\text{prec}$ are the same as in Eqs.~\eqref{SSsolnPrec}, and the S$^4$ term $A_{\mr S^4}^\text{prec}$ is given by Eq.~\eqref{AS4prec}. 
The purely in-plane SS contributions are included in $\tilde{A}_\text{SS}^\text{in\,plane}$ and $\tilde{B}_{p,\text{SS}}^\text{in\,plane}$, which are obtained by writing an ansatz with unknown coefficients and matching it to a 4PN-expanded precessing-spin Hamiltonian with $p_r=0$. We indicate with $\langle..\rangle$ the result of orbit averaging the in-plane spin components, as explained in Sec.~\ref{orbit-average} below. We do not include in-plane S$^4$ corrections for simplicity, and because it is not straightforward to consistently orbit average the S$^4$ terms in the Hamiltonian. 
Thus, the partial precessing-spin Hamiltonian agrees with the full precessing-spin Hamiltonian $H_\text{eff}^\text{prec}$ from Sec.~\ref{sec:HamPrec} when PN expanded to 4PN and to $\Order(S^3)$ (included) for circular orbits.

For the terms $\tilde{A}_\text{SS}^\text{in\,plane}$ and $\tilde{B}_{p,\text{SS}}^\text{in\,plane}$, we obtain
\begin{widetext}
\begin{subequations}
\begin{align}
\tilde{A}_\text{SS}^\text{in\,plane} &= \frac{M}{r^3}\left[2 \left(\bm{n}\cdot\bm{a}_+\right)^2+3 C_+^{n\cdot a^2}\right]
+ \frac{M^2}{r^4} \bigg[\frac{33}{8} \delta  \bm{n}\cdot\bm{a}_+ \bm{n}\cdot\bm{a}_-+\left(-\frac{\nu }{2}-\frac{3}{8}\right) \left(\bm{n}\cdot\bm{a}_-\right)^2+\left(\frac{7 \nu }{4}-\frac{31}{4}\right) \left(\bm{n}\cdot\bm{a}_+\right)^2\nonumber\\
&\qquad
+3 \delta  C_-^{n\cdot a^2}+3 \nu  C_+^{n\cdot a^2}\bigg] 
+ \frac{M^3}{r^5} \bigg[\delta  \left(17 \nu -\frac{1}{4}\right) \bm{n}\cdot\bm{a}_+ \bm{n}\cdot\bm{a}_-+\left(-\frac{41 \nu ^2}{8}+\frac{583 \nu }{32}-\frac{171}{64}\right) \left(\bm{n}\cdot\bm{a}_-\right)^2 \nonumber\\
&\qquad
+\left(\frac{187}{64}-\frac{11 \nu ^2}{8}+\frac{1435 \nu }{96}\right) \left(\bm{n}\cdot\bm{a}_+\right)^2
+\delta  \left(\frac{81 \nu }{16}-\frac{99}{224}\right) C_-^{n\cdot a^2}+\left(\frac{867}{224}-\frac{13 \nu ^2}{16}+\frac{179 \nu }{56}\right) C_+^{n\cdot a^2}\bigg], \\
%%%%
\tilde{B}_{p,\text{SS}}^\text{in\,plane} &= -\frac{\left(\bm{n}\cdot\bm{a}_+\right)^2}{r^2}
+ \frac{M}{r^3} \left[\frac{3}{4} \delta  \bm{n}\cdot\bm{a}_+ \bm{n}\cdot\bm{a}_-+\left(\frac{3}{16}-\frac{3 \nu }{4}\right) \left(\bm{n}\cdot\bm{a}_-\right)^2+\left(-\frac{7 \nu }{4}-\frac{15}{16}\right) \left(\bm{n}\cdot\bm{a}_+\right)^2+(3-3 \nu ) C_+^{n\cdot a^2}\right] \nonumber\\
&\quad
+ \frac{M^2}{r^4} \bigg[\delta  \left(\!-\frac{49 \nu }{4}-\frac{43}{8}\right) \bm{n}\cdot\bm{a}_+ \bm{n}\cdot\bm{a}_-+\left(\frac{19 \nu ^2}{8}-\frac{545 \nu }{32}+\frac{219}{64}\right) \left(\bm{n}\cdot\bm{a}_-\right)^2
+\left(\frac{11 \nu ^2}{8}-\frac{805 \nu }{96}+\frac{125}{64}\right) \left(\bm{n}\cdot\bm{a}_+\right)^2 \nonumber\\
&\qquad
+\delta  \left(\frac{189}{32}-\frac{81 \nu }{16}\right) C_-^{n\cdot a^2}+\left(\frac{13 \nu ^2}{16}-\frac{203 \nu }{8}+\frac{51}{32}\right) C_+^{n\cdot a^2}\bigg],
\end{align}
\end{subequations}
and for the cubic-in-spin term ${G}^\text{pprec}_{a^3}$, we get
\begin{align}
{G}^\text{pprec}_{a^3}&= \bm{L}\cdot \bm{a}_+ \Bigg\lbrace
\frac{L^2}{\mu^2r^3} \left[
\frac{\delta}{2}(\bm{n}\cdot\bm{a}_-)(\bm{n}\cdot\bm{a}_+)
-\frac{\left(\bm{n}\cdot\bm{a}_+\right)^2}{4}
\right]
+\frac{M}{r^2} \Bigg[
-\frac{a_+^2}{4} - \frac{3}{4} \left(\bm{n}\cdot\bm{a}_+\right)^2
+\delta\frac{5}{24} (\bm{a}_+\cdot\bm{a}_-)
\nonumber\\
&\qquad\quad
-\delta\frac{5}{3} (\bm{n}\cdot\bm{a}_-) (\bm{n}\cdot\bm{a}_+)
+\frac{1}{8} a_1^2 \left(3\delta \widetilde{C}_{\rm 1ES^2} +4 \widetilde{C}_{\rm 1BS^3}\right)
+\frac{1}{8} a_2^2 \left(4 \widetilde{C}_{\rm 2BS^3}-3 \widetilde{C}_{\rm 2ES^2} \delta\right) \nonumber\\
&\qquad\quad
-\frac{3}{8} (\bm{a}_1\cdot\bm{a}_2) \left(\widetilde{C}_{\rm 1ES^2} (\delta -3)-\widetilde{C}_{\rm 2ES^2} (\delta +3)\right)
+\frac{3}{4} (\bm{n}\cdot\bm{a}_1) (\bm{n}\cdot\bm{a}_2) \left(\widetilde{C}_{\rm 1ES^2} (2 \delta -7)-\widetilde{C}_{\rm 2ES^2} (2 \delta +7)\right) \nonumber\\
&\qquad\quad
+ \left(\bm{n}\cdot\bm{a}_1\right)^2 \left(-\frac{3}{8} \widetilde{C}_{\rm 1ES^2} (2 \delta +3)-\frac{5 \widetilde{C}_{\rm 1BS^3}}{2}\right)
+\frac{1}{8} \left(\bm{n}\cdot\bm{a}_2\right)^2 \left(\widetilde{C}_{\rm 2ES^2} (6 \delta -9)-20 \widetilde{C}_{\rm 2BS^3}\right)
\Bigg]
\Bigg\rbrace \nonumber\\
&\quad
+\delta \bm{L}\cdot \bm{a}_- \Bigg\lbrace
-\frac{L^2 \left(\bm{n}\cdot\bm{a}_+\right)^2}{4 \mu^2 r^3}
+ \frac{M}{r^2}\Bigg[
\frac{a_+^2}{24} + \frac{5}{12} \left(\bm{n}\cdot\bm{a}_+\right)^2 
+\frac{1}{8} a_1^2 \left(4 \widetilde{C}_{\rm 1BS^3}-3 \widetilde{C}_{\rm 1ES^2}\right)
+\frac{1}{8} a_2^2 \left(3 \widetilde{C}_{\rm 2ES^2}-4 \widetilde{C}_{\rm 2BS^3}\right)\nonumber\\
&\qquad\quad
-\frac{3}{8} (\bm{a}_1\cdot\bm{a}_2) \left(\widetilde{C}_{\rm 1ES^2} (\delta -3)+\widetilde{C}_{\rm 2ES^2} (\delta +3)\right)
+\left(\bm{n}\cdot\bm{a}_1\right)^2 \left(-\frac{3}{8} \widetilde{C}_{\rm 1ES^2} (\delta -6)-\frac{5 \widetilde{C}_{\rm 1BS^3}}{2}\right)\nonumber\\
&\qquad\quad
+\frac{3}{4} (\bm{n}\cdot\bm{a}_1) (\bm{n}\cdot\bm{a}_2) \left(\widetilde{C}_{\rm 1ES^2} (2 \delta -7)+\widetilde{C}_{\rm 2ES^2} (2 \delta +7)\right) 
+\left(\bm{n}\cdot\bm{a}_2\right)^2 \left(\frac{5 \widetilde{C}_{\rm 2BS^3}}{2}-\frac{3}{8} \widetilde{C}_{\rm 2ES^2} (\delta +6)\right)
\!\Bigg]\!\Bigg\rbrace.
\end{align}\\
\end{widetext}

\subsection{Orbit averaging the precessing-spin contributions} 
\label{orbit-average}

To simplify the EOMs, we remove the explicit dependence of the Hamiltonian on the $\bm{n}\cdot\bm{a}_{\mr i}$ terms by taking their orbit average.
Since the spin-precession timescale ($\sim v^{-5}$) is larger than the orbital timescale ($\sim v^{-3}$), orbit-averaging the in-plane spin contributions is expected to provide a good approximation for the dynamics.

We define the unit vectors $(\lNhat,\bm{n},\lamNhat)$ in Cartesian coordinates such that $\lNhat$ is aligned with the $z$-axis, and hence the vector components are given by
\begin{equation}
\begin{aligned}
\lNhat &=  (0, 0, 1),\\
\bm{n} &= (\cos \phi, \sin \phi, 0),  \\
\lamNhat & \equiv \lNhat \cross \bm{n} = (-\sin \phi, \cos \phi, 0),
\end{aligned}
\end{equation}
where $\phi$ is the orbital phase.
When neglecting RR effects, an orbit average yields
\begin{equation}
\begin{aligned}
\langle n^i \rangle &= \frac{1}{2\pi} \int_{0}^{2\pi} n^i \di \phi = 0 = \langle \lambda_{\rm N}^i \rangle, \\
\langle n^i n^j \rangle &= \langle \lambda_{\rm N}^i \lambda_{\rm N}^j \rangle
=  \frac{1}{2} \left(\delta^{ij} - l_{\rm N}^i l_{\rm N}^j\right),
\end{aligned}
\end{equation}
which lead to the following relations for the spin dot products:
\begin{align}
\label{nSAvg}
&\left\langle (\bm{n}\cdot\bm{S}_{\mr i}) \bm{n} \right\rangle = \left\langle (\lamNhat\cdot\bm{S}_{\mr i}) \lamNhat \right\rangle = \frac{1}{2} \left[\bm{S}_{\mr i} - (\lNhat\cdot\bm{S}_{\mr i}) \lNhat\right], \nonumber\\
&\left\langle (\bm{n}\cdot\bm{S}_{\mr i})^2 \right\rangle = 
\left\langle (\lamNhat\cdot\bm{S}_{\mr i})^2 \right\rangle =
\frac{1}{2} \left[S_{\mr i}^2 - (\lNhat\cdot\bm{S}_{\mr i})^2\right], \nonumber\\
&\langle (\bm{n}\cdot\bm{S}_1)(\bm{n}\cdot\bm{S}_2) \rangle = 
\langle (\lamNhat\cdot\bm{S}_1)(\lamNhat\cdot\bm{S}_2) \rangle \nonumber\\
&\qquad\qquad=
\frac{1}{2} \left[\bm{S}_1\cdot\bm{S}_2 - (\lNhat\cdot\bm{S}_1)(\lNhat\cdot\bm{S}_2)\right].
\end{align}

The Hamiltonian from Sec.~\ref{sec:simpHam} depends on $(\bm{n} \cdot \bm{a}_+)^2$, $(\bm{n} \cdot \bm{a}_-)^2$ and $(\bm{n} \cdot \bm{a}_+) (\bm{n} \cdot \bm{a}_-)$, and can be made a function of only $\bm{a}_\pm$ and $\lNhat$ using the following orbit-averaged expressions:
\begin{align}
\label{naAvg}
(\bm{n} \cdot \bm{a}_\pm)^2 &\simeq \frac{1}{2} \left[a_\pm^2 - (\lNhat \cdot\bm{a}_\pm)^2\right], \\
(\bm{n} \cdot \bm{a}_+) (\bm{n} \cdot \bm{a}_-) &\simeq \frac{1}{2} \left[\bm{a}_+\cdot \bm{a}_- - (\lNhat \cdot\bm{a}_+) (\lNhat \cdot\bm{a}_-)\right].\nonumber
\end{align}
When taking the orbit average, we neglect RR since the Hamiltonian encodes the conservative dynamics, and the RR timescale ($\sim v^{-8}$) is much larger than the spin-precession timescale. We account for dissipative effects in the EOMs through the RR force and the orbital-frequency evolution equation, as described below. 

When restricting to binary black holes, the explicit expression of the partial-precessing Hamiltonian as function of $\bm{a}_\pm$ and $\lNhat$ is given in Appendix A of Ref.~\cite{RamosBuadesv5}.

\subsection{Equations of motion}
The ``Newtonian" angular-momentum vector $\bm{L}_{\rm N}$ is perpendicular to the instantaneous orbital plane, since it is defined by
\begin{equation}
\label{LNdef}
\bm{L}_{\rm N} \equiv \mu \bm{r} \cross \bm{\mr v},
\end{equation}
where $\bm{\mr v}\equiv \dot{\bm{r}}$ is the velocity.
We use a co-precessing frame aligned with the orthonormal unit vectors $(\lNhat,\bm{n},\lamNhat)$, with $\lNhat$ being the direction of $\bm{L}_{\rm N}$.
Since $\lNhat$ is perpendicular to $\bm{r}$ and $\bm{\mr v}$, we can write the velocity as
\begin{equation}
\label{defvOmega}
\bm{\mr v} = \dot{r} \bm{n} + r \Omega \lamNhat,
\end{equation}
which can be considered as a definition for the orbital frequency $\Omega$, implying that $\Omega = |\bm{n}\cross\bm{\mr v}|/r$.

The Hamiltonian is expressed in terms of the canonical angular momentum $\bm{L} \equiv \bm{r}\cross\bm{p}$. 
We denote the $\bm{L}$-based unit vectors by $(\lhat,\bm{n},\bm{\lambda})$, where $\lhat$ is the direction of $\bm{L}$ and $\bm{\lambda} \equiv \lhat \cross \bm{n}$, then express the EOMs derived from the Hamiltonian in terms of $\lNhat$.

In the co-precessing frame, the partial precessing-spin dynamics can be approximated by the following EOMs:
\begin{equation}
\label{EOMsAlign}
\begin{aligned}
\dot{r} &= \frac{\partial H_\text{EOB}^\text{pprec}}{\partial p_r}, \quad
&\dot{\phi} &= \frac{\partial H_\text{EOB}^\text{pprec}}{\partial p_\phi}, \\
\dot{p}_r &= -\frac{\partial H_\text{EOB}^\text{pprec}}{\partial r} + \mF_r, \quad
&\dot{p}_\phi &= \mF_\phi,
\end{aligned}
\end{equation}
where $H_\text{EOB}^\text{pprec}$ is related to $H_\text{eff}^\text{pprec}$ from Eq.~\eqref{HeffAnzSimp} through Eq.~\eqref{EOBmap}, and the quantities  $\langle {G}^\text{pprec}_{a^3} \rangle$, $\langle \tilde{A}_\text{SS}^\text{in\,plane}\rangle$ and $\langle \tilde{B}_{p,\text{SS}}^\text{in\,plane}\rangle$ in $H_\text{eff}^\text{pprec}$ can be expressed in terms of $\bm{a}_+$, $\bm{a}_-$ and $\lNhat$ once we replace the $\bm{n}\cdot \bm{a}_\pm$ terms by their orbit average using Eq.~\eqref{naAvg}. The explicit expressions are given in Appendix A of Ref.~\cite{RamosBuadesv5}.

At each time step, we also evolve the PN-expanded equations for the spins and angular momentum, given by
\begin{equation}
\label{EOMsSlN}
\begin{aligned}
\dot{\bm{S}}_{\mr i} &= \bm{\Omega}_{S_{\mr i}} \cross \bm{S}_{\mr i}, \\
\bm{L} &= \bm{L}(\lNhat,v,\bm{S}_{\mr i}), \\
\dotlNhat &= \dotlNhat(\lNhat,v,\bm{S}_{\mr i}),
\end{aligned}
\end{equation}
where $v \equiv (M \Omega)^{1/3}$, and $\bm{\Omega}_{S_{\mr i}} \equiv  \partial H_\text{EOB}^\text{prec}/ \partial \bm{S}_{\mr i}$ is the spin-precession frequency, computed in a PN expansion from the precessing-spin EOB Hamiltonian. 
These equations are derived in the following subsections to NNLO SS in an orbit average for quasi-circular orbits.

Equations~\eqref{EOMsAlign} and~\eqref{EOMsSlN} can be solved simultaneously for the dynamical variables.
Alternatively, they can be decoupled by computing the orbital frequency used in Eqs.~\eqref{EOMsSlN} in a PN expansion, which can be expressed as
\begin{equation}
\label{defdvdt}
\dot{v} = \left[\frac{\dot{E}(v)}{\di E(v)/\di v}\right]_\text{PN-expanded},
\end{equation} 
where $E(v)$ is the energy of the binary system and $\dot{E}(v)$ is the rate of energy loss.

Using the EOB orbital frequency, obtained by solving Eqs.~\eqref{EOMsAlign}, can lead to slightly more accurate results when solving Eqs.~\eqref{EOMsSlN} than the PN-expanded frequency from Eq.~\eqref{defdvdt}. 
However, the \texttt{SEOBNRv5PHM} waveform model~\cite{RamosBuadesv5} uses the PN-expanded frequency since decoupling Eqs.~\eqref{EOMsAlign} and~\eqref{EOMsSlN} makes it possible to use the post-adiabatic approximation~\cite{Damour:2012ky,Nagar:2018gnk,Rettegno:2019tzh,Mihaylov:2021bpf}, which improves the computational efficiency of the model.

In Sec.~\ref{sec:omegadot} below, we obtain $\dot{v}$ in a PN-expanded form, including the NNLO SS contribution that was recently derived in the flux in Ref.~\cite{Cho:2022syn}.
In all equations derived in this Section, we include PN orders up to NNLO SS, which implies different powers in $v$ for each quantity depending on the LO, as summarized in Table~\ref{tab:vOrders}.

Some precessing-spin waveform models, such as Refs.~\cite{Schmidt:2010it,Pan:2013rra,Ossokine:2020kjp}, considered a co-precessing frame adapted to the orbital angular momentum $\bm{L}$, instead of $\bm{L}_{\rm N}$.
Therefore, for completeness, we also provide in Appendix~\ref{app:LEOMs}, the EOMs expressed in terms of $\lhat$.

\begin{table}[ht]
\caption{Orders in $v$ at which the nonspinning, SO and SS contributions first enter $r$, $\bm{L}$, $\dot{\bm{S}}_{\mr i}$, $\dotlNhat$ and $\dot{v}$ for quasi-circular orbits. The last column indicates the highest power in $v$ we include in each quantity.}
\label{tab:vOrders}
\begin{ruledtabular}
\begin{tabular}{c@{\hspace{8pt}}|@{\hspace{-8pt}}ccc@{\hspace{8pt}}|@{\hspace{-8pt}}c}
quantity & LO S$^0$ & LO SO  & LO SS & highest order \\
\hline
$r$ & $v^0$ & $v^3$ & $v^4$ & $v^8$ \\
$\bm{L}$ & $v^{-1}$ & $v^2$ & $v^3$ & $v^7$ \\
$\dot{\bm{S}}_{\mr i}$ & -- & $v^5$ & $v^6$ & $v^{10}$ \\
$\dotlNhat$ & --  & $v^6$ & $v^7$ & $v^{11}$ \\
$\dot{v}$ & $v^9$  & $v^{12}$ & $v^{13}$ & $v^{17}$ \\
\end{tabular}
\end{ruledtabular}
\end{table}

\subsection{Angular momentum vector}
To obtain the angular momentum unit vector $\lhat$ in terms of $\lNhat$, we first use the EOMs~\eqref{EOMsPrecess}, and the definition of $\bm{L}_{\rm N}$ from Eq.~\eqref{LNdef}, to get $\lNhat$ in a PN expansion, i.e.,
\begin{equation}
\label{lNhatEq}
\lNhat \equiv \frac{\bm{L}_{\rm N}}{|\bm{L}_{\rm N}|} = \frac{\mu}{|\bm{L}_{\rm N}|}  \bm{r} \cross \frac{\partial H_\text{EOB}^\text{prec}}{\partial \bm{p}},
\end{equation}
where we use the Hamiltonian before taking the orbit average of the in-plane spin terms.

Then, we specialize to circular orbits, which are defined by $p_r = 0$ and $\dot{p}_r = 0$. 
To obtain $r$ and $L$ as functions of $v$ for circular orbits, we solve
\begin{equation}
\label{circConds}
\begin{aligned}
\frac{\di}{\di t} (\bm{r}\cdot\bm{p}) &= \bm{r}\cdot\dot{\bm{p}} + \dot{\bm{r}}\cdot\bm{p} =0, \\
v^3 &= \frac{M}{r}\,|\bm{n}\cross\bm{\mr v}|,
\end{aligned}
\end{equation}
for $r(\lhat,\bm{\lambda},\bm{n},\bm{S}_{\mr i},v)$ and $L(\lhat,\bm{\lambda},\bm{n},\bm{S}_{\mr i},v)$, perturbatively in a PN expansion, after using the EOMs~\eqref{EOMsPrecess} without RR. 

We substitute that solution for $r$ and $L$ in Eq.~\eqref{lNhatEq}, and replace $\bm{\lambda}$ using
\begin{align}
\bm{S}_{\mr i} &= (\bm{n}\cdot\bm{S}_{\mr i})\bm{n} + (\bm{\lambda}\cdot\bm{S}_{\mr i})\bm{\lambda} + (\lhat\cdot\bm{S}_{\mr i})\lhat \nonumber\\
&= (\bm{n}\cdot\bm{S}_{\mr i})\bm{n} + (\lamNhat\cdot\bm{S}_{\mr i})\lamNhat + (\lNhat\cdot\bm{S}_{\mr i})\lNhat,
\end{align}
which implies that
\begin{equation}
\begin{aligned}
&(\bm{\lambda}\cdot\bm{S}_{\mr i})^2 + (\lhat\cdot\bm{S}_{\mr i})^2 =  (\lamNhat\cdot\bm{S}_{\mr i})^2 + (\lNhat\cdot\bm{S}_{\mr i})^2, \\
&(\bm{\lambda}\cdot\bm{S}_1)(\bm{\lambda}\cdot\bm{S}_2) + (\lhat\cdot\bm{S}_1)(\lhat\cdot\bm{S}_2) \\
&\qquad=  (\lamNhat\cdot\bm{S}_1)(\lamNhat\cdot\bm{S}_2) + (\lNhat\cdot\bm{S}_1)(\lNhat\cdot\bm{S}_2).
\end{aligned}
\end{equation}
That way, the right-hand side of Eq.~\eqref{lNhatEq} only depends on $\lhat$, $\lNhat$, $\lamNhat$, $v$ and the spins.

To solve Eq.~\eqref{lNhatEq} for $\lhat(\lNhat,\lamNhat,\bm{S}_{\mr i},v)$, we expand it in spin, such that
\begin{widetext}
\begin{subequations}
\label{lvslN}
\begin{equation}
\lhat \equiv \lNhat + \lhat_\text{SO} + \lhat_\text{SS},
\end{equation}
since $\lhat$ is in the same direction as $\lNhat$ for nonspinning binaries, while $\lhat_\text{SO}$ and $\lhat_\text{SS}$ are the SO and SS contributions.
Solving order by order in spin, we obtain
\begin{align}
\lhat_\text{SO} &= \frac{(\lamNhat\cdot\bm{S}_1) \lamNhat}{M\mu} \bigg\lbrace
-\frac{v^3}{2}   (3X_2 + \nu)
+ v^5 \left[\frac{\nu ^2}{8}-\frac{9 \nu }{8}+\left(\frac{3 \nu }{2}+\frac{9}{8}\right) X_2\right] \nonumber\\
&\quad\qquad
+ v^7 \left[\frac{\nu ^3}{48}+\frac{9 \nu ^2}{4}-\frac{27 \nu }{16}+\left(-\frac{\nu ^2}{2}+\frac{15 \nu }{4}+\frac{27}{16}\right) X_2\right]
\bigg\rbrace   + 1\leftrightarrow 2, \\
%%%%
\lhat_\text{SS} &= \frac{v^4}{M^2\mu^2} \lamNhat \left[
(\lNhat\cdot\bm{S}_1)(\lamNhat\cdot\bm{S}_1) (X_2 - \nu)
+ \nu (\lNhat\cdot\bm{S}_1)(\lamNhat\cdot\bm{S}_2) \right] \nonumber\\
&\quad
+ \frac{v^6}{M^2\mu^2} \bigg\lbrace
\lamNhat (\lNhat\cdot\bm{S}_1)(\lamNhat\cdot\bm{S}_1) \left[\frac{\nu ^2}{6}+\frac{5 \nu }{2}+\left(-\frac{11 \nu }{3}-\frac{5}{2}\right) X_2\right]
+ \lamNhat (\lNhat\cdot\bm{S}_1)(\lamNhat\cdot\bm{S}_2) \left(-\frac{7 \nu ^2}{6}-4 \nu\right)  \nonumber\\
&\quad\qquad
+ \lNhat (\lamNhat\cdot\bm{S}_1)^2 \left[-\frac{\nu ^2}{8}+\frac{9 \nu }{8}+\left(-\frac{3 \nu }{4}-\frac{9}{8}\right) X_2\right] 
+ \lNhat (\lamNhat\cdot\bm{S}_1)(\lamNhat\cdot\bm{S}_2) \left(-\frac{\nu ^2}{8}-\frac{3}{2} \nu\right)
\bigg\rbrace \nonumber\\
&\quad
+ \frac{v^8}{M^2\mu^2} \bigg\lbrace
\lamNhat (\lNhat\cdot\bm{S}_1)(\lamNhat\cdot\bm{S}_1) \left[-\frac{211 \nu ^3}{144}-\frac{41 \nu ^2}{16}-\frac{9 \nu }{16}+\left(-\frac{479 \nu ^2}{144}+2 \nu +\frac{9}{16}\right) X_2\right]\nonumber\\
&\quad\qquad
+ \lNhat (\lamNhat\cdot\bm{S}_1)^2 \left[\frac{\nu ^3}{16}-\frac{45 \nu ^2}{16}-\frac{27 \nu }{16}+\left(\frac{15 \nu ^2}{16}+\frac{9 \nu }{8}+\frac{27}{16}\right) X_2\right] \nonumber\\
&\quad\qquad
+ \lamNhat (\lNhat\cdot\bm{S}_1)(\lamNhat\cdot\bm{S}_2) \left[
-\frac{179 \nu ^3}{144}+\frac{9 \nu ^2}{32}-\frac{3 \nu }{8}+\left(\frac{3 \nu ^2}{8}+\frac{3 \nu }{2}\right) X_2\right]  \nonumber\\
&\quad\qquad
+ \lNhat (\lamNhat\cdot\bm{S}_1)(\lamNhat\cdot\bm{S}_2) \left[\frac{\nu ^3}{16}+\frac{69 \nu ^2}{32}+\frac{9 \nu }{8}\right]
\bigg\rbrace  + 1 \leftrightarrow 2,
\end{align}
\end{subequations}
which is independent of the spin-quadrupole constants.
Substituting $\lhat(\lNhat)$ in the solution of Eqs.~\eqref{circConds} yields $r(\lNhat,\lamNhat,\bm{n},\bm{S}_{\mr i},v)$ and $L(\lNhat,\lamNhat,\bm{n},\bm{S}_{\mr i},v)$, which are given in Appendix~\ref{app:rLCirclN}. 

Finally, we use these relations to obtain $\bm{L} = L \lhat$ and take its orbit average using Eqs.~\eqref{nSAvg}, leading to
\begin{subequations}
\label{LvslNAvg}
\begin{align}
\bm{L} &\equiv \frac{\mu M }{v}\left(\bar{\bm{L}}_{S^0} + \bar{\bm{L}}_\text{SO} + \bar{\bm{L}}_{S_1S_2} + \bar{\bm{L}}_{S^2} + \bar{\bm{L}}_{S^2\tilde{C}}\right), \\
\bar{\bm{L}}_{S^0} &= \lNhat\bigg\lbrace
1
+\left(\frac{\nu }{6}+\frac{3}{2}\right) v^2
+ \left(\frac{\nu ^2}{24}-\frac{19 \nu }{8}+\frac{27}{8}\right) v^4
+ \left[\frac{7 \nu ^3}{1296}+\frac{31 \nu ^2}{24}+\left(\frac{41 \pi ^2}{24}-\frac{6889}{144}\right) \nu +\frac{135}{16}\right] v^6 \nonumber\\
&\quad
+ v^8 \bigg[-\frac{55 \nu ^4}{31104}-\frac{215 \nu ^3}{1728}+\left(\frac{356035}{3456}-\frac{2255 \pi ^2}{576}\right) \nu ^2+\nu  \left(\frac{98869}{5760}-\frac{128 \gamma_E}{3}-\frac{6455 \pi ^2}{1536}-\frac{256}{3} \ln 2-\frac{128 \ln v}{3}\right) \nonumber\\
&\quad\qquad +\frac{2835}{128}\bigg]
\bigg\rbrace, \\
%%%%
\bar{\bm{L}}_\text{SO} &=
\frac{v^3}{M\mu} \left[(\lNhat\cdot\bm{S}_1) \lNhat \left(-\frac{7 \nu }{12}-\frac{7 X_2}{4}\right)+\bm{S}_1 \left(-\frac{\nu }{4}-\frac{3 X_2}{4}\right)\right] \nonumber\\
&\quad
+ \frac{v^5}{M\mu} \left\lbrace(\lNhat\cdot\bm{S}_1) \lNhat \left[\frac{11 \nu ^2}{144}-\frac{55 \nu }{16}+\left(\frac{55 \nu }{24}-\frac{33}{16}\right) X_2\right]+\bm{S}_1 \left[\frac{\nu ^2}{48}-\frac{15 \nu }{16}+\left(\frac{5 \nu }{8}-\frac{9}{16}\right) X_2\right]\right\rbrace\nonumber\\
&\quad
+ \frac{v^7}{M\mu} \bigg\lbrace(\lNhat\cdot\bm{S}_1) \lNhat \left[\frac{5 \nu ^3}{96}+\frac{275 \nu ^2}{32}-\frac{405 \nu }{32}+\left(-\frac{25 \nu ^2}{32}+\frac{195 \nu }{8}-\frac{135}{32}\right) X_2\right] \nonumber\\
&\quad\qquad
+\bm{S}_1 \left[\frac{\nu ^3}{96}+\frac{55 \nu ^2}{32}-\frac{81 \nu }{32}+\left(-\frac{5 \nu ^2}{32}+\frac{39 \nu }{8}-\frac{27}{32}\right) X_2\right]\bigg\rbrace
+ 1 \leftrightarrow 2, \\
%%%%
\bar{\bm{L}}_{S_1S_2} &= 
\frac{\nu v^4}{(M\mu)^2} \left\lbrace\lNhat \left[2 (\lNhat\cdot\bm{S}_1) (\lNhat\cdot\bm{S}_2)-(\bm{S}_1\cdot\bm{S}_2)\right]+\frac{(\lNhat\cdot\bm{S}_1) \bm{S}_2}{2}+\frac{(\lNhat\cdot\bm{S}_2) \bm{S}_1}{2}\right\rbrace \nonumber\\
&\quad
+ \frac{\nu v^6}{(M\mu)^2} \bigg\lbrace\lNhat \left[(\lNhat\cdot\bm{S}_1) (\lNhat\cdot\bm{S}_2) \left(\frac{13 \nu }{36}-\frac{7}{6}\right)+\frac{2 \nu  (\bm{S}_1\cdot\bm{S}_2)}{3}\right]\nonumber\\
&\quad\qquad
+\bm{S}_2 (\lNhat\cdot\bm{S}_1)\left(\frac{5}{4}-\frac{7 \nu }{24}\right)+\bm{S}_1 (\lNhat\cdot\bm{S}_2) \left(\frac{5}{4}-\frac{7 \nu }{24}\right)\bigg\rbrace \nonumber\\
&\quad
+ \frac{\nu v^8}{(M\mu)^2} \bigg\lbrace\lNhat \left[(\lNhat\cdot\bm{S}_1) (\lNhat\cdot\bm{S}_2) \left(-\frac{361 \nu ^2}{432}+\frac{361 \nu }{288}+\frac{15}{4}\right)+\left(-\frac{5 \nu ^2}{72}-\frac{245 \nu }{24}-\frac{5}{4}\right) (\bm{S}_1\cdot\bm{S}_2)\right] \nonumber\\
&\quad \qquad
+\bm{S}_2 (\lNhat\cdot\bm{S}_1) \left(-\frac{223}{288} \nu ^2-\frac{349 \nu }{64}+\frac{15}{8}\right)
+\bm{S}_1(\lNhat\cdot\bm{S}_2) \left(-\frac{223}{288} \nu ^2-\frac{349 \nu }{64}+\frac{15}{8}\right)\bigg\rbrace,\\
%%%%
\bar{\bm{L}}_{S^2} &= \frac{v^4}{(M\mu)^2} \left\lbrace
\lNhat \left[(\lNhat\cdot\bm{S}_1)^2 \left(X_2-\nu \right)+S_1^2 \left(\frac{\nu }{2}-\frac{X_2}{2}\right)\right]
+(\lNhat\cdot\bm{S}_1) \bm{S}_1 \left(\frac{X_2}{2}-\frac{\nu }{2}\right)\right\rbrace\nonumber\\
&\quad
+ \frac{v^6}{(M\mu)^2} \bigg\lbrace\lNhat \left\lbrace
(\lNhat\cdot\bm{S}_1)^2 \left[\frac{121 \nu ^2}{72}+\frac{35 \nu }{8}+\left(\frac{11 \nu }{2}-\frac{35}{8}\right) X_2\right]
+S_1^2 \left[-\frac{\nu ^2}{3}-\nu +\left(1-\frac{5 \nu }{3}\right) X_2\right]\right\rbrace \nonumber\\
&\quad\qquad
+\bm{S}_1 (\lNhat\cdot\bm{S}_1) \left[\frac{5 \nu ^2}{24}-\frac{11 \nu }{8}+\left(\frac{11}{8}-\frac{\nu }{2}\right) X_2\right]\bigg\rbrace \nonumber\\
&\quad
+ \frac{v^8}{(M\mu)^2} \bigg\lbrace\lNhat (\lNhat\cdot\bm{S}_1)^2 \left[-\frac{505 \nu ^3}{864}+\frac{347 \nu ^2}{96}+\frac{111 \nu }{32}+\left(-\frac{2833 \nu ^2}{288}+\frac{199 \nu }{16}-\frac{111}{32}\right) X_2\right] \nonumber\\
&\quad\qquad
+ \lNhat S_1^2 \left[\frac{5 \nu ^3}{144}+\frac{275 \nu ^2}{48}-\frac{15 \nu }{16}+\left(\frac{295 \nu ^2}{144}-\frac{455 \nu }{48}+\frac{15}{16}\right) X_2\right] \nonumber\\
&\quad\qquad
+\bm{S}_1 (\lNhat\cdot\bm{S}_1) \left[-\frac{235 \nu ^3}{288}+\frac{563 \nu ^2}{96}-\frac{21 \nu }{32}+\left(-\frac{113 \nu ^2}{32}-\frac{7 \nu }{3}+\frac{21}{32}\right) X_2\right]\bigg\rbrace
+ 1 \leftrightarrow 2, \\
%%%%
\bar{\bm{L}}_{S^2\tilde{C}} &=  \frac{\widetilde{C}_{\rm 1ES^2}}{(M\mu)^2} \lNhat \bigg\lbrace
v^4 \left[(\lNhat\cdot\bm{S}_1)^2 \left(\frac{3 X_2}{2}-\frac{3 \nu }{2}\right)+S_1^2 \left(\frac{\nu }{2}-\frac{X_2}{2}\right)\right] \nonumber\\
&\qquad
+ v^6 \left[(\lNhat\cdot\bm{S}_1)^2 \left(\nu ^2-3 \nu +(3 \nu +3) X_2\right)+S_1^2 \left(-\frac{\nu ^2}{3}+\nu +(-\nu -1) X_2\right)\right]\nonumber\\
&\qquad
+ v^8 \bigg[
(\lNhat\cdot\bm{S}_1)^2 \left(-\frac{5 \nu ^3}{48}+\frac{1475 \nu ^2}{112}-\frac{135 \nu }{16}+\left(-\frac{65 \nu ^2}{16}+\frac{55 \nu }{16}+\frac{135}{16}\right) X_2\right) \nonumber\\
&\quad\qquad
+S_1^2 \left(\frac{5 \nu ^3}{144}-\frac{1475 \nu ^2}{336}+\frac{45 \nu }{16}+\left(\frac{65 \nu ^2}{48}-\frac{55 \nu }{48}-\frac{45}{16}\right) X_2\right)
\bigg]
\bigg\rbrace+ 1 \leftrightarrow 2.
\end{align}
\end{subequations}
\end{widetext}
For aligned spins, $L(v)$ is gauge invariant, and our result agrees with the literature (e.g., with Refs.~\cite{Bohe:2012mr,Levi:2014sba,Levi:2016ofk,Cho:2022syn}).
However, for precessing spins, $\bm{L}$ is gauge dependent, and our result disagrees with Refs.~\cite{Bohe:2012mr,Akcay:2020qrj,SpinTaylorNotes}, even at LO SO, because these references used the covariant (Tulczyjew-Dixon) SSC~\cite{Tulczyjew:1959,Dixon:1979}, while we use the canonical Newton-Wigner (NW) SSC~\cite{pryce1948mass,newton1949localized}, since we are working in a Hamiltonian formalism~\cite{Vines:2016unv,Barausse:2009aa}.
Appendix~\ref{app:SSC} shows how to transform between our result and that of Refs.~\cite{Bohe:2012mr,Akcay:2020qrj,SpinTaylorNotes} at LO SO.

\subsection{Spin-evolution equations}
\label{sec:Sdot}
We obtain the spin-precession frequency $\bm{\Omega}_{S_{\mr i}}$ by differentiating the Hamiltonian with respect to the spin vector.
Then, we take the circular-orbit limit by setting $p_r = 0$ and replacing $r$ and $L$ by Eqs.~\eqref{LCirclN} and~\eqref{rCirclN}. Finally, averaging the spin components over an orbit using Eqs.~\eqref{nSAvg} yields
\begin{widetext}
\begin{subequations}
\label{S1dotAvg}
\begin{align}
\dot{\bm{S}}_1 &= \bm{\Omega}_{S_1} \cross \bm{S}_1, \\
\bm{\Omega}_{S_1} &= 
\frac{\lNhat}{M} \bigg\lbrace
v^5 \left(\frac{3 X_2}{2}+\frac{\nu }{2}\right)
+ v^7 \left[\left(\frac{9}{8}-\frac{5 \nu }{4}\right) X_2-\frac{\nu ^2}{24}+\frac{15 \nu }{8}\right] \nonumber\\
&\quad\qquad
+ v^9 \left[\left(\frac{5 \nu ^2}{16}-\frac{39 \nu }{4}+\frac{27}{16}\right) X_2-\frac{\nu ^3}{48}-\frac{55 \nu ^2}{16}+\frac{81 \nu }{16}\right]\!\bigg\rbrace \nonumber\\
&\quad
+ \frac{v^6}{M^2\mu} \left\lbrace
\lNhat \left[\lNhat\cdot\bm{S}_1 \left(\frac{3 \nu }{2}-\frac{3 X_2}{2}\right)-\frac{3  \nu }{2}\lNhat\cdot\bm{S}_2\right]+\frac{\nu}{2}\bm{S}_2
\right\rbrace \nonumber\\
&\quad
+ \frac{v^8}{M^2\mu} \left\lbrace
\lNhat \left[\lNhat\cdot\bm{S}_1 \left(-\frac{17 \nu ^2}{12}-\frac{9 \nu }{4}+\left(\frac{9}{4}-\frac{15 \nu }{4}\right) X_2\right)+\lNhat\cdot\bm{S}_2 \left(\frac{\nu ^2}{12}-\frac{\nu }{2}\right)\right]-\frac{\nu^2}{4}\bm{S}_2
\right\rbrace \nonumber\\
&\quad
+ \frac{v^{10}}{M^2\mu} \bigg\lbrace
\lNhat \bigg[\lNhat\cdot\bm{S}_1 \left(\frac{121 \nu ^3}{144}-\frac{91 \nu ^2}{16}-\frac{27 \nu }{16}+\left(\frac{385 \nu ^2}{48}-\frac{97 \nu }{16}+\frac{27}{16}\right) X_2\right) \nonumber\\
&\quad\qquad
+\lNhat\cdot\bm{S}_2 \left(\frac{103 \nu ^3}{144}+\frac{139 \nu ^2}{48}-\frac{9 \nu }{4}\right)\bigg] 
+\left(\frac{\nu ^3}{48}+\frac{49 \nu ^2}{16}+\frac{3 \nu }{8}\right) \bm{S}_2
\bigg\rbrace, \nonumber\\
&\quad
+\frac{\widetilde{C}_{\rm 1ES^2}}{M^2\mu} \lNhat (\lNhat\cdot\bm{S}_1) \bigg\lbrace
v^6 \left(\frac{3 \nu }{2}-\frac{3 X_2}{2}\right)
+ v^8 \left[-\frac{3 \nu ^2}{4}+\frac{9 \nu }{4}+\left(-\frac{9 \nu }{4}-\frac{9}{4}\right) X_2\right] \nonumber\\
&\quad\qquad
+ v^{10} \left[\frac{\nu ^3}{16}-\frac{885 \nu ^2}{112}+\frac{81 \nu }{16}+\left(\frac{39 \nu ^2}{16}-\frac{33 \nu }{16}-\frac{81}{16}\right) X_2\right]
\bigg\rbrace,
\label{OmegaS1}
\end{align}
\end{subequations}
\end{widetext}
and similarly $\dot{\bm{S}}_2 = \bm{\Omega}_{S_2} \cross \bm{S}_2$, with $\bm{\Omega}_{S_2}$ given by Eq.~\eqref{OmegaS1} after exchanging the two bodies' labels $1\leftrightarrow2$.
The SO and LO SS parts of the spin-precession frequency agree with the orbit-averaged results given by Eqs.~(1)-(5) of Ref.~\cite{Akcay:2020qrj}, but the NLO and NNLO SS terms do not agree with Refs.~\cite{Bohe:2015ana,SpinTaylorNotes} because of the different gauge.

\subsection{Evolution of the orbital frequency}
\label{sec:omegadot}
The evolution equation for the orbital frequency is given by Eq.~\eqref{defdvdt} in terms of the energy loss and the derivative of the binding energy.
The circular-orbit binding energy can be obtained from the Hamiltonian (minus the rest mass) by setting $p_r = 0$, replacing $r$, $L$ and $\lhat$ by Eqs.~\eqref{LCirclN}, \eqref{rCirclN} and \eqref{lvslN}, then taking the orbit average. This leads to
\begin{widetext}
\begin{subequations}
\begin{align}
E(v) &\equiv - \frac{\mu v^2}{2} \left(\bar{E}^{S^0} + \bar{E}_\text{SO} + \bar{E}_{S_1S_2} + \bar{E}_{S^2} + \bar{E}_{S^2\tilde{C}} \right), \\
\bar{E}_{S^0} &= 1
+\left(-\frac{\nu }{12}-\frac{3}{4}\right) v^2
+\left(-\frac{\nu ^2}{24}+\frac{19 \nu }{8}-\frac{27}{8}\right) v^4
+\left[-\frac{35 \nu ^3}{5184}-\frac{155 \nu ^2}{96}+\left(\frac{34445}{576}-\frac{205 \pi ^2}{96}\right) \nu -\frac{675}{64}\right] v^6,\\
%%%%
\bar{E}_\text{SO} &= \frac{\lNhat\cdot\bm{S}_1}{M\mu} \bigg\lbrace
v^3 \left(\frac{2 \nu }{3}+2 X_2\right)
+ v^5 \left[-\frac{\nu ^2}{9}+5 \nu +\left(3-\frac{10 \nu }{3}\right) X_2\right] \nonumber\\
&\quad
+ v^7 \left[-\frac{\nu ^3}{12}-\frac{55 \nu ^2}{4}+\frac{81 \nu }{4}+\left(\frac{5 \nu ^2}{4}-39 \nu +\frac{27}{4}\right) X_2\right]
\bigg\rbrace + 1 \leftrightarrow 2, \\
%%%%
\bar{E}_{S_1S_2} &= \frac{\nu}{(M\mu)^2} \bigg\lbrace v^4 \left[-3(\lNhat\cdot\bm{S}_1)(\lNhat\cdot\bm{S}_2)+(\bm{S}_1\cdot\bm{S}_2)\right]
+ v^6 \left[(\lNhat\cdot\bm{S}_1)(\lNhat\cdot\bm{S}_2) \left(\frac{5 \nu}{18}-\frac{5 }{3}\right)-\frac{5 \nu}{6}(\bm{S}_1\cdot\bm{S}_2)\right] \nonumber\\
&\quad
+ v^8 \left[(\lNhat\cdot\bm{S}_1)(\lNhat\cdot\bm{S}_2) \left(\frac{721 \nu ^2}{216}+\frac{973 \nu}{72}-\frac{21}{2}\right)+(\bm{S}_1\cdot\bm{S}_2)\left(\frac{7 \nu ^2}{72}+\frac{343 \nu}{24}+\frac{7}{4}\right) \right] \bigg\rbrace,\\
%%%%
\bar{E}_{S^2} &= \frac{v^4}{(M\mu)^2} \left[(\lNhat\cdot\bm{S}_1)^2 \left(\frac{3 \nu }{2}-\frac{3 X_2}{2}\right)+S_1^2 \left(\frac{X_2}{2}-\frac{\nu }{2}\right)\right]\nonumber\\
&\quad
+ \frac{v^6}{(M\mu)^2} \bigg\lbrace(\lNhat\cdot\bm{S}_1)^2 \left[-\frac{85 \nu ^2}{36}-\frac{15 \nu }{4}+\left(\frac{15}{4}-\frac{25 \nu }{4}\right) X_2\right] 
+S_1^2 \left[\frac{5 \nu ^2}{12}+\frac{5 \nu }{4}+\left(\frac{25 \nu }{12}-\frac{5}{4}\right) X_2\right]\bigg\rbrace\nonumber\\
&\quad
+ \frac{v^8}{(M\mu)^2} \bigg\lbrace
S_1^2 \left[-\frac{7 \nu ^3}{144}-\frac{385 \nu ^2}{48}+\frac{21 \nu }{16}+\left(-\frac{413 \nu ^2}{144}+\frac{637 \nu }{48}-\frac{21}{16}\right) X_2\right]\nonumber\\
&\qquad\qquad
+(\lNhat\cdot\bm{S}_1)^2 \left[\frac{847 \nu ^3}{432}-\frac{637 \nu ^2}{48}-\frac{63 \nu }{16}+\left(\frac{2695 \nu ^2}{144}-\frac{679 \nu }{48}+\frac{63}{16}\right) X_2\right]\bigg\rbrace
+ 1\leftrightarrow2, \\
%%%%
\bar{E}_{S^2\tilde{C}} &= \frac{\widetilde{C}_{\rm 1ES^2}}{(M\mu)^2} \bigg\lbrace
v^4 \left[(\lNhat\cdot\bm{S}_1)^2 \left(\frac{3 \nu }{2}-\frac{3 X_2}{2}\right)+S_1^2 \left(\frac{X_2}{2}-\frac{\nu }{2}\right)\right] \nonumber\\
&\qquad
+ v^6 \left[(\lNhat\cdot\bm{S}_1)^2 \left(-\frac{5 \nu ^2}{4}+\frac{15 \nu }{4}+\left(-\frac{15 \nu }{4}-\frac{15}{4}\right) X_2\right)+S_1^2 \left(\frac{5 \nu ^2}{12}-\frac{5 \nu }{4}+\left(\frac{5 \nu }{4}+\frac{5}{4}\right) X_2\right)\right]\nonumber\\
&\qquad
+ v^8 \bigg[
S_1^2 \left(-\frac{7 \nu ^3}{144}+\frac{295 \nu ^2}{48}-\frac{63 \nu }{16}+\left(-\frac{91 \nu ^2}{48}+\frac{77 \nu }{48}+\frac{63}{16}\right) X_2\right) \nonumber\\
&\qquad\qquad
+ (\lNhat\cdot\bm{S}_1)^2 \left(\frac{7 \nu ^3}{48}-\frac{295 \nu ^2}{16}+\frac{189 \nu }{16}+\left(\frac{91 \nu ^2}{16}-\frac{77 \nu }{16}-\frac{189}{16}\right) X_2\right)
\bigg]
\bigg\rbrace + 1\leftrightarrow2.
\end{align}
\end{subequations}
\end{widetext}
Note that we did not include the 4PN nonspinning contribution in the binding energy to keep it at the same order as the energy flux, which is known to 3.5PN~\cite{Blanchet:2004ek}.
The nonspinning and SO parts agree with Eqs.~(233) and~(415) of Ref.~\cite{Blanchet:2013haa}, while the SS part agrees in the aligned-spin limit with, e.g., Refs.~\cite{Levi:2014sba,Levi:2016ofk}.

The NNLO SO contribution to the energy flux was derived in Ref.~\cite{Bohe:2013cla}, while the NNLO SS (4PN beyond the LO) contribution was derived in Ref.~\cite{Cho:2022syn}, though the SS tail contribution at 3.5PN was obtained for aligned spins only.
The result in Ref.~\cite{Cho:2022syn} is expressed in terms of gauge-dependent quantities. Therefore, we use their EOMs to obtain the circular-orbit energy flux as a function of $v$, and orbit-average the in-plane spin components, leading to
\begin{widetext}
\begin{subequations}
\label{dEdt}
\begin{align}
\dot{E} &\equiv -\frac{32 \nu^2 v^{10}}{5} \left(\dot{\bar{E}}_{S^0} + \dot{\bar{E}}_\text{SO} + \dot{\bar{E}}_{S_1S_2} + \dot{\bar{E}}_{S^2} + \dot{\bar{E}}_{S^2\tilde{C}}\right), \\
%%%%
\dot{\bar{E}}_{S^0} &= 1 
+ v^2 \left(-\frac{35 \nu }{12}-\frac{1247}{336}\right)
+ 4\pi v^3
+ v^4 \left(\frac{65 \nu ^2}{18}+\frac{9271 \nu }{504}-\frac{44711}{9072}\right)
+ \pi v^5 \left(-\frac{583 \nu }{24}-\frac{8191}{672}\right) \nonumber\\
&\quad
+ v^6 \left[-\frac{775 \nu ^3}{324}-\frac{94403 \nu ^2}{3024}-\frac{134543 \nu }{7776}+\pi ^2 \left(\frac{41 \nu }{48}+\frac{16}{3}\right)-\frac{1712 \ln v}{105}-\frac{1712 \gamma_E}{105}+\frac{6643739519}{69854400}-\frac{3424 \ln 2}{105}\right] \nonumber\\
&\quad
+ \pi v^7 \left(\frac{193385 \nu ^2}{3024}+\frac{214745 \nu }{1728}-\frac{16285}{504}\right), \\
%%%%
\dot{\bar{E}}_\text{SO} &= \frac{\lNhat\cdot\bm{S}_1}{M\mu} \bigg\lbrace
v^3 \left(-\frac{3 \nu }{2}-\frac{5 X_2}{4}\right)
+ v^5 \left(\frac{157 \nu ^2}{18}-\frac{23 \nu }{8}+\left(\frac{43 \nu }{4}-\frac{13}{16}\right) X_2\right)
+ \pi v^6 \left(-\frac{17 \nu }{3}-\frac{31 X_2}{6}\right) \nonumber\\
&\qquad
+ v^7 \left[-\frac{1117 \nu ^3}{54}+\frac{625 \nu ^2}{189}+\frac{180955 \nu }{13608}+\left(-\frac{1501 \nu ^2}{36}+\frac{1849 \nu }{126}+\frac{9535}{336}\right) X_2\right]\nonumber\\
&\qquad
+ \pi v^8 \left[\frac{21241 \nu ^2}{336}-\frac{10069 \nu }{672}+\left(\frac{130583 \nu }{2016}-\frac{7163}{672}\right) X_2\right]
\bigg\rbrace + 1\leftrightarrow 2, \\
%%%%
\dot{\bar{E}}_{S_1S_2} &= \frac{\nu}{(M\mu)^2} \bigg\lbrace v^4 \left[\frac{289}{48} (\lNhat\cdot\bm{S}_1)(\lNhat\cdot\bm{S}_2)-\frac{103}{48}(\bm{S}_1\cdot\bm{S}_2)\right] \nonumber\\
&\quad
+ v^6 \left[
(\lNhat\cdot\bm{S}_1)(\lNhat\cdot\bm{S}_2) \left(-\frac{2023 \nu }{72}-\frac{5647}{168}\right)
+ (\bm{S}_1\cdot\bm{S}_2) \left(\frac{821 \nu }{72}+\frac{2123}{84}\right)
\right] \nonumber\\
&\quad
+ v^8 \bigg[
(\lNhat\cdot\bm{S}_1)(\lNhat\cdot\bm{S}_2)\left(\frac{2161 \nu ^2}{48}+\frac{60241 \nu }{252}+\frac{107771}{1512}\right) 
+ (\bm{S}_1\cdot\bm{S}_2) \left(-\frac{4405 \nu ^2}{144}-\frac{194687 \nu }{1008}-\frac{895429}{9072}\right)
\bigg]\nonumber\\
&\quad
+ \frac{63 \pi}{4} v^7(\lNhat\cdot\bm{S}_1)(\lNhat\cdot\bm{S}_2)\bigg\rbrace,\\
%%%%
\dot{\bar{E}}_{S^2} &= \frac{v^4}{(M\mu)^2} \left[(\lNhat\cdot\bm{S}_1)^2 \left(\frac{287 X_2}{96}-\frac{287 \nu }{96}\right)+S_1^2 \left(\frac{89 \nu }{96}-\frac{89 X_2}{96}\right)\right] \nonumber\\
&\quad
+ \frac{v^6}{(M\mu)^2} \left\lbrace
(\lNhat\cdot\bm{S}_1)^2 \left[\frac{2621 \nu ^2}{144}+\frac{1255 \nu }{56}-\left(\frac{461 \nu }{72}+\frac{1255}{56}\right) X_2\right]
+S_1^2 \left[\left(\frac{185 \nu }{72}+\frac{801}{56}\right) X_2-\frac{727 \nu ^2}{144}-\frac{801 \nu }{56}\right]\right\rbrace \nonumber\\
&\quad
+ \frac{v^8}{(M\mu)^2} \bigg\lbrace
(\lNhat\cdot\bm{S}_1)^2 \left[-\frac{5615 \nu ^3}{96}-\frac{62031 \nu ^2}{448}-\frac{250813 \nu }{6048}+\left(-\frac{11903 \nu ^2}{288}+\frac{202963 \nu }{1344}+\frac{250813}{6048}\right) X_2\right] \nonumber\\
&\quad\qquad
+S_1^2 \left[\frac{3371 \nu ^3}{288}+\frac{406253 \nu ^2}{4032}+\frac{963901 \nu }{18144}+\left(\frac{439 \nu ^2}{96}-\frac{389723 \nu }{4032}-\frac{963901}{18144}\right) X_2\right]
\bigg\rbrace \nonumber\\
&\quad
+ \frac{\pi v^7}{(M\mu)^2} (\lNhat\cdot\bm{S}_1)^2 \left(\frac{65 X_2}{8}-\frac{65 \nu }{8}\right) + 1\leftrightarrow 2, \\
%%%%
\dot{\bar{E}}_{S^2\tilde{C}} &= \frac{\widetilde{C}_{\rm 1ES^2}}{(M\mu)^2} \bigg\lbrace
v^4 \left[(\lNhat\cdot\bm{S}_1)^2 \left(3 X_2-3 \nu \right)+S_1^2 \left(\nu -X_2\right)\right] \nonumber\\
&\qquad
+ v^6 \left[(\lNhat\cdot\bm{S}_1)^2 \left(\frac{129 \nu ^2}{8}+\frac{837 \nu }{112}+\left(-\frac{135 \nu }{16}-\frac{837}{112}\right) X_2\right)+S_1^2 \left(-\frac{43 \nu ^2}{8}-\frac{279 \nu }{112}+\left(\frac{45 \nu }{16}+\frac{279}{112}\right) X_2\right)\right]\nonumber\\
&\qquad
+ v^8 \bigg[
(\lNhat\cdot\bm{S}_1)^2 \left(-\frac{81 \nu ^3}{2}-\frac{41191 \nu ^2}{672}+\frac{74911 \nu }{3024}+\left(-\frac{209 \nu ^2}{48}+\frac{46801 \nu }{672}-\frac{74911}{3024}\right) X_2\right) \nonumber\\
&\qquad\qquad
+ S_1^2 \left(\frac{27 \nu ^3}{2}+\frac{41191 \nu ^2}{2016}-\frac{74911 \nu }{9072}+\left(\frac{209 \nu ^2}{144}-\frac{46801 \nu }{2016}+\frac{74911}{9072}\right) X_2\right)
\bigg]\nonumber\\
&\qquad
+ \pi  v^7 (\lNhat\cdot\bm{S}_1)^2 \left(8 X_2-8 \nu \right)
\bigg\rbrace + 1\leftrightarrow 2,
\end{align}
\end{subequations}
where the SS tail part ($\Order(v^7)$ beyond the LO) is only known for aligned spins, so we expressed it in terms of $\lNhat\cdot\bm{S}_{\mr i}$ as an approximation for the precessing case, which would also depend on $S_1^2$ and $\bm{S}_1\cdot\bm{S}_2$.
 
Inserting $E$ and $\dot{E}$ in Eq.~\eqref{defdvdt} and PN expanding yields
\begin{subequations}
\label{vdot}
\begin{align}
\dot{v} &\equiv \frac{32 \nu  v^9}{5M} \left(\dot{\bar{v}}_{S^0} + \dot{\bar{v}}_\text{SO} + \dot{\bar{v}}_{S_1S_2} + \dot{\bar{v}}_{S^2} + \dot{\bar{v}}_{S^2\tilde{C}}\right),\\
%%%%
\label{vdotS0}
\dot{\bar{v}}_{S^0} &= 1 + \left(-\frac{11 \nu }{4}-\frac{743}{336}\right) v^2
+4 \pi  v^3
+ \left(\frac{59 \nu ^2}{18}+\frac{13661 \nu }{2016}+\frac{34103}{18144}\right) v^4
+ \pi  \left(-\frac{189 \nu }{8}-\frac{4159}{672}\right) v^5 \nonumber\\
&\quad
+ v^6 \left[\frac{541 \nu ^2}{896}-\frac{5605 \nu ^3}{2592}-\frac{56198689 \nu }{217728}+\pi ^2 \left(\frac{451 \nu }{48}+\frac{16}{3}\right)-\frac{1712 \ln v}{105}-\frac{1712 \gamma_E }{105}+\frac{16447322263}{139708800}-\frac{3424 \ln 2}{105}\right] \nonumber\\
&\quad
+ \pi  \left(\frac{91495 \nu ^2}{1512}+\frac{358675 \nu }{6048}-\frac{4415}{4032}\right) v^7,\\
%%%%
\dot{\bar{v}}_\text{SO} &= \frac{\lNhat\cdot\bm{S}_1}{M\mu} \bigg\lbrace
v^3 \left(-\frac{19 \nu }{6}-\frac{25 X_2}{4}\right)
+ v^5 \left[\frac{79 \nu ^2}{6}-\frac{21611 \nu }{1008}+\left(\frac{281 \nu }{8}-\frac{809}{84}\right) X_2\right]
+ \pi  v^6 \left(-\frac{37 \nu }{3}-\frac{151 X_2}{6}\right) \nonumber\\
&\quad
+v^7 \left[-\frac{10819 \nu ^3}{432}+\frac{40289 \nu ^2}{288}-\frac{1932041 \nu }{18144}+\left(-\frac{2903 \nu ^2}{32}+\frac{257023 \nu }{1008}-\frac{1195759}{18144}\right) X_2\right] \nonumber\\
&\quad
+ \pi  v^8 \left[\frac{34303 \nu ^2}{336}-\frac{46957 \nu }{504}+\left(\frac{50483 \nu }{224}-\frac{1665}{28}\right) X_2\right]
\bigg\rbrace + 1\leftrightarrow 2, \\
%%%%
\dot{\bar{v}}_{S_1S_2} &= \frac{\nu}{(M\mu)^2} \bigg\lbrace
v^4 \left[\frac{721 (\lNhat\cdot\bm{S}_1)(\lNhat\cdot\bm{S}_2)}{48}-\frac{247 (\bm{S}_1\cdot\bm{S}_2)}{48}\right]\nonumber\\
&\quad
+ v^6 \left[(\lNhat\cdot\bm{S}_1)(\lNhat\cdot\bm{S}_2) \left(\frac{14433}{224}-\frac{11779 \nu }{288}\right)+\left(\frac{6373 \nu }{288}+\frac{16255}{672}\right) (\bm{S}_1\cdot\bm{S}_2)\right] \nonumber\\
&\quad
+ \pi  v^7 \left[\frac{207 (\lNhat\cdot\bm{S}_1)(\lNhat\cdot\bm{S}_2)}{4}-12 (\bm{S}_1\cdot\bm{S}_2)\right]
+ v^8 \bigg[
\left(-\frac{162541 \nu ^2}{3456}-\frac{195697 \nu }{896}-\frac{9355721}{72576}\right) (\bm{S}_1\cdot\bm{S}_2) \nonumber\\
&\quad\qquad
+ (\lNhat\cdot\bm{S}_1)(\lNhat\cdot\bm{S}_2) \left(\frac{33163 \nu ^2}{3456}-\frac{10150387 \nu }{24192}+\frac{21001565}{24192}\right)
\bigg]\bigg\rbrace, \\
%%%%
\dot{\bar{v}}_{S^2} &= 
\frac{v^4}{(M\mu)^2} \left[(\lNhat\cdot\bm{S}_1)^2 \left(\frac{719 X_2}{96}-\frac{719 \nu }{96}\right)+S_1^2 \left(\frac{233 \nu }{96}-\frac{233 X_2}{96}\right)\right] \nonumber\\
&\quad
+ \frac{v^6}{(M\mu)^2} \bigg\lbrace
(\lNhat\cdot\bm{S}_1)^2 \left[\frac{25373 \nu ^2}{576}+\frac{2185 \nu }{448}+\left(\frac{19423 \nu }{576}-\frac{2185}{448}\right) X_2\right] \nonumber\\
&\quad\qquad
+S_1^2 \left[-\frac{6011 \nu ^2}{576}-\frac{8503 \nu }{448}+\left(\frac{8503}{448}-\frac{1177 \nu }{576}\right) X_2\right]
\bigg\rbrace \nonumber\\
&\quad
+\frac{\pi  v^7}{(M\mu)^2} \left[(\lNhat\cdot\bm{S}_1)^2 \left(\frac{209 X_2}{8}-\frac{209 \nu }{8}\right)+S_1^2 \left(6 \nu -6 X_2\right)\right]\nonumber\\
&\quad
+ \frac{v^8}{(M\mu)^2}\bigg\lbrace
(\lNhat\cdot\bm{S}_1)^2 \left[\left(\frac{11888267}{48384}-\frac{2392243 \nu ^2}{6912}+\frac{4063301 \nu }{16128}\right) X_2-\frac{869429 \nu ^3}{6912}+\frac{14283281 \nu ^2}{48384}-\frac{11888267 \nu }{48384}\right] \nonumber\\
&\quad\qquad
+S_1^2 \left[\frac{138323 \nu ^3}{6912}+\frac{711521 \nu ^2}{5376}+\frac{8207303 \nu }{145152}+\left(\frac{250693 \nu ^2}{6912}-\frac{812353 \nu }{5376}-\frac{8207303}{145152}\right) X_2\right]
\bigg\rbrace + 1\leftrightarrow 2, \\
%%%%
\dot{\bar{v}}_{S^2\tilde{C}} &= \frac{\widetilde{C}_{\rm 1ES^2}}{(M\mu)^2} \bigg\lbrace
v^4 \left[(\lNhat\cdot\bm{S}_1)^2 \left(\frac{15 X_2}{2}-\frac{15 \nu }{2}\right)+S_1^2 \left(\frac{5 \nu }{2}-\frac{5 X_2}{2}\right)\right] \nonumber\\
&\qquad
+ v^6 \left[(\lNhat\cdot\bm{S}_1)^2 \left(\frac{129 \nu ^2}{4}-\frac{1977 \nu }{224}+\left(\frac{1977}{224}-\frac{73 \nu }{16}\right) X_2\right)+S_1^2 \left(-\frac{43 \nu ^2}{4}+\frac{659 \nu }{224}+\left(\frac{73 \nu }{48}-\frac{659}{224}\right) X_2\right)\right]\nonumber\\
&\qquad
+ v^8 \bigg[
(\lNhat\cdot\bm{S}_1)^2 \left(-\frac{1567 \nu ^3}{24}+\frac{29329 \nu ^2}{224}-\frac{597271 \nu }{6048}+\left(-\frac{5675 \nu ^2}{96}-\frac{1517 \nu }{168}+\frac{597271}{6048}\right) X_2\right) \nonumber\\
&\qquad\qquad
+ S_1^2 \left(\frac{1567 \nu ^3}{72}-\frac{29329 \nu ^2}{672}+\frac{597271 \nu }{18144}+\left(\frac{5675 \nu ^2}{288}+\frac{1517 \nu }{504}-\frac{597271}{18144}\right) X_2\right)
\bigg] \nonumber\\
&\qquad
+ \pi  v^7 \left[(\lNhat\cdot\bm{S}_1)^2 \left(26 X_2-26 \nu \right)+S_1^2 \left(6 \nu -6 X_2\right)\right]
\bigg\rbrace + 1\leftrightarrow 2.
\end{align}
\end{subequations}
\end{widetext}
The SO and LO SS parts of $\dot{v}$ agree with, e.g., Eq.~(A1) of Ref.~\cite{Chatziioannou:2013dza}.

\subsection{Evolution of the angular momentum vector}
\label{sec:lNdot}
To obtain the PN expansion for $\dotlNhat$, we start from the equation for the total angular momentum $\bm{J} = \bm{L} + \bm{S}_1 + \bm{S}_2$.
We first neglect RR, and in the following subsection compute the RR contribution. 
Setting $\dot{\bm{J}} = 0$ yields 
\begin{equation}
\label{Jdot0}
\dot{\bm{L}} + \dot{\bm{S}}_1 + \dot{\bm{S}}_2 = 0,
\end{equation}
where $\dot{\bm{S}}_{\mr i}$ is given by Eq.~\eqref{S1dotAvg}, while $\dot{\bm{L}}$ can be computed by taking the time derivative of Eq.~\eqref{LvslNAvg}.

Solving Eq.~\eqref{Jdot0} for $\dotlNhat$ yields\footnote{
To solve Eq.~\eqref{Jdot0}, we split $\dotlNhat$ and $\dot{\bm{S}}_{\mr i}$ into SO and SS contributions, such that $\dotlNhat \equiv \dotlNhat^\text{SO} + \dotlNhat^\text{SS}$ and $\dot{\bm{S}}_{\mr i} \equiv \dot{\bm{S}}_{\mr i}^\text{SO} + \dot{\bm{S}}_{\mr i}^\text{SS}$, then solve order by order in spin for $\dotlNhat^\text{SO}$ and $\dotlNhat^\text{SS}$.
When performing this calculation, several simplifications can be done:
$\dot{\bm{S}}_{\mr i}$ is perpendicular to $\bm{S}_{\mr i}$, leading to $\bm{S}_1\cdot\dot{\bm{S}}_1 = 0 = \bm{S}_2\cdot\dot{\bm{S}}_2$, and since $\dot{\bm{S}}_{\mr i}^\text{SO}$ is perpendicular to $\lNhat$, we get $\lNhat\cdot\dot{\bm{S}}_{\mr i}^\text{SO} = 0$.
}
\begin{widetext}
\begin{subequations}
\begin{align}
\dotlNhat &\equiv \dotlNhat^\text{SO} + \dotlNhat^{S_1S_2} + \dotlNhat^{S^2} + \dotlNhat^{S^2\tilde{C}},\\
\dotlNhat^\text{SO} &= \frac{\lNhat\cross\bm{S}_1}{M^2\mu} \bigg\lbrace
v^6 \left(-\frac{\nu }{2}-\frac{3 X_2}{2}\right)
+ v^8 \left[\frac{\nu ^2}{4}-\frac{9 \nu }{4}+\left(\frac{9 \nu }{4}+\frac{9}{4}\right) X_2\right] \nonumber\\
&\quad\qquad
+ v^{10} \left[-\frac{\nu ^3}{48}+\frac{81 \nu ^2}{16}-\frac{27 \nu }{16}+\left(-\frac{21 \nu ^2}{16}+\frac{63 \nu }{16}+\frac{27}{16}\right) X_2\right]
\bigg\rbrace + 1 \leftrightarrow 2,  \\
%%%%
\dotlNhat^{S_1S_2} &= \frac{\nu}{M^3\mu^2} \bigg\lbrace 
\frac{3}{2} v^7 \left[(\lNhat\cross\bm{S}_1)(\lNhat\cdot\bm{S}_2) + (\lNhat\cross\bm{S}_2)(\lNhat\cdot\bm{S}_1)\right] \nonumber\\
&\qquad
+ v^9 \bigg[
(\lNhat\cross\bm{S}_1)(\lNhat\cdot\bm{S}_2) \left(-\frac{5 \nu }{4}-\frac{15 X_2}{8}-\frac{21}{4}\right)
+ (\lNhat\cross\bm{S}_2)(\lNhat\cdot\bm{S}_1) \left(-\frac{5 \nu }{4}+\frac{15 X_2}{8}-\frac{57}{8}\right) \nonumber\\
&\qquad\qquad
- \frac{5}{8} \delta \lNhat (\lNhat\cdot\bm{S}_1\cross\bm{S}_2)
+ \frac{3}{8} \delta (\bm{S}_1\cross\bm{S}_2)
\bigg]\nonumber\\
&\qquad
+ v^{11} \bigg\lbrace
(\lNhat\cross\bm{S}_1)(\lNhat\cdot\bm{S}_2)\left[-\frac{\nu ^2}{6}+\frac{25 \nu }{4}+\left(\frac{71 \nu }{32}+\frac{9}{32}\right) X_2+\frac{15}{16}\right]
\nonumber\\
&\qquad\qquad
+ (\lNhat\cross\bm{S}_2)(\lNhat\cdot\bm{S}_1) \left[-\frac{\nu ^2}{6}+\frac{271 \nu }{32}+\left(-\frac{71 \nu }{32}-\frac{9}{32}\right) X_2+\frac{39}{32}\right] \nonumber\\
&\qquad\qquad
+ \frac{\delta}{96} (89 \nu -27)\lNhat (\lNhat\cdot\bm{S}_1\cross\bm{S}_2) 
-\frac{9}{32} \delta (2 \nu +1) (\bm{S}_1\cross\bm{S}_2) 
\bigg\rbrace
\bigg\rbrace,\\
%%%%
\dotlNhat^{S^2} &= \frac{(\lNhat\cross\bm{S}_1)(\lNhat\cdot\bm{S}_1)}{M^3\mu^2}
\bigg\lbrace
v^7 \left(\frac{3 X_2}{2}-\frac{3 \nu }{2}\right)
+ v^9 \left[2 \nu ^2+9 \nu +(3 \nu -9) X_2\right] \nonumber\\
&\quad\qquad
+ v^{11} \left[-\frac{23 \nu ^3}{16}-\frac{157 \nu ^2}{16}-\frac{93 \nu }{16}+\left(-\frac{439 \nu ^2}{48}+4 \nu +\frac{93}{16}\right) X_2\right]
\bigg\rbrace + 1 \leftrightarrow 2, \\
%%%%
\dotlNhat^{S^2\tilde{C}} &= \frac{\widetilde{C}_{\rm 1ES^2}}{M^3\mu^2}(\lNhat\cross\bm{S}_1)(\lNhat\cdot\bm{S}_1) \bigg\lbrace
v^7 \left(\frac{3 X_2}{2}-\frac{3 \nu }{2}\right)
+ v^9 \left[\frac{11 \nu ^2}{8}+\frac{9 \nu }{8}+\left(\frac{11 \nu }{4}-\frac{9}{8}\right) X_2\right] \nonumber\\
&\quad\qquad
+ v^{11} \left[-\frac{43 \nu ^3}{96}+\frac{1077 \nu ^2}{224}+\frac{27 \nu }{32}+\left(-\frac{479 \nu ^2}{96}+\frac{3 \nu }{2}-\frac{27}{32}\right) X_2\right]
\bigg\rbrace+ 1 \leftrightarrow 2,
\end{align}
\end{subequations}
\end{widetext}
which agrees up to NLO SO (i.e. to $\Order(v^8)$) with Eq.~(4c) of Ref.~\cite{Akcay:2020qrj} provided that one uses the coefficients of $\bm{L}(\lNhat)$ from Eq.~\eqref{LvslNAvg}, instead of those in Ref.~\cite{Akcay:2020qrj} because of the different SSC.
Note that $\dotlNhat$ has a component parallel to $\lNhat$, which enters at NLO and NNLO S$_1$S$_2$, and is given by
\begin{align}
\dotlNhat\cdot\lNhat &= \frac{\nu \lNhat\cdot\bm{S}_1\cross\bm{S}_2}{M^3\mu^2} 
\left[-\frac{\delta}{4} v^9 
+\frac{\delta}{96} (35 \nu -54) v^{11}\right].
\end{align}

\subsection{Radiation-reaction contribution to $\dotlNhat$}
When computing $\dotlNhat$, RR enters through $\dot{v}$, which is given by Eq.~\eqref{vdot}, and from the nonzero $\dot{\bm{J}}$, which is given by
\begin{align}
\label{JdotVec}
\dot{\bm{J}} &= \dot{\bm{L}} + \dot{\bm{S}}_1 + \dot{\bm{S}}_2 \nonumber\\
&=\dot{\bm{r}} \cross \bm{p} + \bm{r} \cross \dot{\bm{p}} + \dot{\bm{S}}_1 + \dot{\bm{S}}_2\nonumber\\
&= \bm{r}\cross\bm{\mathcal{F}} + \dot{\bm{S}}_1^\text{RR} + \dot{\bm{S}}_2^\text{RR},
\end{align}
where we used the EOMs~\eqref{EOMsPrecess} to relate $\dot{\bm{J}}$ to the RR force $\bm{\mathcal{F}}$.
Since we are working to NNLO SS (i.e. to $\Order(v^{10})$) in $\dot{\bm{L}}$ and $\dot{\bm{S}}_{\mr i}$, we only need $\dot{\bm{J}}$ to $\Order(v^3)$ beyond its LO, which is $\Order(v^7)$, and we can neglect the RR contribution to $\dot{\bm{S}}_{\mr i}$ because it starts at $\Order(v^{11}S^2)$~\cite{Will:2005sn,Maia:2017yok}.

The RR force $\bm{\mathcal{F}}$ for circular orbits in the \texttt{SEOBNR} waveform models is chosen to be in a gauge such that~\cite{Buonanno:2005xu,Ossokine:2020kjp}\footnote{
This relation was derived in Ref.~\cite{Buonanno:2005xu} for precessing spins, and is given at LO SO by Eq.~(3.27) there, which includes an extra term depending on $(\bm{p}\cdot\bm{S}_{\mr i}) \bm{L}$ that averages to zero over an orbit.}
\begin{equation}
\bm{\mathcal{F}} = \frac{\dot{E}}{\Omega L} \bm{p}.
\end{equation} 
Using the energy loss from Eq.~\eqref{dEdt} and expanding to LO SO for circular orbits, we get
\begin{align}
\bm{\mathcal{F}} &\simeq - \frac{32}{5} \nu^2 v^9 \bm{\lambda} \bigg\lbrace
1 + v^2 \left(-\frac{13 \nu }{4}-\frac{1247}{336}\right)
+ 4\pi v^3 \nonumber\\
&\quad
+ \frac{v^3}{M\mu} \left[\lhat\cdot\bm{S}_1 \left(-\frac{4 \nu }{3}-\frac{3 X_2}{4}\right) + 1\leftrightarrow 2\right]
\bigg\rbrace,
\end{align}
where we did not write the $\bm{n}$ component of $\bm{\mathcal{F}}$ since it does not contribute to $\dot{\bm{J}}$ and is proportional to $p_r$.
Then, from Eq.~\eqref{JdotVec}, and using Eq.~\eqref{lvslN} to replace $\lhat=\bm{n}\cross\bm{\lambda}$ by $\lNhat$, we obtain
\begin{align}
\dot{\bm{J}} &= -\frac{32}{5}M\nu^2 v^7 \bigg\lbrace
\lNhat \left[1 + v^2 \left(-\frac{35 \nu }{12}-\frac{1247}{336}\right) + 4\pi v^3\right] \nonumber\\
&\qquad
+ \frac{v^3}{M\mu}\bigg[
\lNhat (\lNhat\cdot\bm{S}_1) \left(-\frac{5}{4} \nu-\frac{1}{2} X_2\right) \nonumber\\
&\qquad\qquad
+ \bm{S}_1 \left(-\frac{\nu }{4}-\frac{3 X_2}{4}\right)
+ 1\leftrightarrow 2
\bigg]
\bigg\rbrace.
\end{align}

Following similar steps as in the previous subsection, except for including $\dot{\bm{J}}$ and $\dot{v}$, we obtain the following RR contribution to $\dotlNhat$:
\begin{align}
\dotlNhat^\text{RR} &= -\frac{64}{5} \frac{v^8}{M} \bigg\lbrace
\nu \lNhat \left[1 + v^2 \left(-\frac{37 \nu }{12}-\frac{1751}{336}\right) + 4\pi v^3\right] \nonumber\\
&\quad
+ \frac{v^3}{M\mu} \bigg[
\lNhat (\lNhat\cdot\bm{S}_1) \left(\frac{9 \nu ^2}{8}-\frac{19 \nu }{12}+5 X_2\nu -\frac{25}{8}X_2 \right) \nonumber\\
&\quad\qquad
+ \bm{S}_1 \left(\frac{\nu ^2}{8}+\frac{3 \nu X_2}{8}\right) 
+ 1\leftrightarrow 2
\bigg]
\bigg\rbrace.
\end{align}
We do not include this RR contribution in the \texttt{SEOBNRv5PHM} waveform model~\cite{RamosBuadesv5}, but we checked that it has a negligible effect on the dynamics.

\section{Conclusions}
\label{sec:conclusions}

In this paper, we derived an aligned-spin Hamiltonian (Sec.~\ref{sec:HamAlign}), which is used in the \texttt{SEOBNRv5HM} waveform model~\cite{Pompiliv5}, and a full precessing-spin Hamiltonian (Sec.~\ref{sec:HamPrec}) that reduces in the test-mass limit to the exact Kerr Hamiltonian for generic orbits. 
The Hamiltonians include the nonspinning part at 4PN order, with partial 5PN and 5.5PN results, in addition to the full 4PN spin information (NNLO SO, NNLO SS, LO S$^3$, LO S$^4$).
The full 5PN spin contributions (NNNLO SO and SS, NLO S$^3$ and S$^4$) to the conservative dynamics are known from the recent work in Refs.~\cite{Antonelli:2020aeb,Antonelli:2020ybz,Mandal:2022nty,Kim:2022pou,Mandal:2022ufb,Kim:2022bwv,Levi:2022dqm,Levi:2022rrq,Levi:2019kgk,Levi:2020lfn}, but we leave their inclusion in the Hamiltonian for future work.
Our results include the spin-multipole constants, and are thus valid for NSs, though one also needs to include dynamical tidal effects, which can be included as was done in \texttt{SEOBNRv4T}~\cite{Steinhoff:2021dsn}.

Furthermore, we derived (in Sec.~\ref{sec:PNEOMs}) a simpler
precessing-spin Hamiltonian, $H_{\rm EOB}^{\rm pprec}$, and PN-expanded EOMs, which orbit average
the in-plane spin components, and are used in the computationally efficient \texttt{SEOBNRv5PHM}
waveform model~\cite{RamosBuadesv5}.  We included in the EOMs the NNLO
SO and SS contributions in a gauge consistent with our EOB Hamiltonian
and the NW (canonical) SSC.  Extending the precessing-spin EOMs to include LO S$^3$ and LO
S$^4$ is straightforward, but the equations become lengthy, and would
likely have a smaller effect than the error introduced due to orbit
averaging the SS contributions.  It would still be interesting to
compute those higher-order spin contributions and quantify their 
effect on the dynamics.  It is also important to extend the RR force
and waveform modes for precessing spins beyond the LO SO and SS
contributions derived in Refs.~\cite{Arun:2008kb,Boyle:2014ioa}.

The results obtained in this paper have contributed to improving the accuracy of \texttt{SEOBNRv5} waveform models, as detailed in Refs.~\cite{Pompiliv5,RamosBuadesv5}.
For example, Ref.~\cite{RamosBuadesv5} demonstrated that using the partially-precessing Hamiltonian $H_{\rm EOB}^{\rm pprec}$, and comparing the waveforms to a set of highly-precessing NR simulations, led to 100\% (86.4\%) of cases with a maximum unfaithfulness below 3\% (1\%), while using the aligned-spin Hamiltonian $H_{\rm EOB}^{\rm align}$ led to 95.8\% (75.4\%) of cases below 3\% (1\%).
Furthermore, we generally find that \texttt{SEOBNRv5} waveform models provide noticeable improvements in accuracy compared to the previous version of the model, \texttt{SEOBNRv4}, and to other \texttt{IMRPhenom} and \texttt{TEOBResumS} models (see for example Fig.~9 of Ref.~\cite{Pompiliv5} and Fig.~4 of Ref.~\cite{RamosBuadesv5}).
These results highlight the importance of including and resumming analytical PN information in waveform models.

\begin{acknowledgments}
We thank Jan Steinhoff,  Maarten van de Meent and Justin Vines for useful discussions. 
We are also grateful to Jan Steinhoff for providing us with a \texttt{Mathematica} file for performing canonical transformations and matching Hamiltonians.
This work made use of the tensor algebra package \texttt{xAct}~\cite{xAct,martin2008xperm} in \texttt{Mathematica}. 

The \texttt{SEOBNRv5} waveform models are publicly available through the python package \texttt{pySEOBNR} \href{https://git.ligo.org/waveforms/software/pyseobnr}{\texttt{git.ligo.org/waveforms/software/pyseobnr}}. Stable versions of \texttt{pySEOBNR} are published through the Python Package Index (PyPI), and can be installed via ~\texttt{pip install pyseobnr}.

Part of M.K.'s work on this paper is supported by Perimeter Institute for Theoretical Physics.
Research at Perimeter Institute is supported in part by the Government of Canada through the Department of Innovation, Science and Economic Development and by the Province of Ontario through the Ministry of Colleges and Universities.\\
\end{acknowledgments}

\appendix
\begin{widetext}
\section{Angular momentum and separation for circular orbits}
\label{app:rLCirclN}
In this Appendix, we write $r$ and $L$ for circular orbits and precessing spins, which are obtained by solving Eqs.~\eqref{circConds} and replacing $\lhat$ by $\lhat(\lNhat)$ from Eq.~\eqref{lvslN}.

For $L(\lNhat,\bm{n},\lamNhat,\bm{S}_{\mr i},v)$, we get
\begin{subequations}
\label{LCirclN}
\begin{align}
L &\equiv \frac{M\mu}{v}\left(\bar{L}_{S^0} + \bar{L}_\text{SO} + \bar{L}_{S_1S_2} + \bar{L}_{S^2} + \bar{L}_{S^2\tilde{C}}\right), \\
%%%%
\label{LCirclNS0}
\bar{L}_{S^0} & = 1
+ v^2 \left(\frac{\nu }{6}+\frac{3}{2}\right)
+ v^4 \left(\frac{\nu ^2}{24}-\frac{19 \nu }{8}+\frac{27}{8}\right)
+ v^6 \left[\frac{7 \nu ^3}{1296}+\frac{31 \nu ^2}{24}+\left(\frac{41 \pi ^2}{24}-\frac{6889}{144}\right) \nu +\frac{135}{16}\right]  \nonumber\\
&\quad
+ v^8 \bigg[\frac{2835}{128}
+\nu  \left(\frac{98869}{5760}-\frac{128 \gamma_E}{3}-\frac{6455 \pi ^2}{1536}-\frac{256 \ln 2}{3}  -\frac{128 \ln v}{3}\right)
+\left(\frac{356035}{3456}-\frac{2255 \pi ^2}{576}\right) \nu ^2 \nonumber\\
&\quad\qquad
-\frac{215 \nu ^3}{1728}-\frac{55 \nu ^4}{31104}\bigg],\\
%%%%
\bar{L}_\text{SO} &= \frac{\lNhat\cdot \bm{S}_1}{M\mu}
\bigg\lbrace
v^3 \left(-\frac{5 \nu }{6}-\frac{5 X_2}{2}\right) 
+ v^5 \left[\frac{7 \nu ^2}{72}-\frac{35 \nu }{8}+\left(\frac{35 \nu }{12}-\frac{21}{8}\right) X_2\right] \nonumber\\
&\qquad
+ v^7 \left[\frac{\nu ^3}{16}+\frac{165 \nu ^2}{16}-\frac{243 \nu }{16}+\left(-\frac{15 \nu ^2}{16}+\frac{117 \nu }{4}-\frac{81}{16}\right) X_2\right]
\bigg\rbrace + 1\leftrightarrow 2, \\
%%%%
\bar{L}_{S_1S_2} &= \frac{\nu}{M^2\mu^2} \bigg\lbrace
2 v^4 \left[(\lNhat\cdot\bm{S}_1)(\lNhat\cdot\bm{S}_2) - (\bm{n}\cdot\bm{S}_1)(\bm{n}\cdot\bm{S}_2)\right] \nonumber\\
&\qquad
+ v^6 \bigg[
(\bm{n}\cdot\bm{S}_1)(\bm{n}\cdot\bm{S}_2) \left(\frac{16}{3}-\frac{2 \nu }{3}\right)
+ (\lNhat\cdot\bm{S}_1)(\lNhat\cdot\bm{S}_2) \left(\frac{4 \nu }{9}+\frac{4}{3}\right)
+ (\lamNhat\cdot\bm{S}_1)(\lamNhat\cdot\bm{S}_2) \left(\frac{9 \nu }{4}-\frac{7}{3}\right)
\bigg]\nonumber\\
&\qquad
+ v^8 \bigg[
(\bm{n}\cdot\bm{S}_1)(\bm{n}\cdot\bm{S}_2) \left(-\frac{205 \nu ^2}{72}-\frac{5315 \nu }{144}+\frac{15}{4}\right)
+ (\lNhat\cdot\bm{S}_1)(\lNhat\cdot\bm{S}_2) \left(-\frac{265 \nu ^2}{108}-\frac{715 \nu }{36}+\frac{25}{4}\right) \nonumber\\
&\quad\qquad
+ (\lamNhat\cdot\bm{S}_1)(\lamNhat\cdot\bm{S}_2) \left(\frac{21 \nu ^2}{8}+\frac{235 \nu }{18}-4\right)
\bigg]
\bigg\rbrace,\\
%%%%
\bar{L}_{S^2}  &= \frac{v^4 }{M^2\mu^2}(\nu -X_2) \left[(\bm{n}\cdot\bm{S}_1)^2 - (\lNhat\cdot\bm{S}_1)^2\right]
+ \frac{v^6}{M^2\mu^2} \bigg\lbrace
(\bm{n}\cdot\bm{S}_1)^2 \left[-\frac{5 \nu ^2}{3}-\frac{23 \nu }{3}+\left(\frac{23}{3}-8 \nu \right) X_2\right] \nonumber\\
&\quad\qquad
+ (\lNhat\cdot\bm{S}_1)^2 \left[\frac{14 \nu ^2}{9}+2 \nu +\left(\frac{10 \nu }{3}-2\right) X_2\right]
+ (\lamNhat\cdot\bm{S}_1)^2 \left[\frac{9 \nu ^2}{8}+\frac{109 \nu }{24}+\left(\frac{65 \nu }{12}-\frac{109}{24}\right) X_2\right]\bigg\rbrace\nonumber\\
&\quad
+ \frac{v^8}{M^2\mu^2} \bigg\lbrace
(\lNhat\cdot\bm{S}_1)^2 \left[-\frac{295 \nu ^3}{216}+\frac{365 \nu ^2}{24}+\frac{15 \nu }{8}+\left(-\frac{815 \nu ^2}{72}+\frac{5 \nu }{8}-\frac{15}{8}\right) X_2\right]\nonumber\\
&\quad\qquad
+ (\bm{n}\cdot\bm{S}_1)^2 \left[-\frac{185 \nu ^3}{144}+\frac{2275 \nu ^2}{144}-\frac{95 \nu }{16}+\left(\frac{175 \nu ^2}{48}-\frac{985 \nu }{36}+\frac{95}{16}\right) X_2\right] \nonumber\\
&\quad\qquad
+(\lamNhat\cdot\bm{S}_1)^2 \left[\frac{21 \nu ^3}{16}-\frac{55 \nu ^2}{36}+\frac{65 \nu }{16}+\left(-\frac{13 \nu ^2}{36}+\frac{1237 \nu }{144}-\frac{65}{16}\right) X_2\right]
\bigg\rbrace + 1\leftrightarrow 2,\\
%%%%
\bar{L}_{S^2\tilde{C}}  &= \frac{\widetilde{C}_{\rm 1ES^2}}{M^2\mu^2} \bigg\lbrace
v^4 \left[(\lNhat\cdot\bm{S}_1)^2 \left(X_2-\nu \right)+(\bm{n}\cdot\bm{S}_1)^2 \left(2 \nu -2 X_2\right)+(\lamNhat\cdot\bm{S}_1)^2 \left(X_2-\nu \right)\right] \nonumber\\
&\qquad
+ v^6 \bigg[
(\lNhat\cdot\bm{S}_1)^2 \left(\frac{2 \nu ^2}{3}-2 \nu +(2 \nu +2) X_2\right)+(\bm{n}\cdot\bm{S}_1)^2 \left(-\frac{4 \nu ^2}{3}+4 \nu +(-4 \nu -4) X_2\right)\nonumber\\
&\qquad\qquad
+(\lamNhat\cdot\bm{S}_1)^2 \left(\frac{2 \nu ^2}{3}-2 \nu +(2 \nu +2) X_2\right)
\bigg] \nonumber\\
&\qquad
+ v^8 \bigg[
(\bm{n}\cdot\bm{S}_1)^2 \left(\frac{5 \nu ^3}{36}-\frac{1475 \nu ^2}{84}+\frac{45 \nu }{4}+\left(\frac{65 \nu ^2}{12}-\frac{55 \nu }{12}-\frac{45}{4}\right) X_2\right)\nonumber\\
&\qquad\qquad
+ (\lNhat\cdot\bm{S}_1)^2 \left(-\frac{5 \nu ^3}{72}+\frac{1475 \nu ^2}{168}-\frac{45 \nu }{8}+\left(-\frac{65 \nu ^2}{24}+\frac{55 \nu }{24}+\frac{45}{8}\right) X_2\right)\nonumber\\
&\qquad\qquad
+ (\lamNhat\cdot\bm{S}_1)^2 \left(-\frac{5 \nu ^3}{72}+\frac{1475 \nu ^2}{168}-\frac{45 \nu }{8}+\left(-\frac{65 \nu ^2}{24}+\frac{55 \nu }{24}+\frac{45}{8}\right) X_2\right)
\bigg]
\bigg\rbrace + 1\leftrightarrow 2,
\end{align}
\end{subequations}
which agrees for aligned spins with, e.g., Eq.~(8.24) of Ref.~\cite{Levi:2014sba} and Eq.~(5.2) of Ref.~\cite{Levi:2016ofk}.

For $r(\lNhat,\bm{n},\lamNhat,\bm{S}_{\mr i},v)$, we obtain
\begin{subequations}
\label{rCirclN}
\begin{align}
r &\equiv \frac{M}{v^2} \left(\bar{r}_{S^0} + \bar{r}_\text{SO} + \bar{r}_{S_1S_2} + \bar{r}_{S^2} + \bar{r}_{S^2\tilde{C}}\right), \\
\bar{r}_{S^0} &= 1 + v^2 \frac{\nu }{3}
+ v^4 \left(\frac{\nu ^2}{9}-\frac{5 \nu }{4}\right)
+ v^6 \left[\frac{2 \nu ^3}{81}+\frac{11 \nu ^2}{12}+\left(\frac{41 \pi ^2}{48}-\frac{1585}{72}\right) \nu\right] \nonumber\\
&\quad
+ v^8 \left[\left(\frac{544}{9}-\frac{451 \pi ^2}{192}\right) \nu ^2+\nu  \left(\frac{153211}{2880}-\frac{64 \gamma_E}{3}-\frac{11375 \pi ^2}{3072}-\frac{128}{3} \ln 2-\frac{64 \ln v}{3}\right)\right], \\
%%%%
\bar{r}_\text{SO} &= \frac{\lNhat\cdot \bm{S}_1}{M\mu} \bigg\lbrace
v^3 \left(-\frac{\nu }{6}-\frac{X_2}{2}\right)
+ v^5 \left[-\frac{19 \nu ^2}{48}+\frac{3 \nu }{16}+\left(\frac{\nu }{8}-\frac{3}{16}\right) X_2\right] \nonumber\\
&\qquad
+ v^7 \left[-\frac{47 \nu ^3}{3456}-\frac{61 \nu ^2}{192}-\frac{5 \nu }{128}+\left(\frac{907 \nu ^2}{1152}+\frac{139 \nu }{32}+\frac{5}{128}\right) X_2\right]
\bigg\rbrace + 1\leftrightarrow 2, \\
%%%%
\bar{r}_{S_1S_2} &= \frac{\nu}{M^2\mu^2}\bigg\lbrace
-2 v^4 (\bm{n}\cdot\bm{S}_1)(\bm{n}\cdot\bm{S}_2) \nonumber\\
&\quad
+ v^6 \bigg[
(\bm{n}\cdot\bm{S}_1)(\bm{n}\cdot\bm{S}_2) \left(\frac{85}{24}-\frac{2 \nu }{3}\right)
+ (\lNhat\cdot\bm{S}_1)(\lNhat\cdot\bm{S}_2) \left(\frac{5 \nu }{18}-\frac{8}{3}\right)
+ (\lamNhat\cdot\bm{S}_1)(\lamNhat\cdot\bm{S}_2) \left(\nu -\frac{8}{3}\right)
\bigg] \nonumber\\
&\quad
+ v^8 \bigg[
(\bm{n}\cdot\bm{S}_1)(\bm{n}\cdot\bm{S}_2) \left(-\frac{59 \nu ^2}{48}-\frac{1115 \nu }{96}-\frac{87}{32}\right)
+ (\lNhat\cdot\bm{S}_1)(\lNhat\cdot\bm{S}_2) \left(\frac{5 \nu ^2}{12}+\frac{55 \nu }{24}+\frac{3}{2}\right) \nonumber\\
&\quad\qquad
+ (\lamNhat\cdot\bm{S}_1)(\lamNhat\cdot\bm{S}_2) \left(\frac{35 \nu ^2}{24}+\frac{183 \nu }{16}+\frac{15}{4}\right)
\bigg]
\bigg\rbrace,\\
%%%%
\bar{r}_{S^2} &= \frac{v^4}{M^2\mu^2} (\nu -X_2) (\bm{n}\cdot\bm{S}_1)^2
+ \frac{v^6}{M^2\mu^2} \bigg\lbrace
(\bm{n}\cdot\bm{S}_1)^2 \left[\frac{\nu ^2}{12}-\frac{49 \nu }{12}+\left(\frac{49}{12}-\frac{25 \nu }{6}\right) X_2\right] \nonumber\\
&\quad\qquad
+ (\lNhat\cdot\bm{S}_1)^2 \left[\frac{29 \nu ^2}{36}+\frac{11 \nu }{4}+\left(\frac{11 \nu }{6}-\frac{11}{4}\right) X_2\right]
+ (\lamNhat\cdot\bm{S}_1)^2 \left[\frac{\nu ^2}{2}+\frac{17 \nu }{6}+\left(\frac{7 \nu }{3}-\frac{17}{6}\right) X_2\right]
\bigg\rbrace \nonumber\\
&\quad
+ \frac{v^8}{M^2\mu^2} \bigg\lbrace
(\bm{n}\cdot\bm{S}_1)^2 \left[-\frac{101 \nu ^3}{32}+\frac{585 \nu ^2}{32}-\frac{67 \nu }{32}+\left(-\frac{595 \nu ^2}{96}-\frac{163 \nu }{8}+\frac{67}{32}\right) X_2\right] \nonumber\\
&\quad\qquad
+ (\lNhat\cdot\bm{S}_1)^2 \left[\frac{5 \nu ^3}{24}-\frac{47 \nu ^2}{24}-3 \nu +\left(-\frac{13 \nu ^2}{8}-\frac{25 \nu }{24}+3\right) X_2\right]\nonumber\\
&\quad\qquad
+ (\lamNhat\cdot\bm{S}_1)^2 \left[\frac{35 \nu ^3}{48}-\frac{105 \nu ^2}{16}-\frac{75 \nu }{16}+\left(\frac{9 \nu ^2}{16}+\frac{15 \nu }{8}+\frac{75}{16}\right) X_2\right]
\bigg\rbrace + 1\leftrightarrow 2,\\
%%%%
\bar{r}_{S^2\tilde{C}} &= \frac{\widetilde{C}_{\rm 1ES^2}}{M^2\mu^2} \bigg\lbrace
v^4 \left[(\lNhat\cdot\bm{S}_1)^2 \left(\frac{X_2}{2}-\frac{\nu }{2}\right)+(\bm{n}\cdot\bm{S}_1)^2 \left(\nu -X_2\right)+(\lamNhat\cdot\bm{S}_1)^2 \left(\frac{X_2}{2}-\frac{\nu }{2}\right)\right] \nonumber\\
&\qquad
+ v^6 \bigg[
(\lNhat\cdot\bm{S}_1)^2 \left(\frac{\nu ^2}{2}+\frac{5 \nu  X_2}{6}\right)+(\bm{n}\cdot\bm{S}_1)^2 \left(\frac{\nu ^2}{2}-\frac{3 \nu }{2}+\left(\frac{3}{2}-\frac{19 \nu }{6}\right) X_2\right)+(\lamNhat\cdot\bm{S}_1)^2 \left(\frac{\nu ^2}{2}+\frac{5 \nu  X_2}{6}\right)
\bigg]\nonumber\\
&\qquad
+ v^8 \bigg[
(\lNhat\cdot\bm{S}_1)^2 \left(\frac{325 \nu ^2}{84}+\left(-2 \nu ^2-\frac{5 \nu }{6}\right) X_2\right)+(\lamNhat\cdot\bm{S}_1)^2 \left(\frac{325 \nu ^2}{84}+\left(-2 \nu ^2-\frac{5 \nu }{6}\right) X_2\right) \nonumber\\
&\qquad\qquad
+ (\bm{n}\cdot\bm{S}_1)^2 \left(-\frac{45 \nu ^3}{32}+\frac{2297 \nu ^2}{672}+\frac{69 \nu }{32}+\left(\frac{11 \nu ^2}{32}-\frac{43 \nu }{12}-\frac{69}{32}\right) X_2\right)
\bigg]
\bigg\rbrace + 1\leftrightarrow 2.
\end{align}
\end{subequations}
~\\
\end{widetext}
~\\

\section{SSC transformation for the angular momentum}
\label{app:SSC}
The transformation from the canonical NW SSC to the covariant SSC is given by the center-of-mass shift~\cite{Kidder:1995zr}
\begin{equation}
{\bm{\mr x}_{\mr i}}_\text{(\rm {NW})} \to {\bm{\mr x}_{\mr i}}_\text{(cov)} + \frac{1}{2 m_{\mr i}} {\bm{\mr v}_{\mr i}}_\text{(cov)}\times \bm{S}_{\mr i},
\end{equation}
where $\bm{\mr x}_{\mr i}$ and $\bm{\mr v}_{\mr i} \equiv \dot{\bm{\mr x}}_{\mr i}$ are the position and velocity vectors of each body, with $\mr i = 1,2$.

In the NW SSC, the vector ${\bm{L}_{\rm N}}_\text{(NW)} = \mu \bm{r}_\text{(NW)} \cross \bm{\mr v}_\text{(NW)}$, where $\bm{r} \equiv \bm{\mr x}_1-\bm{\mr x}_2$ and $\bm{\mr v} \equiv \bm{\mr v}_1 - \bm{\mr v}_2$ are the relative position and velocity, respectively.
Transforming to the covariant SSC leads to
\begin{align}
&{\bm{L}_{\rm N}}_\text{(NW)} = \mu \bm{r}_\text{(cov)} \cross \bm{\mr v}_\text{(cov)} \nonumber\\
&\quad
+ \frac{X_2^2}{2}\bigg[
\left(\frac{M}{r}+ \bm{\mr v}^2\right) \bm{S}_1- (\bm{\mr v}\cdot\bm{S}_1) \bm{\mr v} 
 - \frac{M}{r} (\bm{n}\cdot\bm{S}_1) \bm{n}
 \bigg] \nonumber\\
&\quad
+ \frac{X_1^2}{2}\bigg[
\left(\frac{M}{r}+ \bm{\mr v}^2\right) \bm{S}_2- (\bm{\mr v}\cdot\bm{S}_2) \bm{\mr v} 
 - \frac{M}{r} (\bm{n}\cdot\bm{S}_2) \bm{n}
 \bigg],
\end{align}
where all quantities on the right-hand side are in the covariant SSC, but we dropped the label in the SO part to simplify the notation.
Dividing ${\bm{L}_{\rm N}}_\text{(NW)}$ by its magnitude, using $\bm{r} = r \bm{n}$ and $\bm{\mr v} = r \Omega \lamNhat = r v^3 \lamNhat/M$ for circular orbits, and taking an orbit average yields
\begin{align}
\label{lNhatSSC}
{\lNhat}_\text{(NW)} &= {\lNhat}_\text{(cov)} + \frac{v^3}{2M\mu}
\Big\lbrace 
X_2^2\left[\bm{S}_1 - (\lNhat\cdot\bm{S}_1) \lNhat\right]  \nonumber\\
&\qquad\qquad
+ X_1^2 \left[\bm{S}_2 - (\lNhat\cdot\bm{S}_2) \lNhat \right]
\Big\rbrace.
\end{align}

Our result for $\bm{L}(\lNhat)$ in Eq.~\eqref{LvslNAvg} is in the NW SSC, while Eq.~(A5) of Ref.~\cite{Akcay:2020qrj} is in the covariant SSC; the difference up to LO SO is given by
\begin{align}
\label{LdiffSSC}
&\bm{L}^\text{Eq.\,\eqref{LvslNAvg}}_\text{(NW)} - \bm{L}^\text{Ref.\,\cite{Akcay:2020qrj}}_\text{(cov)} = \nonumber\\
&\qquad
-\frac{v^2}{2}
\Big\lbrace 
X_2^2\left[\bm{S}_1 - (\lNhat\cdot\bm{S}_1) \lNhat\right] \nonumber\\
&\qquad\qquad
+ X_1^2 \left[\bm{S}_2 - (\lNhat\cdot\bm{S}_2) \lNhat \right]
\Big\rbrace + \Order(v^3).
\end{align}
The leading PN order of $\bm{L}(\lNhat)$ in Eq.~\eqref{LvslNAvg} is $\mu M \lNhat/v$, and $v$ is invariant under an SSC transformation; hence, the difference is due to the SSC transformation of $\lNhat$. 
Indeed, we see from Eqs.~\eqref{LdiffSSC} and ~\eqref{lNhatSSC} that accounting for that transformation cancels the difference between our result and that of Ref.~\cite{Akcay:2020qrj}.

\section{Equations of motion in terms of $l$}
\label{app:LEOMs}
In Sec.~\ref{sec:PNEOMs}, we derived the PN-expanded EOMs for precessing spins by using a frame adapted to the vector $\bm{L}_{\rm N}\equiv \mu \bm{r}\cross \bm{v}$.
In this Appendix, we include the corresponding equations in a frame adapted to the orbital angular momentum $\bm{L}\equiv \bm{r}\cross\bm{p}$, since one can define a co-precessing frame to be aligned with $\lhat$, as in the \texttt{SEOBNRv4PHM} waveform model~\cite{Ossokine:2020kjp} for example.

The EOMs in this case are given by Eqs.~\eqref{EOMsAlign} and
\begin{align}
\dot{\bm{S}}_{\mr i} &= \bm{\Omega}_{S_{\mr i}} \cross \bm{S}_{\mr i}, \nonumber\\
\dot{\lhat} &= \dot{\lhat}(\lhat,\bm{S}_{\mr i},v), \\
\dot{v} &= \frac{\dot{E}(v)}{\di E(v)/\di v}. \nonumber
\end{align}
The effective Hamiltonian is the same as in Sec.~\ref{sec:simpHam}, except that we replace the $\bm{n}\cdot\bm{a}_\pm$ terms by the following orbit average
\begin{equation}
\begin{aligned}
(\bm{n} \cdot \bm{a}_\pm)^2 &\simeq \frac{1}{2} \left[a_\pm^2 - (\lhat \cdot\bm{a}_\pm)^2\right], \\
(\bm{n} \cdot \bm{a}_+) (\bm{n} \cdot \bm{a}_-) &\simeq \frac{1}{2} \left[\bm{a}_+\cdot \bm{a}_- - (\lhat \cdot\bm{a}_+) (\lhat \cdot\bm{a}_-)\right],
\end{aligned}
\end{equation}
i.e., in terms of $\lhat$, instead of $\lNhat$ as in Eqs.~\eqref{naAvg}.

The spin-precession frequency $\bm{\Omega}_{S_{\mr i}}$ can be directly computed by taking the derivative of the Hamiltonian with respect to the spin vector, then orbit averaging the in-plane spin components, leading to
\begin{widetext}
\begin{subequations}
\begin{align}
\bm{\Omega}_{S_1} &= 
\frac{\lhat}{M} \bigg\lbrace
v^5 \left(\frac{3 X_2}{2}+\frac{\nu }{2}\right)
+ v^7 \left[\left(\frac{9}{8}-\frac{5 \nu }{4}\right) X_2-\frac{\nu ^2}{24}+\frac{15 \nu }{8}\right] \nonumber\\
&\quad\qquad
+ v^9 \left[\left(\frac{5 \nu ^2}{16}-\frac{39 \nu }{4}+\frac{27}{16}\right) X_2-\frac{\nu ^3}{48}-\frac{55 \nu ^2}{16}+\frac{81 \nu }{16}\right]\!\bigg\rbrace \nonumber\\
&\quad
+ \frac{v^6}{M^2\mu} \left\lbrace
\lhat \left[\lhat\cdot\bm{S}_1 \left(\frac{3 \nu }{2}-\frac{3 X_2}{2}\right)-\frac{3  \nu }{2}\lhat\cdot\bm{S}_2\right]+\frac{\nu}{2}\bm{S}_2
\right\rbrace \nonumber\\
&\quad
+ \frac{v^8}{M^2\mu} \left\lbrace
\lhat \left[\lhat\cdot\bm{S}_1 \left(-\frac{37 \nu ^2}{24}-\frac{9 \nu }{8}+\left(\frac{9}{8}-\frac{9 \nu }{2}\right) X_2\right)
+\lhat\cdot\bm{S}_2 \left(-\frac{\nu ^2}{24}-2 \nu\right)\right]
+ \bm{S}_2 \left(\frac{3 \nu }{2}-\frac{\nu ^2}{8}\right)
\right\rbrace \nonumber\\
&\quad
+ \frac{v^{10}}{M^2\mu} \bigg\lbrace
\lhat \bigg[\lhat\cdot\bm{S}_1 \left(\frac{127 \nu ^3}{144}-\frac{17 \nu ^2}{2}-\frac{27 \nu }{16}+\left(\frac{53 \nu ^2}{6}-\frac{25 \nu }{4}+\frac{27}{16}\right) X_2\right) \nonumber\\
&\quad\qquad
+\lhat\cdot\bm{S}_2 \left(\frac{109 \nu ^3}{144}+\frac{443 \nu ^2}{96}-\frac{27 \nu }{8}\right)\bigg] 
+\left(-\frac{\nu ^3}{48}+\frac{43 \nu ^2}{32}+\frac{3 \nu }{2}\right) \bm{S}_2
\bigg\rbrace, \nonumber\\
&\quad
+\frac{\widetilde{C}_{\rm 1ES^2}}{M^2\mu} \lhat (\lhat\cdot\bm{S}_1) \bigg\lbrace
v^6 \left(\frac{3 \nu }{2}-\frac{3 X_2}{2}\right)
+ v^8 \left[-\frac{3 \nu ^2}{4}+\frac{9 \nu }{4}+\left(-\frac{9 \nu }{4}-\frac{9}{4}\right) X_2\right] \nonumber\\
&\quad\qquad
+ v^{10} \left[\frac{\nu ^3}{16}-\frac{885 \nu ^2}{112}+\frac{81 \nu }{16}+\left(\frac{39 \nu ^2}{16}-\frac{33 \nu }{16}-\frac{81}{16}\right) X_2\right]
\bigg\rbrace,
\end{align}
\end{subequations}
and similarly for $\bm{\Omega}_{S_2}$.

To compute the equation for $\dot{\lhat}$, we first need the orbit-averaged angular momentum, which can be obtained by solving Eqs.~\eqref{circConds} for $r$ and $L$, leading to
\begin{subequations}
\begin{align}
L &\equiv \frac{M\mu}{v}\left(\bar{L}^{S^0} + \bar{L}_\text{SO} + \bar{L}_{S_1S_2} + \bar{L}_{S^2} + \bar{L}_{S^2\tilde{C}}\right), \\
%%%%
\bar{L}_\text{SO} &= \frac{\lhat\cdot \bm{S}_1}{M\mu}
\bigg\lbrace
v^3 \left(-\frac{5 \nu }{6}-\frac{5 X_2}{2}\right) 
+ v^5 \left[\frac{7 \nu ^2}{72}-\frac{35 \nu }{8}+\left(\frac{35 \nu }{12}-\frac{21}{8}\right) X_2\right] \nonumber\\
&\qquad
+ v^7 \left[\frac{\nu ^3}{16}+\frac{165 \nu ^2}{16}-\frac{243 \nu }{16}+\left(-\frac{15 \nu ^2}{16}+\frac{117 \nu }{4}-\frac{81}{16}\right) X_2\right]
\bigg\rbrace + 1\leftrightarrow 2, \\
%%%%
\bar{L}_{S_1S_2} &= \frac{\nu}{M^2\mu^2} \bigg\lbrace
v^4 \left[3(\lhat\cdot\bm{S}_1)(\lhat\cdot\bm{S}_2) - \bm{S}_1\cdot\bm{S}_2\right]
+ v^6 \left[
(\lhat\cdot\bm{S}_1)(\lhat\cdot\bm{S}_2)\left(\frac{5 \nu }{72}+\frac{29}{6}\right)
+ \bm{S}_1\cdot\bm{S}_2 \left(\frac{3 \nu }{8}-\frac{7}{2}\right)
\right]\nonumber\\
&\qquad
+ v^8 \left[
(\lhat\cdot\bm{S}_1)(\lhat\cdot\bm{S}_2)\left(-\frac{539 \nu ^2}{216}-\frac{4181 \nu }{288}+\frac{99}{8}\right)
+ \bm{S}_1\cdot\bm{S}_2 \left(\frac{\nu ^2}{24}-\frac{171 \nu }{32}-\frac{49}{8}\right)
\right]
\bigg\rbrace,\\
%%%%
\bar{L}_{S^2}  &= \frac{v^4 }{M^2\mu^2} \left[(\lhat\cdot\bm{S}_1)^2 \left(\frac{3 X_2}{2}-\frac{3 \nu }{2}\right) + \bm{S}_1^2 \left(\frac{\nu }{2}-\frac{X_2}{2}\right) \right] \nonumber\\
&\quad
+ \frac{v^6}{M^2\mu^2} \left\lbrace
(\lhat\cdot\bm{S}_1)^2 \left[\frac{293 \nu ^2}{144}+\frac{27 \nu }{16}+\left(\frac{47 \nu }{8}-\frac{27}{16}\right) X_2\right]
+ \bm{S}_1^2 \left[-\frac{23 \nu ^2}{48}+\nu  \left(\frac{5}{16}-\frac{61 X_2}{24}\right)-\frac{5 X_2}{16}\right] \right\rbrace\nonumber\\
&\quad
+ \frac{v^8}{M^2\mu^2} \bigg\lbrace
(\lhat\cdot\bm{S}_1)^2 \left[-\frac{629 \nu ^3}{432}+\frac{1315 \nu ^2}{96}+\frac{9 \nu }{4}+\left(-\frac{4189 \nu ^2}{288}+\frac{1039 \nu }{96}-\frac{9}{4}\right) X_2\right] \nonumber\\
&\quad\qquad
+ \bm{S}_1^2 \left[\frac{13 \nu ^3}{144}+\frac{145 \nu ^2}{96}-\frac{3 \nu }{8}+\left(\frac{929 \nu ^2}{288}-\frac{979 \nu }{96}+\frac{3}{8}\right) X_2\right]
\bigg\rbrace + 1\leftrightarrow 2,\\
%%%%
\bar{L}_{S^2\tilde{C}}  &= \frac{\widetilde{C}_{\rm 1ES^2}}{M^2\mu^2} \bigg\lbrace
v^4 \left[(\lhat\cdot\bm{S}_1)^2 \left(\frac{3 X_2}{2}-\frac{3 \nu }{2}\right)
+ \bm{S}_1^2 \left(\frac{\nu }{2}-\frac{X_2}{2}\right)\right] \nonumber\\
&\qquad
+ v^6 \bigg[
(\lhat\cdot\bm{S}_1)^2 \left(\nu ^2+3 \nu  \left(X_2-1\right)+3 X_2\right)
+ \bm{S}_1^2 \left(-\frac{\nu ^2}{3}+\nu  \left(1-X_2\right)-X_2\right)
\bigg] \nonumber\\
&\qquad
+ v^8 \bigg[
(\lhat\cdot\bm{S}_1)^2 \left(-\frac{5 \nu ^3}{48}+\nu ^2 \left(\frac{1475}{112}-\frac{65 X_2}{16}\right)+\nu  \left(\frac{55 X_2}{16}-\frac{135}{16}\right)+\frac{135 X_2}{16}\right) \nonumber\\
&\qquad\qquad
+ \bm{S}_1^2 \left(\frac{5 \nu ^3}{144}+\nu ^2 \left(\frac{65 X_2}{48}-\frac{1475}{336}\right)+\nu  \left(\frac{45}{16}-\frac{55 X_2}{48}\right)-\frac{45 X_2}{16}\right)
\bigg]
\bigg\rbrace + 1\leftrightarrow 2,
\end{align}
\end{subequations}
where the nonspinning part is the same as in Eq.~\eqref{LCirclNS0}.
Taking the time derivative of $\bm{L}=L \lhat$, then solving $\dot{\bm{J}} = \dot{\bm{L}} + \dot{\bm{S}}_1 + \dot{\bm{S}}_2$ for $\dot{\lhat}$ yields
\begin{subequations}
\begin{align}
\dot{\lhat} &\equiv \dot{\lhat}_\text{SO} + \dot{\lhat}_{S_1S_2} + \dot{\lhat}_{S^2} + \dot{\lhat}_{S^2\tilde{C}} + \dot{\lhat}_\text{RR},\\
\dot{\lhat}_\text{SO} &= \frac{\lhat\cross\bm{S}_1}{M^2\mu} \bigg\lbrace
v^6 \left(-\frac{\nu }{2}-\frac{3 X_2}{2}\right)
+ v^8 \left[\frac{\nu ^2}{8}-\frac{9 \nu }{8}+\left(\frac{3 \nu }{2}+\frac{9}{8}\right) X_2\right] \nonumber\\
&\quad\qquad
+ v^{10} \left[\frac{\nu ^3}{48}+\frac{9 \nu ^2}{4}-\frac{27 \nu }{16}+\left(-\frac{\nu ^2}{2}+\frac{15 \nu }{4}+\frac{27}{16}\right) X_2\right]
\bigg\rbrace + 1 \leftrightarrow 2,  \\
%%%%
\dot{\lhat}_{S_1S_2} &= \frac{\nu}{M^3\mu^2} \bigg\lbrace 
\frac{3}{2} v^7 \left[(\lhat\cross\bm{S}_1)(\lhat\cdot\bm{S}_2) + (\lhat\cross\bm{S}_2)(\lhat\cdot\bm{S}_1)\right] \nonumber\\
&\qquad
+ v^9 \left[
(\lhat\cross\bm{S}_1)(\lhat\cdot\bm{S}_2) \left(-\frac{5 \nu }{8}-\frac{21}{4}\right)
+ (\lhat\cross\bm{S}_2)(\lhat\cdot\bm{S}_1) \left(-\frac{5 \nu }{8}-\frac{21}{4}\right)
-\frac{\delta}{4} \lhat (\lhat\cdot\bm{S}_1\cross\bm{S}_2) 
\right]\nonumber\\
&\qquad
+ v^{11} \bigg\lbrace
(\lhat\cross\bm{S}_1)(\lhat\cdot\bm{S}_2)\left[-\frac{9 \nu ^2}{16}+\frac{241 \nu }{32}+\left(-\frac{3 \nu }{8}-\frac{3}{2}\right) X_2+\frac{15}{16}\right]
\nonumber\\
&\qquad\qquad
+ (\lhat\cross\bm{S}_2)(\lhat\cdot\bm{S}_1) \left[-\frac{9 \nu ^2}{16}+\frac{229 \nu }{32}+\left(\frac{3 \nu }{8}+\frac{3}{2}\right) X_2-\frac{9}{16}\right]
+ \frac{\delta}{48} (22 \nu +27) \lhat (\lhat\cdot\bm{S}_1\cross\bm{S}_2)
\bigg\rbrace
\bigg\rbrace,\\
%%%%
\dot{\lhat}_{S^2} &= \frac{(\lhat\cross\bm{S}_1)(\lhat\cdot\bm{S}_1)}{M^3\mu^2}
\bigg\lbrace
v^7 \left(\frac{3 X_2}{2}-\frac{3 \nu }{2}\right)
+ v^9 \left[\frac{11 \nu ^2}{8}+\frac{57 \nu }{8}+\left(\frac{7 \nu }{4}-\frac{57}{8}\right) X_2\right] \nonumber\\
&\quad\qquad
+ v^{11} \left[-\frac{43 \nu ^3}{48}-\frac{153 \nu ^2}{16}-\frac{45 \nu }{16}+\left(-\frac{289 \nu ^2}{48}+\frac{27 \nu }{4}+\frac{45}{16}\right) X_2\right]
\bigg\rbrace + 1 \leftrightarrow 2, \\
%%%%
\dot{\lhat}_{S^2\tilde{C}} &= \frac{\widetilde{C}_{\rm 1ES^2}}{M^3\mu^2}(\lhat\cross\bm{S}_1)(\lhat\cdot\bm{S}_1) \bigg\lbrace
v^7 \left(\frac{3 X_2}{2}-\frac{3 \nu }{2}\right)
+ v^9 \left(\nu ^2+2 \nu  X_2\right) \nonumber\\
&\quad\qquad
+ v^{11} \left[-\frac{\nu ^3}{6}+\frac{159 \nu ^2}{56}+\left(\frac{21 \nu }{8}-\frac{17 \nu ^2}{6}\right) X_2\right]
\bigg\rbrace + 1 \leftrightarrow 2,\\
%%%%
\dot{\lhat}_\text{RR} &=  -\frac{64}{5} \frac{v^8}{M}\lhat \bigg\lbrace
\nu + v^2 \nu \left(-\frac{37 \nu }{12}-\frac{1751}{336}\right) + 4\pi \nu v^3
+ \frac{v^3}{M\mu} \bigg[
(\lhat\cdot\bm{S}_1) \left(\frac{7 \nu ^2}{4}-\frac{19 \nu }{12}+\left(\frac{55 \nu }{8}-\frac{25}{8}\right) X_2 \right) \nonumber\\
&\quad\qquad
+ (\lhat\cdot\bm{S}_2) \left(\frac{7 \nu ^2}{4}-\frac{19 \nu }{12}+\left(\frac{55 \nu }{8}-\frac{25}{8}\right) X_1\right)
\bigg]
\bigg\rbrace.
\end{align}
\end{subequations}

The evolution of the orbital frequency can be obtained by replacing $\lNhat$ in Eq.~\eqref{vdot} by its PN expansion in terms of $\lhat$, resulting in
\begin{subequations}
\begin{align}
\dot{v} &\equiv \frac{32 \nu  v^9}{5M} \left(\dot{\bar{v}}_{S^0} + \dot{\bar{v}}_\text{SO} + \dot{\bar{v}}_{S_1S_2} + \dot{\bar{v}}_{S^2} + \dot{\bar{v}}_{S^2\tilde{C}}\right),\\
%%%%
\dot{\bar{v}}_\text{SO} &= \frac{\lhat\cdot\bm{S}_1}{M\mu} \bigg\lbrace
v^3 \left(-\frac{19 \nu }{6}-\frac{25 X_2}{4}\right)
+ v^5 \left[\frac{79 \nu ^2}{6}-\frac{21611 \nu }{1008}+\left(\frac{281 \nu }{8}-\frac{809}{84}\right) X_2\right]
+ \pi  v^6 \left(-\frac{37 \nu }{3}-\frac{151 X_2}{6}\right) \nonumber\\
&\quad
+v^7 \left[-\frac{10819 \nu ^3}{432}+\frac{40289 \nu ^2}{288}-\frac{1932041 \nu }{18144}+\left(-\frac{2903 \nu ^2}{32}+\frac{257023 \nu }{1008}-\frac{1195759}{18144}\right) X_2\right] \nonumber\\
&\quad
+ \pi  v^8 \left[\frac{34303 \nu ^2}{336}-\frac{46957 \nu }{504}+\left(\frac{50483 \nu }{224}-\frac{1665}{28}\right) X_2\right]
\bigg\rbrace + 1\leftrightarrow 2, \\
%%%%
\dot{\bar{v}}_{S_1S_2} &= \frac{\nu v^4}{M^2\mu^2} \left[\frac{721}{48}(\lhat\cdot\bm{S}_1)(\lhat\cdot\bm{S}_2)-\frac{247 }{48}(\bm{S}_1\cdot\bm{S}_2)\right]\nonumber\\
&\quad
+ \frac{\nu v^6}{M^2\mu^2} \left[(\lhat\cdot\bm{S}_1)(\lhat\cdot\bm{S}_2) \left(\frac{17415}{224}-\frac{11323 \nu }{288}\right)
+\left(\frac{5917 \nu }{288}+\frac{7309}{672}\right) (\bm{S}_1\cdot\bm{S}_2)\right] \nonumber\\
&\quad
+ \frac{\nu \pi  v^7}{M^2\mu^2} \left[\frac{207}{4}(\lhat\cdot\bm{S}_1)(\lhat\cdot\bm{S}_2)-12 (\bm{S}_1\cdot\bm{S}_2)\right]
+ \frac{\nu v^8}{M^2\mu^2} \bigg[
\left(-\frac{138421 \nu ^2}{3456}-\frac{1203227 \nu }{8064}-\frac{1623071}{10368}\right) (\bm{S}_1\cdot\bm{S}_2) \nonumber\\
&\quad\qquad
+ (\lhat\cdot\bm{S}_1)(\lhat\cdot\bm{S}_2) \left(\frac{9043 \nu ^2}{3456}-\frac{11824525 \nu }{24192}+\frac{21670157}{24192}\right)
\bigg], \\
%%%%
\dot{\bar{v}}_{S^2} &= 
\frac{v^4}{M^2\mu^2} \left[(\lhat\cdot\bm{S}_1)^2 \left(\frac{719 X_2}{96}-\frac{719 \nu }{96}\right)+S_1^2 \left(\frac{233 \nu }{96}-\frac{233 X_2}{96}\right)\right] \nonumber\\
&\quad
+ \frac{v^6}{M^2\mu^2} \bigg\lbrace
(\lhat\cdot\bm{S}_1)^2 \left[\frac{25829 \nu ^2}{576}+\frac{85 \nu }{448}+\left(\frac{21691 \nu }{576}-\frac{85}{448}\right) X_2\right] \nonumber\\
&\qquad
+S_1^2 \left[-\frac{6467 \nu ^2}{576}-\frac{6403 \nu }{448}+\left(\frac{6403}{448}-\frac{3445 \nu }{576}\right) X_2\right]
\bigg\rbrace \nonumber\\
&\quad
+\frac{\pi  v^7}{M^2\mu^2} \left[(\lhat\cdot\bm{S}_1)^2 \left(\frac{209 X_2}{8}-\frac{209 \nu }{8}\right)+S_1^2 \left(6 \nu -6 X_2\right)\right]\nonumber\\
&\quad
+ \frac{v^8}{M^2\mu^2}\bigg\lbrace
(\lhat\cdot\bm{S}_1)^2 \left[-\frac{893549 \nu ^3}{6912}+\frac{16130213 \nu ^2}{48384}-\frac{12067655 \nu }{48384}+\left(-\frac{2540311 \nu ^2}{6912}+\frac{3888965 \nu }{16128}+\frac{12067655}{48384}\right) X_2\right] \nonumber\\
&\qquad
+S_1^2 \left[\frac{162443 \nu ^3}{6912}+\frac{1518919 \nu ^2}{16128}+\frac{8745467 \nu }{145152}+\left(\frac{398761 \nu ^2}{6912}-\frac{754241 \nu }{5376}-\frac{8745467}{145152}\right) X_2\right]
\!\bigg\rbrace + 1\leftrightarrow 2, \\
%%%%
\dot{\bar{v}}_{S^2\tilde{C}} &= \frac{\widetilde{C}_{\rm 1ES^2}}{M^2\mu^2} \bigg\lbrace
v^4 \left[(\lhat\cdot\bm{S}_1)^2 \left(\frac{15 X_2}{2}-\frac{15 \nu }{2}\right)+S_1^2 \left(\frac{5 \nu }{2}-\frac{5 X_2}{2}\right)\right] \nonumber\\
&\qquad
+ v^6 \left[(\lhat\cdot\bm{S}_1)^2 \left(\frac{129 \nu ^2}{4}-\frac{1977 \nu }{224}+\left(\frac{1977}{224}-\frac{73 \nu }{16}\right) X_2\right)+S_1^2 \left(-\frac{43 \nu ^2}{4}+\frac{659 \nu }{224}+\left(\frac{73 \nu }{48}-\frac{659}{224}\right) X_2\right)\right]\nonumber\\
&\qquad
+ v^8 \bigg[
(\lhat\cdot\bm{S}_1)^2 \left(-\frac{1567 \nu ^3}{24}+\frac{29329 \nu ^2}{224}-\frac{597271 \nu }{6048}+\left(-\frac{5675 \nu ^2}{96}-\frac{1517 \nu }{168}+\frac{597271}{6048}\right) X_2\right) \nonumber\\
&\qquad\qquad
+ S_1^2 \left(\frac{1567 \nu ^3}{72}-\frac{29329 \nu ^2}{672}+\frac{597271 \nu }{18144}+\left(\frac{5675 \nu ^2}{288}+\frac{1517 \nu }{504}-\frac{597271}{18144}\right) X_2\right)
\bigg] \nonumber\\
&\qquad
+ \pi  v^7 \left[(\lhat\cdot\bm{S}_1)^2 \left(26 X_2-26 \nu \right)+S_1^2 \left(6 \nu -6 X_2\right)\right]
\bigg\rbrace + 1\leftrightarrow 2,
\end{align}
\end{subequations}
where the nonspinning part is the same as in Eq.~\eqref{vdotS0}.
\end{widetext}

\bibliography{../references}

\begin{thebibliography}{247}%
\makeatletter
\providecommand \@ifxundefined [1]{%
 \@ifx{#1\undefined}
}%
\providecommand \@ifnum [1]{%
 \ifnum #1\expandafter \@firstoftwo
 \else \expandafter \@secondoftwo
 \fi
}%
\providecommand \@ifx [1]{%
 \ifx #1\expandafter \@firstoftwo
 \else \expandafter \@secondoftwo
 \fi
}%
\providecommand \natexlab [1]{#1}%
\providecommand \enquote  [1]{``#1''}%
\providecommand \bibnamefont  [1]{#1}%
\providecommand \bibfnamefont [1]{#1}%
\providecommand \citenamefont [1]{#1}%
\providecommand \href@noop [0]{\@secondoftwo}%
\providecommand \href [0]{\begingroup \@sanitize@url \@href}%
\providecommand \@href[1]{\@@startlink{#1}\@@href}%
\providecommand \@@href[1]{\endgroup#1\@@endlink}%
\providecommand \@sanitize@url [0]{\catcode `\\12\catcode `\$12\catcode
  `\&12\catcode `\#12\catcode `\^12\catcode `\_12\catcode `\%12\relax}%
\providecommand \@@startlink[1]{}%
\providecommand \@@endlink[0]{}%
\providecommand \url  [0]{\begingroup\@sanitize@url \@url }%
\providecommand \@url [1]{\endgroup\@href {#1}{\urlprefix }}%
\providecommand \urlprefix  [0]{URL }%
\providecommand \Eprint [0]{\href }%
\providecommand \doibase [0]{https://doi.org/}%
\providecommand \selectlanguage [0]{\@gobble}%
\providecommand \bibinfo  [0]{\@secondoftwo}%
\providecommand \bibfield  [0]{\@secondoftwo}%
\providecommand \translation [1]{[#1]}%
\providecommand \BibitemOpen [0]{}%
\providecommand \bibitemStop [0]{}%
\providecommand \bibitemNoStop [0]{.\EOS\space}%
\providecommand \EOS [0]{\spacefactor3000\relax}%
\providecommand \BibitemShut  [1]{\csname bibitem#1\endcsname}%
\let\auto@bib@innerbib\@empty
%</preamble>
\bibitem [{\citenamefont {Abbott}\ \emph
  {et~al.}(2019{\natexlab{a}})\citenamefont {Abbott} \emph
  {et~al.}}]{LIGOScientific:2018mvr}%
  \BibitemOpen
  \bibfield  {author} {\bibinfo {author} {\bibfnamefont {B.~P.}\ \bibnamefont
  {Abbott}} \emph {et~al.} (\bibinfo {collaboration} {LIGO Scientific,
  Virgo}),\ }\bibfield  {title} {\bibinfo {title} {{GWTC-1: A
  Gravitational-Wave Transient Catalog of Compact Binary Mergers Observed by
  LIGO and Virgo during the First and Second Observing Runs}},\ }\href
  {https://doi.org/10.1103/PhysRevX.9.031040} {\bibfield  {journal} {\bibinfo
  {journal} {Phys. Rev. X}\ }\textbf {\bibinfo {volume} {9}},\ \bibinfo {pages}
  {031040} (\bibinfo {year} {2019}{\natexlab{a}})},\ \Eprint
  {https://arxiv.org/abs/1811.12907} {arXiv:1811.12907 [astro-ph.HE]}
  \BibitemShut {NoStop}%
\bibitem [{\citenamefont {Abbott}\ \emph
  {et~al.}(2021{\natexlab{a}})\citenamefont {Abbott} \emph
  {et~al.}}]{LIGOScientific:2020ibl}%
  \BibitemOpen
  \bibfield  {author} {\bibinfo {author} {\bibfnamefont {R.}~\bibnamefont
  {Abbott}} \emph {et~al.} (\bibinfo {collaboration} {LIGO Scientific,
  Virgo}),\ }\bibfield  {title} {\bibinfo {title} {{GWTC-2: Compact Binary
  Coalescences Observed by LIGO and Virgo During the First Half of the Third
  Observing Run}},\ }\href {https://doi.org/10.1103/PhysRevX.11.021053}
  {\bibfield  {journal} {\bibinfo  {journal} {Phys. Rev. X}\ }\textbf {\bibinfo
  {volume} {11}},\ \bibinfo {pages} {021053} (\bibinfo {year}
  {2021}{\natexlab{a}})},\ \Eprint {https://arxiv.org/abs/2010.14527}
  {arXiv:2010.14527 [gr-qc]} \BibitemShut {NoStop}%
\bibitem [{\citenamefont {Abbott}\ \emph
  {et~al.}(2021{\natexlab{b}})\citenamefont {Abbott} \emph
  {et~al.}}]{LIGOScientific:2021djp}%
  \BibitemOpen
  \bibfield  {author} {\bibinfo {author} {\bibfnamefont {R.}~\bibnamefont
  {Abbott}} \emph {et~al.} (\bibinfo {collaboration} {LIGO Scientific, VIRGO,
  KAGRA}),\ }\bibfield  {title} {\bibinfo {title} {{GWTC-3: Compact Binary
  Coalescences Observed by LIGO and Virgo During the Second Part of the Third
  Observing Run}},\ }\href@noop {} {\  (\bibinfo {year}
  {2021}{\natexlab{b}})},\ \Eprint {https://arxiv.org/abs/2111.03606}
  {arXiv:2111.03606 [gr-qc]} \BibitemShut {NoStop}%
\bibitem [{\citenamefont {Aasi}\ \emph {et~al.}(2015)\citenamefont {Aasi} \emph
  {et~al.}}]{LIGOScientific:2014pky}%
  \BibitemOpen
  \bibfield  {author} {\bibinfo {author} {\bibfnamefont {J.}~\bibnamefont
  {Aasi}} \emph {et~al.} (\bibinfo {collaboration} {LIGO Scientific}),\
  }\bibfield  {title} {\bibinfo {title} {{Advanced LIGO}},\ }\href
  {https://doi.org/10.1088/0264-9381/32/7/074001} {\bibfield  {journal}
  {\bibinfo  {journal} {Class. Quant. Grav.}\ }\textbf {\bibinfo {volume}
  {32}},\ \bibinfo {pages} {074001} (\bibinfo {year} {2015})},\ \Eprint
  {https://arxiv.org/abs/1411.4547} {arXiv:1411.4547 [gr-qc]} \BibitemShut
  {NoStop}%
\bibitem [{\citenamefont {Acernese}\ \emph {et~al.}(2015)\citenamefont
  {Acernese} \emph {et~al.}}]{VIRGO:2014yos}%
  \BibitemOpen
  \bibfield  {author} {\bibinfo {author} {\bibfnamefont {F.}~\bibnamefont
  {Acernese}} \emph {et~al.} (\bibinfo {collaboration} {VIRGO}),\ }\bibfield
  {title} {\bibinfo {title} {{Advanced Virgo: a second-generation
  interferometric gravitational wave detector}},\ }\href
  {https://doi.org/10.1088/0264-9381/32/2/024001} {\bibfield  {journal}
  {\bibinfo  {journal} {Class. Quant. Grav.}\ }\textbf {\bibinfo {volume}
  {32}},\ \bibinfo {pages} {024001} (\bibinfo {year} {2015})},\ \Eprint
  {https://arxiv.org/abs/1408.3978} {arXiv:1408.3978 [gr-qc]} \BibitemShut
  {NoStop}%
\bibitem [{\citenamefont {Abbott}\ \emph
  {et~al.}(2019{\natexlab{b}})\citenamefont {Abbott} \emph
  {et~al.}}]{LIGOScientific:2018jsj}%
  \BibitemOpen
  \bibfield  {author} {\bibinfo {author} {\bibfnamefont {B.~P.}\ \bibnamefont
  {Abbott}} \emph {et~al.} (\bibinfo {collaboration} {LIGO Scientific,
  Virgo}),\ }\bibfield  {title} {\bibinfo {title} {{Binary Black Hole
  Population Properties Inferred from the First and Second Observing Runs of
  Advanced LIGO and Advanced Virgo}},\ }\href
  {https://doi.org/10.3847/2041-8213/ab3800} {\bibfield  {journal} {\bibinfo
  {journal} {Astrophys. J. Lett.}\ }\textbf {\bibinfo {volume} {882}},\
  \bibinfo {pages} {L24} (\bibinfo {year} {2019}{\natexlab{b}})},\ \Eprint
  {https://arxiv.org/abs/1811.12940} {arXiv:1811.12940 [astro-ph.HE]}
  \BibitemShut {NoStop}%
\bibitem [{\citenamefont {Abbott}\ \emph
  {et~al.}(2021{\natexlab{c}})\citenamefont {Abbott} \emph
  {et~al.}}]{LIGOScientific:2020kqk}%
  \BibitemOpen
  \bibfield  {author} {\bibinfo {author} {\bibfnamefont {R.}~\bibnamefont
  {Abbott}} \emph {et~al.} (\bibinfo {collaboration} {LIGO Scientific,
  Virgo}),\ }\bibfield  {title} {\bibinfo {title} {{Population Properties of
  Compact Objects from the Second LIGO-Virgo Gravitational-Wave Transient
  Catalog}},\ }\href {https://doi.org/10.3847/2041-8213/abe949} {\bibfield
  {journal} {\bibinfo  {journal} {Astrophys. J. Lett.}\ }\textbf {\bibinfo
  {volume} {913}},\ \bibinfo {pages} {L7} (\bibinfo {year}
  {2021}{\natexlab{c}})},\ \Eprint {https://arxiv.org/abs/2010.14533}
  {arXiv:2010.14533 [astro-ph.HE]} \BibitemShut {NoStop}%
\bibitem [{\citenamefont {Abbott}\ \emph
  {et~al.}(2021{\natexlab{d}})\citenamefont {Abbott} \emph
  {et~al.}}]{LIGOScientific:2021psn}%
  \BibitemOpen
  \bibfield  {author} {\bibinfo {author} {\bibfnamefont {R.}~\bibnamefont
  {Abbott}} \emph {et~al.} (\bibinfo {collaboration} {LIGO Scientific, VIRGO,
  KAGRA}),\ }\bibfield  {title} {\bibinfo {title} {{The population of merging
  compact binaries inferred using gravitational waves through GWTC-3}},\
  }\href@noop {} {\  (\bibinfo {year} {2021}{\natexlab{d}})},\ \Eprint
  {https://arxiv.org/abs/2111.03634} {arXiv:2111.03634 [astro-ph.HE]}
  \BibitemShut {NoStop}%
\bibitem [{\citenamefont {P\"urrer}\ and\ \citenamefont
  {Haster}(2020)}]{Purrer:2019jcp}%
  \BibitemOpen
  \bibfield  {author} {\bibinfo {author} {\bibfnamefont {M.}~\bibnamefont
  {P\"urrer}}\ and\ \bibinfo {author} {\bibfnamefont {C.-J.}\ \bibnamefont
  {Haster}},\ }\bibfield  {title} {\bibinfo {title} {{Gravitational waveform
  accuracy requirements for future ground-based detectors}},\ }\href
  {https://doi.org/10.1103/PhysRevResearch.2.023151} {\bibfield  {journal}
  {\bibinfo  {journal} {Phys. Rev. Res.}\ }\textbf {\bibinfo {volume} {2}},\
  \bibinfo {pages} {023151} (\bibinfo {year} {2020})},\ \Eprint
  {https://arxiv.org/abs/1912.10055} {arXiv:1912.10055 [gr-qc]} \BibitemShut
  {NoStop}%
\bibitem [{\citenamefont {Abbott}\ \emph {et~al.}(2018)\citenamefont {Abbott}
  \emph {et~al.}}]{KAGRA:2013rdx}%
  \BibitemOpen
  \bibfield  {author} {\bibinfo {author} {\bibfnamefont {B.~P.}\ \bibnamefont
  {Abbott}} \emph {et~al.} (\bibinfo {collaboration} {KAGRA, LIGO Scientific,
  Virgo, VIRGO}),\ }\bibfield  {title} {\bibinfo {title} {{Prospects for
  observing and localizing gravitational-wave transients with Advanced LIGO,
  Advanced Virgo and KAGRA}},\ }\href
  {https://doi.org/10.1007/s41114-020-00026-9} {\bibfield  {journal} {\bibinfo
  {journal} {Living Rev. Rel.}\ }\textbf {\bibinfo {volume} {21}},\ \bibinfo
  {pages} {3} (\bibinfo {year} {2018})},\ \Eprint
  {https://arxiv.org/abs/1304.0670} {arXiv:1304.0670 [gr-qc]} \BibitemShut
  {NoStop}%
\bibitem [{\citenamefont {Amaro-Seoane}\ \emph {et~al.}(2017)\citenamefont
  {Amaro-Seoane} \emph {et~al.}}]{LISA:2017pwj}%
  \BibitemOpen
  \bibfield  {author} {\bibinfo {author} {\bibfnamefont {P.}~\bibnamefont
  {Amaro-Seoane}} \emph {et~al.} (\bibinfo {collaboration} {LISA}),\ }\bibfield
   {title} {\bibinfo {title} {{Laser Interferometer Space Antenna}},\
  }\href@noop {} {\  (\bibinfo {year} {2017})},\ \Eprint
  {https://arxiv.org/abs/1702.00786} {arXiv:1702.00786 [astro-ph.IM]}
  \BibitemShut {NoStop}%
\bibitem [{\citenamefont {Punturo}\ \emph {et~al.}(2010)\citenamefont {Punturo}
  \emph {et~al.}}]{Punturo:2010zz}%
  \BibitemOpen
  \bibfield  {author} {\bibinfo {author} {\bibfnamefont {M.}~\bibnamefont
  {Punturo}} \emph {et~al.},\ }\bibfield  {title} {\bibinfo {title} {{The
  Einstein Telescope: A third-generation gravitational wave observatory}},\
  }\href {https://doi.org/10.1088/0264-9381/27/19/194002} {\bibfield  {journal}
  {\bibinfo  {journal} {Class. Quant. Grav.}\ }\textbf {\bibinfo {volume}
  {27}},\ \bibinfo {pages} {194002} (\bibinfo {year} {2010})}\BibitemShut
  {NoStop}%
\bibitem [{\citenamefont {Reitze}\ \emph {et~al.}(2019)\citenamefont {Reitze}
  \emph {et~al.}}]{Reitze:2019iox}%
  \BibitemOpen
  \bibfield  {author} {\bibinfo {author} {\bibfnamefont {D.}~\bibnamefont
  {Reitze}} \emph {et~al.},\ }\bibfield  {title} {\bibinfo {title} {{Cosmic
  Explorer: The U.S. Contribution to Gravitational-Wave Astronomy beyond
  LIGO}},\ }\href@noop {} {\bibfield  {journal} {\bibinfo  {journal} {Bull. Am.
  Astron. Soc.}\ }\textbf {\bibinfo {volume} {51}},\ \bibinfo {pages} {035}
  (\bibinfo {year} {2019})},\ \Eprint {https://arxiv.org/abs/1907.04833}
  {arXiv:1907.04833 [astro-ph.IM]} \BibitemShut {NoStop}%
\bibitem [{\citenamefont {Evans}\ \emph {et~al.}(2021)\citenamefont {Evans}
  \emph {et~al.}}]{Evans:2021gyd}%
  \BibitemOpen
  \bibfield  {author} {\bibinfo {author} {\bibfnamefont {M.}~\bibnamefont
  {Evans}} \emph {et~al.},\ }\bibfield  {title} {\bibinfo {title} {{A Horizon
  Study for Cosmic Explorer: Science, Observatories, and Community}},\
  }\href@noop {} {\  (\bibinfo {year} {2021})},\ \Eprint
  {https://arxiv.org/abs/2109.09882} {arXiv:2109.09882 [astro-ph.IM]}
  \BibitemShut {NoStop}%
\bibitem [{\citenamefont {Pretorius}(2005)}]{Pretorius:2005gq}%
  \BibitemOpen
  \bibfield  {author} {\bibinfo {author} {\bibfnamefont {F.}~\bibnamefont
  {Pretorius}},\ }\bibfield  {title} {\bibinfo {title} {{Evolution of binary
  black hole spacetimes}},\ }\href
  {https://doi.org/10.1103/PhysRevLett.95.121101} {\bibfield  {journal}
  {\bibinfo  {journal} {Phys. Rev. Lett.}\ }\textbf {\bibinfo {volume} {95}},\
  \bibinfo {pages} {121101} (\bibinfo {year} {2005})},\ \Eprint
  {https://arxiv.org/abs/gr-qc/0507014} {arXiv:gr-qc/0507014} \BibitemShut
  {NoStop}%
\bibitem [{\citenamefont {Campanelli}\ \emph {et~al.}(2006)\citenamefont
  {Campanelli}, \citenamefont {Lousto}, \citenamefont {Marronetti},\ and\
  \citenamefont {Zlochower}}]{Campanelli:2005dd}%
  \BibitemOpen
  \bibfield  {author} {\bibinfo {author} {\bibfnamefont {M.}~\bibnamefont
  {Campanelli}}, \bibinfo {author} {\bibfnamefont {C.~O.}\ \bibnamefont
  {Lousto}}, \bibinfo {author} {\bibfnamefont {P.}~\bibnamefont {Marronetti}},\
  and\ \bibinfo {author} {\bibfnamefont {Y.}~\bibnamefont {Zlochower}},\
  }\bibfield  {title} {\bibinfo {title} {{Accurate evolutions of orbiting
  black-hole binaries without excision}},\ }\href
  {https://doi.org/10.1103/PhysRevLett.96.111101} {\bibfield  {journal}
  {\bibinfo  {journal} {Phys. Rev. Lett.}\ }\textbf {\bibinfo {volume} {96}},\
  \bibinfo {pages} {111101} (\bibinfo {year} {2006})},\ \Eprint
  {https://arxiv.org/abs/gr-qc/0511048} {arXiv:gr-qc/0511048} \BibitemShut
  {NoStop}%
\bibitem [{\citenamefont {Baker}\ \emph {et~al.}(2006)\citenamefont {Baker},
  \citenamefont {Centrella}, \citenamefont {Choi}, \citenamefont {Koppitz},\
  and\ \citenamefont {van Meter}}]{Baker:2005vv}%
  \BibitemOpen
  \bibfield  {author} {\bibinfo {author} {\bibfnamefont {J.~G.}\ \bibnamefont
  {Baker}}, \bibinfo {author} {\bibfnamefont {J.}~\bibnamefont {Centrella}},
  \bibinfo {author} {\bibfnamefont {D.-I.}\ \bibnamefont {Choi}}, \bibinfo
  {author} {\bibfnamefont {M.}~\bibnamefont {Koppitz}},\ and\ \bibinfo {author}
  {\bibfnamefont {J.}~\bibnamefont {van Meter}},\ }\bibfield  {title} {\bibinfo
  {title} {{Gravitational wave extraction from an inspiraling configuration of
  merging black holes}},\ }\href
  {https://doi.org/10.1103/PhysRevLett.96.111102} {\bibfield  {journal}
  {\bibinfo  {journal} {Phys. Rev. Lett.}\ }\textbf {\bibinfo {volume} {96}},\
  \bibinfo {pages} {111102} (\bibinfo {year} {2006})},\ \Eprint
  {https://arxiv.org/abs/gr-qc/0511103} {arXiv:gr-qc/0511103} \BibitemShut
  {NoStop}%
\bibitem [{\citenamefont {Futamase}\ and\ \citenamefont
  {Itoh}(2007)}]{Futamase:2007zz}%
  \BibitemOpen
  \bibfield  {author} {\bibinfo {author} {\bibfnamefont {T.}~\bibnamefont
  {Futamase}}\ and\ \bibinfo {author} {\bibfnamefont {Y.}~\bibnamefont
  {Itoh}},\ }\bibfield  {title} {\bibinfo {title} {{The post-Newtonian
  approximation for relativistic compact binaries}},\ }\href
  {https://doi.org/10.12942/lrr-2007-2} {\bibfield  {journal} {\bibinfo
  {journal} {Living Rev. Rel.}\ }\textbf {\bibinfo {volume} {10}},\ \bibinfo
  {pages} {2} (\bibinfo {year} {2007})}\BibitemShut {NoStop}%
\bibitem [{\citenamefont {Blanchet}(2014)}]{Blanchet:2013haa}%
  \BibitemOpen
  \bibfield  {author} {\bibinfo {author} {\bibfnamefont {L.}~\bibnamefont
  {Blanchet}},\ }\bibfield  {title} {\bibinfo {title} {{Gravitational Radiation
  from Post-Newtonian Sources and Inspiralling Compact Binaries}},\ }\href
  {https://doi.org/10.12942/lrr-2014-2} {\bibfield  {journal} {\bibinfo
  {journal} {Living Rev. Rel.}\ }\textbf {\bibinfo {volume} {17}},\ \bibinfo
  {pages} {2} (\bibinfo {year} {2014})},\ \Eprint
  {https://arxiv.org/abs/1310.1528} {arXiv:1310.1528 [gr-qc]} \BibitemShut
  {NoStop}%
\bibitem [{\citenamefont {Sch\"afer}\ and\ \citenamefont
  {Jaranowski}(2018)}]{Schafer:2018kuf}%
  \BibitemOpen
  \bibfield  {author} {\bibinfo {author} {\bibfnamefont {G.}~\bibnamefont
  {Sch\"afer}}\ and\ \bibinfo {author} {\bibfnamefont {P.}~\bibnamefont
  {Jaranowski}},\ }\bibfield  {title} {\bibinfo {title} {{Hamiltonian
  formulation of general relativity and post-Newtonian dynamics of compact
  binaries}},\ }\href {https://doi.org/10.1007/s41114-018-0016-5} {\bibfield
  {journal} {\bibinfo  {journal} {Living Rev. Rel.}\ }\textbf {\bibinfo
  {volume} {21}},\ \bibinfo {pages} {7} (\bibinfo {year} {2018})},\ \Eprint
  {https://arxiv.org/abs/1805.07240} {arXiv:1805.07240 [gr-qc]} \BibitemShut
  {NoStop}%
\bibitem [{\citenamefont {Levi}\ and\ \citenamefont
  {Steinhoff}(2015{\natexlab{a}})}]{Levi:2015msa}%
  \BibitemOpen
  \bibfield  {author} {\bibinfo {author} {\bibfnamefont {M.}~\bibnamefont
  {Levi}}\ and\ \bibinfo {author} {\bibfnamefont {J.}~\bibnamefont
  {Steinhoff}},\ }\bibfield  {title} {\bibinfo {title} {{Spinning gravitating
  objects in the effective field theory in the post-Newtonian scheme}},\ }\href
  {https://doi.org/10.1007/JHEP09(2015)219} {\bibfield  {journal} {\bibinfo
  {journal} {JHEP}\ }\textbf {\bibinfo {volume} {09}},\ \bibinfo {pages}
  {219}},\ \Eprint {https://arxiv.org/abs/1501.04956} {arXiv:1501.04956
  [gr-qc]} \BibitemShut {NoStop}%
\bibitem [{\citenamefont {Porto}(2016)}]{Porto:2016pyg}%
  \BibitemOpen
  \bibfield  {author} {\bibinfo {author} {\bibfnamefont {R.~A.}\ \bibnamefont
  {Porto}},\ }\bibfield  {title} {\bibinfo {title} {{The effective field
  theorist\textquoteright{}s approach to gravitational dynamics}},\ }\href
  {https://doi.org/10.1016/j.physrep.2016.04.003} {\bibfield  {journal}
  {\bibinfo  {journal} {Phys. Rept.}\ }\textbf {\bibinfo {volume} {633}},\
  \bibinfo {pages} {1} (\bibinfo {year} {2016})},\ \Eprint
  {https://arxiv.org/abs/1601.04914} {arXiv:1601.04914 [hep-th]} \BibitemShut
  {NoStop}%
\bibitem [{\citenamefont {Levi}(2020)}]{Levi:2018nxp}%
  \BibitemOpen
  \bibfield  {author} {\bibinfo {author} {\bibfnamefont {M.}~\bibnamefont
  {Levi}},\ }\bibfield  {title} {\bibinfo {title} {{Effective Field Theories of
  Post-Newtonian Gravity: A comprehensive review}},\ }\href
  {https://doi.org/10.1088/1361-6633/ab12bc} {\bibfield  {journal} {\bibinfo
  {journal} {Rept. Prog. Phys.}\ }\textbf {\bibinfo {volume} {83}},\ \bibinfo
  {pages} {075901} (\bibinfo {year} {2020})},\ \Eprint
  {https://arxiv.org/abs/1807.01699} {arXiv:1807.01699 [hep-th]} \BibitemShut
  {NoStop}%
\bibitem [{\citenamefont {Isoyama}\ \emph {et~al.}(2020)\citenamefont
  {Isoyama}, \citenamefont {Sturani},\ and\ \citenamefont
  {Nakano}}]{Isoyama:2020lls}%
  \BibitemOpen
  \bibfield  {author} {\bibinfo {author} {\bibfnamefont {S.}~\bibnamefont
  {Isoyama}}, \bibinfo {author} {\bibfnamefont {R.}~\bibnamefont {Sturani}},\
  and\ \bibinfo {author} {\bibfnamefont {H.}~\bibnamefont {Nakano}},\
  }\bibfield  {title} {\bibinfo {title} {{Post-Newtonian templates for
  gravitational waves from compact binary inspirals}},\ }\href@noop {} {\
  (\bibinfo {year} {2020})},\ \Eprint {https://arxiv.org/abs/2012.01350}
  {arXiv:2012.01350 [gr-qc]} \BibitemShut {NoStop}%
\bibitem [{\citenamefont {Poisson}\ and\ \citenamefont
  {Will}(1995)}]{Poisson:1995ef}%
  \BibitemOpen
  \bibfield  {author} {\bibinfo {author} {\bibfnamefont {E.}~\bibnamefont
  {Poisson}}\ and\ \bibinfo {author} {\bibfnamefont {C.~M.}\ \bibnamefont
  {Will}},\ }\bibfield  {title} {\bibinfo {title} {{Gravitational waves from
  inspiraling compact binaries: Parameter estimation using second postNewtonian
  wave forms}},\ }\href {https://doi.org/10.1103/PhysRevD.52.848} {\bibfield
  {journal} {\bibinfo  {journal} {Phys. Rev. D}\ }\textbf {\bibinfo {volume}
  {52}},\ \bibinfo {pages} {848} (\bibinfo {year} {1995})},\ \Eprint
  {https://arxiv.org/abs/gr-qc/9502040} {arXiv:gr-qc/9502040} \BibitemShut
  {NoStop}%
\bibitem [{\citenamefont {Damour}\ \emph {et~al.}(1998)\citenamefont {Damour},
  \citenamefont {Iyer},\ and\ \citenamefont {Sathyaprakash}}]{Damour:1997ub}%
  \BibitemOpen
  \bibfield  {author} {\bibinfo {author} {\bibfnamefont {T.}~\bibnamefont
  {Damour}}, \bibinfo {author} {\bibfnamefont {B.~R.}\ \bibnamefont {Iyer}},\
  and\ \bibinfo {author} {\bibfnamefont {B.~S.}\ \bibnamefont
  {Sathyaprakash}},\ }\bibfield  {title} {\bibinfo {title} {{Improved filters
  for gravitational waves from inspiralling compact binaries}},\ }\href
  {https://doi.org/10.1103/PhysRevD.57.885} {\bibfield  {journal} {\bibinfo
  {journal} {Phys. Rev. D}\ }\textbf {\bibinfo {volume} {57}},\ \bibinfo
  {pages} {885} (\bibinfo {year} {1998})},\ \Eprint
  {https://arxiv.org/abs/gr-qc/9708034} {arXiv:gr-qc/9708034} \BibitemShut
  {NoStop}%
\bibitem [{\citenamefont {Droz}\ \emph {et~al.}(1999)\citenamefont {Droz},
  \citenamefont {Knapp}, \citenamefont {Poisson},\ and\ \citenamefont
  {Owen}}]{Droz:1999qx}%
  \BibitemOpen
  \bibfield  {author} {\bibinfo {author} {\bibfnamefont {S.}~\bibnamefont
  {Droz}}, \bibinfo {author} {\bibfnamefont {D.~J.}\ \bibnamefont {Knapp}},
  \bibinfo {author} {\bibfnamefont {E.}~\bibnamefont {Poisson}},\ and\ \bibinfo
  {author} {\bibfnamefont {B.~J.}\ \bibnamefont {Owen}},\ }\bibfield  {title}
  {\bibinfo {title} {{Gravitational waves from inspiraling compact binaries:
  Validity of the stationary phase approximation to the Fourier transform}},\
  }\href {https://doi.org/10.1103/PhysRevD.59.124016} {\bibfield  {journal}
  {\bibinfo  {journal} {Phys. Rev. D}\ }\textbf {\bibinfo {volume} {59}},\
  \bibinfo {pages} {124016} (\bibinfo {year} {1999})},\ \Eprint
  {https://arxiv.org/abs/gr-qc/9901076} {arXiv:gr-qc/9901076} \BibitemShut
  {NoStop}%
\bibitem [{\citenamefont {Damour}\ \emph
  {et~al.}(2000{\natexlab{a}})\citenamefont {Damour}, \citenamefont {Iyer},\
  and\ \citenamefont {Sathyaprakash}}]{Damour:2000gg}%
  \BibitemOpen
  \bibfield  {author} {\bibinfo {author} {\bibfnamefont {T.}~\bibnamefont
  {Damour}}, \bibinfo {author} {\bibfnamefont {B.~R.}\ \bibnamefont {Iyer}},\
  and\ \bibinfo {author} {\bibfnamefont {B.~S.}\ \bibnamefont
  {Sathyaprakash}},\ }\bibfield  {title} {\bibinfo {title} {{Frequency domain P
  approximant filters for time truncated inspiral gravitational wave signals
  from compact binaries}},\ }\href {https://doi.org/10.1103/PhysRevD.62.084036}
  {\bibfield  {journal} {\bibinfo  {journal} {Phys. Rev. D}\ }\textbf {\bibinfo
  {volume} {62}},\ \bibinfo {pages} {084036} (\bibinfo {year}
  {2000}{\natexlab{a}})},\ \Eprint {https://arxiv.org/abs/gr-qc/0001023}
  {arXiv:gr-qc/0001023} \BibitemShut {NoStop}%
\bibitem [{\citenamefont {Damour}\ \emph {et~al.}(2001)\citenamefont {Damour},
  \citenamefont {Iyer},\ and\ \citenamefont {Sathyaprakash}}]{Damour:2000zb}%
  \BibitemOpen
  \bibfield  {author} {\bibinfo {author} {\bibfnamefont {T.}~\bibnamefont
  {Damour}}, \bibinfo {author} {\bibfnamefont {B.~R.}\ \bibnamefont {Iyer}},\
  and\ \bibinfo {author} {\bibfnamefont {B.~S.}\ \bibnamefont
  {Sathyaprakash}},\ }\bibfield  {title} {\bibinfo {title} {{A Comparison of
  search templates for gravitational waves from binary inspiral}},\ }\href
  {https://doi.org/10.1103/PhysRevD.63.044023} {\bibfield  {journal} {\bibinfo
  {journal} {Phys. Rev. D}\ }\textbf {\bibinfo {volume} {63}},\ \bibinfo
  {pages} {044023} (\bibinfo {year} {2001})},\ \bibinfo {note} {[Erratum:
  Phys.Rev.D 72, 029902 (2005)]},\ \Eprint
  {https://arxiv.org/abs/gr-qc/0010009} {arXiv:gr-qc/0010009} \BibitemShut
  {NoStop}%
\bibitem [{\citenamefont {Buonanno}\ \emph
  {et~al.}(2003{\natexlab{a}})\citenamefont {Buonanno}, \citenamefont {Chen},\
  and\ \citenamefont {Vallisneri}}]{Buonanno:2002ft}%
  \BibitemOpen
  \bibfield  {author} {\bibinfo {author} {\bibfnamefont {A.}~\bibnamefont
  {Buonanno}}, \bibinfo {author} {\bibfnamefont {Y.-b.}\ \bibnamefont {Chen}},\
  and\ \bibinfo {author} {\bibfnamefont {M.}~\bibnamefont {Vallisneri}},\
  }\bibfield  {title} {\bibinfo {title} {{Detection template families for
  gravitational waves from the final stages of binary--black-hole inspirals:
  Nonspinning case}},\ }\href {https://doi.org/10.1103/PhysRevD.67.024016}
  {\bibfield  {journal} {\bibinfo  {journal} {Phys. Rev. D}\ }\textbf {\bibinfo
  {volume} {67}},\ \bibinfo {pages} {024016} (\bibinfo {year}
  {2003}{\natexlab{a}})},\ \bibinfo {note} {[Erratum: Phys.Rev.D 74, 029903
  (2006)]},\ \Eprint {https://arxiv.org/abs/gr-qc/0205122}
  {arXiv:gr-qc/0205122} \BibitemShut {NoStop}%
\bibitem [{\citenamefont {Buonanno}\ \emph
  {et~al.}(2003{\natexlab{b}})\citenamefont {Buonanno}, \citenamefont {Chen},\
  and\ \citenamefont {Vallisneri}}]{Buonanno:2002fy}%
  \BibitemOpen
  \bibfield  {author} {\bibinfo {author} {\bibfnamefont {A.}~\bibnamefont
  {Buonanno}}, \bibinfo {author} {\bibfnamefont {Y.-b.}\ \bibnamefont {Chen}},\
  and\ \bibinfo {author} {\bibfnamefont {M.}~\bibnamefont {Vallisneri}},\
  }\bibfield  {title} {\bibinfo {title} {{Detecting gravitational waves from
  precessing binaries of spinning compact objects: Adiabatic limit}},\ }\href
  {https://doi.org/10.1103/PhysRevD.67.104025} {\bibfield  {journal} {\bibinfo
  {journal} {Phys. Rev. D}\ }\textbf {\bibinfo {volume} {67}},\ \bibinfo
  {pages} {104025} (\bibinfo {year} {2003}{\natexlab{b}})},\ \bibinfo {note}
  {[Erratum: Phys.Rev.D 74, 029904 (2006)]},\ \Eprint
  {https://arxiv.org/abs/gr-qc/0211087} {arXiv:gr-qc/0211087} \BibitemShut
  {NoStop}%
\bibitem [{\citenamefont {Damour}\ \emph {et~al.}(2002)\citenamefont {Damour},
  \citenamefont {Iyer},\ and\ \citenamefont {Sathyaprakash}}]{Damour:2002kr}%
  \BibitemOpen
  \bibfield  {author} {\bibinfo {author} {\bibfnamefont {T.}~\bibnamefont
  {Damour}}, \bibinfo {author} {\bibfnamefont {B.~R.}\ \bibnamefont {Iyer}},\
  and\ \bibinfo {author} {\bibfnamefont {B.~S.}\ \bibnamefont
  {Sathyaprakash}},\ }\bibfield  {title} {\bibinfo {title} {{A Comparison of
  search templates for gravitational waves from binary inspiral - 3.5PN
  update}},\ }\href {https://doi.org/10.1103/PhysRevD.66.027502} {\bibfield
  {journal} {\bibinfo  {journal} {Phys. Rev. D}\ }\textbf {\bibinfo {volume}
  {66}},\ \bibinfo {pages} {027502} (\bibinfo {year} {2002})},\ \Eprint
  {https://arxiv.org/abs/gr-qc/0207021} {arXiv:gr-qc/0207021} \BibitemShut
  {NoStop}%
\bibitem [{\citenamefont {Arun}\ \emph {et~al.}(2005)\citenamefont {Arun},
  \citenamefont {Iyer}, \citenamefont {Sathyaprakash},\ and\ \citenamefont
  {Sundararajan}}]{Arun:2004hn}%
  \BibitemOpen
  \bibfield  {author} {\bibinfo {author} {\bibfnamefont {K.~G.}\ \bibnamefont
  {Arun}}, \bibinfo {author} {\bibfnamefont {B.~R.}\ \bibnamefont {Iyer}},
  \bibinfo {author} {\bibfnamefont {B.~S.}\ \bibnamefont {Sathyaprakash}},\
  and\ \bibinfo {author} {\bibfnamefont {P.~A.}\ \bibnamefont {Sundararajan}},\
  }\bibfield  {title} {\bibinfo {title} {{Parameter estimation of inspiralling
  compact binaries using 3.5 post-Newtonian gravitational wave phasing: The
  Non-spinning case}},\ }\href {https://doi.org/10.1103/PhysRevD.71.084008}
  {\bibfield  {journal} {\bibinfo  {journal} {Phys. Rev. D}\ }\textbf {\bibinfo
  {volume} {71}},\ \bibinfo {pages} {084008} (\bibinfo {year} {2005})},\
  \bibinfo {note} {[Erratum: Phys.Rev.D 72, 069903 (2005)]},\ \Eprint
  {https://arxiv.org/abs/gr-qc/0411146} {arXiv:gr-qc/0411146} \BibitemShut
  {NoStop}%
\bibitem [{\citenamefont {Gopakumar}\ \emph {et~al.}(2008)\citenamefont
  {Gopakumar}, \citenamefont {Hannam}, \citenamefont {Husa},\ and\
  \citenamefont {Bruegmann}}]{Gopakumar:2007vh}%
  \BibitemOpen
  \bibfield  {author} {\bibinfo {author} {\bibfnamefont {A.}~\bibnamefont
  {Gopakumar}}, \bibinfo {author} {\bibfnamefont {M.}~\bibnamefont {Hannam}},
  \bibinfo {author} {\bibfnamefont {S.}~\bibnamefont {Husa}},\ and\ \bibinfo
  {author} {\bibfnamefont {B.}~\bibnamefont {Bruegmann}},\ }\bibfield  {title}
  {\bibinfo {title} {{Comparison between numerical relativity and a new class
  of post-Newtonian gravitational-wave phase evolutions: The Non-spinning
  equal-mass case}},\ }\href {https://doi.org/10.1103/PhysRevD.78.064026}
  {\bibfield  {journal} {\bibinfo  {journal} {Phys. Rev. D}\ }\textbf {\bibinfo
  {volume} {78}},\ \bibinfo {pages} {064026} (\bibinfo {year} {2008})},\
  \Eprint {https://arxiv.org/abs/0712.3737} {arXiv:0712.3737 [gr-qc]}
  \BibitemShut {NoStop}%
\bibitem [{\citenamefont {Boyle}\ \emph {et~al.}(2007)\citenamefont {Boyle},
  \citenamefont {Brown}, \citenamefont {Kidder}, \citenamefont {Mroue},
  \citenamefont {Pfeiffer}, \citenamefont {Scheel}, \citenamefont {Cook},\ and\
  \citenamefont {Teukolsky}}]{Boyle:2007ft}%
  \BibitemOpen
  \bibfield  {author} {\bibinfo {author} {\bibfnamefont {M.}~\bibnamefont
  {Boyle}}, \bibinfo {author} {\bibfnamefont {D.~A.}\ \bibnamefont {Brown}},
  \bibinfo {author} {\bibfnamefont {L.~E.}\ \bibnamefont {Kidder}}, \bibinfo
  {author} {\bibfnamefont {A.~H.}\ \bibnamefont {Mroue}}, \bibinfo {author}
  {\bibfnamefont {H.~P.}\ \bibnamefont {Pfeiffer}}, \bibinfo {author}
  {\bibfnamefont {M.~A.}\ \bibnamefont {Scheel}}, \bibinfo {author}
  {\bibfnamefont {G.~B.}\ \bibnamefont {Cook}},\ and\ \bibinfo {author}
  {\bibfnamefont {S.~A.}\ \bibnamefont {Teukolsky}},\ }\bibfield  {title}
  {\bibinfo {title} {{High-accuracy comparison of numerical relativity
  simulations with post-Newtonian expansions}},\ }\href
  {https://doi.org/10.1103/PhysRevD.76.124038} {\bibfield  {journal} {\bibinfo
  {journal} {Phys. Rev. D}\ }\textbf {\bibinfo {volume} {76}},\ \bibinfo
  {pages} {124038} (\bibinfo {year} {2007})},\ \Eprint
  {https://arxiv.org/abs/0710.0158} {arXiv:0710.0158 [gr-qc]} \BibitemShut
  {NoStop}%
\bibitem [{\citenamefont {Buonanno}\ \emph
  {et~al.}(2009{\natexlab{a}})\citenamefont {Buonanno}, \citenamefont {Iyer},
  \citenamefont {Ochsner}, \citenamefont {Pan},\ and\ \citenamefont
  {Sathyaprakash}}]{Buonanno:2009zt}%
  \BibitemOpen
  \bibfield  {author} {\bibinfo {author} {\bibfnamefont {A.}~\bibnamefont
  {Buonanno}}, \bibinfo {author} {\bibfnamefont {B.}~\bibnamefont {Iyer}},
  \bibinfo {author} {\bibfnamefont {E.}~\bibnamefont {Ochsner}}, \bibinfo
  {author} {\bibfnamefont {Y.}~\bibnamefont {Pan}},\ and\ \bibinfo {author}
  {\bibfnamefont {B.~S.}\ \bibnamefont {Sathyaprakash}},\ }\bibfield  {title}
  {\bibinfo {title} {{Comparison of post-Newtonian templates for compact binary
  inspiral signals in gravitational-wave detectors}},\ }\href
  {https://doi.org/10.1103/PhysRevD.80.084043} {\bibfield  {journal} {\bibinfo
  {journal} {Phys. Rev. D}\ }\textbf {\bibinfo {volume} {80}},\ \bibinfo
  {pages} {084043} (\bibinfo {year} {2009}{\natexlab{a}})},\ \Eprint
  {https://arxiv.org/abs/0907.0700} {arXiv:0907.0700 [gr-qc]} \BibitemShut
  {NoStop}%
\bibitem [{\citenamefont {Ajith}(2011)}]{Ajith:2011ec}%
  \BibitemOpen
  \bibfield  {author} {\bibinfo {author} {\bibfnamefont {P.}~\bibnamefont
  {Ajith}},\ }\bibfield  {title} {\bibinfo {title} {{Addressing the spin
  question in gravitational-wave searches: Waveform templates for inspiralling
  compact binaries with nonprecessing spins}},\ }\href
  {https://doi.org/10.1103/PhysRevD.84.084037} {\bibfield  {journal} {\bibinfo
  {journal} {Phys. Rev. D}\ }\textbf {\bibinfo {volume} {84}},\ \bibinfo
  {pages} {084037} (\bibinfo {year} {2011})},\ \Eprint
  {https://arxiv.org/abs/1107.1267} {arXiv:1107.1267 [gr-qc]} \BibitemShut
  {NoStop}%
\bibitem [{\citenamefont {Klein}\ \emph {et~al.}(2013)\citenamefont {Klein},
  \citenamefont {Cornish},\ and\ \citenamefont {Yunes}}]{Klein:2013qda}%
  \BibitemOpen
  \bibfield  {author} {\bibinfo {author} {\bibfnamefont {A.}~\bibnamefont
  {Klein}}, \bibinfo {author} {\bibfnamefont {N.}~\bibnamefont {Cornish}},\
  and\ \bibinfo {author} {\bibfnamefont {N.}~\bibnamefont {Yunes}},\ }\bibfield
   {title} {\bibinfo {title} {{Gravitational waveforms for precessing,
  quasicircular binaries via multiple scale analysis and uniform asymptotics:
  The near spin alignment case}},\ }\href
  {https://doi.org/10.1103/PhysRevD.88.124015} {\bibfield  {journal} {\bibinfo
  {journal} {Phys. Rev. D}\ }\textbf {\bibinfo {volume} {88}},\ \bibinfo
  {pages} {124015} (\bibinfo {year} {2013})},\ \Eprint
  {https://arxiv.org/abs/1305.1932} {arXiv:1305.1932 [gr-qc]} \BibitemShut
  {NoStop}%
\bibitem [{\citenamefont {Chatziioannou}\ \emph {et~al.}(2013)\citenamefont
  {Chatziioannou}, \citenamefont {Klein}, \citenamefont {Yunes},\ and\
  \citenamefont {Cornish}}]{Chatziioannou:2013dza}%
  \BibitemOpen
  \bibfield  {author} {\bibinfo {author} {\bibfnamefont {K.}~\bibnamefont
  {Chatziioannou}}, \bibinfo {author} {\bibfnamefont {A.}~\bibnamefont
  {Klein}}, \bibinfo {author} {\bibfnamefont {N.}~\bibnamefont {Yunes}},\ and\
  \bibinfo {author} {\bibfnamefont {N.}~\bibnamefont {Cornish}},\ }\bibfield
  {title} {\bibinfo {title} {{Gravitational Waveforms for Precessing,
  Quasicircular Compact Binaries with Multiple Scale Analysis: Small Spin
  Expansion}},\ }\href {https://doi.org/10.1103/PhysRevD.88.063011} {\bibfield
  {journal} {\bibinfo  {journal} {Phys. Rev. D}\ }\textbf {\bibinfo {volume}
  {88}},\ \bibinfo {pages} {063011} (\bibinfo {year} {2013})},\ \Eprint
  {https://arxiv.org/abs/1307.4418} {arXiv:1307.4418 [gr-qc]} \BibitemShut
  {NoStop}%
\bibitem [{\citenamefont {Chatziioannou}\ \emph {et~al.}(2017)\citenamefont
  {Chatziioannou}, \citenamefont {Klein}, \citenamefont {Yunes},\ and\
  \citenamefont {Cornish}}]{Chatziioannou:2017tdw}%
  \BibitemOpen
  \bibfield  {author} {\bibinfo {author} {\bibfnamefont {K.}~\bibnamefont
  {Chatziioannou}}, \bibinfo {author} {\bibfnamefont {A.}~\bibnamefont
  {Klein}}, \bibinfo {author} {\bibfnamefont {N.}~\bibnamefont {Yunes}},\ and\
  \bibinfo {author} {\bibfnamefont {N.}~\bibnamefont {Cornish}},\ }\bibfield
  {title} {\bibinfo {title} {{Constructing Gravitational Waves from Generic
  Spin-Precessing Compact Binary Inspirals}},\ }\href
  {https://doi.org/10.1103/PhysRevD.95.104004} {\bibfield  {journal} {\bibinfo
  {journal} {Phys. Rev. D}\ }\textbf {\bibinfo {volume} {95}},\ \bibinfo
  {pages} {104004} (\bibinfo {year} {2017})},\ \Eprint
  {https://arxiv.org/abs/1703.03967} {arXiv:1703.03967 [gr-qc]} \BibitemShut
  {NoStop}%
\bibitem [{\citenamefont {Mishra}\ \emph {et~al.}(2016)\citenamefont {Mishra},
  \citenamefont {Kela}, \citenamefont {Arun},\ and\ \citenamefont
  {Faye}}]{Mishra:2016whh}%
  \BibitemOpen
  \bibfield  {author} {\bibinfo {author} {\bibfnamefont {C.~K.}\ \bibnamefont
  {Mishra}}, \bibinfo {author} {\bibfnamefont {A.}~\bibnamefont {Kela}},
  \bibinfo {author} {\bibfnamefont {K.~G.}\ \bibnamefont {Arun}},\ and\
  \bibinfo {author} {\bibfnamefont {G.}~\bibnamefont {Faye}},\ }\bibfield
  {title} {\bibinfo {title} {{Ready-to-use post-Newtonian gravitational
  waveforms for binary black holes with nonprecessing spins: An update}},\
  }\href {https://doi.org/10.1103/PhysRevD.93.084054} {\bibfield  {journal}
  {\bibinfo  {journal} {Phys. Rev. D}\ }\textbf {\bibinfo {volume} {93}},\
  \bibinfo {pages} {084054} (\bibinfo {year} {2016})},\ \Eprint
  {https://arxiv.org/abs/1601.05588} {arXiv:1601.05588 [gr-qc]} \BibitemShut
  {NoStop}%
\bibitem [{\citenamefont {Isoyama}\ and\ \citenamefont
  {Nakano}(2018)}]{Isoyama:2017tbp}%
  \BibitemOpen
  \bibfield  {author} {\bibinfo {author} {\bibfnamefont {S.}~\bibnamefont
  {Isoyama}}\ and\ \bibinfo {author} {\bibfnamefont {H.}~\bibnamefont
  {Nakano}},\ }\bibfield  {title} {\bibinfo {title} {{Post-Newtonian templates
  for binary black-hole inspirals: the effect of the horizon fluxes and the
  secular change in the black-hole masses and spins}},\ }\href
  {https://doi.org/10.1088/1361-6382/aa96c5} {\bibfield  {journal} {\bibinfo
  {journal} {Class. Quant. Grav.}\ }\textbf {\bibinfo {volume} {35}},\ \bibinfo
  {pages} {024001} (\bibinfo {year} {2018})},\ \Eprint
  {https://arxiv.org/abs/1705.03869} {arXiv:1705.03869 [gr-qc]} \BibitemShut
  {NoStop}%
\bibitem [{\citenamefont {Moore}\ \emph {et~al.}(2016)\citenamefont {Moore},
  \citenamefont {Favata}, \citenamefont {Arun},\ and\ \citenamefont
  {Mishra}}]{Moore:2016qxz}%
  \BibitemOpen
  \bibfield  {author} {\bibinfo {author} {\bibfnamefont {B.}~\bibnamefont
  {Moore}}, \bibinfo {author} {\bibfnamefont {M.}~\bibnamefont {Favata}},
  \bibinfo {author} {\bibfnamefont {K.~G.}\ \bibnamefont {Arun}},\ and\
  \bibinfo {author} {\bibfnamefont {C.~K.}\ \bibnamefont {Mishra}},\ }\bibfield
   {title} {\bibinfo {title} {{Gravitational-wave phasing for low-eccentricity
  inspiralling compact binaries to 3PN order}},\ }\href
  {https://doi.org/10.1103/PhysRevD.93.124061} {\bibfield  {journal} {\bibinfo
  {journal} {Phys. Rev. D}\ }\textbf {\bibinfo {volume} {93}},\ \bibinfo
  {pages} {124061} (\bibinfo {year} {2016})},\ \Eprint
  {https://arxiv.org/abs/1605.00304} {arXiv:1605.00304 [gr-qc]} \BibitemShut
  {NoStop}%
\bibitem [{\citenamefont {Moore}\ \emph {et~al.}(2018)\citenamefont {Moore},
  \citenamefont {Robson}, \citenamefont {Loutrel},\ and\ \citenamefont
  {Yunes}}]{Moore:2018kvz}%
  \BibitemOpen
  \bibfield  {author} {\bibinfo {author} {\bibfnamefont {B.}~\bibnamefont
  {Moore}}, \bibinfo {author} {\bibfnamefont {T.}~\bibnamefont {Robson}},
  \bibinfo {author} {\bibfnamefont {N.}~\bibnamefont {Loutrel}},\ and\ \bibinfo
  {author} {\bibfnamefont {N.}~\bibnamefont {Yunes}},\ }\bibfield  {title}
  {\bibinfo {title} {{Towards a Fourier domain waveform for non-spinning
  binaries with arbitrary eccentricity}},\ }\href
  {https://doi.org/10.1088/1361-6382/aaea00} {\bibfield  {journal} {\bibinfo
  {journal} {Class. Quant. Grav.}\ }\textbf {\bibinfo {volume} {35}},\ \bibinfo
  {pages} {235006} (\bibinfo {year} {2018})},\ \Eprint
  {https://arxiv.org/abs/1807.07163} {arXiv:1807.07163 [gr-qc]} \BibitemShut
  {NoStop}%
\bibitem [{\citenamefont {Moore}\ and\ \citenamefont
  {Yunes}(2019)}]{Moore:2019xkm}%
  \BibitemOpen
  \bibfield  {author} {\bibinfo {author} {\bibfnamefont {B.}~\bibnamefont
  {Moore}}\ and\ \bibinfo {author} {\bibfnamefont {N.}~\bibnamefont {Yunes}},\
  }\bibfield  {title} {\bibinfo {title} {{A 3PN Fourier Domain Waveform for
  Non-Spinning Binaries with Moderate Eccentricity}},\ }\href
  {https://doi.org/10.1088/1361-6382/ab3778} {\bibfield  {journal} {\bibinfo
  {journal} {Class. Quant. Grav.}\ }\textbf {\bibinfo {volume} {36}},\ \bibinfo
  {pages} {185003} (\bibinfo {year} {2019})},\ \Eprint
  {https://arxiv.org/abs/1903.05203} {arXiv:1903.05203 [gr-qc]} \BibitemShut
  {NoStop}%
\bibitem [{\citenamefont {Blackman}\ \emph {et~al.}(2015)\citenamefont
  {Blackman}, \citenamefont {Field}, \citenamefont {Galley}, \citenamefont
  {Szil\'agyi}, \citenamefont {Scheel}, \citenamefont {Tiglio},\ and\
  \citenamefont {Hemberger}}]{Blackman:2015pia}%
  \BibitemOpen
  \bibfield  {author} {\bibinfo {author} {\bibfnamefont {J.}~\bibnamefont
  {Blackman}}, \bibinfo {author} {\bibfnamefont {S.~E.}\ \bibnamefont {Field}},
  \bibinfo {author} {\bibfnamefont {C.~R.}\ \bibnamefont {Galley}}, \bibinfo
  {author} {\bibfnamefont {B.}~\bibnamefont {Szil\'agyi}}, \bibinfo {author}
  {\bibfnamefont {M.~A.}\ \bibnamefont {Scheel}}, \bibinfo {author}
  {\bibfnamefont {M.}~\bibnamefont {Tiglio}},\ and\ \bibinfo {author}
  {\bibfnamefont {D.~A.}\ \bibnamefont {Hemberger}},\ }\bibfield  {title}
  {\bibinfo {title} {{Fast and Accurate Prediction of Numerical Relativity
  Waveforms from Binary Black Hole Coalescences Using Surrogate Models}},\
  }\href {https://doi.org/10.1103/PhysRevLett.115.121102} {\bibfield  {journal}
  {\bibinfo  {journal} {Phys. Rev. Lett.}\ }\textbf {\bibinfo {volume} {115}},\
  \bibinfo {pages} {121102} (\bibinfo {year} {2015})},\ \Eprint
  {https://arxiv.org/abs/1502.07758} {arXiv:1502.07758 [gr-qc]} \BibitemShut
  {NoStop}%
\bibitem [{\citenamefont {Blackman}\ \emph
  {et~al.}(2017{\natexlab{a}})\citenamefont {Blackman}, \citenamefont {Field},
  \citenamefont {Scheel}, \citenamefont {Galley}, \citenamefont {Hemberger},
  \citenamefont {Schmidt},\ and\ \citenamefont {Smith}}]{Blackman:2017dfb}%
  \BibitemOpen
  \bibfield  {author} {\bibinfo {author} {\bibfnamefont {J.}~\bibnamefont
  {Blackman}}, \bibinfo {author} {\bibfnamefont {S.~E.}\ \bibnamefont {Field}},
  \bibinfo {author} {\bibfnamefont {M.~A.}\ \bibnamefont {Scheel}}, \bibinfo
  {author} {\bibfnamefont {C.~R.}\ \bibnamefont {Galley}}, \bibinfo {author}
  {\bibfnamefont {D.~A.}\ \bibnamefont {Hemberger}}, \bibinfo {author}
  {\bibfnamefont {P.}~\bibnamefont {Schmidt}},\ and\ \bibinfo {author}
  {\bibfnamefont {R.}~\bibnamefont {Smith}},\ }\bibfield  {title} {\bibinfo
  {title} {{A Surrogate Model of Gravitational Waveforms from Numerical
  Relativity Simulations of Precessing Binary Black Hole Mergers}},\ }\href
  {https://doi.org/10.1103/PhysRevD.95.104023} {\bibfield  {journal} {\bibinfo
  {journal} {Phys. Rev. D}\ }\textbf {\bibinfo {volume} {95}},\ \bibinfo
  {pages} {104023} (\bibinfo {year} {2017}{\natexlab{a}})},\ \Eprint
  {https://arxiv.org/abs/1701.00550} {arXiv:1701.00550 [gr-qc]} \BibitemShut
  {NoStop}%
\bibitem [{\citenamefont {Blackman}\ \emph
  {et~al.}(2017{\natexlab{b}})\citenamefont {Blackman}, \citenamefont {Field},
  \citenamefont {Scheel}, \citenamefont {Galley}, \citenamefont {Ott},
  \citenamefont {Boyle}, \citenamefont {Kidder}, \citenamefont {Pfeiffer},\
  and\ \citenamefont {Szil\'agyi}}]{Blackman:2017pcm}%
  \BibitemOpen
  \bibfield  {author} {\bibinfo {author} {\bibfnamefont {J.}~\bibnamefont
  {Blackman}}, \bibinfo {author} {\bibfnamefont {S.~E.}\ \bibnamefont {Field}},
  \bibinfo {author} {\bibfnamefont {M.~A.}\ \bibnamefont {Scheel}}, \bibinfo
  {author} {\bibfnamefont {C.~R.}\ \bibnamefont {Galley}}, \bibinfo {author}
  {\bibfnamefont {C.~D.}\ \bibnamefont {Ott}}, \bibinfo {author} {\bibfnamefont
  {M.}~\bibnamefont {Boyle}}, \bibinfo {author} {\bibfnamefont {L.~E.}\
  \bibnamefont {Kidder}}, \bibinfo {author} {\bibfnamefont {H.~P.}\
  \bibnamefont {Pfeiffer}},\ and\ \bibinfo {author} {\bibfnamefont
  {B.}~\bibnamefont {Szil\'agyi}},\ }\bibfield  {title} {\bibinfo {title}
  {{Numerical relativity waveform surrogate model for generically precessing
  binary black hole mergers}},\ }\href
  {https://doi.org/10.1103/PhysRevD.96.024058} {\bibfield  {journal} {\bibinfo
  {journal} {Phys. Rev. D}\ }\textbf {\bibinfo {volume} {96}},\ \bibinfo
  {pages} {024058} (\bibinfo {year} {2017}{\natexlab{b}})},\ \Eprint
  {https://arxiv.org/abs/1705.07089} {arXiv:1705.07089 [gr-qc]} \BibitemShut
  {NoStop}%
\bibitem [{\citenamefont {Varma}\ \emph
  {et~al.}(2019{\natexlab{a}})\citenamefont {Varma}, \citenamefont {Field},
  \citenamefont {Scheel}, \citenamefont {Blackman}, \citenamefont {Kidder},\
  and\ \citenamefont {Pfeiffer}}]{Varma:2018mmi}%
  \BibitemOpen
  \bibfield  {author} {\bibinfo {author} {\bibfnamefont {V.}~\bibnamefont
  {Varma}}, \bibinfo {author} {\bibfnamefont {S.~E.}\ \bibnamefont {Field}},
  \bibinfo {author} {\bibfnamefont {M.~A.}\ \bibnamefont {Scheel}}, \bibinfo
  {author} {\bibfnamefont {J.}~\bibnamefont {Blackman}}, \bibinfo {author}
  {\bibfnamefont {L.~E.}\ \bibnamefont {Kidder}},\ and\ \bibinfo {author}
  {\bibfnamefont {H.~P.}\ \bibnamefont {Pfeiffer}},\ }\bibfield  {title}
  {\bibinfo {title} {{Surrogate model of hybridized numerical relativity binary
  black hole waveforms}},\ }\href {https://doi.org/10.1103/PhysRevD.99.064045}
  {\bibfield  {journal} {\bibinfo  {journal} {Phys. Rev. D}\ }\textbf {\bibinfo
  {volume} {99}},\ \bibinfo {pages} {064045} (\bibinfo {year}
  {2019}{\natexlab{a}})},\ \Eprint {https://arxiv.org/abs/1812.07865}
  {arXiv:1812.07865 [gr-qc]} \BibitemShut {NoStop}%
\bibitem [{\citenamefont {Varma}\ \emph
  {et~al.}(2019{\natexlab{b}})\citenamefont {Varma}, \citenamefont {Field},
  \citenamefont {Scheel}, \citenamefont {Blackman}, \citenamefont {Gerosa},
  \citenamefont {Stein}, \citenamefont {Kidder},\ and\ \citenamefont
  {Pfeiffer}}]{Varma:2019csw}%
  \BibitemOpen
  \bibfield  {author} {\bibinfo {author} {\bibfnamefont {V.}~\bibnamefont
  {Varma}}, \bibinfo {author} {\bibfnamefont {S.~E.}\ \bibnamefont {Field}},
  \bibinfo {author} {\bibfnamefont {M.~A.}\ \bibnamefont {Scheel}}, \bibinfo
  {author} {\bibfnamefont {J.}~\bibnamefont {Blackman}}, \bibinfo {author}
  {\bibfnamefont {D.}~\bibnamefont {Gerosa}}, \bibinfo {author} {\bibfnamefont
  {L.~C.}\ \bibnamefont {Stein}}, \bibinfo {author} {\bibfnamefont {L.~E.}\
  \bibnamefont {Kidder}},\ and\ \bibinfo {author} {\bibfnamefont {H.~P.}\
  \bibnamefont {Pfeiffer}},\ }\bibfield  {title} {\bibinfo {title} {{Surrogate
  models for precessing binary black hole simulations with unequal masses}},\
  }\href {https://doi.org/10.1103/PhysRevResearch.1.033015} {\bibfield
  {journal} {\bibinfo  {journal} {Phys. Rev. Research.}\ }\textbf {\bibinfo
  {volume} {1}},\ \bibinfo {pages} {033015} (\bibinfo {year}
  {2019}{\natexlab{b}})},\ \Eprint {https://arxiv.org/abs/1905.09300}
  {arXiv:1905.09300 [gr-qc]} \BibitemShut {NoStop}%
\bibitem [{\citenamefont {Williams}\ \emph {et~al.}(2020)\citenamefont
  {Williams}, \citenamefont {Heng}, \citenamefont {Gair}, \citenamefont
  {Clark},\ and\ \citenamefont {Khamesra}}]{Williams:2019vub}%
  \BibitemOpen
  \bibfield  {author} {\bibinfo {author} {\bibfnamefont {D.}~\bibnamefont
  {Williams}}, \bibinfo {author} {\bibfnamefont {I.~S.}\ \bibnamefont {Heng}},
  \bibinfo {author} {\bibfnamefont {J.}~\bibnamefont {Gair}}, \bibinfo {author}
  {\bibfnamefont {J.~A.}\ \bibnamefont {Clark}},\ and\ \bibinfo {author}
  {\bibfnamefont {B.}~\bibnamefont {Khamesra}},\ }\bibfield  {title} {\bibinfo
  {title} {{Precessing numerical relativity waveform surrogate model for binary
  black holes: A Gaussian process regression approach}},\ }\href
  {https://doi.org/10.1103/PhysRevD.101.063011} {\bibfield  {journal} {\bibinfo
   {journal} {Phys. Rev. D}\ }\textbf {\bibinfo {volume} {101}},\ \bibinfo
  {pages} {063011} (\bibinfo {year} {2020})},\ \Eprint
  {https://arxiv.org/abs/1903.09204} {arXiv:1903.09204 [gr-qc]} \BibitemShut
  {NoStop}%
\bibitem [{\citenamefont {Rifat}\ \emph {et~al.}(2020)\citenamefont {Rifat},
  \citenamefont {Field}, \citenamefont {Khanna},\ and\ \citenamefont
  {Varma}}]{Rifat:2019ltp}%
  \BibitemOpen
  \bibfield  {author} {\bibinfo {author} {\bibfnamefont {N.~E.~M.}\
  \bibnamefont {Rifat}}, \bibinfo {author} {\bibfnamefont {S.~E.}\ \bibnamefont
  {Field}}, \bibinfo {author} {\bibfnamefont {G.}~\bibnamefont {Khanna}},\ and\
  \bibinfo {author} {\bibfnamefont {V.}~\bibnamefont {Varma}},\ }\bibfield
  {title} {\bibinfo {title} {{Surrogate model for gravitational wave signals
  from comparable and large-mass-ratio black hole binaries}},\ }\href
  {https://doi.org/10.1103/PhysRevD.101.081502} {\bibfield  {journal} {\bibinfo
   {journal} {Phys. Rev. D}\ }\textbf {\bibinfo {volume} {101}},\ \bibinfo
  {pages} {081502} (\bibinfo {year} {2020})},\ \Eprint
  {https://arxiv.org/abs/1910.10473} {arXiv:1910.10473 [gr-qc]} \BibitemShut
  {NoStop}%
\bibitem [{\citenamefont {Islam}\ \emph {et~al.}(2021)\citenamefont {Islam},
  \citenamefont {Varma}, \citenamefont {Lodman}, \citenamefont {Field},
  \citenamefont {Khanna}, \citenamefont {Scheel}, \citenamefont {Pfeiffer},
  \citenamefont {Gerosa},\ and\ \citenamefont {Kidder}}]{Islam:2021mha}%
  \BibitemOpen
  \bibfield  {author} {\bibinfo {author} {\bibfnamefont {T.}~\bibnamefont
  {Islam}}, \bibinfo {author} {\bibfnamefont {V.}~\bibnamefont {Varma}},
  \bibinfo {author} {\bibfnamefont {J.}~\bibnamefont {Lodman}}, \bibinfo
  {author} {\bibfnamefont {S.~E.}\ \bibnamefont {Field}}, \bibinfo {author}
  {\bibfnamefont {G.}~\bibnamefont {Khanna}}, \bibinfo {author} {\bibfnamefont
  {M.~A.}\ \bibnamefont {Scheel}}, \bibinfo {author} {\bibfnamefont {H.~P.}\
  \bibnamefont {Pfeiffer}}, \bibinfo {author} {\bibfnamefont {D.}~\bibnamefont
  {Gerosa}},\ and\ \bibinfo {author} {\bibfnamefont {L.~E.}\ \bibnamefont
  {Kidder}},\ }\bibfield  {title} {\bibinfo {title} {{Eccentric binary black
  hole surrogate models for the gravitational waveform and remnant properties:
  comparable mass, nonspinning case}},\ }\href
  {https://doi.org/10.1103/PhysRevD.103.064022} {\bibfield  {journal} {\bibinfo
   {journal} {Phys. Rev. D}\ }\textbf {\bibinfo {volume} {103}},\ \bibinfo
  {pages} {064022} (\bibinfo {year} {2021})},\ \Eprint
  {https://arxiv.org/abs/2101.11798} {arXiv:2101.11798 [gr-qc]} \BibitemShut
  {NoStop}%
\bibitem [{\citenamefont {Islam}\ \emph {et~al.}(2022)\citenamefont {Islam},
  \citenamefont {Field}, \citenamefont {Hughes}, \citenamefont {Khanna},
  \citenamefont {Varma}, \citenamefont {Giesler}, \citenamefont {Scheel},
  \citenamefont {Kidder},\ and\ \citenamefont {Pfeiffer}}]{Islam:2022laz}%
  \BibitemOpen
  \bibfield  {author} {\bibinfo {author} {\bibfnamefont {T.}~\bibnamefont
  {Islam}}, \bibinfo {author} {\bibfnamefont {S.~E.}\ \bibnamefont {Field}},
  \bibinfo {author} {\bibfnamefont {S.~A.}\ \bibnamefont {Hughes}}, \bibinfo
  {author} {\bibfnamefont {G.}~\bibnamefont {Khanna}}, \bibinfo {author}
  {\bibfnamefont {V.}~\bibnamefont {Varma}}, \bibinfo {author} {\bibfnamefont
  {M.}~\bibnamefont {Giesler}}, \bibinfo {author} {\bibfnamefont {M.~A.}\
  \bibnamefont {Scheel}}, \bibinfo {author} {\bibfnamefont {L.~E.}\
  \bibnamefont {Kidder}},\ and\ \bibinfo {author} {\bibfnamefont {H.~P.}\
  \bibnamefont {Pfeiffer}},\ }\bibfield  {title} {\bibinfo {title} {{Surrogate
  model for gravitational wave signals from nonspinning, comparable-to
  large-mass-ratio black hole binaries built on black hole perturbation theory
  waveforms calibrated to numerical relativity}},\ }\href
  {https://doi.org/10.1103/PhysRevD.106.104025} {\bibfield  {journal} {\bibinfo
   {journal} {Phys. Rev. D}\ }\textbf {\bibinfo {volume} {106}},\ \bibinfo
  {pages} {104025} (\bibinfo {year} {2022})},\ \Eprint
  {https://arxiv.org/abs/2204.01972} {arXiv:2204.01972 [gr-qc]} \BibitemShut
  {NoStop}%
\bibitem [{\citenamefont {Gonzalez}\ \emph {et~al.}(2009)\citenamefont
  {Gonzalez}, \citenamefont {Sperhake},\ and\ \citenamefont
  {Bruegmann}}]{Gonzalez:2008bi}%
  \BibitemOpen
  \bibfield  {author} {\bibinfo {author} {\bibfnamefont {J.~A.}\ \bibnamefont
  {Gonzalez}}, \bibinfo {author} {\bibfnamefont {U.}~\bibnamefont {Sperhake}},\
  and\ \bibinfo {author} {\bibfnamefont {B.}~\bibnamefont {Bruegmann}},\
  }\bibfield  {title} {\bibinfo {title} {{Black-hole binary simulations: The
  Mass ratio 10:1}},\ }\href {https://doi.org/10.1103/PhysRevD.79.124006}
  {\bibfield  {journal} {\bibinfo  {journal} {Phys. Rev. D}\ }\textbf {\bibinfo
  {volume} {79}},\ \bibinfo {pages} {124006} (\bibinfo {year} {2009})},\
  \Eprint {https://arxiv.org/abs/0811.3952} {arXiv:0811.3952 [gr-qc]}
  \BibitemShut {NoStop}%
\bibitem [{\citenamefont {Buchman}\ \emph {et~al.}(2012)\citenamefont
  {Buchman}, \citenamefont {Pfeiffer}, \citenamefont {Scheel},\ and\
  \citenamefont {Szilagyi}}]{Buchman:2012dw}%
  \BibitemOpen
  \bibfield  {author} {\bibinfo {author} {\bibfnamefont {L.~T.}\ \bibnamefont
  {Buchman}}, \bibinfo {author} {\bibfnamefont {H.~P.}\ \bibnamefont
  {Pfeiffer}}, \bibinfo {author} {\bibfnamefont {M.~A.}\ \bibnamefont
  {Scheel}},\ and\ \bibinfo {author} {\bibfnamefont {B.}~\bibnamefont
  {Szilagyi}},\ }\bibfield  {title} {\bibinfo {title} {{Simulations of
  non-equal mass black hole binaries with spectral methods}},\ }\href
  {https://doi.org/10.1103/PhysRevD.86.084033} {\bibfield  {journal} {\bibinfo
  {journal} {Phys. Rev. D}\ }\textbf {\bibinfo {volume} {86}},\ \bibinfo
  {pages} {084033} (\bibinfo {year} {2012})},\ \Eprint
  {https://arxiv.org/abs/1206.3015} {arXiv:1206.3015 [gr-qc]} \BibitemShut
  {NoStop}%
\bibitem [{\citenamefont {Chu}\ \emph {et~al.}(2009)\citenamefont {Chu},
  \citenamefont {Pfeiffer},\ and\ \citenamefont {Scheel}}]{Chu:2009md}%
  \BibitemOpen
  \bibfield  {author} {\bibinfo {author} {\bibfnamefont {T.}~\bibnamefont
  {Chu}}, \bibinfo {author} {\bibfnamefont {H.~P.}\ \bibnamefont {Pfeiffer}},\
  and\ \bibinfo {author} {\bibfnamefont {M.~A.}\ \bibnamefont {Scheel}},\
  }\bibfield  {title} {\bibinfo {title} {{High accuracy simulations of black
  hole binaries: Spins anti-aligned with the orbital angular momentum}},\
  }\href {https://doi.org/10.1103/PhysRevD.80.124051} {\bibfield  {journal}
  {\bibinfo  {journal} {Phys. Rev. D}\ }\textbf {\bibinfo {volume} {80}},\
  \bibinfo {pages} {124051} (\bibinfo {year} {2009})},\ \Eprint
  {https://arxiv.org/abs/0909.1313} {arXiv:0909.1313 [gr-qc]} \BibitemShut
  {NoStop}%
\bibitem [{\citenamefont {Mroue}\ \emph {et~al.}(2013)\citenamefont {Mroue}
  \emph {et~al.}}]{Mroue:2013xna}%
  \BibitemOpen
  \bibfield  {author} {\bibinfo {author} {\bibfnamefont {A.~H.}\ \bibnamefont
  {Mroue}} \emph {et~al.},\ }\bibfield  {title} {\bibinfo {title} {{Catalog of
  174 Binary Black Hole Simulations for Gravitational Wave Astronomy}},\ }\href
  {https://doi.org/10.1103/PhysRevLett.111.241104} {\bibfield  {journal}
  {\bibinfo  {journal} {Phys. Rev. Lett.}\ }\textbf {\bibinfo {volume} {111}},\
  \bibinfo {pages} {241104} (\bibinfo {year} {2013})},\ \Eprint
  {https://arxiv.org/abs/1304.6077} {arXiv:1304.6077 [gr-qc]} \BibitemShut
  {NoStop}%
\bibitem [{\citenamefont {Boyle}\ \emph {et~al.}(2019)\citenamefont {Boyle}
  \emph {et~al.}}]{Boyle:2019kee}%
  \BibitemOpen
  \bibfield  {author} {\bibinfo {author} {\bibfnamefont {M.}~\bibnamefont
  {Boyle}} \emph {et~al.},\ }\bibfield  {title} {\bibinfo {title} {{The SXS
  Collaboration catalog of binary black hole simulations}},\ }\href
  {https://doi.org/10.1088/1361-6382/ab34e2} {\bibfield  {journal} {\bibinfo
  {journal} {Class. Quant. Grav.}\ }\textbf {\bibinfo {volume} {36}},\ \bibinfo
  {pages} {195006} (\bibinfo {year} {2019})},\ \Eprint
  {https://arxiv.org/abs/1904.04831} {arXiv:1904.04831 [gr-qc]} \BibitemShut
  {NoStop}%
\bibitem [{\citenamefont {Healy}\ \emph {et~al.}(2017)\citenamefont {Healy},
  \citenamefont {Lousto}, \citenamefont {Zlochower},\ and\ \citenamefont
  {Campanelli}}]{Healy:2017psd}%
  \BibitemOpen
  \bibfield  {author} {\bibinfo {author} {\bibfnamefont {J.}~\bibnamefont
  {Healy}}, \bibinfo {author} {\bibfnamefont {C.~O.}\ \bibnamefont {Lousto}},
  \bibinfo {author} {\bibfnamefont {Y.}~\bibnamefont {Zlochower}},\ and\
  \bibinfo {author} {\bibfnamefont {M.}~\bibnamefont {Campanelli}},\ }\bibfield
   {title} {\bibinfo {title} {{The RIT binary black hole simulations
  catalog}},\ }\href {https://doi.org/10.1088/1361-6382/aa91b1} {\bibfield
  {journal} {\bibinfo  {journal} {Class. Quant. Grav.}\ }\textbf {\bibinfo
  {volume} {34}},\ \bibinfo {pages} {224001} (\bibinfo {year} {2017})},\
  \Eprint {https://arxiv.org/abs/1703.03423} {arXiv:1703.03423 [gr-qc]}
  \BibitemShut {NoStop}%
\bibitem [{\citenamefont {Healy}\ and\ \citenamefont
  {Lousto}(2022)}]{Healy:2022wdn}%
  \BibitemOpen
  \bibfield  {author} {\bibinfo {author} {\bibfnamefont {J.}~\bibnamefont
  {Healy}}\ and\ \bibinfo {author} {\bibfnamefont {C.~O.}\ \bibnamefont
  {Lousto}},\ }\bibfield  {title} {\bibinfo {title} {{Fourth RIT binary black
  hole simulations catalog: Extension to eccentric orbits}},\ }\href
  {https://doi.org/10.1103/PhysRevD.105.124010} {\bibfield  {journal} {\bibinfo
   {journal} {Phys. Rev. D}\ }\textbf {\bibinfo {volume} {105}},\ \bibinfo
  {pages} {124010} (\bibinfo {year} {2022})},\ \Eprint
  {https://arxiv.org/abs/2202.00018} {arXiv:2202.00018 [gr-qc]} \BibitemShut
  {NoStop}%
\bibitem [{\citenamefont {Jani}\ \emph {et~al.}(2016)\citenamefont {Jani},
  \citenamefont {Healy}, \citenamefont {Clark}, \citenamefont {London},
  \citenamefont {Laguna},\ and\ \citenamefont {Shoemaker}}]{Jani:2016wkt}%
  \BibitemOpen
  \bibfield  {author} {\bibinfo {author} {\bibfnamefont {K.}~\bibnamefont
  {Jani}}, \bibinfo {author} {\bibfnamefont {J.}~\bibnamefont {Healy}},
  \bibinfo {author} {\bibfnamefont {J.~A.}\ \bibnamefont {Clark}}, \bibinfo
  {author} {\bibfnamefont {L.}~\bibnamefont {London}}, \bibinfo {author}
  {\bibfnamefont {P.}~\bibnamefont {Laguna}},\ and\ \bibinfo {author}
  {\bibfnamefont {D.}~\bibnamefont {Shoemaker}},\ }\bibfield  {title} {\bibinfo
  {title} {{Georgia Tech Catalog of Gravitational Waveforms}},\ }\href
  {https://doi.org/10.1088/0264-9381/33/20/204001} {\bibfield  {journal}
  {\bibinfo  {journal} {Class. Quant. Grav.}\ }\textbf {\bibinfo {volume}
  {33}},\ \bibinfo {pages} {204001} (\bibinfo {year} {2016})},\ \Eprint
  {https://arxiv.org/abs/1605.03204} {arXiv:1605.03204 [gr-qc]} \BibitemShut
  {NoStop}%
\bibitem [{\citenamefont {Foucart}\ \emph {et~al.}(2019)\citenamefont {Foucart}
  \emph {et~al.}}]{Foucart:2018lhe}%
  \BibitemOpen
  \bibfield  {author} {\bibinfo {author} {\bibfnamefont {F.}~\bibnamefont
  {Foucart}} \emph {et~al.},\ }\bibfield  {title} {\bibinfo {title}
  {{Gravitational waveforms from spectral Einstein code simulations: Neutron
  star-neutron star and low-mass black hole-neutron star binaries}},\ }\href
  {https://doi.org/10.1103/PhysRevD.99.044008} {\bibfield  {journal} {\bibinfo
  {journal} {Phys. Rev. D}\ }\textbf {\bibinfo {volume} {99}},\ \bibinfo
  {pages} {044008} (\bibinfo {year} {2019})},\ \Eprint
  {https://arxiv.org/abs/1812.06988} {arXiv:1812.06988 [gr-qc]} \BibitemShut
  {NoStop}%
\bibitem [{\citenamefont {Pan}\ \emph {et~al.}(2008)\citenamefont {Pan},
  \citenamefont {Buonanno}, \citenamefont {Baker}, \citenamefont {Centrella},
  \citenamefont {Kelly}, \citenamefont {McWilliams}, \citenamefont
  {Pretorius},\ and\ \citenamefont {van Meter}}]{Pan:2007nw}%
  \BibitemOpen
  \bibfield  {author} {\bibinfo {author} {\bibfnamefont {Y.}~\bibnamefont
  {Pan}}, \bibinfo {author} {\bibfnamefont {A.}~\bibnamefont {Buonanno}},
  \bibinfo {author} {\bibfnamefont {J.~G.}\ \bibnamefont {Baker}}, \bibinfo
  {author} {\bibfnamefont {J.}~\bibnamefont {Centrella}}, \bibinfo {author}
  {\bibfnamefont {B.~J.}\ \bibnamefont {Kelly}}, \bibinfo {author}
  {\bibfnamefont {S.~T.}\ \bibnamefont {McWilliams}}, \bibinfo {author}
  {\bibfnamefont {F.}~\bibnamefont {Pretorius}},\ and\ \bibinfo {author}
  {\bibfnamefont {J.~R.}\ \bibnamefont {van Meter}},\ }\bibfield  {title}
  {\bibinfo {title} {{A Data-analysis driven comparison of analytic and
  numerical coalescing binary waveforms: Nonspinning case}},\ }\href
  {https://doi.org/10.1103/PhysRevD.77.024014} {\bibfield  {journal} {\bibinfo
  {journal} {Phys. Rev. D}\ }\textbf {\bibinfo {volume} {77}},\ \bibinfo
  {pages} {024014} (\bibinfo {year} {2008})},\ \Eprint
  {https://arxiv.org/abs/0704.1964} {arXiv:0704.1964 [gr-qc]} \BibitemShut
  {NoStop}%
\bibitem [{\citenamefont {Ajith}\ \emph {et~al.}(2007)\citenamefont {Ajith}
  \emph {et~al.}}]{Ajith:2007qp}%
  \BibitemOpen
  \bibfield  {author} {\bibinfo {author} {\bibfnamefont {P.}~\bibnamefont
  {Ajith}} \emph {et~al.},\ }\bibfield  {title} {\bibinfo {title}
  {{Phenomenological template family for black-hole coalescence waveforms}},\
  }\href {https://doi.org/10.1088/0264-9381/24/19/S31} {\bibfield  {journal}
  {\bibinfo  {journal} {Class. Quant. Grav.}\ }\textbf {\bibinfo {volume}
  {24}},\ \bibinfo {pages} {S689} (\bibinfo {year} {2007})},\ \Eprint
  {https://arxiv.org/abs/0704.3764} {arXiv:0704.3764 [gr-qc]} \BibitemShut
  {NoStop}%
\bibitem [{\citenamefont {Ajith}\ \emph {et~al.}(2011)\citenamefont {Ajith}
  \emph {et~al.}}]{Ajith:2009bn}%
  \BibitemOpen
  \bibfield  {author} {\bibinfo {author} {\bibfnamefont {P.}~\bibnamefont
  {Ajith}} \emph {et~al.},\ }\bibfield  {title} {\bibinfo {title}
  {{Inspiral-merger-ringdown waveforms for black-hole binaries with
  non-precessing spins}},\ }\href
  {https://doi.org/10.1103/PhysRevLett.106.241101} {\bibfield  {journal}
  {\bibinfo  {journal} {Phys. Rev. Lett.}\ }\textbf {\bibinfo {volume} {106}},\
  \bibinfo {pages} {241101} (\bibinfo {year} {2011})},\ \Eprint
  {https://arxiv.org/abs/0909.2867} {arXiv:0909.2867 [gr-qc]} \BibitemShut
  {NoStop}%
\bibitem [{\citenamefont {Santamaria}\ \emph {et~al.}(2010)\citenamefont
  {Santamaria} \emph {et~al.}}]{Santamaria:2010yb}%
  \BibitemOpen
  \bibfield  {author} {\bibinfo {author} {\bibfnamefont {L.}~\bibnamefont
  {Santamaria}} \emph {et~al.},\ }\bibfield  {title} {\bibinfo {title}
  {{Matching post-Newtonian and numerical relativity waveforms: systematic
  errors and a new phenomenological model for non-precessing black hole
  binaries}},\ }\href {https://doi.org/10.1103/PhysRevD.82.064016} {\bibfield
  {journal} {\bibinfo  {journal} {Phys. Rev. D}\ }\textbf {\bibinfo {volume}
  {82}},\ \bibinfo {pages} {064016} (\bibinfo {year} {2010})},\ \Eprint
  {https://arxiv.org/abs/1005.3306} {arXiv:1005.3306 [gr-qc]} \BibitemShut
  {NoStop}%
\bibitem [{\citenamefont {Hannam}\ \emph {et~al.}(2014)\citenamefont {Hannam},
  \citenamefont {Schmidt}, \citenamefont {Boh\'e}, \citenamefont {Haegel},
  \citenamefont {Husa}, \citenamefont {Ohme}, \citenamefont {Pratten},\ and\
  \citenamefont {P\"urrer}}]{Hannam:2013oca}%
  \BibitemOpen
  \bibfield  {author} {\bibinfo {author} {\bibfnamefont {M.}~\bibnamefont
  {Hannam}}, \bibinfo {author} {\bibfnamefont {P.}~\bibnamefont {Schmidt}},
  \bibinfo {author} {\bibfnamefont {A.}~\bibnamefont {Boh\'e}}, \bibinfo
  {author} {\bibfnamefont {L.}~\bibnamefont {Haegel}}, \bibinfo {author}
  {\bibfnamefont {S.}~\bibnamefont {Husa}}, \bibinfo {author} {\bibfnamefont
  {F.}~\bibnamefont {Ohme}}, \bibinfo {author} {\bibfnamefont {G.}~\bibnamefont
  {Pratten}},\ and\ \bibinfo {author} {\bibfnamefont {M.}~\bibnamefont
  {P\"urrer}},\ }\bibfield  {title} {\bibinfo {title} {{Simple Model of
  Complete Precessing Black-Hole-Binary Gravitational Waveforms}},\ }\href
  {https://doi.org/10.1103/PhysRevLett.113.151101} {\bibfield  {journal}
  {\bibinfo  {journal} {Phys. Rev. Lett.}\ }\textbf {\bibinfo {volume} {113}},\
  \bibinfo {pages} {151101} (\bibinfo {year} {2014})},\ \Eprint
  {https://arxiv.org/abs/1308.3271} {arXiv:1308.3271 [gr-qc]} \BibitemShut
  {NoStop}%
\bibitem [{\citenamefont {Husa}\ \emph {et~al.}(2016)\citenamefont {Husa},
  \citenamefont {Khan}, \citenamefont {Hannam}, \citenamefont {P\"urrer},
  \citenamefont {Ohme}, \citenamefont {Jim\'enez~Forteza},\ and\ \citenamefont
  {Boh\'e}}]{Husa:2015iqa}%
  \BibitemOpen
  \bibfield  {author} {\bibinfo {author} {\bibfnamefont {S.}~\bibnamefont
  {Husa}}, \bibinfo {author} {\bibfnamefont {S.}~\bibnamefont {Khan}}, \bibinfo
  {author} {\bibfnamefont {M.}~\bibnamefont {Hannam}}, \bibinfo {author}
  {\bibfnamefont {M.}~\bibnamefont {P\"urrer}}, \bibinfo {author}
  {\bibfnamefont {F.}~\bibnamefont {Ohme}}, \bibinfo {author} {\bibfnamefont
  {X.}~\bibnamefont {Jim\'enez~Forteza}},\ and\ \bibinfo {author}
  {\bibfnamefont {A.}~\bibnamefont {Boh\'e}},\ }\bibfield  {title} {\bibinfo
  {title} {{Frequency-domain gravitational waves from nonprecessing black-hole
  binaries. I. New numerical waveforms and anatomy of the signal}},\ }\href
  {https://doi.org/10.1103/PhysRevD.93.044006} {\bibfield  {journal} {\bibinfo
  {journal} {Phys. Rev. D}\ }\textbf {\bibinfo {volume} {93}},\ \bibinfo
  {pages} {044006} (\bibinfo {year} {2016})},\ \Eprint
  {https://arxiv.org/abs/1508.07250} {arXiv:1508.07250 [gr-qc]} \BibitemShut
  {NoStop}%
\bibitem [{\citenamefont {Khan}\ \emph {et~al.}(2016)\citenamefont {Khan},
  \citenamefont {Husa}, \citenamefont {Hannam}, \citenamefont {Ohme},
  \citenamefont {P\"urrer}, \citenamefont {Jim\'enez~Forteza},\ and\
  \citenamefont {Boh\'e}}]{Khan:2015jqa}%
  \BibitemOpen
  \bibfield  {author} {\bibinfo {author} {\bibfnamefont {S.}~\bibnamefont
  {Khan}}, \bibinfo {author} {\bibfnamefont {S.}~\bibnamefont {Husa}}, \bibinfo
  {author} {\bibfnamefont {M.}~\bibnamefont {Hannam}}, \bibinfo {author}
  {\bibfnamefont {F.}~\bibnamefont {Ohme}}, \bibinfo {author} {\bibfnamefont
  {M.}~\bibnamefont {P\"urrer}}, \bibinfo {author} {\bibfnamefont
  {X.}~\bibnamefont {Jim\'enez~Forteza}},\ and\ \bibinfo {author}
  {\bibfnamefont {A.}~\bibnamefont {Boh\'e}},\ }\bibfield  {title} {\bibinfo
  {title} {{Frequency-domain gravitational waves from nonprecessing black-hole
  binaries. II. A phenomenological model for the advanced detector era}},\
  }\href {https://doi.org/10.1103/PhysRevD.93.044007} {\bibfield  {journal}
  {\bibinfo  {journal} {Phys. Rev. D}\ }\textbf {\bibinfo {volume} {93}},\
  \bibinfo {pages} {044007} (\bibinfo {year} {2016})},\ \Eprint
  {https://arxiv.org/abs/1508.07253} {arXiv:1508.07253 [gr-qc]} \BibitemShut
  {NoStop}%
\bibitem [{\citenamefont {London}\ \emph {et~al.}(2018)\citenamefont {London},
  \citenamefont {Khan}, \citenamefont {Fauchon-Jones}, \citenamefont
  {Garc\'\i{}a}, \citenamefont {Hannam}, \citenamefont {Husa}, \citenamefont
  {Jim\'enez-Forteza}, \citenamefont {Kalaghatgi}, \citenamefont {Ohme},\ and\
  \citenamefont {Pannarale}}]{London:2017bcn}%
  \BibitemOpen
  \bibfield  {author} {\bibinfo {author} {\bibfnamefont {L.}~\bibnamefont
  {London}}, \bibinfo {author} {\bibfnamefont {S.}~\bibnamefont {Khan}},
  \bibinfo {author} {\bibfnamefont {E.}~\bibnamefont {Fauchon-Jones}}, \bibinfo
  {author} {\bibfnamefont {C.}~\bibnamefont {Garc\'\i{}a}}, \bibinfo {author}
  {\bibfnamefont {M.}~\bibnamefont {Hannam}}, \bibinfo {author} {\bibfnamefont
  {S.}~\bibnamefont {Husa}}, \bibinfo {author} {\bibfnamefont {X.}~\bibnamefont
  {Jim\'enez-Forteza}}, \bibinfo {author} {\bibfnamefont {C.}~\bibnamefont
  {Kalaghatgi}}, \bibinfo {author} {\bibfnamefont {F.}~\bibnamefont {Ohme}},\
  and\ \bibinfo {author} {\bibfnamefont {F.}~\bibnamefont {Pannarale}},\
  }\bibfield  {title} {\bibinfo {title} {{First higher-multipole model of
  gravitational waves from spinning and coalescing black-hole binaries}},\
  }\href {https://doi.org/10.1103/PhysRevLett.120.161102} {\bibfield  {journal}
  {\bibinfo  {journal} {Phys. Rev. Lett.}\ }\textbf {\bibinfo {volume} {120}},\
  \bibinfo {pages} {161102} (\bibinfo {year} {2018})},\ \Eprint
  {https://arxiv.org/abs/1708.00404} {arXiv:1708.00404 [gr-qc]} \BibitemShut
  {NoStop}%
\bibitem [{\citenamefont {Khan}\ \emph {et~al.}(2019)\citenamefont {Khan},
  \citenamefont {Chatziioannou}, \citenamefont {Hannam},\ and\ \citenamefont
  {Ohme}}]{Khan:2018fmp}%
  \BibitemOpen
  \bibfield  {author} {\bibinfo {author} {\bibfnamefont {S.}~\bibnamefont
  {Khan}}, \bibinfo {author} {\bibfnamefont {K.}~\bibnamefont {Chatziioannou}},
  \bibinfo {author} {\bibfnamefont {M.}~\bibnamefont {Hannam}},\ and\ \bibinfo
  {author} {\bibfnamefont {F.}~\bibnamefont {Ohme}},\ }\bibfield  {title}
  {\bibinfo {title} {{Phenomenological model for the gravitational-wave signal
  from precessing binary black holes with two-spin effects}},\ }\href
  {https://doi.org/10.1103/PhysRevD.100.024059} {\bibfield  {journal} {\bibinfo
   {journal} {Phys. Rev. D}\ }\textbf {\bibinfo {volume} {100}},\ \bibinfo
  {pages} {024059} (\bibinfo {year} {2019})},\ \Eprint
  {https://arxiv.org/abs/1809.10113} {arXiv:1809.10113 [gr-qc]} \BibitemShut
  {NoStop}%
\bibitem [{\citenamefont {Khan}\ \emph {et~al.}(2020)\citenamefont {Khan},
  \citenamefont {Ohme}, \citenamefont {Chatziioannou},\ and\ \citenamefont
  {Hannam}}]{Khan:2019kot}%
  \BibitemOpen
  \bibfield  {author} {\bibinfo {author} {\bibfnamefont {S.}~\bibnamefont
  {Khan}}, \bibinfo {author} {\bibfnamefont {F.}~\bibnamefont {Ohme}}, \bibinfo
  {author} {\bibfnamefont {K.}~\bibnamefont {Chatziioannou}},\ and\ \bibinfo
  {author} {\bibfnamefont {M.}~\bibnamefont {Hannam}},\ }\bibfield  {title}
  {\bibinfo {title} {{Including higher order multipoles in gravitational-wave
  models for precessing binary black holes}},\ }\href
  {https://doi.org/10.1103/PhysRevD.101.024056} {\bibfield  {journal} {\bibinfo
   {journal} {Phys. Rev. D}\ }\textbf {\bibinfo {volume} {101}},\ \bibinfo
  {pages} {024056} (\bibinfo {year} {2020})},\ \Eprint
  {https://arxiv.org/abs/1911.06050} {arXiv:1911.06050 [gr-qc]} \BibitemShut
  {NoStop}%
\bibitem [{\citenamefont {Dietrich}\ \emph {et~al.}(2019)\citenamefont
  {Dietrich}, \citenamefont {Samajdar}, \citenamefont {Khan}, \citenamefont
  {Johnson-McDaniel}, \citenamefont {Dudi},\ and\ \citenamefont
  {Tichy}}]{Dietrich:2019kaq}%
  \BibitemOpen
  \bibfield  {author} {\bibinfo {author} {\bibfnamefont {T.}~\bibnamefont
  {Dietrich}}, \bibinfo {author} {\bibfnamefont {A.}~\bibnamefont {Samajdar}},
  \bibinfo {author} {\bibfnamefont {S.}~\bibnamefont {Khan}}, \bibinfo {author}
  {\bibfnamefont {N.~K.}\ \bibnamefont {Johnson-McDaniel}}, \bibinfo {author}
  {\bibfnamefont {R.}~\bibnamefont {Dudi}},\ and\ \bibinfo {author}
  {\bibfnamefont {W.}~\bibnamefont {Tichy}},\ }\bibfield  {title} {\bibinfo
  {title} {{Improving the NRTidal model for binary neutron star systems}},\
  }\href {https://doi.org/10.1103/PhysRevD.100.044003} {\bibfield  {journal}
  {\bibinfo  {journal} {Phys. Rev. D}\ }\textbf {\bibinfo {volume} {100}},\
  \bibinfo {pages} {044003} (\bibinfo {year} {2019})},\ \Eprint
  {https://arxiv.org/abs/1905.06011} {arXiv:1905.06011 [gr-qc]} \BibitemShut
  {NoStop}%
\bibitem [{\citenamefont {Pratten}\ \emph {et~al.}(2020)\citenamefont
  {Pratten}, \citenamefont {Husa}, \citenamefont {Garcia-Quiros}, \citenamefont
  {Colleoni}, \citenamefont {Ramos-Buades}, \citenamefont {Estelles},\ and\
  \citenamefont {Jaume}}]{Pratten:2020fqn}%
  \BibitemOpen
  \bibfield  {author} {\bibinfo {author} {\bibfnamefont {G.}~\bibnamefont
  {Pratten}}, \bibinfo {author} {\bibfnamefont {S.}~\bibnamefont {Husa}},
  \bibinfo {author} {\bibfnamefont {C.}~\bibnamefont {Garcia-Quiros}}, \bibinfo
  {author} {\bibfnamefont {M.}~\bibnamefont {Colleoni}}, \bibinfo {author}
  {\bibfnamefont {A.}~\bibnamefont {Ramos-Buades}}, \bibinfo {author}
  {\bibfnamefont {H.}~\bibnamefont {Estelles}},\ and\ \bibinfo {author}
  {\bibfnamefont {R.}~\bibnamefont {Jaume}},\ }\bibfield  {title} {\bibinfo
  {title} {{Setting the cornerstone for a family of models for gravitational
  waves from compact binaries: The dominant harmonic for nonprecessing
  quasicircular black holes}},\ }\href
  {https://doi.org/10.1103/PhysRevD.102.064001} {\bibfield  {journal} {\bibinfo
   {journal} {Phys. Rev. D}\ }\textbf {\bibinfo {volume} {102}},\ \bibinfo
  {pages} {064001} (\bibinfo {year} {2020})},\ \Eprint
  {https://arxiv.org/abs/2001.11412} {arXiv:2001.11412 [gr-qc]} \BibitemShut
  {NoStop}%
\bibitem [{\citenamefont {Pratten}\ \emph {et~al.}(2021)\citenamefont {Pratten}
  \emph {et~al.}}]{Pratten:2020ceb}%
  \BibitemOpen
  \bibfield  {author} {\bibinfo {author} {\bibfnamefont {G.}~\bibnamefont
  {Pratten}} \emph {et~al.},\ }\bibfield  {title} {\bibinfo {title}
  {{Computationally efficient models for the dominant and subdominant harmonic
  modes of precessing binary black holes}},\ }\href
  {https://doi.org/10.1103/PhysRevD.103.104056} {\bibfield  {journal} {\bibinfo
   {journal} {Phys. Rev. D}\ }\textbf {\bibinfo {volume} {103}},\ \bibinfo
  {pages} {104056} (\bibinfo {year} {2021})},\ \Eprint
  {https://arxiv.org/abs/2004.06503} {arXiv:2004.06503 [gr-qc]} \BibitemShut
  {NoStop}%
\bibitem [{\citenamefont {Garc\'\i{}a-Quir\'os}\ \emph
  {et~al.}(2020)\citenamefont {Garc\'\i{}a-Quir\'os}, \citenamefont {Colleoni},
  \citenamefont {Husa}, \citenamefont {Estell\'es}, \citenamefont {Pratten},
  \citenamefont {Ramos-Buades}, \citenamefont {Mateu-Lucena},\ and\
  \citenamefont {Jaume}}]{Garcia-Quiros:2020qpx}%
  \BibitemOpen
  \bibfield  {author} {\bibinfo {author} {\bibfnamefont {C.}~\bibnamefont
  {Garc\'\i{}a-Quir\'os}}, \bibinfo {author} {\bibfnamefont {M.}~\bibnamefont
  {Colleoni}}, \bibinfo {author} {\bibfnamefont {S.}~\bibnamefont {Husa}},
  \bibinfo {author} {\bibfnamefont {H.}~\bibnamefont {Estell\'es}}, \bibinfo
  {author} {\bibfnamefont {G.}~\bibnamefont {Pratten}}, \bibinfo {author}
  {\bibfnamefont {A.}~\bibnamefont {Ramos-Buades}}, \bibinfo {author}
  {\bibfnamefont {M.}~\bibnamefont {Mateu-Lucena}},\ and\ \bibinfo {author}
  {\bibfnamefont {R.}~\bibnamefont {Jaume}},\ }\bibfield  {title} {\bibinfo
  {title} {{Multimode frequency-domain model for the gravitational wave signal
  from nonprecessing black-hole binaries}},\ }\href
  {https://doi.org/10.1103/PhysRevD.102.064002} {\bibfield  {journal} {\bibinfo
   {journal} {Phys. Rev. D}\ }\textbf {\bibinfo {volume} {102}},\ \bibinfo
  {pages} {064002} (\bibinfo {year} {2020})},\ \Eprint
  {https://arxiv.org/abs/2001.10914} {arXiv:2001.10914 [gr-qc]} \BibitemShut
  {NoStop}%
\bibitem [{\citenamefont {Estell\'es}\ \emph {et~al.}(2021)\citenamefont
  {Estell\'es}, \citenamefont {Ramos-Buades}, \citenamefont {Husa},
  \citenamefont {Garc\'\i{}a-Quir\'os}, \citenamefont {Colleoni}, \citenamefont
  {Haegel},\ and\ \citenamefont {Jaume}}]{Estelles:2020osj}%
  \BibitemOpen
  \bibfield  {author} {\bibinfo {author} {\bibfnamefont {H.}~\bibnamefont
  {Estell\'es}}, \bibinfo {author} {\bibfnamefont {A.}~\bibnamefont
  {Ramos-Buades}}, \bibinfo {author} {\bibfnamefont {S.}~\bibnamefont {Husa}},
  \bibinfo {author} {\bibfnamefont {C.}~\bibnamefont {Garc\'\i{}a-Quir\'os}},
  \bibinfo {author} {\bibfnamefont {M.}~\bibnamefont {Colleoni}}, \bibinfo
  {author} {\bibfnamefont {L.}~\bibnamefont {Haegel}},\ and\ \bibinfo {author}
  {\bibfnamefont {R.}~\bibnamefont {Jaume}},\ }\bibfield  {title} {\bibinfo
  {title} {{Phenomenological time domain model for dominant quadrupole
  gravitational wave signal of coalescing binary black holes}},\ }\href
  {https://doi.org/10.1103/PhysRevD.103.124060} {\bibfield  {journal} {\bibinfo
   {journal} {Phys. Rev. D}\ }\textbf {\bibinfo {volume} {103}},\ \bibinfo
  {pages} {124060} (\bibinfo {year} {2021})},\ \Eprint
  {https://arxiv.org/abs/2004.08302} {arXiv:2004.08302 [gr-qc]} \BibitemShut
  {NoStop}%
\bibitem [{\citenamefont {Estell\'es}\ \emph
  {et~al.}(2022{\natexlab{a}})\citenamefont {Estell\'es}, \citenamefont {Husa},
  \citenamefont {Colleoni}, \citenamefont {Keitel}, \citenamefont
  {Mateu-Lucena}, \citenamefont {Garc\'\i{}a-Quir\'os}, \citenamefont
  {Ramos-Buades},\ and\ \citenamefont {Borchers}}]{Estelles:2020twz}%
  \BibitemOpen
  \bibfield  {author} {\bibinfo {author} {\bibfnamefont {H.}~\bibnamefont
  {Estell\'es}}, \bibinfo {author} {\bibfnamefont {S.}~\bibnamefont {Husa}},
  \bibinfo {author} {\bibfnamefont {M.}~\bibnamefont {Colleoni}}, \bibinfo
  {author} {\bibfnamefont {D.}~\bibnamefont {Keitel}}, \bibinfo {author}
  {\bibfnamefont {M.}~\bibnamefont {Mateu-Lucena}}, \bibinfo {author}
  {\bibfnamefont {C.}~\bibnamefont {Garc\'\i{}a-Quir\'os}}, \bibinfo {author}
  {\bibfnamefont {A.}~\bibnamefont {Ramos-Buades}},\ and\ \bibinfo {author}
  {\bibfnamefont {A.}~\bibnamefont {Borchers}},\ }\bibfield  {title} {\bibinfo
  {title} {{Time-domain phenomenological model of gravitational-wave
  subdominant harmonics for quasicircular nonprecessing binary black hole
  coalescences}},\ }\href {https://doi.org/10.1103/PhysRevD.105.084039}
  {\bibfield  {journal} {\bibinfo  {journal} {Phys. Rev. D}\ }\textbf {\bibinfo
  {volume} {105}},\ \bibinfo {pages} {084039} (\bibinfo {year}
  {2022}{\natexlab{a}})},\ \Eprint {https://arxiv.org/abs/2012.11923}
  {arXiv:2012.11923 [gr-qc]} \BibitemShut {NoStop}%
\bibitem [{\citenamefont {Estell\'es}\ \emph
  {et~al.}(2022{\natexlab{b}})\citenamefont {Estell\'es}, \citenamefont
  {Colleoni}, \citenamefont {Garc\'\i{}a-Quir\'os}, \citenamefont {Husa},
  \citenamefont {Keitel}, \citenamefont {Mateu-Lucena}, \citenamefont
  {Planas},\ and\ \citenamefont {Ramos-Buades}}]{Estelles:2021gvs}%
  \BibitemOpen
  \bibfield  {author} {\bibinfo {author} {\bibfnamefont {H.}~\bibnamefont
  {Estell\'es}}, \bibinfo {author} {\bibfnamefont {M.}~\bibnamefont
  {Colleoni}}, \bibinfo {author} {\bibfnamefont {C.}~\bibnamefont
  {Garc\'\i{}a-Quir\'os}}, \bibinfo {author} {\bibfnamefont {S.}~\bibnamefont
  {Husa}}, \bibinfo {author} {\bibfnamefont {D.}~\bibnamefont {Keitel}},
  \bibinfo {author} {\bibfnamefont {M.}~\bibnamefont {Mateu-Lucena}}, \bibinfo
  {author} {\bibfnamefont {M.~d.~L.}\ \bibnamefont {Planas}},\ and\ \bibinfo
  {author} {\bibfnamefont {A.}~\bibnamefont {Ramos-Buades}},\ }\bibfield
  {title} {\bibinfo {title} {{New twists in compact binary waveform modeling: A
  fast time-domain model for precession}},\ }\href
  {https://doi.org/10.1103/PhysRevD.105.084040} {\bibfield  {journal} {\bibinfo
   {journal} {Phys. Rev. D}\ }\textbf {\bibinfo {volume} {105}},\ \bibinfo
  {pages} {084040} (\bibinfo {year} {2022}{\natexlab{b}})},\ \Eprint
  {https://arxiv.org/abs/2105.05872} {arXiv:2105.05872 [gr-qc]} \BibitemShut
  {NoStop}%
\bibitem [{\citenamefont {Hamilton}\ \emph {et~al.}(2021)\citenamefont
  {Hamilton}, \citenamefont {London}, \citenamefont {Thompson}, \citenamefont
  {Fauchon-Jones}, \citenamefont {Hannam}, \citenamefont {Kalaghatgi},
  \citenamefont {Khan}, \citenamefont {Pannarale},\ and\ \citenamefont
  {Vano-Vinuales}}]{Hamilton:2021pkf}%
  \BibitemOpen
  \bibfield  {author} {\bibinfo {author} {\bibfnamefont {E.}~\bibnamefont
  {Hamilton}}, \bibinfo {author} {\bibfnamefont {L.}~\bibnamefont {London}},
  \bibinfo {author} {\bibfnamefont {J.~E.}\ \bibnamefont {Thompson}}, \bibinfo
  {author} {\bibfnamefont {E.}~\bibnamefont {Fauchon-Jones}}, \bibinfo {author}
  {\bibfnamefont {M.}~\bibnamefont {Hannam}}, \bibinfo {author} {\bibfnamefont
  {C.}~\bibnamefont {Kalaghatgi}}, \bibinfo {author} {\bibfnamefont
  {S.}~\bibnamefont {Khan}}, \bibinfo {author} {\bibfnamefont {F.}~\bibnamefont
  {Pannarale}},\ and\ \bibinfo {author} {\bibfnamefont {A.}~\bibnamefont
  {Vano-Vinuales}},\ }\bibfield  {title} {\bibinfo {title} {{Model of
  gravitational waves from precessing black-hole binaries through merger and
  ringdown}},\ }\href {https://doi.org/10.1103/PhysRevD.104.124027} {\bibfield
  {journal} {\bibinfo  {journal} {Phys. Rev. D}\ }\textbf {\bibinfo {volume}
  {104}},\ \bibinfo {pages} {124027} (\bibinfo {year} {2021})},\ \Eprint
  {https://arxiv.org/abs/2107.08876} {arXiv:2107.08876 [gr-qc]} \BibitemShut
  {NoStop}%
\bibitem [{\citenamefont {Buonanno}\ and\ \citenamefont
  {Damour}(1999)}]{Buonanno:1998gg}%
  \BibitemOpen
  \bibfield  {author} {\bibinfo {author} {\bibfnamefont {A.}~\bibnamefont
  {Buonanno}}\ and\ \bibinfo {author} {\bibfnamefont {T.}~\bibnamefont
  {Damour}},\ }\bibfield  {title} {\bibinfo {title} {{Effective one-body
  approach to general relativistic two-body dynamics}},\ }\href
  {https://doi.org/10.1103/PhysRevD.59.084006} {\bibfield  {journal} {\bibinfo
  {journal} {Phys. Rev. D}\ }\textbf {\bibinfo {volume} {59}},\ \bibinfo
  {pages} {084006} (\bibinfo {year} {1999})},\ \Eprint
  {https://arxiv.org/abs/gr-qc/9811091} {arXiv:gr-qc/9811091} \BibitemShut
  {NoStop}%
\bibitem [{\citenamefont {Buonanno}\ and\ \citenamefont
  {Damour}(2000)}]{Buonanno:2000ef}%
  \BibitemOpen
  \bibfield  {author} {\bibinfo {author} {\bibfnamefont {A.}~\bibnamefont
  {Buonanno}}\ and\ \bibinfo {author} {\bibfnamefont {T.}~\bibnamefont
  {Damour}},\ }\bibfield  {title} {\bibinfo {title} {{Transition from inspiral
  to plunge in binary black hole coalescences}},\ }\href
  {https://doi.org/10.1103/PhysRevD.62.064015} {\bibfield  {journal} {\bibinfo
  {journal} {Phys. Rev. D}\ }\textbf {\bibinfo {volume} {62}},\ \bibinfo
  {pages} {064015} (\bibinfo {year} {2000})},\ \Eprint
  {https://arxiv.org/abs/gr-qc/0001013} {arXiv:gr-qc/0001013} \BibitemShut
  {NoStop}%
\bibitem [{\citenamefont {Damour}\ \emph
  {et~al.}(2000{\natexlab{b}})\citenamefont {Damour}, \citenamefont
  {Jaranowski},\ and\ \citenamefont {Schaefer}}]{Damour:2000we}%
  \BibitemOpen
  \bibfield  {author} {\bibinfo {author} {\bibfnamefont {T.}~\bibnamefont
  {Damour}}, \bibinfo {author} {\bibfnamefont {P.}~\bibnamefont {Jaranowski}},\
  and\ \bibinfo {author} {\bibfnamefont {G.}~\bibnamefont {Schaefer}},\
  }\bibfield  {title} {\bibinfo {title} {{On the determination of the last
  stable orbit for circular general relativistic binaries at the third
  postNewtonian approximation}},\ }\href
  {https://doi.org/10.1103/PhysRevD.62.084011} {\bibfield  {journal} {\bibinfo
  {journal} {Phys. Rev. D}\ }\textbf {\bibinfo {volume} {62}},\ \bibinfo
  {pages} {084011} (\bibinfo {year} {2000}{\natexlab{b}})},\ \Eprint
  {https://arxiv.org/abs/gr-qc/0005034} {arXiv:gr-qc/0005034} \BibitemShut
  {NoStop}%
\bibitem [{\citenamefont {Damour}(2001)}]{Damour:2001tu}%
  \BibitemOpen
  \bibfield  {author} {\bibinfo {author} {\bibfnamefont {T.}~\bibnamefont
  {Damour}},\ }\bibfield  {title} {\bibinfo {title} {{Coalescence of two
  spinning black holes: an effective one-body approach}},\ }\href
  {https://doi.org/10.1103/PhysRevD.64.124013} {\bibfield  {journal} {\bibinfo
  {journal} {Phys. Rev. D}\ }\textbf {\bibinfo {volume} {64}},\ \bibinfo
  {pages} {124013} (\bibinfo {year} {2001})},\ \Eprint
  {https://arxiv.org/abs/gr-qc/0103018} {arXiv:gr-qc/0103018} \BibitemShut
  {NoStop}%
\bibitem [{\citenamefont {Buonanno}\ \emph {et~al.}(2006)\citenamefont
  {Buonanno}, \citenamefont {Chen},\ and\ \citenamefont
  {Damour}}]{Buonanno:2005xu}%
  \BibitemOpen
  \bibfield  {author} {\bibinfo {author} {\bibfnamefont {A.}~\bibnamefont
  {Buonanno}}, \bibinfo {author} {\bibfnamefont {Y.}~\bibnamefont {Chen}},\
  and\ \bibinfo {author} {\bibfnamefont {T.}~\bibnamefont {Damour}},\
  }\bibfield  {title} {\bibinfo {title} {{Transition from inspiral to plunge in
  precessing binaries of spinning black holes}},\ }\href
  {https://doi.org/10.1103/PhysRevD.74.104005} {\bibfield  {journal} {\bibinfo
  {journal} {Phys. Rev. D}\ }\textbf {\bibinfo {volume} {74}},\ \bibinfo
  {pages} {104005} (\bibinfo {year} {2006})},\ \Eprint
  {https://arxiv.org/abs/gr-qc/0508067} {arXiv:gr-qc/0508067} \BibitemShut
  {NoStop}%
\bibitem [{\citenamefont {Buonanno}\ \emph {et~al.}(2007)\citenamefont
  {Buonanno}, \citenamefont {Pan}, \citenamefont {Baker}, \citenamefont
  {Centrella}, \citenamefont {Kelly}, \citenamefont {McWilliams},\ and\
  \citenamefont {van Meter}}]{Buonanno:2007pf}%
  \BibitemOpen
  \bibfield  {author} {\bibinfo {author} {\bibfnamefont {A.}~\bibnamefont
  {Buonanno}}, \bibinfo {author} {\bibfnamefont {Y.}~\bibnamefont {Pan}},
  \bibinfo {author} {\bibfnamefont {J.~G.}\ \bibnamefont {Baker}}, \bibinfo
  {author} {\bibfnamefont {J.}~\bibnamefont {Centrella}}, \bibinfo {author}
  {\bibfnamefont {B.~J.}\ \bibnamefont {Kelly}}, \bibinfo {author}
  {\bibfnamefont {S.~T.}\ \bibnamefont {McWilliams}},\ and\ \bibinfo {author}
  {\bibfnamefont {J.~R.}\ \bibnamefont {van Meter}},\ }\bibfield  {title}
  {\bibinfo {title} {{Toward faithful templates for non-spinning binary black
  holes using the effective-one-body approach}},\ }\href
  {https://doi.org/10.1103/PhysRevD.76.104049} {\bibfield  {journal} {\bibinfo
  {journal} {Phys. Rev. D}\ }\textbf {\bibinfo {volume} {76}},\ \bibinfo
  {pages} {104049} (\bibinfo {year} {2007})},\ \Eprint
  {https://arxiv.org/abs/0706.3732} {arXiv:0706.3732 [gr-qc]} \BibitemShut
  {NoStop}%
\bibitem [{\citenamefont {Damour}\ and\ \citenamefont
  {Nagar}(2008)}]{Damour:2007yf}%
  \BibitemOpen
  \bibfield  {author} {\bibinfo {author} {\bibfnamefont {T.}~\bibnamefont
  {Damour}}\ and\ \bibinfo {author} {\bibfnamefont {A.}~\bibnamefont {Nagar}},\
  }\bibfield  {title} {\bibinfo {title} {{Comparing Effective-One-Body
  gravitational waveforms to accurate numerical data}},\ }\href
  {https://doi.org/10.1103/PhysRevD.77.024043} {\bibfield  {journal} {\bibinfo
  {journal} {Phys. Rev. D}\ }\textbf {\bibinfo {volume} {77}},\ \bibinfo
  {pages} {024043} (\bibinfo {year} {2008})},\ \Eprint
  {https://arxiv.org/abs/0711.2628} {arXiv:0711.2628 [gr-qc]} \BibitemShut
  {NoStop}%
\bibitem [{\citenamefont {Damour}\ \emph {et~al.}(2009)\citenamefont {Damour},
  \citenamefont {Iyer},\ and\ \citenamefont {Nagar}}]{Damour:2008gu}%
  \BibitemOpen
  \bibfield  {author} {\bibinfo {author} {\bibfnamefont {T.}~\bibnamefont
  {Damour}}, \bibinfo {author} {\bibfnamefont {B.~R.}\ \bibnamefont {Iyer}},\
  and\ \bibinfo {author} {\bibfnamefont {A.}~\bibnamefont {Nagar}},\ }\bibfield
   {title} {\bibinfo {title} {{Improved resummation of post-Newtonian
  multipolar waveforms from circularized compact binaries}},\ }\href
  {https://doi.org/10.1103/PhysRevD.79.064004} {\bibfield  {journal} {\bibinfo
  {journal} {Phys. Rev. D}\ }\textbf {\bibinfo {volume} {79}},\ \bibinfo
  {pages} {064004} (\bibinfo {year} {2009})},\ \Eprint
  {https://arxiv.org/abs/0811.2069} {arXiv:0811.2069 [gr-qc]} \BibitemShut
  {NoStop}%
\bibitem [{\citenamefont {Buonanno}\ \emph
  {et~al.}(2009{\natexlab{b}})\citenamefont {Buonanno}, \citenamefont {Pan},
  \citenamefont {Pfeiffer}, \citenamefont {Scheel}, \citenamefont {Buchman},\
  and\ \citenamefont {Kidder}}]{Buonanno:2009qa}%
  \BibitemOpen
  \bibfield  {author} {\bibinfo {author} {\bibfnamefont {A.}~\bibnamefont
  {Buonanno}}, \bibinfo {author} {\bibfnamefont {Y.}~\bibnamefont {Pan}},
  \bibinfo {author} {\bibfnamefont {H.~P.}\ \bibnamefont {Pfeiffer}}, \bibinfo
  {author} {\bibfnamefont {M.~A.}\ \bibnamefont {Scheel}}, \bibinfo {author}
  {\bibfnamefont {L.~T.}\ \bibnamefont {Buchman}},\ and\ \bibinfo {author}
  {\bibfnamefont {L.~E.}\ \bibnamefont {Kidder}},\ }\bibfield  {title}
  {\bibinfo {title} {{Effective-one-body waveforms calibrated to numerical
  relativity simulations: Coalescence of non-spinning, equal-mass black
  holes}},\ }\href {https://doi.org/10.1103/PhysRevD.79.124028} {\bibfield
  {journal} {\bibinfo  {journal} {Phys. Rev. D}\ }\textbf {\bibinfo {volume}
  {79}},\ \bibinfo {pages} {124028} (\bibinfo {year} {2009}{\natexlab{b}})},\
  \Eprint {https://arxiv.org/abs/0902.0790} {arXiv:0902.0790 [gr-qc]}
  \BibitemShut {NoStop}%
\bibitem [{\citenamefont {Pan}\ \emph {et~al.}(2011)\citenamefont {Pan},
  \citenamefont {Buonanno}, \citenamefont {Boyle}, \citenamefont {Buchman},
  \citenamefont {Kidder}, \citenamefont {Pfeiffer},\ and\ \citenamefont
  {Scheel}}]{Pan:2011gk}%
  \BibitemOpen
  \bibfield  {author} {\bibinfo {author} {\bibfnamefont {Y.}~\bibnamefont
  {Pan}}, \bibinfo {author} {\bibfnamefont {A.}~\bibnamefont {Buonanno}},
  \bibinfo {author} {\bibfnamefont {M.}~\bibnamefont {Boyle}}, \bibinfo
  {author} {\bibfnamefont {L.~T.}\ \bibnamefont {Buchman}}, \bibinfo {author}
  {\bibfnamefont {L.~E.}\ \bibnamefont {Kidder}}, \bibinfo {author}
  {\bibfnamefont {H.~P.}\ \bibnamefont {Pfeiffer}},\ and\ \bibinfo {author}
  {\bibfnamefont {M.~A.}\ \bibnamefont {Scheel}},\ }\bibfield  {title}
  {\bibinfo {title} {{Inspiral-merger-ringdown multipolar waveforms of
  nonspinning black-hole binaries using the effective-one-body formalism}},\
  }\href {https://doi.org/10.1103/PhysRevD.84.124052} {\bibfield  {journal}
  {\bibinfo  {journal} {Phys. Rev. D}\ }\textbf {\bibinfo {volume} {84}},\
  \bibinfo {pages} {124052} (\bibinfo {year} {2011})},\ \Eprint
  {https://arxiv.org/abs/1106.1021} {arXiv:1106.1021 [gr-qc]} \BibitemShut
  {NoStop}%
\bibitem [{\citenamefont {Damour}\ \emph {et~al.}(2013)\citenamefont {Damour},
  \citenamefont {Nagar},\ and\ \citenamefont {Bernuzzi}}]{Damour:2012ky}%
  \BibitemOpen
  \bibfield  {author} {\bibinfo {author} {\bibfnamefont {T.}~\bibnamefont
  {Damour}}, \bibinfo {author} {\bibfnamefont {A.}~\bibnamefont {Nagar}},\ and\
  \bibinfo {author} {\bibfnamefont {S.}~\bibnamefont {Bernuzzi}},\ }\bibfield
  {title} {\bibinfo {title} {{Improved effective-one-body description of
  coalescing nonspinning black-hole binaries and its numerical-relativity
  completion}},\ }\href {https://doi.org/10.1103/PhysRevD.87.084035} {\bibfield
   {journal} {\bibinfo  {journal} {Phys. Rev. D}\ }\textbf {\bibinfo {volume}
  {87}},\ \bibinfo {pages} {084035} (\bibinfo {year} {2013})},\ \Eprint
  {https://arxiv.org/abs/1212.4357} {arXiv:1212.4357 [gr-qc]} \BibitemShut
  {NoStop}%
\bibitem [{\citenamefont {Damour}\ \emph {et~al.}(2015)\citenamefont {Damour},
  \citenamefont {Jaranowski},\ and\ \citenamefont
  {Sch\"afer}}]{Damour:2015isa}%
  \BibitemOpen
  \bibfield  {author} {\bibinfo {author} {\bibfnamefont {T.}~\bibnamefont
  {Damour}}, \bibinfo {author} {\bibfnamefont {P.}~\bibnamefont {Jaranowski}},\
  and\ \bibinfo {author} {\bibfnamefont {G.}~\bibnamefont {Sch\"afer}},\
  }\bibfield  {title} {\bibinfo {title} {{Fourth post-Newtonian effective
  one-body dynamics}},\ }\href {https://doi.org/10.1103/PhysRevD.91.084024}
  {\bibfield  {journal} {\bibinfo  {journal} {Phys. Rev. D}\ }\textbf {\bibinfo
  {volume} {91}},\ \bibinfo {pages} {084024} (\bibinfo {year} {2015})},\
  \Eprint {https://arxiv.org/abs/1502.07245} {arXiv:1502.07245 [gr-qc]}
  \BibitemShut {NoStop}%
\bibitem [{\citenamefont {Nagar}\ \emph
  {et~al.}(2020{\natexlab{a}})\citenamefont {Nagar}, \citenamefont {Pratten},
  \citenamefont {Riemenschneider},\ and\ \citenamefont
  {Gamba}}]{Nagar:2019wds}%
  \BibitemOpen
  \bibfield  {author} {\bibinfo {author} {\bibfnamefont {A.}~\bibnamefont
  {Nagar}}, \bibinfo {author} {\bibfnamefont {G.}~\bibnamefont {Pratten}},
  \bibinfo {author} {\bibfnamefont {G.}~\bibnamefont {Riemenschneider}},\ and\
  \bibinfo {author} {\bibfnamefont {R.}~\bibnamefont {Gamba}},\ }\bibfield
  {title} {\bibinfo {title} {{Multipolar effective one body model for
  nonspinning black hole binaries}},\ }\href
  {https://doi.org/10.1103/PhysRevD.101.024041} {\bibfield  {journal} {\bibinfo
   {journal} {Phys. Rev. D}\ }\textbf {\bibinfo {volume} {101}},\ \bibinfo
  {pages} {024041} (\bibinfo {year} {2020}{\natexlab{a}})},\ \Eprint
  {https://arxiv.org/abs/1904.09550} {arXiv:1904.09550 [gr-qc]} \BibitemShut
  {NoStop}%
\bibitem [{\citenamefont {Damour}\ \emph
  {et~al.}(2008{\natexlab{a}})\citenamefont {Damour}, \citenamefont {Nagar},
  \citenamefont {Dorband}, \citenamefont {Pollney},\ and\ \citenamefont
  {Rezzolla}}]{Damour:2007vq}%
  \BibitemOpen
  \bibfield  {author} {\bibinfo {author} {\bibfnamefont {T.}~\bibnamefont
  {Damour}}, \bibinfo {author} {\bibfnamefont {A.}~\bibnamefont {Nagar}},
  \bibinfo {author} {\bibfnamefont {E.~N.}\ \bibnamefont {Dorband}}, \bibinfo
  {author} {\bibfnamefont {D.}~\bibnamefont {Pollney}},\ and\ \bibinfo {author}
  {\bibfnamefont {L.}~\bibnamefont {Rezzolla}},\ }\bibfield  {title} {\bibinfo
  {title} {{Faithful Effective-One-Body waveforms of equal-mass coalescing
  black-hole binaries}},\ }\href {https://doi.org/10.1103/PhysRevD.77.084017}
  {\bibfield  {journal} {\bibinfo  {journal} {Phys. Rev. D}\ }\textbf {\bibinfo
  {volume} {77}},\ \bibinfo {pages} {084017} (\bibinfo {year}
  {2008}{\natexlab{a}})},\ \Eprint {https://arxiv.org/abs/0712.3003}
  {arXiv:0712.3003 [gr-qc]} \BibitemShut {NoStop}%
\bibitem [{\citenamefont {Damour}\ \emph
  {et~al.}(2008{\natexlab{b}})\citenamefont {Damour}, \citenamefont
  {Jaranowski},\ and\ \citenamefont {Schaefer}}]{Damour:2008qf}%
  \BibitemOpen
  \bibfield  {author} {\bibinfo {author} {\bibfnamefont {T.}~\bibnamefont
  {Damour}}, \bibinfo {author} {\bibfnamefont {P.}~\bibnamefont {Jaranowski}},\
  and\ \bibinfo {author} {\bibfnamefont {G.}~\bibnamefont {Schaefer}},\
  }\bibfield  {title} {\bibinfo {title} {{Effective one body approach to the
  dynamics of two spinning black holes with next-to-leading order spin-orbit
  coupling}},\ }\href {https://doi.org/10.1103/PhysRevD.78.024009} {\bibfield
  {journal} {\bibinfo  {journal} {Phys. Rev. D}\ }\textbf {\bibinfo {volume}
  {78}},\ \bibinfo {pages} {024009} (\bibinfo {year} {2008}{\natexlab{b}})},\
  \Eprint {https://arxiv.org/abs/0803.0915} {arXiv:0803.0915 [gr-qc]}
  \BibitemShut {NoStop}%
\bibitem [{\citenamefont {Pan}\ \emph {et~al.}(2010)\citenamefont {Pan},
  \citenamefont {Buonanno}, \citenamefont {Buchman}, \citenamefont {Chu},
  \citenamefont {Kidder}, \citenamefont {Pfeiffer},\ and\ \citenamefont
  {Scheel}}]{Pan:2009wj}%
  \BibitemOpen
  \bibfield  {author} {\bibinfo {author} {\bibfnamefont {Y.}~\bibnamefont
  {Pan}}, \bibinfo {author} {\bibfnamefont {A.}~\bibnamefont {Buonanno}},
  \bibinfo {author} {\bibfnamefont {L.~T.}\ \bibnamefont {Buchman}}, \bibinfo
  {author} {\bibfnamefont {T.}~\bibnamefont {Chu}}, \bibinfo {author}
  {\bibfnamefont {L.~E.}\ \bibnamefont {Kidder}}, \bibinfo {author}
  {\bibfnamefont {H.~P.}\ \bibnamefont {Pfeiffer}},\ and\ \bibinfo {author}
  {\bibfnamefont {M.~A.}\ \bibnamefont {Scheel}},\ }\bibfield  {title}
  {\bibinfo {title} {{Effective-one-body waveforms calibrated to numerical
  relativity simulations: coalescence of non-precessing, spinning, equal-mass
  black holes}},\ }\href {https://doi.org/10.1103/PhysRevD.81.084041}
  {\bibfield  {journal} {\bibinfo  {journal} {Phys. Rev. D}\ }\textbf {\bibinfo
  {volume} {81}},\ \bibinfo {pages} {084041} (\bibinfo {year} {2010})},\
  \Eprint {https://arxiv.org/abs/0912.3466} {arXiv:0912.3466 [gr-qc]}
  \BibitemShut {NoStop}%
\bibitem [{\citenamefont {Damour}\ \emph
  {et~al.}(2008{\natexlab{c}})\citenamefont {Damour}, \citenamefont {Nagar},
  \citenamefont {Hannam}, \citenamefont {Husa},\ and\ \citenamefont
  {Bruegmann}}]{Damour:2008te}%
  \BibitemOpen
  \bibfield  {author} {\bibinfo {author} {\bibfnamefont {T.}~\bibnamefont
  {Damour}}, \bibinfo {author} {\bibfnamefont {A.}~\bibnamefont {Nagar}},
  \bibinfo {author} {\bibfnamefont {M.}~\bibnamefont {Hannam}}, \bibinfo
  {author} {\bibfnamefont {S.}~\bibnamefont {Husa}},\ and\ \bibinfo {author}
  {\bibfnamefont {B.}~\bibnamefont {Bruegmann}},\ }\bibfield  {title} {\bibinfo
  {title} {{Accurate Effective-One-Body waveforms of inspiralling and
  coalescing black-hole binaries}},\ }\href
  {https://doi.org/10.1103/PhysRevD.78.044039} {\bibfield  {journal} {\bibinfo
  {journal} {Phys. Rev. D}\ }\textbf {\bibinfo {volume} {78}},\ \bibinfo
  {pages} {044039} (\bibinfo {year} {2008}{\natexlab{c}})},\ \Eprint
  {https://arxiv.org/abs/0803.3162} {arXiv:0803.3162 [gr-qc]} \BibitemShut
  {NoStop}%
\bibitem [{\citenamefont {Barausse}\ and\ \citenamefont
  {Buonanno}(2010)}]{Barausse:2009xi}%
  \BibitemOpen
  \bibfield  {author} {\bibinfo {author} {\bibfnamefont {E.}~\bibnamefont
  {Barausse}}\ and\ \bibinfo {author} {\bibfnamefont {A.}~\bibnamefont
  {Buonanno}},\ }\bibfield  {title} {\bibinfo {title} {{An Improved
  effective-one-body Hamiltonian for spinning black-hole binaries}},\ }\href
  {https://doi.org/10.1103/PhysRevD.81.084024} {\bibfield  {journal} {\bibinfo
  {journal} {Phys. Rev. D}\ }\textbf {\bibinfo {volume} {81}},\ \bibinfo
  {pages} {084024} (\bibinfo {year} {2010})},\ \Eprint
  {https://arxiv.org/abs/0912.3517} {arXiv:0912.3517 [gr-qc]} \BibitemShut
  {NoStop}%
\bibitem [{\citenamefont {Barausse}\ and\ \citenamefont
  {Buonanno}(2011)}]{Barausse:2011ys}%
  \BibitemOpen
  \bibfield  {author} {\bibinfo {author} {\bibfnamefont {E.}~\bibnamefont
  {Barausse}}\ and\ \bibinfo {author} {\bibfnamefont {A.}~\bibnamefont
  {Buonanno}},\ }\bibfield  {title} {\bibinfo {title} {{Extending the
  effective-one-body Hamiltonian of black-hole binaries to include
  next-to-next-to-leading spin-orbit couplings}},\ }\href
  {https://doi.org/10.1103/PhysRevD.84.104027} {\bibfield  {journal} {\bibinfo
  {journal} {Phys. Rev. D}\ }\textbf {\bibinfo {volume} {84}},\ \bibinfo
  {pages} {104027} (\bibinfo {year} {2011})},\ \Eprint
  {https://arxiv.org/abs/1107.2904} {arXiv:1107.2904 [gr-qc]} \BibitemShut
  {NoStop}%
\bibitem [{\citenamefont {Nagar}(2011)}]{Nagar:2011fx}%
  \BibitemOpen
  \bibfield  {author} {\bibinfo {author} {\bibfnamefont {A.}~\bibnamefont
  {Nagar}},\ }\bibfield  {title} {\bibinfo {title} {{Effective one body
  Hamiltonian of two spinning black-holes with next-to-next-to-leading order
  spin-orbit coupling}},\ }\href {https://doi.org/10.1103/PhysRevD.84.084028}
  {\bibfield  {journal} {\bibinfo  {journal} {Phys. Rev. D}\ }\textbf {\bibinfo
  {volume} {84}},\ \bibinfo {pages} {084028} (\bibinfo {year} {2011})},\
  \bibinfo {note} {[Erratum: Phys.Rev.D 88, 089901 (2013)]},\ \Eprint
  {https://arxiv.org/abs/1106.4349} {arXiv:1106.4349 [gr-qc]} \BibitemShut
  {NoStop}%
\bibitem [{\citenamefont {Damour}\ and\ \citenamefont
  {Nagar}(2014)}]{Damour:2014sva}%
  \BibitemOpen
  \bibfield  {author} {\bibinfo {author} {\bibfnamefont {T.}~\bibnamefont
  {Damour}}\ and\ \bibinfo {author} {\bibfnamefont {A.}~\bibnamefont {Nagar}},\
  }\bibfield  {title} {\bibinfo {title} {{New effective-one-body description of
  coalescing nonprecessing spinning black-hole binaries}},\ }\href
  {https://doi.org/10.1103/PhysRevD.90.044018} {\bibfield  {journal} {\bibinfo
  {journal} {Phys. Rev. D}\ }\textbf {\bibinfo {volume} {90}},\ \bibinfo
  {pages} {044018} (\bibinfo {year} {2014})},\ \Eprint
  {https://arxiv.org/abs/1406.6913} {arXiv:1406.6913 [gr-qc]} \BibitemShut
  {NoStop}%
\bibitem [{\citenamefont {Balmelli}\ and\ \citenamefont
  {Damour}(2015)}]{Balmelli:2015zsa}%
  \BibitemOpen
  \bibfield  {author} {\bibinfo {author} {\bibfnamefont {S.}~\bibnamefont
  {Balmelli}}\ and\ \bibinfo {author} {\bibfnamefont {T.}~\bibnamefont
  {Damour}},\ }\bibfield  {title} {\bibinfo {title} {{New effective-one-body
  Hamiltonian with next-to-leading order spin-spin coupling}},\ }\href
  {https://doi.org/10.1103/PhysRevD.92.124022} {\bibfield  {journal} {\bibinfo
  {journal} {Phys. Rev. D}\ }\textbf {\bibinfo {volume} {92}},\ \bibinfo
  {pages} {124022} (\bibinfo {year} {2015})},\ \Eprint
  {https://arxiv.org/abs/1509.08135} {arXiv:1509.08135 [gr-qc]} \BibitemShut
  {NoStop}%
\bibitem [{\citenamefont {Khalil}\ \emph {et~al.}(2020)\citenamefont {Khalil},
  \citenamefont {Steinhoff}, \citenamefont {Vines},\ and\ \citenamefont
  {Buonanno}}]{Khalil:2020mmr}%
  \BibitemOpen
  \bibfield  {author} {\bibinfo {author} {\bibfnamefont {M.}~\bibnamefont
  {Khalil}}, \bibinfo {author} {\bibfnamefont {J.}~\bibnamefont {Steinhoff}},
  \bibinfo {author} {\bibfnamefont {J.}~\bibnamefont {Vines}},\ and\ \bibinfo
  {author} {\bibfnamefont {A.}~\bibnamefont {Buonanno}},\ }\bibfield  {title}
  {\bibinfo {title} {{Fourth post-Newtonian effective-one-body Hamiltonians
  with generic spins}},\ }\href {https://doi.org/10.1103/PhysRevD.101.104034}
  {\bibfield  {journal} {\bibinfo  {journal} {Phys. Rev. D}\ }\textbf {\bibinfo
  {volume} {101}},\ \bibinfo {pages} {104034} (\bibinfo {year} {2020})},\
  \Eprint {https://arxiv.org/abs/2003.04469} {arXiv:2003.04469 [gr-qc]}
  \BibitemShut {NoStop}%
\bibitem [{\citenamefont {Taracchini}\ \emph {et~al.}(2012)\citenamefont
  {Taracchini}, \citenamefont {Pan}, \citenamefont {Buonanno}, \citenamefont
  {Barausse}, \citenamefont {Boyle}, \citenamefont {Chu}, \citenamefont
  {Lovelace}, \citenamefont {Pfeiffer},\ and\ \citenamefont
  {Scheel}}]{Taracchini:2012ig}%
  \BibitemOpen
  \bibfield  {author} {\bibinfo {author} {\bibfnamefont {A.}~\bibnamefont
  {Taracchini}}, \bibinfo {author} {\bibfnamefont {Y.}~\bibnamefont {Pan}},
  \bibinfo {author} {\bibfnamefont {A.}~\bibnamefont {Buonanno}}, \bibinfo
  {author} {\bibfnamefont {E.}~\bibnamefont {Barausse}}, \bibinfo {author}
  {\bibfnamefont {M.}~\bibnamefont {Boyle}}, \bibinfo {author} {\bibfnamefont
  {T.}~\bibnamefont {Chu}}, \bibinfo {author} {\bibfnamefont {G.}~\bibnamefont
  {Lovelace}}, \bibinfo {author} {\bibfnamefont {H.~P.}\ \bibnamefont
  {Pfeiffer}},\ and\ \bibinfo {author} {\bibfnamefont {M.~A.}\ \bibnamefont
  {Scheel}},\ }\bibfield  {title} {\bibinfo {title} {{Prototype
  effective-one-body model for nonprecessing spinning inspiral-merger-ringdown
  waveforms}},\ }\href {https://doi.org/10.1103/PhysRevD.86.024011} {\bibfield
  {journal} {\bibinfo  {journal} {Phys. Rev. D}\ }\textbf {\bibinfo {volume}
  {86}},\ \bibinfo {pages} {024011} (\bibinfo {year} {2012})},\ \Eprint
  {https://arxiv.org/abs/1202.0790} {arXiv:1202.0790 [gr-qc]} \BibitemShut
  {NoStop}%
\bibitem [{\citenamefont {Taracchini}\ \emph {et~al.}(2014)\citenamefont
  {Taracchini} \emph {et~al.}}]{Taracchini:2013rva}%
  \BibitemOpen
  \bibfield  {author} {\bibinfo {author} {\bibfnamefont {A.}~\bibnamefont
  {Taracchini}} \emph {et~al.},\ }\bibfield  {title} {\bibinfo {title}
  {{Effective-one-body model for black-hole binaries with generic mass ratios
  and spins}},\ }\href {https://doi.org/10.1103/PhysRevD.89.061502} {\bibfield
  {journal} {\bibinfo  {journal} {Phys. Rev. D}\ }\textbf {\bibinfo {volume}
  {89}},\ \bibinfo {pages} {061502} (\bibinfo {year} {2014})},\ \Eprint
  {https://arxiv.org/abs/1311.2544} {arXiv:1311.2544 [gr-qc]} \BibitemShut
  {NoStop}%
\bibitem [{\citenamefont {Boh\'e}\ \emph {et~al.}(2017)\citenamefont {Boh\'e}
  \emph {et~al.}}]{Bohe:2016gbl}%
  \BibitemOpen
  \bibfield  {author} {\bibinfo {author} {\bibfnamefont {A.}~\bibnamefont
  {Boh\'e}} \emph {et~al.},\ }\bibfield  {title} {\bibinfo {title} {{Improved
  effective-one-body model of spinning, nonprecessing binary black holes for
  the era of gravitational-wave astrophysics with advanced detectors}},\ }\href
  {https://doi.org/10.1103/PhysRevD.95.044028} {\bibfield  {journal} {\bibinfo
  {journal} {Phys. Rev. D}\ }\textbf {\bibinfo {volume} {95}},\ \bibinfo
  {pages} {044028} (\bibinfo {year} {2017})},\ \Eprint
  {https://arxiv.org/abs/1611.03703} {arXiv:1611.03703 [gr-qc]} \BibitemShut
  {NoStop}%
\bibitem [{\citenamefont {Cotesta}\ \emph {et~al.}(2018)\citenamefont
  {Cotesta}, \citenamefont {Buonanno}, \citenamefont {Boh\'e}, \citenamefont
  {Taracchini}, \citenamefont {Hinder},\ and\ \citenamefont
  {Ossokine}}]{Cotesta:2018fcv}%
  \BibitemOpen
  \bibfield  {author} {\bibinfo {author} {\bibfnamefont {R.}~\bibnamefont
  {Cotesta}}, \bibinfo {author} {\bibfnamefont {A.}~\bibnamefont {Buonanno}},
  \bibinfo {author} {\bibfnamefont {A.}~\bibnamefont {Boh\'e}}, \bibinfo
  {author} {\bibfnamefont {A.}~\bibnamefont {Taracchini}}, \bibinfo {author}
  {\bibfnamefont {I.}~\bibnamefont {Hinder}},\ and\ \bibinfo {author}
  {\bibfnamefont {S.}~\bibnamefont {Ossokine}},\ }\bibfield  {title} {\bibinfo
  {title} {{Enriching the Symphony of Gravitational Waves from Binary Black
  Holes by Tuning Higher Harmonics}},\ }\href
  {https://doi.org/10.1103/PhysRevD.98.084028} {\bibfield  {journal} {\bibinfo
  {journal} {Phys. Rev. D}\ }\textbf {\bibinfo {volume} {98}},\ \bibinfo
  {pages} {084028} (\bibinfo {year} {2018})},\ \Eprint
  {https://arxiv.org/abs/1803.10701} {arXiv:1803.10701 [gr-qc]} \BibitemShut
  {NoStop}%
\bibitem [{\citenamefont {Pan}\ \emph {et~al.}(2014)\citenamefont {Pan},
  \citenamefont {Buonanno}, \citenamefont {Taracchini}, \citenamefont {Kidder},
  \citenamefont {Mrou\'e}, \citenamefont {Pfeiffer}, \citenamefont {Scheel},\
  and\ \citenamefont {Szil\'agyi}}]{Pan:2013rra}%
  \BibitemOpen
  \bibfield  {author} {\bibinfo {author} {\bibfnamefont {Y.}~\bibnamefont
  {Pan}}, \bibinfo {author} {\bibfnamefont {A.}~\bibnamefont {Buonanno}},
  \bibinfo {author} {\bibfnamefont {A.}~\bibnamefont {Taracchini}}, \bibinfo
  {author} {\bibfnamefont {L.~E.}\ \bibnamefont {Kidder}}, \bibinfo {author}
  {\bibfnamefont {A.~H.}\ \bibnamefont {Mrou\'e}}, \bibinfo {author}
  {\bibfnamefont {H.~P.}\ \bibnamefont {Pfeiffer}}, \bibinfo {author}
  {\bibfnamefont {M.~A.}\ \bibnamefont {Scheel}},\ and\ \bibinfo {author}
  {\bibfnamefont {B.}~\bibnamefont {Szil\'agyi}},\ }\bibfield  {title}
  {\bibinfo {title} {{Inspiral-merger-ringdown waveforms of spinning,
  precessing black-hole binaries in the effective-one-body formalism}},\ }\href
  {https://doi.org/10.1103/PhysRevD.89.084006} {\bibfield  {journal} {\bibinfo
  {journal} {Phys. Rev. D}\ }\textbf {\bibinfo {volume} {89}},\ \bibinfo
  {pages} {084006} (\bibinfo {year} {2014})},\ \Eprint
  {https://arxiv.org/abs/1307.6232} {arXiv:1307.6232 [gr-qc]} \BibitemShut
  {NoStop}%
\bibitem [{\citenamefont {Babak}\ \emph {et~al.}(2017)\citenamefont {Babak},
  \citenamefont {Taracchini},\ and\ \citenamefont {Buonanno}}]{Babak:2016tgq}%
  \BibitemOpen
  \bibfield  {author} {\bibinfo {author} {\bibfnamefont {S.}~\bibnamefont
  {Babak}}, \bibinfo {author} {\bibfnamefont {A.}~\bibnamefont {Taracchini}},\
  and\ \bibinfo {author} {\bibfnamefont {A.}~\bibnamefont {Buonanno}},\
  }\bibfield  {title} {\bibinfo {title} {{Validating the effective-one-body
  model of spinning, precessing binary black holes against numerical
  relativity}},\ }\href {https://doi.org/10.1103/PhysRevD.95.024010} {\bibfield
   {journal} {\bibinfo  {journal} {Phys. Rev. D}\ }\textbf {\bibinfo {volume}
  {95}},\ \bibinfo {pages} {024010} (\bibinfo {year} {2017})},\ \Eprint
  {https://arxiv.org/abs/1607.05661} {arXiv:1607.05661 [gr-qc]} \BibitemShut
  {NoStop}%
\bibitem [{\citenamefont {Ossokine}\ \emph {et~al.}(2020)\citenamefont
  {Ossokine} \emph {et~al.}}]{Ossokine:2020kjp}%
  \BibitemOpen
  \bibfield  {author} {\bibinfo {author} {\bibfnamefont {S.}~\bibnamefont
  {Ossokine}} \emph {et~al.},\ }\bibfield  {title} {\bibinfo {title}
  {{Multipolar Effective-One-Body Waveforms for Precessing Binary Black Holes:
  Construction and Validation}},\ }\href
  {https://doi.org/10.1103/PhysRevD.102.044055} {\bibfield  {journal} {\bibinfo
   {journal} {Phys. Rev. D}\ }\textbf {\bibinfo {volume} {102}},\ \bibinfo
  {pages} {044055} (\bibinfo {year} {2020})},\ \Eprint
  {https://arxiv.org/abs/2004.09442} {arXiv:2004.09442 [gr-qc]} \BibitemShut
  {NoStop}%
\bibitem [{\citenamefont {Nagar}\ \emph {et~al.}(2019)\citenamefont {Nagar},
  \citenamefont {Messina}, \citenamefont {Rettegno}, \citenamefont {Bini},
  \citenamefont {Damour}, \citenamefont {Geralico}, \citenamefont {Akcay},\
  and\ \citenamefont {Bernuzzi}}]{Nagar:2018plt}%
  \BibitemOpen
  \bibfield  {author} {\bibinfo {author} {\bibfnamefont {A.}~\bibnamefont
  {Nagar}}, \bibinfo {author} {\bibfnamefont {F.}~\bibnamefont {Messina}},
  \bibinfo {author} {\bibfnamefont {P.}~\bibnamefont {Rettegno}}, \bibinfo
  {author} {\bibfnamefont {D.}~\bibnamefont {Bini}}, \bibinfo {author}
  {\bibfnamefont {T.}~\bibnamefont {Damour}}, \bibinfo {author} {\bibfnamefont
  {A.}~\bibnamefont {Geralico}}, \bibinfo {author} {\bibfnamefont
  {S.}~\bibnamefont {Akcay}},\ and\ \bibinfo {author} {\bibfnamefont
  {S.}~\bibnamefont {Bernuzzi}},\ }\bibfield  {title} {\bibinfo {title}
  {{Nonlinear-in-spin effects in effective-one-body waveform models of
  spin-aligned, inspiralling, neutron star binaries}},\ }\href
  {https://doi.org/10.1103/PhysRevD.99.044007} {\bibfield  {journal} {\bibinfo
  {journal} {Phys. Rev. D}\ }\textbf {\bibinfo {volume} {99}},\ \bibinfo
  {pages} {044007} (\bibinfo {year} {2019})},\ \Eprint
  {https://arxiv.org/abs/1812.07923} {arXiv:1812.07923 [gr-qc]} \BibitemShut
  {NoStop}%
\bibitem [{\citenamefont {Nagar}\ \emph {et~al.}(2018)\citenamefont {Nagar}
  \emph {et~al.}}]{Nagar:2018zoe}%
  \BibitemOpen
  \bibfield  {author} {\bibinfo {author} {\bibfnamefont {A.}~\bibnamefont
  {Nagar}} \emph {et~al.},\ }\bibfield  {title} {\bibinfo {title} {{Time-domain
  effective-one-body gravitational waveforms for coalescing compact binaries
  with nonprecessing spins, tides and self-spin effects}},\ }\href
  {https://doi.org/10.1103/PhysRevD.98.104052} {\bibfield  {journal} {\bibinfo
  {journal} {Phys. Rev. D}\ }\textbf {\bibinfo {volume} {98}},\ \bibinfo
  {pages} {104052} (\bibinfo {year} {2018})},\ \Eprint
  {https://arxiv.org/abs/1806.01772} {arXiv:1806.01772 [gr-qc]} \BibitemShut
  {NoStop}%
\bibitem [{\citenamefont {Akcay}\ \emph {et~al.}(2021)\citenamefont {Akcay},
  \citenamefont {Gamba},\ and\ \citenamefont {Bernuzzi}}]{Akcay:2020qrj}%
  \BibitemOpen
  \bibfield  {author} {\bibinfo {author} {\bibfnamefont {S.}~\bibnamefont
  {Akcay}}, \bibinfo {author} {\bibfnamefont {R.}~\bibnamefont {Gamba}},\ and\
  \bibinfo {author} {\bibfnamefont {S.}~\bibnamefont {Bernuzzi}},\ }\bibfield
  {title} {\bibinfo {title} {{Hybrid post-Newtonian effective-one-body scheme
  for spin-precessing compact-binary waveforms up to merger}},\ }\href
  {https://doi.org/10.1103/PhysRevD.103.024014} {\bibfield  {journal} {\bibinfo
   {journal} {Phys. Rev. D}\ }\textbf {\bibinfo {volume} {103}},\ \bibinfo
  {pages} {024014} (\bibinfo {year} {2021})},\ \Eprint
  {https://arxiv.org/abs/2005.05338} {arXiv:2005.05338 [gr-qc]} \BibitemShut
  {NoStop}%
\bibitem [{\citenamefont {Gamba}\ \emph {et~al.}(2022)\citenamefont {Gamba},
  \citenamefont {Ak\c{c}ay}, \citenamefont {Bernuzzi},\ and\ \citenamefont
  {Williams}}]{Gamba:2021ydi}%
  \BibitemOpen
  \bibfield  {author} {\bibinfo {author} {\bibfnamefont {R.}~\bibnamefont
  {Gamba}}, \bibinfo {author} {\bibfnamefont {S.}~\bibnamefont {Ak\c{c}ay}},
  \bibinfo {author} {\bibfnamefont {S.}~\bibnamefont {Bernuzzi}},\ and\
  \bibinfo {author} {\bibfnamefont {J.}~\bibnamefont {Williams}},\ }\bibfield
  {title} {\bibinfo {title} {{Effective-one-body waveforms for precessing
  coalescing compact binaries with post-Newtonian twist}},\ }\href
  {https://doi.org/10.1103/PhysRevD.106.024020} {\bibfield  {journal} {\bibinfo
   {journal} {Phys. Rev. D}\ }\textbf {\bibinfo {volume} {106}},\ \bibinfo
  {pages} {024020} (\bibinfo {year} {2022})},\ \Eprint
  {https://arxiv.org/abs/2111.03675} {arXiv:2111.03675 [gr-qc]} \BibitemShut
  {NoStop}%
\bibitem [{\citenamefont {Bini}\ and\ \citenamefont
  {Damour}(2012)}]{Bini:2012ji}%
  \BibitemOpen
  \bibfield  {author} {\bibinfo {author} {\bibfnamefont {D.}~\bibnamefont
  {Bini}}\ and\ \bibinfo {author} {\bibfnamefont {T.}~\bibnamefont {Damour}},\
  }\bibfield  {title} {\bibinfo {title} {{Gravitational radiation reaction
  along general orbits in the effective one-body formalism}},\ }\href
  {https://doi.org/10.1103/PhysRevD.86.124012} {\bibfield  {journal} {\bibinfo
  {journal} {Phys. Rev. D}\ }\textbf {\bibinfo {volume} {86}},\ \bibinfo
  {pages} {124012} (\bibinfo {year} {2012})},\ \Eprint
  {https://arxiv.org/abs/1210.2834} {arXiv:1210.2834 [gr-qc]} \BibitemShut
  {NoStop}%
\bibitem [{\citenamefont {Hinderer}\ and\ \citenamefont
  {Babak}(2017)}]{Hinderer:2017jcs}%
  \BibitemOpen
  \bibfield  {author} {\bibinfo {author} {\bibfnamefont {T.}~\bibnamefont
  {Hinderer}}\ and\ \bibinfo {author} {\bibfnamefont {S.}~\bibnamefont
  {Babak}},\ }\bibfield  {title} {\bibinfo {title} {{Foundations of an
  effective-one-body model for coalescing binaries on eccentric orbits}},\
  }\href {https://doi.org/10.1103/PhysRevD.96.104048} {\bibfield  {journal}
  {\bibinfo  {journal} {Phys. Rev. D}\ }\textbf {\bibinfo {volume} {96}},\
  \bibinfo {pages} {104048} (\bibinfo {year} {2017})},\ \Eprint
  {https://arxiv.org/abs/1707.08426} {arXiv:1707.08426 [gr-qc]} \BibitemShut
  {NoStop}%
\bibitem [{\citenamefont {Nagar}\ \emph {et~al.}(2021)\citenamefont {Nagar},
  \citenamefont {Bonino},\ and\ \citenamefont {Rettegno}}]{Nagar:2021gss}%
  \BibitemOpen
  \bibfield  {author} {\bibinfo {author} {\bibfnamefont {A.}~\bibnamefont
  {Nagar}}, \bibinfo {author} {\bibfnamefont {A.}~\bibnamefont {Bonino}},\ and\
  \bibinfo {author} {\bibfnamefont {P.}~\bibnamefont {Rettegno}},\ }\bibfield
  {title} {\bibinfo {title} {{Effective one-body multipolar waveform model for
  spin-aligned, quasicircular, eccentric, hyperbolic black hole binaries}},\
  }\href {https://doi.org/10.1103/PhysRevD.103.104021} {\bibfield  {journal}
  {\bibinfo  {journal} {Phys. Rev. D}\ }\textbf {\bibinfo {volume} {103}},\
  \bibinfo {pages} {104021} (\bibinfo {year} {2021})},\ \Eprint
  {https://arxiv.org/abs/2101.08624} {arXiv:2101.08624 [gr-qc]} \BibitemShut
  {NoStop}%
\bibitem [{\citenamefont {Khalil}\ \emph {et~al.}(2021)\citenamefont {Khalil},
  \citenamefont {Buonanno}, \citenamefont {Steinhoff},\ and\ \citenamefont
  {Vines}}]{Khalil:2021txt}%
  \BibitemOpen
  \bibfield  {author} {\bibinfo {author} {\bibfnamefont {M.}~\bibnamefont
  {Khalil}}, \bibinfo {author} {\bibfnamefont {A.}~\bibnamefont {Buonanno}},
  \bibinfo {author} {\bibfnamefont {J.}~\bibnamefont {Steinhoff}},\ and\
  \bibinfo {author} {\bibfnamefont {J.}~\bibnamefont {Vines}},\ }\bibfield
  {title} {\bibinfo {title} {{Radiation-reaction force and multipolar waveforms
  for eccentric, spin-aligned binaries in the effective-one-body formalism}},\
  }\href {https://doi.org/10.1103/PhysRevD.104.024046} {\bibfield  {journal}
  {\bibinfo  {journal} {Phys. Rev. D}\ }\textbf {\bibinfo {volume} {104}},\
  \bibinfo {pages} {024046} (\bibinfo {year} {2021})},\ \Eprint
  {https://arxiv.org/abs/2104.11705} {arXiv:2104.11705 [gr-qc]} \BibitemShut
  {NoStop}%
\bibitem [{\citenamefont {Ramos-Buades}\ \emph {et~al.}(2022)\citenamefont
  {Ramos-Buades}, \citenamefont {Buonanno}, \citenamefont {Khalil},\ and\
  \citenamefont {Ossokine}}]{Ramos-Buades:2021adz}%
  \BibitemOpen
  \bibfield  {author} {\bibinfo {author} {\bibfnamefont {A.}~\bibnamefont
  {Ramos-Buades}}, \bibinfo {author} {\bibfnamefont {A.}~\bibnamefont
  {Buonanno}}, \bibinfo {author} {\bibfnamefont {M.}~\bibnamefont {Khalil}},\
  and\ \bibinfo {author} {\bibfnamefont {S.}~\bibnamefont {Ossokine}},\
  }\bibfield  {title} {\bibinfo {title} {{Effective-one-body multipolar
  waveforms for eccentric binary black holes with nonprecessing spins}},\
  }\href {https://doi.org/10.1103/PhysRevD.105.044035} {\bibfield  {journal}
  {\bibinfo  {journal} {Phys. Rev. D}\ }\textbf {\bibinfo {volume} {105}},\
  \bibinfo {pages} {044035} (\bibinfo {year} {2022})},\ \Eprint
  {https://arxiv.org/abs/2112.06952} {arXiv:2112.06952 [gr-qc]} \BibitemShut
  {NoStop}%
\bibitem [{\citenamefont {Albanesi}\ \emph {et~al.}(2022)\citenamefont
  {Albanesi}, \citenamefont {Placidi}, \citenamefont {Nagar}, \citenamefont
  {Orselli},\ and\ \citenamefont {Bernuzzi}}]{Albanesi:2022xge}%
  \BibitemOpen
  \bibfield  {author} {\bibinfo {author} {\bibfnamefont {S.}~\bibnamefont
  {Albanesi}}, \bibinfo {author} {\bibfnamefont {A.}~\bibnamefont {Placidi}},
  \bibinfo {author} {\bibfnamefont {A.}~\bibnamefont {Nagar}}, \bibinfo
  {author} {\bibfnamefont {M.}~\bibnamefont {Orselli}},\ and\ \bibinfo {author}
  {\bibfnamefont {S.}~\bibnamefont {Bernuzzi}},\ }\bibfield  {title} {\bibinfo
  {title} {{New avenue for accurate analytical waveforms and fluxes for
  eccentric compact binaries}},\ }\href
  {https://doi.org/10.1103/PhysRevD.105.L121503} {\bibfield  {journal}
  {\bibinfo  {journal} {Phys. Rev. D}\ }\textbf {\bibinfo {volume} {105}},\
  \bibinfo {pages} {L121503} (\bibinfo {year} {2022})},\ \Eprint
  {https://arxiv.org/abs/2203.16286} {arXiv:2203.16286 [gr-qc]} \BibitemShut
  {NoStop}%
\bibitem [{\citenamefont {Bernuzzi}\ \emph {et~al.}(2015)\citenamefont
  {Bernuzzi}, \citenamefont {Nagar}, \citenamefont {Dietrich},\ and\
  \citenamefont {Damour}}]{Bernuzzi:2014owa}%
  \BibitemOpen
  \bibfield  {author} {\bibinfo {author} {\bibfnamefont {S.}~\bibnamefont
  {Bernuzzi}}, \bibinfo {author} {\bibfnamefont {A.}~\bibnamefont {Nagar}},
  \bibinfo {author} {\bibfnamefont {T.}~\bibnamefont {Dietrich}},\ and\
  \bibinfo {author} {\bibfnamefont {T.}~\bibnamefont {Damour}},\ }\bibfield
  {title} {\bibinfo {title} {{Modeling the Dynamics of Tidally Interacting
  Binary Neutron Stars up to the Merger}},\ }\href
  {https://doi.org/10.1103/PhysRevLett.114.161103} {\bibfield  {journal}
  {\bibinfo  {journal} {Phys. Rev. Lett.}\ }\textbf {\bibinfo {volume} {114}},\
  \bibinfo {pages} {161103} (\bibinfo {year} {2015})},\ \Eprint
  {https://arxiv.org/abs/1412.4553} {arXiv:1412.4553 [gr-qc]} \BibitemShut
  {NoStop}%
\bibitem [{\citenamefont {Akcay}\ \emph {et~al.}(2019)\citenamefont {Akcay},
  \citenamefont {Bernuzzi}, \citenamefont {Messina}, \citenamefont {Nagar},
  \citenamefont {Ortiz},\ and\ \citenamefont {Rettegno}}]{Akcay:2018yyh}%
  \BibitemOpen
  \bibfield  {author} {\bibinfo {author} {\bibfnamefont {S.}~\bibnamefont
  {Akcay}}, \bibinfo {author} {\bibfnamefont {S.}~\bibnamefont {Bernuzzi}},
  \bibinfo {author} {\bibfnamefont {F.}~\bibnamefont {Messina}}, \bibinfo
  {author} {\bibfnamefont {A.}~\bibnamefont {Nagar}}, \bibinfo {author}
  {\bibfnamefont {N.}~\bibnamefont {Ortiz}},\ and\ \bibinfo {author}
  {\bibfnamefont {P.}~\bibnamefont {Rettegno}},\ }\bibfield  {title} {\bibinfo
  {title} {{Effective-one-body multipolar waveform for tidally interacting
  binary neutron stars up to merger}},\ }\href
  {https://doi.org/10.1103/PhysRevD.99.044051} {\bibfield  {journal} {\bibinfo
  {journal} {Phys. Rev. D}\ }\textbf {\bibinfo {volume} {99}},\ \bibinfo
  {pages} {044051} (\bibinfo {year} {2019})},\ \Eprint
  {https://arxiv.org/abs/1812.02744} {arXiv:1812.02744 [gr-qc]} \BibitemShut
  {NoStop}%
\bibitem [{\citenamefont {Steinhoff}\ \emph {et~al.}(2021)\citenamefont
  {Steinhoff}, \citenamefont {Hinderer}, \citenamefont {Dietrich},\ and\
  \citenamefont {Foucart}}]{Steinhoff:2021dsn}%
  \BibitemOpen
  \bibfield  {author} {\bibinfo {author} {\bibfnamefont {J.}~\bibnamefont
  {Steinhoff}}, \bibinfo {author} {\bibfnamefont {T.}~\bibnamefont {Hinderer}},
  \bibinfo {author} {\bibfnamefont {T.}~\bibnamefont {Dietrich}},\ and\
  \bibinfo {author} {\bibfnamefont {F.}~\bibnamefont {Foucart}},\ }\bibfield
  {title} {\bibinfo {title} {{Spin effects on neutron star fundamental-mode
  dynamical tides: Phenomenology and comparison to numerical simulations}},\
  }\href {https://doi.org/10.1103/PhysRevResearch.3.033129} {\bibfield
  {journal} {\bibinfo  {journal} {Phys. Rev. Res.}\ }\textbf {\bibinfo {volume}
  {3}},\ \bibinfo {pages} {033129} (\bibinfo {year} {2021})},\ \Eprint
  {https://arxiv.org/abs/2103.06100} {arXiv:2103.06100 [gr-qc]} \BibitemShut
  {NoStop}%
\bibitem [{\citenamefont {Matas}\ \emph {et~al.}(2020)\citenamefont {Matas}
  \emph {et~al.}}]{Matas:2020wab}%
  \BibitemOpen
  \bibfield  {author} {\bibinfo {author} {\bibfnamefont {A.}~\bibnamefont
  {Matas}} \emph {et~al.},\ }\bibfield  {title} {\bibinfo {title}
  {{Aligned-spin neutron-star\textendash{}black-hole waveform model based on
  the effective-one-body approach and numerical-relativity simulations}},\
  }\href {https://doi.org/10.1103/PhysRevD.102.043023} {\bibfield  {journal}
  {\bibinfo  {journal} {Phys. Rev. D}\ }\textbf {\bibinfo {volume} {102}},\
  \bibinfo {pages} {043023} (\bibinfo {year} {2020})},\ \Eprint
  {https://arxiv.org/abs/2004.10001} {arXiv:2004.10001 [gr-qc]} \BibitemShut
  {NoStop}%
\bibitem [{\citenamefont {Gonzalez}\ \emph {et~al.}(2022)\citenamefont
  {Gonzalez}, \citenamefont {Gamba}, \citenamefont {Breschi}, \citenamefont
  {Zappa}, \citenamefont {Carullo}, \citenamefont {Bernuzzi},\ and\
  \citenamefont {Nagar}}]{Gonzalez:2022prs}%
  \BibitemOpen
  \bibfield  {author} {\bibinfo {author} {\bibfnamefont {A.}~\bibnamefont
  {Gonzalez}}, \bibinfo {author} {\bibfnamefont {R.}~\bibnamefont {Gamba}},
  \bibinfo {author} {\bibfnamefont {M.}~\bibnamefont {Breschi}}, \bibinfo
  {author} {\bibfnamefont {F.}~\bibnamefont {Zappa}}, \bibinfo {author}
  {\bibfnamefont {G.}~\bibnamefont {Carullo}}, \bibinfo {author} {\bibfnamefont
  {S.}~\bibnamefont {Bernuzzi}},\ and\ \bibinfo {author} {\bibfnamefont
  {A.}~\bibnamefont {Nagar}},\ }\bibfield  {title} {\bibinfo {title}
  {{Numerical-Relativity-Informed Effective-One-Body model for
  Black-Hole-Neutron-Star Mergers with Higher Modes and Spin Precession}},\
  }\href@noop {} {\  (\bibinfo {year} {2022})},\ \Eprint
  {https://arxiv.org/abs/2212.03909} {arXiv:2212.03909 [gr-qc]} \BibitemShut
  {NoStop}%
\bibitem [{\citenamefont {Damour}(2016)}]{Damour:2016gwp}%
  \BibitemOpen
  \bibfield  {author} {\bibinfo {author} {\bibfnamefont {T.}~\bibnamefont
  {Damour}},\ }\bibfield  {title} {\bibinfo {title} {{Gravitational scattering,
  post-Minkowskian approximation and Effective One-Body theory}},\ }\href
  {https://doi.org/10.1103/PhysRevD.94.104015} {\bibfield  {journal} {\bibinfo
  {journal} {Phys. Rev. D}\ }\textbf {\bibinfo {volume} {94}},\ \bibinfo
  {pages} {104015} (\bibinfo {year} {2016})},\ \Eprint
  {https://arxiv.org/abs/1609.00354} {arXiv:1609.00354 [gr-qc]} \BibitemShut
  {NoStop}%
\bibitem [{\citenamefont {Damour}(2018)}]{Damour:2017zjx}%
  \BibitemOpen
  \bibfield  {author} {\bibinfo {author} {\bibfnamefont {T.}~\bibnamefont
  {Damour}},\ }\bibfield  {title} {\bibinfo {title} {{High-energy gravitational
  scattering and the general relativistic two-body problem}},\ }\href
  {https://doi.org/10.1103/PhysRevD.97.044038} {\bibfield  {journal} {\bibinfo
  {journal} {Phys. Rev. D}\ }\textbf {\bibinfo {volume} {97}},\ \bibinfo
  {pages} {044038} (\bibinfo {year} {2018})},\ \Eprint
  {https://arxiv.org/abs/1710.10599} {arXiv:1710.10599 [gr-qc]} \BibitemShut
  {NoStop}%
\bibitem [{\citenamefont {Antonelli}\ \emph {et~al.}(2019)\citenamefont
  {Antonelli}, \citenamefont {Buonanno}, \citenamefont {Steinhoff},
  \citenamefont {van~de Meent},\ and\ \citenamefont
  {Vines}}]{Antonelli:2019ytb}%
  \BibitemOpen
  \bibfield  {author} {\bibinfo {author} {\bibfnamefont {A.}~\bibnamefont
  {Antonelli}}, \bibinfo {author} {\bibfnamefont {A.}~\bibnamefont {Buonanno}},
  \bibinfo {author} {\bibfnamefont {J.}~\bibnamefont {Steinhoff}}, \bibinfo
  {author} {\bibfnamefont {M.}~\bibnamefont {van~de Meent}},\ and\ \bibinfo
  {author} {\bibfnamefont {J.}~\bibnamefont {Vines}},\ }\bibfield  {title}
  {\bibinfo {title} {{Energetics of two-body Hamiltonians in post-Minkowskian
  gravity}},\ }\href {https://doi.org/10.1103/PhysRevD.99.104004} {\bibfield
  {journal} {\bibinfo  {journal} {Phys. Rev. D}\ }\textbf {\bibinfo {volume}
  {99}},\ \bibinfo {pages} {104004} (\bibinfo {year} {2019})},\ \Eprint
  {https://arxiv.org/abs/1901.07102} {arXiv:1901.07102 [gr-qc]} \BibitemShut
  {NoStop}%
\bibitem [{\citenamefont {Damgaard}\ and\ \citenamefont
  {Vanhove}(2021)}]{Damgaard:2021rnk}%
  \BibitemOpen
  \bibfield  {author} {\bibinfo {author} {\bibfnamefont {P.~H.}\ \bibnamefont
  {Damgaard}}\ and\ \bibinfo {author} {\bibfnamefont {P.}~\bibnamefont
  {Vanhove}},\ }\bibfield  {title} {\bibinfo {title} {{Remodeling the effective
  one-body formalism in post-Minkowskian gravity}},\ }\href
  {https://doi.org/10.1103/PhysRevD.104.104029} {\bibfield  {journal} {\bibinfo
   {journal} {Phys. Rev. D}\ }\textbf {\bibinfo {volume} {104}},\ \bibinfo
  {pages} {104029} (\bibinfo {year} {2021})},\ \Eprint
  {https://arxiv.org/abs/2108.11248} {arXiv:2108.11248 [hep-th]} \BibitemShut
  {NoStop}%
\bibitem [{\citenamefont {Khalil}\ \emph {et~al.}(2022)\citenamefont {Khalil},
  \citenamefont {Buonanno}, \citenamefont {Steinhoff},\ and\ \citenamefont
  {Vines}}]{Khalil:2022ylj}%
  \BibitemOpen
  \bibfield  {author} {\bibinfo {author} {\bibfnamefont {M.}~\bibnamefont
  {Khalil}}, \bibinfo {author} {\bibfnamefont {A.}~\bibnamefont {Buonanno}},
  \bibinfo {author} {\bibfnamefont {J.}~\bibnamefont {Steinhoff}},\ and\
  \bibinfo {author} {\bibfnamefont {J.}~\bibnamefont {Vines}},\ }\bibfield
  {title} {\bibinfo {title} {{Energetics and scattering of gravitational
  two-body systems at fourth post-Minkowskian order}},\ }\href
  {https://doi.org/10.1103/PhysRevD.106.024042} {\bibfield  {journal} {\bibinfo
   {journal} {Phys. Rev. D}\ }\textbf {\bibinfo {volume} {106}},\ \bibinfo
  {pages} {024042} (\bibinfo {year} {2022})},\ \Eprint
  {https://arxiv.org/abs/2204.05047} {arXiv:2204.05047 [gr-qc]} \BibitemShut
  {NoStop}%
\bibitem [{\citenamefont {Damour}\ and\ \citenamefont
  {Rettegno}(2022)}]{Damour:2022ybd}%
  \BibitemOpen
  \bibfield  {author} {\bibinfo {author} {\bibfnamefont {T.}~\bibnamefont
  {Damour}}\ and\ \bibinfo {author} {\bibfnamefont {P.}~\bibnamefont
  {Rettegno}},\ }\bibfield  {title} {\bibinfo {title} {{Strong-field scattering
  of two black holes: Numerical Relativity meets Post-Minkowskian gravity}},\
  }\href@noop {} {\  (\bibinfo {year} {2022})},\ \Eprint
  {https://arxiv.org/abs/2211.01399} {arXiv:2211.01399 [gr-qc]} \BibitemShut
  {NoStop}%
\bibitem [{\citenamefont {Damour}(2010)}]{Damour:2009sm}%
  \BibitemOpen
  \bibfield  {author} {\bibinfo {author} {\bibfnamefont {T.}~\bibnamefont
  {Damour}},\ }\bibfield  {title} {\bibinfo {title} {{Gravitational Self Force
  in a Schwarzschild Background and the Effective One Body Formalism}},\ }\href
  {https://doi.org/10.1103/PhysRevD.81.024017} {\bibfield  {journal} {\bibinfo
  {journal} {Phys. Rev. D}\ }\textbf {\bibinfo {volume} {81}},\ \bibinfo
  {pages} {024017} (\bibinfo {year} {2010})},\ \Eprint
  {https://arxiv.org/abs/0910.5533} {arXiv:0910.5533 [gr-qc]} \BibitemShut
  {NoStop}%
\bibitem [{\citenamefont {Yunes}\ \emph {et~al.}(2010)\citenamefont {Yunes},
  \citenamefont {Buonanno}, \citenamefont {Hughes}, \citenamefont
  {Coleman~Miller},\ and\ \citenamefont {Pan}}]{Yunes:2009ef}%
  \BibitemOpen
  \bibfield  {author} {\bibinfo {author} {\bibfnamefont {N.}~\bibnamefont
  {Yunes}}, \bibinfo {author} {\bibfnamefont {A.}~\bibnamefont {Buonanno}},
  \bibinfo {author} {\bibfnamefont {S.~A.}\ \bibnamefont {Hughes}}, \bibinfo
  {author} {\bibfnamefont {M.}~\bibnamefont {Coleman~Miller}},\ and\ \bibinfo
  {author} {\bibfnamefont {Y.}~\bibnamefont {Pan}},\ }\bibfield  {title}
  {\bibinfo {title} {{Modeling Extreme Mass Ratio Inspirals within the
  Effective-One-Body Approach}},\ }\href
  {https://doi.org/10.1103/PhysRevLett.104.091102} {\bibfield  {journal}
  {\bibinfo  {journal} {Phys. Rev. Lett.}\ }\textbf {\bibinfo {volume} {104}},\
  \bibinfo {pages} {091102} (\bibinfo {year} {2010})},\ \Eprint
  {https://arxiv.org/abs/0909.4263} {arXiv:0909.4263 [gr-qc]} \BibitemShut
  {NoStop}%
\bibitem [{\citenamefont {Yunes}\ \emph {et~al.}(2011)\citenamefont {Yunes},
  \citenamefont {Buonanno}, \citenamefont {Hughes}, \citenamefont {Pan},
  \citenamefont {Barausse}, \citenamefont {Miller},\ and\ \citenamefont
  {Throwe}}]{Yunes:2010zj}%
  \BibitemOpen
  \bibfield  {author} {\bibinfo {author} {\bibfnamefont {N.}~\bibnamefont
  {Yunes}}, \bibinfo {author} {\bibfnamefont {A.}~\bibnamefont {Buonanno}},
  \bibinfo {author} {\bibfnamefont {S.~A.}\ \bibnamefont {Hughes}}, \bibinfo
  {author} {\bibfnamefont {Y.}~\bibnamefont {Pan}}, \bibinfo {author}
  {\bibfnamefont {E.}~\bibnamefont {Barausse}}, \bibinfo {author}
  {\bibfnamefont {M.~C.}\ \bibnamefont {Miller}},\ and\ \bibinfo {author}
  {\bibfnamefont {W.}~\bibnamefont {Throwe}},\ }\bibfield  {title} {\bibinfo
  {title} {{Extreme Mass-Ratio Inspirals in the Effective-One-Body Approach:
  Quasi-Circular, Equatorial Orbits around a Spinning Black Hole}},\ }\href
  {https://doi.org/10.1103/PhysRevD.83.044044} {\bibfield  {journal} {\bibinfo
  {journal} {Phys. Rev. D}\ }\textbf {\bibinfo {volume} {83}},\ \bibinfo
  {pages} {044044} (\bibinfo {year} {2011})},\ \bibinfo {note} {[Erratum:
  Phys.Rev.D 88, 109904 (2013)]},\ \Eprint {https://arxiv.org/abs/1009.6013}
  {arXiv:1009.6013 [gr-qc]} \BibitemShut {NoStop}%
\bibitem [{\citenamefont {Barausse}\ \emph {et~al.}(2012)\citenamefont
  {Barausse}, \citenamefont {Buonanno},\ and\ \citenamefont
  {Le~Tiec}}]{Barausse:2011dq}%
  \BibitemOpen
  \bibfield  {author} {\bibinfo {author} {\bibfnamefont {E.}~\bibnamefont
  {Barausse}}, \bibinfo {author} {\bibfnamefont {A.}~\bibnamefont {Buonanno}},\
  and\ \bibinfo {author} {\bibfnamefont {A.}~\bibnamefont {Le~Tiec}},\
  }\bibfield  {title} {\bibinfo {title} {{The complete non-spinning
  effective-one-body metric at linear order in the mass ratio}},\ }\href
  {https://doi.org/10.1103/PhysRevD.85.064010} {\bibfield  {journal} {\bibinfo
  {journal} {Phys. Rev. D}\ }\textbf {\bibinfo {volume} {85}},\ \bibinfo
  {pages} {064010} (\bibinfo {year} {2012})},\ \Eprint
  {https://arxiv.org/abs/1111.5610} {arXiv:1111.5610 [gr-qc]} \BibitemShut
  {NoStop}%
\bibitem [{\citenamefont {Akcay}\ \emph {et~al.}(2012)\citenamefont {Akcay},
  \citenamefont {Barack}, \citenamefont {Damour},\ and\ \citenamefont
  {Sago}}]{Akcay:2012ea}%
  \BibitemOpen
  \bibfield  {author} {\bibinfo {author} {\bibfnamefont {S.}~\bibnamefont
  {Akcay}}, \bibinfo {author} {\bibfnamefont {L.}~\bibnamefont {Barack}},
  \bibinfo {author} {\bibfnamefont {T.}~\bibnamefont {Damour}},\ and\ \bibinfo
  {author} {\bibfnamefont {N.}~\bibnamefont {Sago}},\ }\bibfield  {title}
  {\bibinfo {title} {{Gravitational self-force and the effective-one-body
  formalism between the innermost stable circular orbit and the light ring}},\
  }\href {https://doi.org/10.1103/PhysRevD.86.104041} {\bibfield  {journal}
  {\bibinfo  {journal} {Phys. Rev. D}\ }\textbf {\bibinfo {volume} {86}},\
  \bibinfo {pages} {104041} (\bibinfo {year} {2012})},\ \Eprint
  {https://arxiv.org/abs/1209.0964} {arXiv:1209.0964 [gr-qc]} \BibitemShut
  {NoStop}%
\bibitem [{\citenamefont {Antonelli}\ \emph
  {et~al.}(2020{\natexlab{a}})\citenamefont {Antonelli}, \citenamefont {van~de
  Meent}, \citenamefont {Buonanno}, \citenamefont {Steinhoff},\ and\
  \citenamefont {Vines}}]{Antonelli:2019fmq}%
  \BibitemOpen
  \bibfield  {author} {\bibinfo {author} {\bibfnamefont {A.}~\bibnamefont
  {Antonelli}}, \bibinfo {author} {\bibfnamefont {M.}~\bibnamefont {van~de
  Meent}}, \bibinfo {author} {\bibfnamefont {A.}~\bibnamefont {Buonanno}},
  \bibinfo {author} {\bibfnamefont {J.}~\bibnamefont {Steinhoff}},\ and\
  \bibinfo {author} {\bibfnamefont {J.}~\bibnamefont {Vines}},\ }\bibfield
  {title} {\bibinfo {title} {{Quasicircular inspirals and plunges from
  nonspinning effective-one-body Hamiltonians with gravitational self-force
  information}},\ }\href {https://doi.org/10.1103/PhysRevD.101.024024}
  {\bibfield  {journal} {\bibinfo  {journal} {Phys. Rev. D}\ }\textbf {\bibinfo
  {volume} {101}},\ \bibinfo {pages} {024024} (\bibinfo {year}
  {2020}{\natexlab{a}})},\ \Eprint {https://arxiv.org/abs/1907.11597}
  {arXiv:1907.11597 [gr-qc]} \BibitemShut {NoStop}%
\bibitem [{\citenamefont {Nagar}\ and\ \citenamefont
  {Albanesi}(2022)}]{Nagar:2022fep}%
  \BibitemOpen
  \bibfield  {author} {\bibinfo {author} {\bibfnamefont {A.}~\bibnamefont
  {Nagar}}\ and\ \bibinfo {author} {\bibfnamefont {S.}~\bibnamefont
  {Albanesi}},\ }\bibfield  {title} {\bibinfo {title} {{Toward a gravitational
  self-force-informed effective-one-body waveform model for nonprecessing,
  eccentric, large-mass-ratio inspirals}},\ }\href
  {https://doi.org/10.1103/PhysRevD.106.064049} {\bibfield  {journal} {\bibinfo
   {journal} {Phys. Rev. D}\ }\textbf {\bibinfo {volume} {106}},\ \bibinfo
  {pages} {064049} (\bibinfo {year} {2022})},\ \Eprint
  {https://arxiv.org/abs/2207.14002} {arXiv:2207.14002 [gr-qc]} \BibitemShut
  {NoStop}%
\bibitem [{\citenamefont {Field}\ \emph {et~al.}(2014)\citenamefont {Field},
  \citenamefont {Galley}, \citenamefont {Hesthaven}, \citenamefont {Kaye},\
  and\ \citenamefont {Tiglio}}]{Field:2013cfa}%
  \BibitemOpen
  \bibfield  {author} {\bibinfo {author} {\bibfnamefont {S.~E.}\ \bibnamefont
  {Field}}, \bibinfo {author} {\bibfnamefont {C.~R.}\ \bibnamefont {Galley}},
  \bibinfo {author} {\bibfnamefont {J.~S.}\ \bibnamefont {Hesthaven}}, \bibinfo
  {author} {\bibfnamefont {J.}~\bibnamefont {Kaye}},\ and\ \bibinfo {author}
  {\bibfnamefont {M.}~\bibnamefont {Tiglio}},\ }\bibfield  {title} {\bibinfo
  {title} {{Fast prediction and evaluation of gravitational waveforms using
  surrogate models}},\ }\href {https://doi.org/10.1103/PhysRevX.4.031006}
  {\bibfield  {journal} {\bibinfo  {journal} {Phys. Rev. X}\ }\textbf {\bibinfo
  {volume} {4}},\ \bibinfo {pages} {031006} (\bibinfo {year} {2014})},\ \Eprint
  {https://arxiv.org/abs/1308.3565} {arXiv:1308.3565 [gr-qc]} \BibitemShut
  {NoStop}%
\bibitem [{\citenamefont {P\"urrer}(2014)}]{Purrer:2014fza}%
  \BibitemOpen
  \bibfield  {author} {\bibinfo {author} {\bibfnamefont {M.}~\bibnamefont
  {P\"urrer}},\ }\bibfield  {title} {\bibinfo {title} {{Frequency domain
  reduced order models for gravitational waves from aligned-spin compact
  binaries}},\ }\href {https://doi.org/10.1088/0264-9381/31/19/195010}
  {\bibfield  {journal} {\bibinfo  {journal} {Class. Quant. Grav.}\ }\textbf
  {\bibinfo {volume} {31}},\ \bibinfo {pages} {195010} (\bibinfo {year}
  {2014})},\ \Eprint {https://arxiv.org/abs/1402.4146} {arXiv:1402.4146
  [gr-qc]} \BibitemShut {NoStop}%
\bibitem [{\citenamefont {P\"urrer}(2016)}]{Purrer:2015tud}%
  \BibitemOpen
  \bibfield  {author} {\bibinfo {author} {\bibfnamefont {M.}~\bibnamefont
  {P\"urrer}},\ }\bibfield  {title} {\bibinfo {title} {{Frequency domain
  reduced order model of aligned-spin effective-one-body waveforms with generic
  mass-ratios and spins}},\ }\href {https://doi.org/10.1103/PhysRevD.93.064041}
  {\bibfield  {journal} {\bibinfo  {journal} {Phys. Rev. D}\ }\textbf {\bibinfo
  {volume} {93}},\ \bibinfo {pages} {064041} (\bibinfo {year} {2016})},\
  \Eprint {https://arxiv.org/abs/1512.02248} {arXiv:1512.02248 [gr-qc]}
  \BibitemShut {NoStop}%
\bibitem [{\citenamefont {Lackey}\ \emph {et~al.}(2017)\citenamefont {Lackey},
  \citenamefont {Bernuzzi}, \citenamefont {Galley}, \citenamefont {Meidam},\
  and\ \citenamefont {Van Den~Broeck}}]{Lackey:2016krb}%
  \BibitemOpen
  \bibfield  {author} {\bibinfo {author} {\bibfnamefont {B.~D.}\ \bibnamefont
  {Lackey}}, \bibinfo {author} {\bibfnamefont {S.}~\bibnamefont {Bernuzzi}},
  \bibinfo {author} {\bibfnamefont {C.~R.}\ \bibnamefont {Galley}}, \bibinfo
  {author} {\bibfnamefont {J.}~\bibnamefont {Meidam}},\ and\ \bibinfo {author}
  {\bibfnamefont {C.}~\bibnamefont {Van Den~Broeck}},\ }\bibfield  {title}
  {\bibinfo {title} {{Effective-one-body waveforms for binary neutron stars
  using surrogate models}},\ }\href
  {https://doi.org/10.1103/PhysRevD.95.104036} {\bibfield  {journal} {\bibinfo
  {journal} {Phys. Rev. D}\ }\textbf {\bibinfo {volume} {95}},\ \bibinfo
  {pages} {104036} (\bibinfo {year} {2017})},\ \Eprint
  {https://arxiv.org/abs/1610.04742} {arXiv:1610.04742 [gr-qc]} \BibitemShut
  {NoStop}%
\bibitem [{\citenamefont {Lackey}\ \emph {et~al.}(2019)\citenamefont {Lackey},
  \citenamefont {P\"urrer}, \citenamefont {Taracchini},\ and\ \citenamefont
  {Marsat}}]{Lackey:2018zvw}%
  \BibitemOpen
  \bibfield  {author} {\bibinfo {author} {\bibfnamefont {B.~D.}\ \bibnamefont
  {Lackey}}, \bibinfo {author} {\bibfnamefont {M.}~\bibnamefont {P\"urrer}},
  \bibinfo {author} {\bibfnamefont {A.}~\bibnamefont {Taracchini}},\ and\
  \bibinfo {author} {\bibfnamefont {S.}~\bibnamefont {Marsat}},\ }\bibfield
  {title} {\bibinfo {title} {{Surrogate model for an aligned-spin effective one
  body waveform model of binary neutron star inspirals using Gaussian process
  regression}},\ }\href {https://doi.org/10.1103/PhysRevD.100.024002}
  {\bibfield  {journal} {\bibinfo  {journal} {Phys. Rev. D}\ }\textbf {\bibinfo
  {volume} {100}},\ \bibinfo {pages} {024002} (\bibinfo {year} {2019})},\
  \Eprint {https://arxiv.org/abs/1812.08643} {arXiv:1812.08643 [gr-qc]}
  \BibitemShut {NoStop}%
\bibitem [{\citenamefont {Cotesta}\ \emph {et~al.}(2020)\citenamefont
  {Cotesta}, \citenamefont {Marsat},\ and\ \citenamefont
  {P\"urrer}}]{Cotesta:2020qhw}%
  \BibitemOpen
  \bibfield  {author} {\bibinfo {author} {\bibfnamefont {R.}~\bibnamefont
  {Cotesta}}, \bibinfo {author} {\bibfnamefont {S.}~\bibnamefont {Marsat}},\
  and\ \bibinfo {author} {\bibfnamefont {M.}~\bibnamefont {P\"urrer}},\
  }\bibfield  {title} {\bibinfo {title} {{Frequency domain reduced order model
  of aligned-spin effective-one-body waveforms with higher-order modes}},\
  }\href {https://doi.org/10.1103/PhysRevD.101.124040} {\bibfield  {journal}
  {\bibinfo  {journal} {Phys. Rev. D}\ }\textbf {\bibinfo {volume} {101}},\
  \bibinfo {pages} {124040} (\bibinfo {year} {2020})},\ \Eprint
  {https://arxiv.org/abs/2003.12079} {arXiv:2003.12079 [gr-qc]} \BibitemShut
  {NoStop}%
\bibitem [{\citenamefont {Gadre}\ \emph {et~al.}(2022)\citenamefont {Gadre},
  \citenamefont {P\"urrer}, \citenamefont {Field}, \citenamefont {Ossokine},\
  and\ \citenamefont {Varma}}]{Gadre:2022sed}%
  \BibitemOpen
  \bibfield  {author} {\bibinfo {author} {\bibfnamefont {B.}~\bibnamefont
  {Gadre}}, \bibinfo {author} {\bibfnamefont {M.}~\bibnamefont {P\"urrer}},
  \bibinfo {author} {\bibfnamefont {S.~E.}\ \bibnamefont {Field}}, \bibinfo
  {author} {\bibfnamefont {S.}~\bibnamefont {Ossokine}},\ and\ \bibinfo
  {author} {\bibfnamefont {V.}~\bibnamefont {Varma}},\ }\bibfield  {title}
  {\bibinfo {title} {{A fully precessing higher-mode surrogate model of
  effective-one-body waveforms}},\ }\href@noop {} {\  (\bibinfo {year}
  {2022})},\ \Eprint {https://arxiv.org/abs/2203.00381} {arXiv:2203.00381
  [gr-qc]} \BibitemShut {NoStop}%
\bibitem [{\citenamefont {Tissino}\ \emph {et~al.}(2022)\citenamefont
  {Tissino}, \citenamefont {Carullo}, \citenamefont {Breschi}, \citenamefont
  {Gamba}, \citenamefont {Schmidt},\ and\ \citenamefont
  {Bernuzzi}}]{Tissino:2022thn}%
  \BibitemOpen
  \bibfield  {author} {\bibinfo {author} {\bibfnamefont {J.}~\bibnamefont
  {Tissino}}, \bibinfo {author} {\bibfnamefont {G.}~\bibnamefont {Carullo}},
  \bibinfo {author} {\bibfnamefont {M.}~\bibnamefont {Breschi}}, \bibinfo
  {author} {\bibfnamefont {R.}~\bibnamefont {Gamba}}, \bibinfo {author}
  {\bibfnamefont {S.}~\bibnamefont {Schmidt}},\ and\ \bibinfo {author}
  {\bibfnamefont {S.}~\bibnamefont {Bernuzzi}},\ }\bibfield  {title} {\bibinfo
  {title} {{Combining effective-one-body accuracy and reduced-order-quadrature
  speed for binary neutron star merger parameter estimation with machine
  learning}},\ }\href@noop {} {\  (\bibinfo {year} {2022})},\ \Eprint
  {https://arxiv.org/abs/2210.15684} {arXiv:2210.15684 [gr-qc]} \BibitemShut
  {NoStop}%
\bibitem [{\citenamefont {Khan}\ and\ \citenamefont
  {Green}(2021)}]{Khan:2020fso}%
  \BibitemOpen
  \bibfield  {author} {\bibinfo {author} {\bibfnamefont {S.}~\bibnamefont
  {Khan}}\ and\ \bibinfo {author} {\bibfnamefont {R.}~\bibnamefont {Green}},\
  }\bibfield  {title} {\bibinfo {title} {{Gravitational-wave surrogate models
  powered by artificial neural networks}},\ }\href
  {https://doi.org/10.1103/PhysRevD.103.064015} {\bibfield  {journal} {\bibinfo
   {journal} {Phys. Rev. D}\ }\textbf {\bibinfo {volume} {103}},\ \bibinfo
  {pages} {064015} (\bibinfo {year} {2021})},\ \Eprint
  {https://arxiv.org/abs/2008.12932} {arXiv:2008.12932 [gr-qc]} \BibitemShut
  {NoStop}%
\bibitem [{\citenamefont {Thomas}\ \emph {et~al.}(2022)\citenamefont {Thomas},
  \citenamefont {Pratten},\ and\ \citenamefont {Schmidt}}]{Thomas:2022rmc}%
  \BibitemOpen
  \bibfield  {author} {\bibinfo {author} {\bibfnamefont {L.~M.}\ \bibnamefont
  {Thomas}}, \bibinfo {author} {\bibfnamefont {G.}~\bibnamefont {Pratten}},\
  and\ \bibinfo {author} {\bibfnamefont {P.}~\bibnamefont {Schmidt}},\
  }\bibfield  {title} {\bibinfo {title} {{Accelerating multimodal gravitational
  waveforms from precessing compact binaries with artificial neural
  networks}},\ }\href {https://doi.org/10.1103/PhysRevD.106.104029} {\bibfield
  {journal} {\bibinfo  {journal} {Phys. Rev. D}\ }\textbf {\bibinfo {volume}
  {106}},\ \bibinfo {pages} {104029} (\bibinfo {year} {2022})},\ \Eprint
  {https://arxiv.org/abs/2205.14066} {arXiv:2205.14066 [gr-qc]} \BibitemShut
  {NoStop}%
\bibitem [{\citenamefont {Nagar}\ \emph
  {et~al.}(2020{\natexlab{b}})\citenamefont {Nagar}, \citenamefont
  {Riemenschneider}, \citenamefont {Pratten}, \citenamefont {Rettegno},\ and\
  \citenamefont {Messina}}]{Nagar:2020pcj}%
  \BibitemOpen
  \bibfield  {author} {\bibinfo {author} {\bibfnamefont {A.}~\bibnamefont
  {Nagar}}, \bibinfo {author} {\bibfnamefont {G.}~\bibnamefont
  {Riemenschneider}}, \bibinfo {author} {\bibfnamefont {G.}~\bibnamefont
  {Pratten}}, \bibinfo {author} {\bibfnamefont {P.}~\bibnamefont {Rettegno}},\
  and\ \bibinfo {author} {\bibfnamefont {F.}~\bibnamefont {Messina}},\
  }\bibfield  {title} {\bibinfo {title} {{Multipolar effective one body
  waveform model for spin-aligned black hole binaries}},\ }\href
  {https://doi.org/10.1103/PhysRevD.102.024077} {\bibfield  {journal} {\bibinfo
   {journal} {Phys. Rev. D}\ }\textbf {\bibinfo {volume} {102}},\ \bibinfo
  {pages} {024077} (\bibinfo {year} {2020}{\natexlab{b}})},\ \Eprint
  {https://arxiv.org/abs/2001.09082} {arXiv:2001.09082 [gr-qc]} \BibitemShut
  {NoStop}%
\bibitem [{LVK()}]{LVKO4}%
  \BibitemOpen
  \href@noop {} {}\bibinfo {howpublished}
  {\url{https://observing.docs.ligo.org/plan/}},\ \bibinfo {note} {accessed:
  2023-02-25}\BibitemShut {NoStop}%
\bibitem [{\citenamefont {Mihaylov}\ \emph {et~al.}(2023)\citenamefont
  {Mihaylov}, \citenamefont {Ossokine}, \citenamefont {Buonanno}, \citenamefont
  {Estelles}, \citenamefont {Pompili}, \citenamefont {P\"urrer},\ and\
  \citenamefont {Ramos-Buades}}]{Mihaylovv5}%
  \BibitemOpen
  \bibfield  {author} {\bibinfo {author} {\bibfnamefont {D.~P.}\ \bibnamefont
  {Mihaylov}}, \bibinfo {author} {\bibfnamefont {S.}~\bibnamefont {Ossokine}},
  \bibinfo {author} {\bibfnamefont {A.}~\bibnamefont {Buonanno}}, \bibinfo
  {author} {\bibfnamefont {H.}~\bibnamefont {Estelles}}, \bibinfo {author}
  {\bibfnamefont {L.}~\bibnamefont {Pompili}}, \bibinfo {author} {\bibfnamefont
  {M.}~\bibnamefont {P\"urrer}},\ and\ \bibinfo {author} {\bibfnamefont
  {A.}~\bibnamefont {Ramos-Buades}},\ }\bibfield  {title} {\bibinfo {title}
  {{pySEOBNR: a software package for the next generation of effective-one-body
  multipolar waveform models}},\ }\href@noop {} {\  (\bibinfo {year} {2023})},\
  \Eprint {https://arxiv.org/abs/2303.18203} {arXiv:2303.18203 [gr-qc]}
  \BibitemShut {NoStop}%
\bibitem [{\citenamefont {Pompili}\ \emph {et~al.}(2023)\citenamefont
  {Pompili}, \citenamefont {Buonanno}, \citenamefont {Estell\'es},
  \citenamefont {Khalil}, \citenamefont {van~de Meent}, \citenamefont
  {Mihaylov}, \citenamefont {Ossokine}, \citenamefont {P{\"u}rrer},
  \citenamefont {Ramos-Buades}, \citenamefont {Kumar~Mehta}, \citenamefont
  {Cotesta}, \citenamefont {Marsat}, \citenamefont {Boyle}, \citenamefont
  {Kidder}, \citenamefont {Pfeiffer}, \citenamefont {Scheel}, \citenamefont
  {R{\"u}ter}, \citenamefont {Vu}, \citenamefont {Dudi}, \citenamefont {Ma},
  \citenamefont {Mitman}, \citenamefont {Melchor}, \citenamefont {Thomas},\
  and\ \citenamefont {Sanchez}}]{Pompiliv5}%
  \BibitemOpen
  \bibfield  {author} {\bibinfo {author} {\bibfnamefont {L.}~\bibnamefont
  {Pompili}}, \bibinfo {author} {\bibfnamefont {A.}~\bibnamefont {Buonanno}},
  \bibinfo {author} {\bibfnamefont {H.}~\bibnamefont {Estell\'es}}, \bibinfo
  {author} {\bibfnamefont {M.}~\bibnamefont {Khalil}}, \bibinfo {author}
  {\bibfnamefont {M.}~\bibnamefont {van~de Meent}}, \bibinfo {author}
  {\bibfnamefont {D.}~\bibnamefont {Mihaylov}}, \bibinfo {author}
  {\bibfnamefont {S.}~\bibnamefont {Ossokine}}, \bibinfo {author}
  {\bibfnamefont {M.}~\bibnamefont {P{\"u}rrer}}, \bibinfo {author}
  {\bibfnamefont {A.}~\bibnamefont {Ramos-Buades}}, \bibinfo {author}
  {\bibfnamefont {A.}~\bibnamefont {Kumar~Mehta}}, \bibinfo {author}
  {\bibfnamefont {R.}~\bibnamefont {Cotesta}}, \bibinfo {author} {\bibfnamefont
  {S.}~\bibnamefont {Marsat}}, \bibinfo {author} {\bibfnamefont
  {M.}~\bibnamefont {Boyle}}, \bibinfo {author} {\bibfnamefont {L.~E.}\
  \bibnamefont {Kidder}}, \bibinfo {author} {\bibfnamefont {H.~P.}\
  \bibnamefont {Pfeiffer}}, \bibinfo {author} {\bibfnamefont {M.~A.}\
  \bibnamefont {Scheel}}, \bibinfo {author} {\bibfnamefont {H.~R.}\
  \bibnamefont {R{\"u}ter}}, \bibinfo {author} {\bibfnamefont {N.}~\bibnamefont
  {Vu}}, \bibinfo {author} {\bibfnamefont {R.}~\bibnamefont {Dudi}}, \bibinfo
  {author} {\bibfnamefont {S.}~\bibnamefont {Ma}}, \bibinfo {author}
  {\bibfnamefont {K.}~\bibnamefont {Mitman}}, \bibinfo {author} {\bibfnamefont
  {D.}~\bibnamefont {Melchor}}, \bibinfo {author} {\bibfnamefont
  {S.}~\bibnamefont {Thomas}},\ and\ \bibinfo {author} {\bibfnamefont
  {J.}~\bibnamefont {Sanchez}},\ }\bibfield  {title} {\bibinfo {title} {{Laying
  the foundation of the effective-one-body waveform models SEOBNRv5: improved
  accuracy and efficiency for spinning non-precessing binary black holes}},\
  }\href@noop {} {\  (\bibinfo {year} {2023})},\ \Eprint
  {https://arxiv.org/abs/2303.18039} {arXiv:2303.18039 [gr-qc]} \BibitemShut
  {NoStop}%
\bibitem [{\citenamefont {Ramos-Buades}\ \emph {et~al.}(2023)\citenamefont
  {Ramos-Buades}, \citenamefont {Buonanno}, \citenamefont {Estell\'es},
  \citenamefont {Khalil}, \citenamefont {Mihaylov}, \citenamefont {Ossokine},
  \citenamefont {Pompili},\ and\ \citenamefont {Shiferaw}}]{RamosBuadesv5}%
  \BibitemOpen
  \bibfield  {author} {\bibinfo {author} {\bibfnamefont {A.}~\bibnamefont
  {Ramos-Buades}}, \bibinfo {author} {\bibfnamefont {A.}~\bibnamefont
  {Buonanno}}, \bibinfo {author} {\bibfnamefont {H.}~\bibnamefont
  {Estell\'es}}, \bibinfo {author} {\bibfnamefont {M.}~\bibnamefont {Khalil}},
  \bibinfo {author} {\bibfnamefont {D.~P.}\ \bibnamefont {Mihaylov}}, \bibinfo
  {author} {\bibfnamefont {S.}~\bibnamefont {Ossokine}}, \bibinfo {author}
  {\bibfnamefont {L.}~\bibnamefont {Pompili}},\ and\ \bibinfo {author}
  {\bibfnamefont {M.}~\bibnamefont {Shiferaw}},\ }\bibfield  {title} {\bibinfo
  {title} {{SEOBNRv5PHM: Next generation of accurate and efficient multipolar
  precessing-spin effective-one-body waveforms for binary black holes}},\
  }\href@noop {} {\  (\bibinfo {year} {2023})},\ \Eprint
  {https://arxiv.org/abs/2303.18046} {arXiv:2303.18046 [gr-qc]} \BibitemShut
  {NoStop}%
\bibitem [{\citenamefont {van~de Meent}\ \emph {et~al.}(2023)\citenamefont
  {van~de Meent}, \citenamefont {Buonanno}, \citenamefont {Mihaylov},
  \citenamefont {Ossokine}, \citenamefont {Pompili}, \citenamefont {Warburton},
  \citenamefont {Pound}, \citenamefont {Wardell}, \citenamefont {Durkan},\ and\
  \citenamefont {Miller}}]{VandeMeentv5}%
  \BibitemOpen
  \bibfield  {author} {\bibinfo {author} {\bibfnamefont {M.}~\bibnamefont
  {van~de Meent}}, \bibinfo {author} {\bibfnamefont {A.}~\bibnamefont
  {Buonanno}}, \bibinfo {author} {\bibfnamefont {D.~P.}\ \bibnamefont
  {Mihaylov}}, \bibinfo {author} {\bibfnamefont {S.}~\bibnamefont {Ossokine}},
  \bibinfo {author} {\bibfnamefont {L.}~\bibnamefont {Pompili}}, \bibinfo
  {author} {\bibfnamefont {N.}~\bibnamefont {Warburton}}, \bibinfo {author}
  {\bibfnamefont {A.}~\bibnamefont {Pound}}, \bibinfo {author} {\bibfnamefont
  {B.}~\bibnamefont {Wardell}}, \bibinfo {author} {\bibfnamefont
  {L.}~\bibnamefont {Durkan}},\ and\ \bibinfo {author} {\bibfnamefont
  {J.}~\bibnamefont {Miller}},\ }\bibfield  {title} {\bibinfo {title}
  {{Enhancing the SEOBNRv5 effective-one-body waveform model with second-order
  gravitational self-force fluxes}},\ }\href@noop {} {\  (\bibinfo {year}
  {2023})},\ \Eprint {https://arxiv.org/abs/2303.18026} {arXiv:2303.18026
  [gr-qc]} \BibitemShut {NoStop}%
\bibitem [{\citenamefont {Knowles}\ \emph {et~al.}(2018)\citenamefont
  {Knowles}, \citenamefont {Devine}, \citenamefont {Buch}, \citenamefont
  {Bilgili}, \citenamefont {Adams}, \citenamefont {Etienne},\ and\
  \citenamefont {Mcwilliams}}]{Knowles:2018hqq}%
  \BibitemOpen
  \bibfield  {author} {\bibinfo {author} {\bibfnamefont {T.~D.}\ \bibnamefont
  {Knowles}}, \bibinfo {author} {\bibfnamefont {C.}~\bibnamefont {Devine}},
  \bibinfo {author} {\bibfnamefont {D.~A.}\ \bibnamefont {Buch}}, \bibinfo
  {author} {\bibfnamefont {S.~A.}\ \bibnamefont {Bilgili}}, \bibinfo {author}
  {\bibfnamefont {T.~R.}\ \bibnamefont {Adams}}, \bibinfo {author}
  {\bibfnamefont {Z.~B.}\ \bibnamefont {Etienne}},\ and\ \bibinfo {author}
  {\bibfnamefont {S.~T.}\ \bibnamefont {Mcwilliams}},\ }\bibfield  {title}
  {\bibinfo {title} {{Improving performance of SEOBNRv3 by $\sim$300x}},\
  }\href {https://doi.org/10.1088/1361-6382/aacb8c} {\bibfield  {journal}
  {\bibinfo  {journal} {Class. Quant. Grav.}\ }\textbf {\bibinfo {volume}
  {35}},\ \bibinfo {pages} {155003} (\bibinfo {year} {2018})},\ \Eprint
  {https://arxiv.org/abs/1803.06346} {arXiv:1803.06346 [gr-qc]} \BibitemShut
  {NoStop}%
\bibitem [{\citenamefont {Devine}\ \emph {et~al.}(2016)\citenamefont {Devine},
  \citenamefont {Etienne},\ and\ \citenamefont {McWilliams}}]{Devine:2016ovp}%
  \BibitemOpen
  \bibfield  {author} {\bibinfo {author} {\bibfnamefont {C.}~\bibnamefont
  {Devine}}, \bibinfo {author} {\bibfnamefont {Z.~B.}\ \bibnamefont
  {Etienne}},\ and\ \bibinfo {author} {\bibfnamefont {S.~T.}\ \bibnamefont
  {McWilliams}},\ }\bibfield  {title} {\bibinfo {title} {{Optimizing spinning
  time-domain gravitational waveforms for Advanced LIGO data analysis}},\
  }\href {https://doi.org/10.1088/0264-9381/33/12/125025} {\bibfield  {journal}
  {\bibinfo  {journal} {Class. Quant. Grav.}\ }\textbf {\bibinfo {volume}
  {33}},\ \bibinfo {pages} {125025} (\bibinfo {year} {2016})},\ \Eprint
  {https://arxiv.org/abs/1601.03393} {arXiv:1601.03393 [astro-ph.HE]}
  \BibitemShut {NoStop}%
\bibitem [{\citenamefont {Sturani}()}]{SpinTaylorNotes}%
  \BibitemOpen
  \bibfield  {author} {\bibinfo {author} {\bibfnamefont {R.}~\bibnamefont
  {Sturani}},\ }\href@noop {} {\bibinfo {title} {Note on the derivation of the
  angular momentum and spin precessing equations in spintaylor codes}},\
  \bibinfo {howpublished}
  {\url{https://dcc.ligo.org/public/0122/T1500554/023/dLdS.pdf}}\BibitemShut
  {NoStop}%
\bibitem [{\citenamefont {Apostolatos}\ \emph {et~al.}(1994)\citenamefont
  {Apostolatos}, \citenamefont {Cutler}, \citenamefont {Sussman},\ and\
  \citenamefont {Thorne}}]{Apostolatos:1994mx}%
  \BibitemOpen
  \bibfield  {author} {\bibinfo {author} {\bibfnamefont {T.~A.}\ \bibnamefont
  {Apostolatos}}, \bibinfo {author} {\bibfnamefont {C.}~\bibnamefont {Cutler}},
  \bibinfo {author} {\bibfnamefont {G.~J.}\ \bibnamefont {Sussman}},\ and\
  \bibinfo {author} {\bibfnamefont {K.~S.}\ \bibnamefont {Thorne}},\ }\bibfield
   {title} {\bibinfo {title} {{Spin induced orbital precession and its
  modulation of the gravitational wave forms from merging binaries}},\ }\href
  {https://doi.org/10.1103/PhysRevD.49.6274} {\bibfield  {journal} {\bibinfo
  {journal} {Phys. Rev. D}\ }\textbf {\bibinfo {volume} {49}},\ \bibinfo
  {pages} {6274} (\bibinfo {year} {1994})}\BibitemShut {NoStop}%
\bibitem [{\citenamefont {Boyle}\ \emph {et~al.}(2011)\citenamefont {Boyle},
  \citenamefont {Owen},\ and\ \citenamefont {Pfeiffer}}]{Boyle:2011gg}%
  \BibitemOpen
  \bibfield  {author} {\bibinfo {author} {\bibfnamefont {M.}~\bibnamefont
  {Boyle}}, \bibinfo {author} {\bibfnamefont {R.}~\bibnamefont {Owen}},\ and\
  \bibinfo {author} {\bibfnamefont {H.~P.}\ \bibnamefont {Pfeiffer}},\
  }\bibfield  {title} {\bibinfo {title} {{A geometric approach to the
  precession of compact binaries}},\ }\href
  {https://doi.org/10.1103/PhysRevD.84.124011} {\bibfield  {journal} {\bibinfo
  {journal} {Phys. Rev. D}\ }\textbf {\bibinfo {volume} {84}},\ \bibinfo
  {pages} {124011} (\bibinfo {year} {2011})},\ \Eprint
  {https://arxiv.org/abs/1110.2965} {arXiv:1110.2965 [gr-qc]} \BibitemShut
  {NoStop}%
\bibitem [{\citenamefont {Schmidt}\ \emph {et~al.}(2012)\citenamefont
  {Schmidt}, \citenamefont {Hannam},\ and\ \citenamefont
  {Husa}}]{Schmidt:2012rh}%
  \BibitemOpen
  \bibfield  {author} {\bibinfo {author} {\bibfnamefont {P.}~\bibnamefont
  {Schmidt}}, \bibinfo {author} {\bibfnamefont {M.}~\bibnamefont {Hannam}},\
  and\ \bibinfo {author} {\bibfnamefont {S.}~\bibnamefont {Husa}},\ }\bibfield
  {title} {\bibinfo {title} {{Towards models of gravitational waveforms from
  generic binaries: A simple approximate mapping between precessing and
  non-precessing inspiral signals}},\ }\href
  {https://doi.org/10.1103/PhysRevD.86.104063} {\bibfield  {journal} {\bibinfo
  {journal} {Phys. Rev. D}\ }\textbf {\bibinfo {volume} {86}},\ \bibinfo
  {pages} {104063} (\bibinfo {year} {2012})},\ \Eprint
  {https://arxiv.org/abs/1207.3088} {arXiv:1207.3088 [gr-qc]} \BibitemShut
  {NoStop}%
\bibitem [{\citenamefont {Boh\'e}\ \emph {et~al.}(2013)\citenamefont {Boh\'e},
  \citenamefont {Marsat},\ and\ \citenamefont {Blanchet}}]{Bohe:2013cla}%
  \BibitemOpen
  \bibfield  {author} {\bibinfo {author} {\bibfnamefont {A.}~\bibnamefont
  {Boh\'e}}, \bibinfo {author} {\bibfnamefont {S.}~\bibnamefont {Marsat}},\
  and\ \bibinfo {author} {\bibfnamefont {L.}~\bibnamefont {Blanchet}},\
  }\bibfield  {title} {\bibinfo {title} {{Next-to-next-to-leading order
  spin\textendash{}orbit effects in the gravitational wave flux and orbital
  phasing of compact binaries}},\ }\href
  {https://doi.org/10.1088/0264-9381/30/13/135009} {\bibfield  {journal}
  {\bibinfo  {journal} {Class. Quant. Grav.}\ }\textbf {\bibinfo {volume}
  {30}},\ \bibinfo {pages} {135009} (\bibinfo {year} {2013})},\ \Eprint
  {https://arxiv.org/abs/1303.7412} {arXiv:1303.7412 [gr-qc]} \BibitemShut
  {NoStop}%
\bibitem [{anc()}]{ancmaterial}%
  \BibitemOpen
  \href@noop {} {\ }\bibinfo {note} {The ancillary files \texttt{Hamiltonian.m}
  and \texttt{PNexp\_EOMs.m} contain the precessing-spin Hamiltonians and
  PN-expanded equations of motion derived in this paper.}\BibitemShut {Stop}%
\bibitem [{\citenamefont {Bini}\ \emph {et~al.}(2019)\citenamefont {Bini},
  \citenamefont {Damour},\ and\ \citenamefont {Geralico}}]{Bini:2019nra}%
  \BibitemOpen
  \bibfield  {author} {\bibinfo {author} {\bibfnamefont {D.}~\bibnamefont
  {Bini}}, \bibinfo {author} {\bibfnamefont {T.}~\bibnamefont {Damour}},\ and\
  \bibinfo {author} {\bibfnamefont {A.}~\bibnamefont {Geralico}},\ }\bibfield
  {title} {\bibinfo {title} {{Novel approach to binary dynamics: application to
  the fifth post-Newtonian level}},\ }\href
  {https://doi.org/10.1103/PhysRevLett.123.231104} {\bibfield  {journal}
  {\bibinfo  {journal} {Phys. Rev. Lett.}\ }\textbf {\bibinfo {volume} {123}},\
  \bibinfo {pages} {231104} (\bibinfo {year} {2019})},\ \Eprint
  {https://arxiv.org/abs/1909.02375} {arXiv:1909.02375 [gr-qc]} \BibitemShut
  {NoStop}%
\bibitem [{\citenamefont {Bini}\ \emph
  {et~al.}(2020{\natexlab{a}})\citenamefont {Bini}, \citenamefont {Damour},\
  and\ \citenamefont {Geralico}}]{Bini:2020wpo}%
  \BibitemOpen
  \bibfield  {author} {\bibinfo {author} {\bibfnamefont {D.}~\bibnamefont
  {Bini}}, \bibinfo {author} {\bibfnamefont {T.}~\bibnamefont {Damour}},\ and\
  \bibinfo {author} {\bibfnamefont {A.}~\bibnamefont {Geralico}},\ }\bibfield
  {title} {\bibinfo {title} {{Binary dynamics at the fifth and fifth-and-a-half
  post-Newtonian orders}},\ }\href
  {https://doi.org/10.1103/PhysRevD.102.024062} {\bibfield  {journal} {\bibinfo
   {journal} {Phys. Rev. D}\ }\textbf {\bibinfo {volume} {102}},\ \bibinfo
  {pages} {024062} (\bibinfo {year} {2020}{\natexlab{a}})},\ \Eprint
  {https://arxiv.org/abs/2003.11891} {arXiv:2003.11891 [gr-qc]} \BibitemShut
  {NoStop}%
\bibitem [{\citenamefont {Bl\"umlein}\ \emph {et~al.}(2021)\citenamefont
  {Bl\"umlein}, \citenamefont {Maier}, \citenamefont {Marquard},\ and\
  \citenamefont {Sch\"afer}}]{Blumlein:2021txe}%
  \BibitemOpen
  \bibfield  {author} {\bibinfo {author} {\bibfnamefont {J.}~\bibnamefont
  {Bl\"umlein}}, \bibinfo {author} {\bibfnamefont {A.}~\bibnamefont {Maier}},
  \bibinfo {author} {\bibfnamefont {P.}~\bibnamefont {Marquard}},\ and\
  \bibinfo {author} {\bibfnamefont {G.}~\bibnamefont {Sch\"afer}},\ }\bibfield
  {title} {\bibinfo {title} {{The fifth-order post-Newtonian Hamiltonian
  dynamics of two-body systems from an effective field theory approach}},\
  }\href@noop {} {\  (\bibinfo {year} {2021})},\ \Eprint
  {https://arxiv.org/abs/2110.13822} {arXiv:2110.13822 [gr-qc]} \BibitemShut
  {NoStop}%
\bibitem [{\citenamefont {Bini}\ \emph
  {et~al.}(2020{\natexlab{b}})\citenamefont {Bini}, \citenamefont {Damour},\
  and\ \citenamefont {Geralico}}]{Bini:2020nsb}%
  \BibitemOpen
  \bibfield  {author} {\bibinfo {author} {\bibfnamefont {D.}~\bibnamefont
  {Bini}}, \bibinfo {author} {\bibfnamefont {T.}~\bibnamefont {Damour}},\ and\
  \bibinfo {author} {\bibfnamefont {A.}~\bibnamefont {Geralico}},\ }\bibfield
  {title} {\bibinfo {title} {{Sixth post-Newtonian local-in-time dynamics of
  binary systems}},\ }\href {https://doi.org/10.1103/PhysRevD.102.024061}
  {\bibfield  {journal} {\bibinfo  {journal} {Phys. Rev. D}\ }\textbf {\bibinfo
  {volume} {102}},\ \bibinfo {pages} {024061} (\bibinfo {year}
  {2020}{\natexlab{b}})},\ \Eprint {https://arxiv.org/abs/2004.05407}
  {arXiv:2004.05407 [gr-qc]} \BibitemShut {NoStop}%
\bibitem [{\citenamefont {Bini}\ \emph
  {et~al.}(2020{\natexlab{c}})\citenamefont {Bini}, \citenamefont {Damour},\
  and\ \citenamefont {Geralico}}]{Bini:2020hmy}%
  \BibitemOpen
  \bibfield  {author} {\bibinfo {author} {\bibfnamefont {D.}~\bibnamefont
  {Bini}}, \bibinfo {author} {\bibfnamefont {T.}~\bibnamefont {Damour}},\ and\
  \bibinfo {author} {\bibfnamefont {A.}~\bibnamefont {Geralico}},\ }\bibfield
  {title} {\bibinfo {title} {{Sixth post-Newtonian nonlocal-in-time dynamics of
  binary systems}},\ }\href {https://doi.org/10.1103/PhysRevD.102.084047}
  {\bibfield  {journal} {\bibinfo  {journal} {Phys. Rev. D}\ }\textbf {\bibinfo
  {volume} {102}},\ \bibinfo {pages} {084047} (\bibinfo {year}
  {2020}{\natexlab{c}})},\ \Eprint {https://arxiv.org/abs/2007.11239}
  {arXiv:2007.11239 [gr-qc]} \BibitemShut {NoStop}%
\bibitem [{\citenamefont {Balmelli}\ and\ \citenamefont
  {Jetzer}(2013)}]{Balmelli:2013zna}%
  \BibitemOpen
  \bibfield  {author} {\bibinfo {author} {\bibfnamefont {S.}~\bibnamefont
  {Balmelli}}\ and\ \bibinfo {author} {\bibfnamefont {P.}~\bibnamefont
  {Jetzer}},\ }\bibfield  {title} {\bibinfo {title} {{Effective-one-body
  Hamiltonian with next-to-leading order spin-spin coupling for two
  nonprecessing black holes with aligned spins}},\ }\href
  {https://doi.org/10.1103/PhysRevD.87.124036} {\bibfield  {journal} {\bibinfo
  {journal} {Phys. Rev. D}\ }\textbf {\bibinfo {volume} {87}},\ \bibinfo
  {pages} {124036} (\bibinfo {year} {2013})},\ \bibinfo {note} {[Erratum:
  Phys.Rev.D 90, 089905 (2014)]},\ \Eprint {https://arxiv.org/abs/1305.5674}
  {arXiv:1305.5674 [gr-qc]} \BibitemShut {NoStop}%
\bibitem [{\citenamefont {Balmelli}\ and\ \citenamefont
  {Jetzer}(2015)}]{Balmelli:2015lva}%
  \BibitemOpen
  \bibfield  {author} {\bibinfo {author} {\bibfnamefont {S.}~\bibnamefont
  {Balmelli}}\ and\ \bibinfo {author} {\bibfnamefont {P.}~\bibnamefont
  {Jetzer}},\ }\bibfield  {title} {\bibinfo {title} {{Effective-one-body
  Hamiltonian with next-to-leading order spin-spin coupling}},\ }\href
  {https://doi.org/10.1103/PhysRevD.91.064011} {\bibfield  {journal} {\bibinfo
  {journal} {Phys. Rev. D}\ }\textbf {\bibinfo {volume} {91}},\ \bibinfo
  {pages} {064011} (\bibinfo {year} {2015})},\ \Eprint
  {https://arxiv.org/abs/1502.01343} {arXiv:1502.01343 [gr-qc]} \BibitemShut
  {NoStop}%
\bibitem [{\citenamefont {Antonelli}\ \emph
  {et~al.}(2020{\natexlab{b}})\citenamefont {Antonelli}, \citenamefont
  {Kavanagh}, \citenamefont {Khalil}, \citenamefont {Steinhoff},\ and\
  \citenamefont {Vines}}]{Antonelli:2020aeb}%
  \BibitemOpen
  \bibfield  {author} {\bibinfo {author} {\bibfnamefont {A.}~\bibnamefont
  {Antonelli}}, \bibinfo {author} {\bibfnamefont {C.}~\bibnamefont {Kavanagh}},
  \bibinfo {author} {\bibfnamefont {M.}~\bibnamefont {Khalil}}, \bibinfo
  {author} {\bibfnamefont {J.}~\bibnamefont {Steinhoff}},\ and\ \bibinfo
  {author} {\bibfnamefont {J.}~\bibnamefont {Vines}},\ }\bibfield  {title}
  {\bibinfo {title} {{Gravitational spin-orbit coupling through
  third-subleading post-Newtonian order: from first-order self-force to
  arbitrary mass ratios}},\ }\href
  {https://doi.org/10.1103/PhysRevLett.125.011103} {\bibfield  {journal}
  {\bibinfo  {journal} {Phys. Rev. Lett.}\ }\textbf {\bibinfo {volume} {125}},\
  \bibinfo {pages} {011103} (\bibinfo {year} {2020}{\natexlab{b}})},\ \Eprint
  {https://arxiv.org/abs/2003.11391} {arXiv:2003.11391 [gr-qc]} \BibitemShut
  {NoStop}%
\bibitem [{\citenamefont {Antonelli}\ \emph
  {et~al.}(2020{\natexlab{c}})\citenamefont {Antonelli}, \citenamefont
  {Kavanagh}, \citenamefont {Khalil}, \citenamefont {Steinhoff},\ and\
  \citenamefont {Vines}}]{Antonelli:2020ybz}%
  \BibitemOpen
  \bibfield  {author} {\bibinfo {author} {\bibfnamefont {A.}~\bibnamefont
  {Antonelli}}, \bibinfo {author} {\bibfnamefont {C.}~\bibnamefont {Kavanagh}},
  \bibinfo {author} {\bibfnamefont {M.}~\bibnamefont {Khalil}}, \bibinfo
  {author} {\bibfnamefont {J.}~\bibnamefont {Steinhoff}},\ and\ \bibinfo
  {author} {\bibfnamefont {J.}~\bibnamefont {Vines}},\ }\bibfield  {title}
  {\bibinfo {title} {{Gravitational spin-orbit and aligned spin$_1$-spin$_2$
  couplings through third-subleading post-Newtonian orders}},\ }\href
  {https://doi.org/10.1103/PhysRevD.102.124024} {\bibfield  {journal} {\bibinfo
   {journal} {Phys. Rev. D}\ }\textbf {\bibinfo {volume} {102}},\ \bibinfo
  {pages} {124024} (\bibinfo {year} {2020}{\natexlab{c}})},\ \Eprint
  {https://arxiv.org/abs/2010.02018} {arXiv:2010.02018 [gr-qc]} \BibitemShut
  {NoStop}%
\bibitem [{\citenamefont {Barausse}\ \emph {et~al.}(2009)\citenamefont
  {Barausse}, \citenamefont {Racine},\ and\ \citenamefont
  {Buonanno}}]{Barausse:2009aa}%
  \BibitemOpen
  \bibfield  {author} {\bibinfo {author} {\bibfnamefont {E.}~\bibnamefont
  {Barausse}}, \bibinfo {author} {\bibfnamefont {E.}~\bibnamefont {Racine}},\
  and\ \bibinfo {author} {\bibfnamefont {A.}~\bibnamefont {Buonanno}},\
  }\bibfield  {title} {\bibinfo {title} {{Hamiltonian of a spinning
  test-particle in curved spacetime}},\ }\href
  {https://doi.org/10.1103/PhysRevD.85.069904} {\bibfield  {journal} {\bibinfo
  {journal} {Phys. Rev. D}\ }\textbf {\bibinfo {volume} {80}},\ \bibinfo
  {pages} {104025} (\bibinfo {year} {2009})},\ \bibinfo {note} {[Erratum:
  Phys.Rev.D 85, 069904 (2012)]},\ \Eprint {https://arxiv.org/abs/0907.4745}
  {arXiv:0907.4745 [gr-qc]} \BibitemShut {NoStop}%
\bibitem [{\citenamefont {Vines}\ \emph {et~al.}(2016)\citenamefont {Vines},
  \citenamefont {Kunst}, \citenamefont {Steinhoff},\ and\ \citenamefont
  {Hinderer}}]{Vines:2016unv}%
  \BibitemOpen
  \bibfield  {author} {\bibinfo {author} {\bibfnamefont {J.}~\bibnamefont
  {Vines}}, \bibinfo {author} {\bibfnamefont {D.}~\bibnamefont {Kunst}},
  \bibinfo {author} {\bibfnamefont {J.}~\bibnamefont {Steinhoff}},\ and\
  \bibinfo {author} {\bibfnamefont {T.}~\bibnamefont {Hinderer}},\ }\bibfield
  {title} {\bibinfo {title} {{Canonical Hamiltonian for an extended test body
  in curved spacetime: To quadratic order in spin}},\ }\href
  {https://doi.org/10.1103/PhysRevD.104.029902} {\bibfield  {journal} {\bibinfo
   {journal} {Phys. Rev. D}\ }\textbf {\bibinfo {volume} {93}},\ \bibinfo
  {pages} {103008} (\bibinfo {year} {2016})},\ \bibinfo {note} {[Erratum:
  Phys.Rev.D 104, 029902 (2021)]},\ \Eprint {https://arxiv.org/abs/1601.07529}
  {arXiv:1601.07529 [gr-qc]} \BibitemShut {NoStop}%
\bibitem [{\citenamefont {{LIGO Scientific Collaboration}}(2018)}]{lalsuite}%
  \BibitemOpen
  \bibfield  {author} {\bibinfo {author} {\bibnamefont {{LIGO Scientific
  Collaboration}}},\ }\href {https://doi.org/10.7935/GT1W-FZ16} {\bibinfo
  {title} {{LIGO} {A}lgorithm {L}ibrary - {LALS}uite}},\ \bibinfo
  {howpublished} {free software (GPL)} (\bibinfo {year} {2018})\BibitemShut
  {NoStop}%
\bibitem [{\citenamefont {Vines}\ and\ \citenamefont
  {Steinhoff}(2018)}]{Vines:2016qwa}%
  \BibitemOpen
  \bibfield  {author} {\bibinfo {author} {\bibfnamefont {J.}~\bibnamefont
  {Vines}}\ and\ \bibinfo {author} {\bibfnamefont {J.}~\bibnamefont
  {Steinhoff}},\ }\bibfield  {title} {\bibinfo {title} {{Spin-multipole effects
  in binary black holes and the test-body limit}},\ }\href
  {https://doi.org/10.1103/PhysRevD.97.064010} {\bibfield  {journal} {\bibinfo
  {journal} {Phys. Rev. D}\ }\textbf {\bibinfo {volume} {97}},\ \bibinfo
  {pages} {064010} (\bibinfo {year} {2018})},\ \Eprint
  {https://arxiv.org/abs/1606.08832} {arXiv:1606.08832 [gr-qc]} \BibitemShut
  {NoStop}%
\bibitem [{\citenamefont {Ohta}\ \emph {et~al.}(1974)\citenamefont {Ohta},
  \citenamefont {Okamura}, \citenamefont {Hiida},\ and\ \citenamefont
  {Kimura}}]{Ohta:1974pq}%
  \BibitemOpen
  \bibfield  {author} {\bibinfo {author} {\bibfnamefont {T.}~\bibnamefont
  {Ohta}}, \bibinfo {author} {\bibfnamefont {H.}~\bibnamefont {Okamura}},
  \bibinfo {author} {\bibfnamefont {K.}~\bibnamefont {Hiida}},\ and\ \bibinfo
  {author} {\bibfnamefont {T.}~\bibnamefont {Kimura}},\ }\bibfield  {title}
  {\bibinfo {title} {{Higher order gravitational potential for many-body
  system}},\ }\href {https://doi.org/10.1143/PTP.51.1220} {\bibfield  {journal}
  {\bibinfo  {journal} {Prog. Theor. Phys.}\ }\textbf {\bibinfo {volume}
  {51}},\ \bibinfo {pages} {1220} (\bibinfo {year} {1974})}\BibitemShut
  {NoStop}%
\bibitem [{\citenamefont {Damour}\ and\ \citenamefont
  {Sch\"afer}(1985)}]{Damour:1985mt}%
  \BibitemOpen
  \bibfield  {author} {\bibinfo {author} {\bibfnamefont {T.}~\bibnamefont
  {Damour}}\ and\ \bibinfo {author} {\bibfnamefont {G.}~\bibnamefont
  {Sch\"afer}},\ }\bibfield  {title} {\bibinfo {title} {{Lagrangians for$n$
  point masses at the second post-Newtonian approximation of general
  relativity}},\ }\href {https://doi.org/10.1007/BF00773685} {\bibfield
  {journal} {\bibinfo  {journal} {Gen. Rel. Grav.}\ }\textbf {\bibinfo {volume}
  {17}},\ \bibinfo {pages} {879} (\bibinfo {year} {1985})}\BibitemShut
  {NoStop}%
\bibitem [{\citenamefont {Blanchet}\ and\ \citenamefont
  {Faye}(2001)}]{Blanchet:2000ub}%
  \BibitemOpen
  \bibfield  {author} {\bibinfo {author} {\bibfnamefont {L.}~\bibnamefont
  {Blanchet}}\ and\ \bibinfo {author} {\bibfnamefont {G.}~\bibnamefont
  {Faye}},\ }\bibfield  {title} {\bibinfo {title} {{General relativistic
  dynamics of compact binaries at the third postNewtonian order}},\ }\href
  {https://doi.org/10.1103/PhysRevD.63.062005} {\bibfield  {journal} {\bibinfo
  {journal} {Phys. Rev. D}\ }\textbf {\bibinfo {volume} {63}},\ \bibinfo
  {pages} {062005} (\bibinfo {year} {2001})},\ \Eprint
  {https://arxiv.org/abs/gr-qc/0007051} {arXiv:gr-qc/0007051} \BibitemShut
  {NoStop}%
\bibitem [{\citenamefont {Blanchet}\ and\ \citenamefont
  {Faye}(2000)}]{Blanchet:2000nv}%
  \BibitemOpen
  \bibfield  {author} {\bibinfo {author} {\bibfnamefont {L.}~\bibnamefont
  {Blanchet}}\ and\ \bibinfo {author} {\bibfnamefont {G.}~\bibnamefont
  {Faye}},\ }\bibfield  {title} {\bibinfo {title} {{Equations of motion of
  point particle binaries at the third postNewtonian order}},\ }\href
  {https://doi.org/10.1016/S0375-9601(00)00360-1} {\bibfield  {journal}
  {\bibinfo  {journal} {Phys. Lett. A}\ }\textbf {\bibinfo {volume} {271}},\
  \bibinfo {pages} {58} (\bibinfo {year} {2000})},\ \Eprint
  {https://arxiv.org/abs/gr-qc/0004009} {arXiv:gr-qc/0004009} \BibitemShut
  {NoStop}%
\bibitem [{\citenamefont {Jaranowski}\ and\ \citenamefont
  {Schaefer}(1998)}]{Jaranowski:1997ky}%
  \BibitemOpen
  \bibfield  {author} {\bibinfo {author} {\bibfnamefont {P.}~\bibnamefont
  {Jaranowski}}\ and\ \bibinfo {author} {\bibfnamefont {G.}~\bibnamefont
  {Schaefer}},\ }\bibfield  {title} {\bibinfo {title} {{Third postNewtonian
  higher order ADM Hamilton dynamics for two-body point mass systems}},\ }\href
  {https://doi.org/10.1103/PhysRevD.57.7274} {\bibfield  {journal} {\bibinfo
  {journal} {Phys. Rev. D}\ }\textbf {\bibinfo {volume} {57}},\ \bibinfo
  {pages} {7274} (\bibinfo {year} {1998})},\ \bibinfo {note} {[Erratum:
  Phys.Rev.D 63, 029902 (2001)]},\ \Eprint
  {https://arxiv.org/abs/gr-qc/9712075} {arXiv:gr-qc/9712075} \BibitemShut
  {NoStop}%
\bibitem [{\citenamefont {Jaranowski}\ and\ \citenamefont
  {Schaefer}(2000)}]{Jaranowski:1999qd}%
  \BibitemOpen
  \bibfield  {author} {\bibinfo {author} {\bibfnamefont {P.}~\bibnamefont
  {Jaranowski}}\ and\ \bibinfo {author} {\bibfnamefont {G.}~\bibnamefont
  {Schaefer}},\ }\bibfield  {title} {\bibinfo {title} {{The Binary black hole
  dynamics at the third postNewtonian order in the orbital motion}},\ }\href
  {https://doi.org/10.1002/(SICI)1521-3889(200005)9:3/5<378::AID-ANDP378>3.0.CO;2-M}
  {\bibfield  {journal} {\bibinfo  {journal} {Annalen Phys.}\ }\textbf
  {\bibinfo {volume} {9}},\ \bibinfo {pages} {378} (\bibinfo {year} {2000})},\
  \Eprint {https://arxiv.org/abs/gr-qc/0003054} {arXiv:gr-qc/0003054}
  \BibitemShut {NoStop}%
\bibitem [{\citenamefont {Itoh}\ and\ \citenamefont
  {Futamase}(2003)}]{Itoh:2003fy}%
  \BibitemOpen
  \bibfield  {author} {\bibinfo {author} {\bibfnamefont {Y.}~\bibnamefont
  {Itoh}}\ and\ \bibinfo {author} {\bibfnamefont {T.}~\bibnamefont
  {Futamase}},\ }\bibfield  {title} {\bibinfo {title} {{New derivation of a
  third postNewtonian equation of motion for relativistic compact binaries
  without ambiguity}},\ }\href {https://doi.org/10.1103/PhysRevD.68.121501}
  {\bibfield  {journal} {\bibinfo  {journal} {Phys. Rev. D}\ }\textbf {\bibinfo
  {volume} {68}},\ \bibinfo {pages} {121501} (\bibinfo {year} {2003})},\
  \Eprint {https://arxiv.org/abs/gr-qc/0310028} {arXiv:gr-qc/0310028}
  \BibitemShut {NoStop}%
\bibitem [{\citenamefont {Damour}\ \emph {et~al.}(2014)\citenamefont {Damour},
  \citenamefont {Jaranowski},\ and\ \citenamefont
  {Sch\"afer}}]{Damour:2014jta}%
  \BibitemOpen
  \bibfield  {author} {\bibinfo {author} {\bibfnamefont {T.}~\bibnamefont
  {Damour}}, \bibinfo {author} {\bibfnamefont {P.}~\bibnamefont {Jaranowski}},\
  and\ \bibinfo {author} {\bibfnamefont {G.}~\bibnamefont {Sch\"afer}},\
  }\bibfield  {title} {\bibinfo {title} {{Nonlocal-in-time action for the
  fourth post-Newtonian conservative dynamics of two-body systems}},\ }\href
  {https://doi.org/10.1103/PhysRevD.89.064058} {\bibfield  {journal} {\bibinfo
  {journal} {Phys. Rev. D}\ }\textbf {\bibinfo {volume} {89}},\ \bibinfo
  {pages} {064058} (\bibinfo {year} {2014})},\ \Eprint
  {https://arxiv.org/abs/1401.4548} {arXiv:1401.4548 [gr-qc]} \BibitemShut
  {NoStop}%
\bibitem [{\citenamefont {Damour}\ \emph {et~al.}(2016)\citenamefont {Damour},
  \citenamefont {Jaranowski},\ and\ \citenamefont
  {Sch\"afer}}]{Damour:2016abl}%
  \BibitemOpen
  \bibfield  {author} {\bibinfo {author} {\bibfnamefont {T.}~\bibnamefont
  {Damour}}, \bibinfo {author} {\bibfnamefont {P.}~\bibnamefont {Jaranowski}},\
  and\ \bibinfo {author} {\bibfnamefont {G.}~\bibnamefont {Sch\"afer}},\
  }\bibfield  {title} {\bibinfo {title} {{Conservative dynamics of two-body
  systems at the fourth post-Newtonian approximation of general relativity}},\
  }\href {https://doi.org/10.1103/PhysRevD.93.084014} {\bibfield  {journal}
  {\bibinfo  {journal} {Phys. Rev. D}\ }\textbf {\bibinfo {volume} {93}},\
  \bibinfo {pages} {084014} (\bibinfo {year} {2016})},\ \Eprint
  {https://arxiv.org/abs/1601.01283} {arXiv:1601.01283 [gr-qc]} \BibitemShut
  {NoStop}%
\bibitem [{\citenamefont {Bernard}\ \emph {et~al.}(2018)\citenamefont
  {Bernard}, \citenamefont {Blanchet}, \citenamefont {Faye},\ and\
  \citenamefont {Marchand}}]{Bernard:2017ktp}%
  \BibitemOpen
  \bibfield  {author} {\bibinfo {author} {\bibfnamefont {L.}~\bibnamefont
  {Bernard}}, \bibinfo {author} {\bibfnamefont {L.}~\bibnamefont {Blanchet}},
  \bibinfo {author} {\bibfnamefont {G.}~\bibnamefont {Faye}},\ and\ \bibinfo
  {author} {\bibfnamefont {T.}~\bibnamefont {Marchand}},\ }\bibfield  {title}
  {\bibinfo {title} {{Center-of-Mass Equations of Motion and Conserved
  Integrals of Compact Binary Systems at the Fourth Post-Newtonian Order}},\
  }\href {https://doi.org/10.1103/PhysRevD.97.044037} {\bibfield  {journal}
  {\bibinfo  {journal} {Phys. Rev. D}\ }\textbf {\bibinfo {volume} {97}},\
  \bibinfo {pages} {044037} (\bibinfo {year} {2018})},\ \Eprint
  {https://arxiv.org/abs/1711.00283} {arXiv:1711.00283 [gr-qc]} \BibitemShut
  {NoStop}%
\bibitem [{\citenamefont {Marchand}\ \emph {et~al.}(2018)\citenamefont
  {Marchand}, \citenamefont {Bernard}, \citenamefont {Blanchet},\ and\
  \citenamefont {Faye}}]{Marchand:2017pir}%
  \BibitemOpen
  \bibfield  {author} {\bibinfo {author} {\bibfnamefont {T.}~\bibnamefont
  {Marchand}}, \bibinfo {author} {\bibfnamefont {L.}~\bibnamefont {Bernard}},
  \bibinfo {author} {\bibfnamefont {L.}~\bibnamefont {Blanchet}},\ and\
  \bibinfo {author} {\bibfnamefont {G.}~\bibnamefont {Faye}},\ }\bibfield
  {title} {\bibinfo {title} {{Ambiguity-Free Completion of the Equations of
  Motion of Compact Binary Systems at the Fourth Post-Newtonian Order}},\
  }\href {https://doi.org/10.1103/PhysRevD.97.044023} {\bibfield  {journal}
  {\bibinfo  {journal} {Phys. Rev. D}\ }\textbf {\bibinfo {volume} {97}},\
  \bibinfo {pages} {044023} (\bibinfo {year} {2018})},\ \Eprint
  {https://arxiv.org/abs/1707.09289} {arXiv:1707.09289 [gr-qc]} \BibitemShut
  {NoStop}%
\bibitem [{\citenamefont {Foffa}\ and\ \citenamefont
  {Sturani}(2019)}]{Foffa:2019rdf}%
  \BibitemOpen
  \bibfield  {author} {\bibinfo {author} {\bibfnamefont {S.}~\bibnamefont
  {Foffa}}\ and\ \bibinfo {author} {\bibfnamefont {R.}~\bibnamefont
  {Sturani}},\ }\bibfield  {title} {\bibinfo {title} {{Conservative dynamics of
  binary systems to fourth Post-Newtonian order in the EFT approach I:
  Regularized Lagrangian}},\ }\href
  {https://doi.org/10.1103/PhysRevD.100.024047} {\bibfield  {journal} {\bibinfo
   {journal} {Phys. Rev. D}\ }\textbf {\bibinfo {volume} {100}},\ \bibinfo
  {pages} {024047} (\bibinfo {year} {2019})},\ \Eprint
  {https://arxiv.org/abs/1903.05113} {arXiv:1903.05113 [gr-qc]} \BibitemShut
  {NoStop}%
\bibitem [{\citenamefont {Foffa}\ \emph {et~al.}(2019)\citenamefont {Foffa},
  \citenamefont {Porto}, \citenamefont {Rothstein},\ and\ \citenamefont
  {Sturani}}]{Foffa:2019yfl}%
  \BibitemOpen
  \bibfield  {author} {\bibinfo {author} {\bibfnamefont {S.}~\bibnamefont
  {Foffa}}, \bibinfo {author} {\bibfnamefont {R.~A.}\ \bibnamefont {Porto}},
  \bibinfo {author} {\bibfnamefont {I.}~\bibnamefont {Rothstein}},\ and\
  \bibinfo {author} {\bibfnamefont {R.}~\bibnamefont {Sturani}},\ }\bibfield
  {title} {\bibinfo {title} {{Conservative dynamics of binary systems to fourth
  Post-Newtonian order in the EFT approach II: Renormalized Lagrangian}},\
  }\href {https://doi.org/10.1103/PhysRevD.100.024048} {\bibfield  {journal}
  {\bibinfo  {journal} {Phys. Rev. D}\ }\textbf {\bibinfo {volume} {100}},\
  \bibinfo {pages} {024048} (\bibinfo {year} {2019})},\ \Eprint
  {https://arxiv.org/abs/1903.05118} {arXiv:1903.05118 [gr-qc]} \BibitemShut
  {NoStop}%
\bibitem [{\citenamefont {Bl\"umlein}\ \emph {et~al.}(2020)\citenamefont
  {Bl\"umlein}, \citenamefont {Maier}, \citenamefont {Marquard},\ and\
  \citenamefont {Sch\"afer}}]{Blumlein:2020pog}%
  \BibitemOpen
  \bibfield  {author} {\bibinfo {author} {\bibfnamefont {J.}~\bibnamefont
  {Bl\"umlein}}, \bibinfo {author} {\bibfnamefont {A.}~\bibnamefont {Maier}},
  \bibinfo {author} {\bibfnamefont {P.}~\bibnamefont {Marquard}},\ and\
  \bibinfo {author} {\bibfnamefont {G.}~\bibnamefont {Sch\"afer}},\ }\bibfield
  {title} {\bibinfo {title} {{Fourth post-Newtonian Hamiltonian dynamics of
  two-body systems from an effective field theory approach}},\ }\href
  {https://doi.org/10.1016/j.nuclphysb.2020.115041} {\bibfield  {journal}
  {\bibinfo  {journal} {Nucl. Phys. B}\ }\textbf {\bibinfo {volume} {955}},\
  \bibinfo {pages} {115041} (\bibinfo {year} {2020})},\ \Eprint
  {https://arxiv.org/abs/2003.01692} {arXiv:2003.01692 [gr-qc]} \BibitemShut
  {NoStop}%
\bibitem [{\citenamefont {Tulczyjew}(1959)}]{Tulczyjew:1959}%
  \BibitemOpen
  \bibfield  {author} {\bibinfo {author} {\bibfnamefont {W.}~\bibnamefont
  {Tulczyjew}},\ }\bibfield  {title} {\bibinfo {title} {{Equations of motion of
  rotating bodies in general relativity theory}},\ }\href@noop {} {\bibfield
  {journal} {\bibinfo  {journal} {Acta Phys. Polon.}\ }\textbf {\bibinfo
  {volume} {18}},\ \bibinfo {pages} {37} (\bibinfo {year} {1959})},\ \bibinfo
  {note} {[Erratum: Acta Phys. Polon. \textbf{18}, 534 (1959)]}\BibitemShut
  {NoStop}%
\bibitem [{\citenamefont {Tagoshi}\ \emph {et~al.}(2001)\citenamefont
  {Tagoshi}, \citenamefont {Ohashi},\ and\ \citenamefont
  {Owen}}]{Tagoshi:2000zg}%
  \BibitemOpen
  \bibfield  {author} {\bibinfo {author} {\bibfnamefont {H.}~\bibnamefont
  {Tagoshi}}, \bibinfo {author} {\bibfnamefont {A.}~\bibnamefont {Ohashi}},\
  and\ \bibinfo {author} {\bibfnamefont {B.~J.}\ \bibnamefont {Owen}},\
  }\bibfield  {title} {\bibinfo {title} {{Gravitational field and equations of
  motion of spinning compact binaries to 2.5 postNewtonian order}},\ }\href
  {https://doi.org/10.1103/PhysRevD.63.044006} {\bibfield  {journal} {\bibinfo
  {journal} {Phys. Rev. D}\ }\textbf {\bibinfo {volume} {63}},\ \bibinfo
  {pages} {044006} (\bibinfo {year} {2001})},\ \Eprint
  {https://arxiv.org/abs/gr-qc/0010014} {arXiv:gr-qc/0010014} \BibitemShut
  {NoStop}%
\bibitem [{\citenamefont {Porto}(2006)}]{Porto:2005ac}%
  \BibitemOpen
  \bibfield  {author} {\bibinfo {author} {\bibfnamefont {R.~A.}\ \bibnamefont
  {Porto}},\ }\bibfield  {title} {\bibinfo {title} {{Post-Newtonian corrections
  to the motion of spinning bodies in NRGR}},\ }\href
  {https://doi.org/10.1103/PhysRevD.73.104031} {\bibfield  {journal} {\bibinfo
  {journal} {Phys. Rev. D}\ }\textbf {\bibinfo {volume} {73}},\ \bibinfo
  {pages} {104031} (\bibinfo {year} {2006})},\ \Eprint
  {https://arxiv.org/abs/gr-qc/0511061} {arXiv:gr-qc/0511061} \BibitemShut
  {NoStop}%
\bibitem [{\citenamefont {Faye}\ \emph {et~al.}(2006)\citenamefont {Faye},
  \citenamefont {Blanchet},\ and\ \citenamefont {Buonanno}}]{Faye:2006gx}%
  \BibitemOpen
  \bibfield  {author} {\bibinfo {author} {\bibfnamefont {G.}~\bibnamefont
  {Faye}}, \bibinfo {author} {\bibfnamefont {L.}~\bibnamefont {Blanchet}},\
  and\ \bibinfo {author} {\bibfnamefont {A.}~\bibnamefont {Buonanno}},\
  }\bibfield  {title} {\bibinfo {title} {{Higher-order spin effects in the
  dynamics of compact binaries. I. Equations of motion}},\ }\href
  {https://doi.org/10.1103/PhysRevD.74.104033} {\bibfield  {journal} {\bibinfo
  {journal} {Phys. Rev. D}\ }\textbf {\bibinfo {volume} {74}},\ \bibinfo
  {pages} {104033} (\bibinfo {year} {2006})},\ \Eprint
  {https://arxiv.org/abs/gr-qc/0605139} {arXiv:gr-qc/0605139} \BibitemShut
  {NoStop}%
\bibitem [{\citenamefont {Damour}\ \emph
  {et~al.}(2008{\natexlab{d}})\citenamefont {Damour}, \citenamefont
  {Jaranowski},\ and\ \citenamefont {Schaefer}}]{Damour:2007nc}%
  \BibitemOpen
  \bibfield  {author} {\bibinfo {author} {\bibfnamefont {T.}~\bibnamefont
  {Damour}}, \bibinfo {author} {\bibfnamefont {P.}~\bibnamefont {Jaranowski}},\
  and\ \bibinfo {author} {\bibfnamefont {G.}~\bibnamefont {Schaefer}},\
  }\bibfield  {title} {\bibinfo {title} {{Hamiltonian of two spinning compact
  bodies with next-to-leading order gravitational spin-orbit coupling}},\
  }\href {https://doi.org/10.1103/PhysRevD.77.064032} {\bibfield  {journal}
  {\bibinfo  {journal} {Phys. Rev. D}\ }\textbf {\bibinfo {volume} {77}},\
  \bibinfo {pages} {064032} (\bibinfo {year} {2008}{\natexlab{d}})},\ \Eprint
  {https://arxiv.org/abs/0711.1048} {arXiv:0711.1048 [gr-qc]} \BibitemShut
  {NoStop}%
\bibitem [{\citenamefont {Steinhoff}\ \emph
  {et~al.}(2008{\natexlab{a}})\citenamefont {Steinhoff}, \citenamefont
  {Schaefer},\ and\ \citenamefont {Hergt}}]{Steinhoff:2008zr}%
  \BibitemOpen
  \bibfield  {author} {\bibinfo {author} {\bibfnamefont {J.}~\bibnamefont
  {Steinhoff}}, \bibinfo {author} {\bibfnamefont {G.}~\bibnamefont
  {Schaefer}},\ and\ \bibinfo {author} {\bibfnamefont {S.}~\bibnamefont
  {Hergt}},\ }\bibfield  {title} {\bibinfo {title} {{ADM canonical formalism
  for gravitating spinning objects}},\ }\href
  {https://doi.org/10.1103/PhysRevD.77.104018} {\bibfield  {journal} {\bibinfo
  {journal} {Phys. Rev. D}\ }\textbf {\bibinfo {volume} {77}},\ \bibinfo
  {pages} {104018} (\bibinfo {year} {2008}{\natexlab{a}})},\ \Eprint
  {https://arxiv.org/abs/0805.3136} {arXiv:0805.3136 [gr-qc]} \BibitemShut
  {NoStop}%
\bibitem [{\citenamefont {Hartung}\ and\ \citenamefont
  {Steinhoff}(2011{\natexlab{a}})}]{Hartung:2011te}%
  \BibitemOpen
  \bibfield  {author} {\bibinfo {author} {\bibfnamefont {J.}~\bibnamefont
  {Hartung}}\ and\ \bibinfo {author} {\bibfnamefont {J.}~\bibnamefont
  {Steinhoff}},\ }\bibfield  {title} {\bibinfo {title}
  {{Next-to-next-to-leading order post-Newtonian spin-orbit Hamiltonian for
  self-gravitating binaries}},\ }\href {https://doi.org/10.1002/andp.201100094}
  {\bibfield  {journal} {\bibinfo  {journal} {Annalen Phys.}\ }\textbf
  {\bibinfo {volume} {523}},\ \bibinfo {pages} {783} (\bibinfo {year}
  {2011}{\natexlab{a}})},\ \Eprint {https://arxiv.org/abs/1104.3079}
  {arXiv:1104.3079 [gr-qc]} \BibitemShut {NoStop}%
\bibitem [{\citenamefont {Perrodin}(2010)}]{Perrodin:2010dy}%
  \BibitemOpen
  \bibfield  {author} {\bibinfo {author} {\bibfnamefont {D.~L.}\ \bibnamefont
  {Perrodin}},\ }\bibfield  {title} {\bibinfo {title} {{Subleading Spin-Orbit
  Correction to the Newtonian Potential in Effective Field Theory Formalism}},\
  }in\ \href {https://doi.org/10.1142/9789814374552_0041} {\emph {\bibinfo
  {booktitle} {{12th Marcel Grossmann Meeting on General Relativity}}}}\
  (\bibinfo {year} {2010})\ pp.\ \bibinfo {pages} {725--727},\ \Eprint
  {https://arxiv.org/abs/1005.0634} {arXiv:1005.0634 [gr-qc]} \BibitemShut
  {NoStop}%
\bibitem [{\citenamefont {Porto}(2010)}]{Porto:2010tr}%
  \BibitemOpen
  \bibfield  {author} {\bibinfo {author} {\bibfnamefont {R.~A.}\ \bibnamefont
  {Porto}},\ }\bibfield  {title} {\bibinfo {title} {{Next to leading order
  spin-orbit effects in the motion of inspiralling compact binaries}},\ }\href
  {https://doi.org/10.1088/0264-9381/27/20/205001} {\bibfield  {journal}
  {\bibinfo  {journal} {Class. Quant. Grav.}\ }\textbf {\bibinfo {volume}
  {27}},\ \bibinfo {pages} {205001} (\bibinfo {year} {2010})},\ \Eprint
  {https://arxiv.org/abs/1005.5730} {arXiv:1005.5730 [gr-qc]} \BibitemShut
  {NoStop}%
\bibitem [{\citenamefont {Hartung}\ \emph {et~al.}(2013)\citenamefont
  {Hartung}, \citenamefont {Steinhoff},\ and\ \citenamefont
  {Schafer}}]{Hartung:2013dza}%
  \BibitemOpen
  \bibfield  {author} {\bibinfo {author} {\bibfnamefont {J.}~\bibnamefont
  {Hartung}}, \bibinfo {author} {\bibfnamefont {J.}~\bibnamefont {Steinhoff}},\
  and\ \bibinfo {author} {\bibfnamefont {G.}~\bibnamefont {Schafer}},\
  }\bibfield  {title} {\bibinfo {title} {{Next-to-next-to-leading order
  post-Newtonian linear-in-spin binary Hamiltonians}},\ }\href
  {https://doi.org/10.1002/andp.201200271} {\bibfield  {journal} {\bibinfo
  {journal} {Annalen Phys.}\ }\textbf {\bibinfo {volume} {525}},\ \bibinfo
  {pages} {359} (\bibinfo {year} {2013})},\ \Eprint
  {https://arxiv.org/abs/1302.6723} {arXiv:1302.6723 [gr-qc]} \BibitemShut
  {NoStop}%
\bibitem [{\citenamefont {Marsat}\ \emph {et~al.}(2013)\citenamefont {Marsat},
  \citenamefont {Bohe}, \citenamefont {Faye},\ and\ \citenamefont
  {Blanchet}}]{Marsat:2012fn}%
  \BibitemOpen
  \bibfield  {author} {\bibinfo {author} {\bibfnamefont {S.}~\bibnamefont
  {Marsat}}, \bibinfo {author} {\bibfnamefont {A.}~\bibnamefont {Bohe}},
  \bibinfo {author} {\bibfnamefont {G.}~\bibnamefont {Faye}},\ and\ \bibinfo
  {author} {\bibfnamefont {L.}~\bibnamefont {Blanchet}},\ }\bibfield  {title}
  {\bibinfo {title} {{Next-to-next-to-leading order spin-orbit effects in the
  equations of motion of compact binary systems}},\ }\href
  {https://doi.org/10.1088/0264-9381/30/5/055007} {\bibfield  {journal}
  {\bibinfo  {journal} {Class. Quant. Grav.}\ }\textbf {\bibinfo {volume}
  {30}},\ \bibinfo {pages} {055007} (\bibinfo {year} {2013})},\ \Eprint
  {https://arxiv.org/abs/1210.4143} {arXiv:1210.4143 [gr-qc]} \BibitemShut
  {NoStop}%
\bibitem [{\citenamefont {Bohe}\ \emph {et~al.}(2013)\citenamefont {Bohe},
  \citenamefont {Marsat}, \citenamefont {Faye},\ and\ \citenamefont
  {Blanchet}}]{Bohe:2012mr}%
  \BibitemOpen
  \bibfield  {author} {\bibinfo {author} {\bibfnamefont {A.}~\bibnamefont
  {Bohe}}, \bibinfo {author} {\bibfnamefont {S.}~\bibnamefont {Marsat}},
  \bibinfo {author} {\bibfnamefont {G.}~\bibnamefont {Faye}},\ and\ \bibinfo
  {author} {\bibfnamefont {L.}~\bibnamefont {Blanchet}},\ }\bibfield  {title}
  {\bibinfo {title} {{Next-to-next-to-leading order spin-orbit effects in the
  near-zone metric and precession equations of compact binaries}},\ }\href
  {https://doi.org/10.1088/0264-9381/30/7/075017} {\bibfield  {journal}
  {\bibinfo  {journal} {Class. Quant. Grav.}\ }\textbf {\bibinfo {volume}
  {30}},\ \bibinfo {pages} {075017} (\bibinfo {year} {2013})},\ \Eprint
  {https://arxiv.org/abs/1212.5520} {arXiv:1212.5520 [gr-qc]} \BibitemShut
  {NoStop}%
\bibitem [{\citenamefont {Levi}\ and\ \citenamefont
  {Steinhoff}(2016{\natexlab{a}})}]{Levi:2015uxa}%
  \BibitemOpen
  \bibfield  {author} {\bibinfo {author} {\bibfnamefont {M.}~\bibnamefont
  {Levi}}\ and\ \bibinfo {author} {\bibfnamefont {J.}~\bibnamefont
  {Steinhoff}},\ }\bibfield  {title} {\bibinfo {title}
  {{Next-to-next-to-leading order gravitational spin-orbit coupling via the
  effective field theory for spinning objects in the post-Newtonian scheme}},\
  }\href {https://doi.org/10.1088/1475-7516/2016/01/011} {\bibfield  {journal}
  {\bibinfo  {journal} {JCAP}\ }\textbf {\bibinfo {volume} {01}},\ \bibinfo
  {pages} {011}},\ \Eprint {https://arxiv.org/abs/1506.05056} {arXiv:1506.05056
  [gr-qc]} \BibitemShut {NoStop}%
\bibitem [{\citenamefont {Levi}\ \emph
  {et~al.}(2021{\natexlab{a}})\citenamefont {Levi}, \citenamefont {Mcleod},\
  and\ \citenamefont {Von~Hippel}}]{Levi:2020kvb}%
  \BibitemOpen
  \bibfield  {author} {\bibinfo {author} {\bibfnamefont {M.}~\bibnamefont
  {Levi}}, \bibinfo {author} {\bibfnamefont {A.~J.}\ \bibnamefont {Mcleod}},\
  and\ \bibinfo {author} {\bibfnamefont {M.}~\bibnamefont {Von~Hippel}},\
  }\bibfield  {title} {\bibinfo {title} {{N$^{3}$LO gravitational spin-orbit
  coupling at order G$^{4}$}},\ }\href
  {https://doi.org/10.1007/JHEP07(2021)115} {\bibfield  {journal} {\bibinfo
  {journal} {JHEP}\ }\textbf {\bibinfo {volume} {07}},\ \bibinfo {pages}
  {115}},\ \Eprint {https://arxiv.org/abs/2003.02827} {arXiv:2003.02827
  [hep-th]} \BibitemShut {NoStop}%
\bibitem [{\citenamefont {Barker}\ and\ \citenamefont
  {O'Connell}(1975)}]{Barker:1975ae}%
  \BibitemOpen
  \bibfield  {author} {\bibinfo {author} {\bibfnamefont {B.~M.}\ \bibnamefont
  {Barker}}\ and\ \bibinfo {author} {\bibfnamefont {R.~F.}\ \bibnamefont
  {O'Connell}},\ }\bibfield  {title} {\bibinfo {title} {{Gravitational Two-Body
  Problem with Arbitrary Masses, Spins, and Quadrupole Moments}},\ }\href
  {https://doi.org/10.1103/PhysRevD.12.329} {\bibfield  {journal} {\bibinfo
  {journal} {Phys. Rev. D}\ }\textbf {\bibinfo {volume} {12}},\ \bibinfo
  {pages} {329} (\bibinfo {year} {1975})}\BibitemShut {NoStop}%
\bibitem [{\citenamefont {Barker}\ and\ \citenamefont
  {O'Connell}(1979)}]{Barker:OConnell:1979}%
  \BibitemOpen
  \bibfield  {author} {\bibinfo {author} {\bibfnamefont {B.~M.}\ \bibnamefont
  {Barker}}\ and\ \bibinfo {author} {\bibfnamefont {R.~F.}\ \bibnamefont
  {O'Connell}},\ }\bibfield  {title} {\bibinfo {title} {The gravitational
  interaction: Spin, rotation, and quantum effects---a review},\ }\href
  {https://doi.org/10.1007/BF00756587} {\bibfield  {journal} {\bibinfo
  {journal} {Gen. Relativ. Gravit.}\ }\textbf {\bibinfo {volume} {11}},\
  \bibinfo {pages} {149} (\bibinfo {year} {1979})}\BibitemShut {NoStop}%
\bibitem [{\citenamefont {D'Eath}(1975)}]{DEath:1975wqz}%
  \BibitemOpen
  \bibfield  {author} {\bibinfo {author} {\bibfnamefont {P.~D.}\ \bibnamefont
  {D'Eath}},\ }\bibfield  {title} {\bibinfo {title} {{Interaction of two black
  holes in the slow-motion limit}},\ }\href
  {https://doi.org/10.1103/PhysRevD.12.2183} {\bibfield  {journal} {\bibinfo
  {journal} {Phys. Rev. D}\ }\textbf {\bibinfo {volume} {12}},\ \bibinfo
  {pages} {2183} (\bibinfo {year} {1975})}\BibitemShut {NoStop}%
\bibitem [{\citenamefont {Hergt}\ and\ \citenamefont
  {Schaefer}(2008)}]{Hergt:2008jn}%
  \BibitemOpen
  \bibfield  {author} {\bibinfo {author} {\bibfnamefont {S.}~\bibnamefont
  {Hergt}}\ and\ \bibinfo {author} {\bibfnamefont {G.}~\bibnamefont
  {Schaefer}},\ }\bibfield  {title} {\bibinfo {title} {{Higher-order-in-spin
  interaction Hamiltonians for binary black holes from Poincare invariance}},\
  }\href {https://doi.org/10.1103/PhysRevD.78.124004} {\bibfield  {journal}
  {\bibinfo  {journal} {Phys. Rev. D}\ }\textbf {\bibinfo {volume} {78}},\
  \bibinfo {pages} {124004} (\bibinfo {year} {2008})},\ \Eprint
  {https://arxiv.org/abs/0809.2208} {arXiv:0809.2208 [gr-qc]} \BibitemShut
  {NoStop}%
\bibitem [{\citenamefont {Porto}\ and\ \citenamefont
  {Rothstein}(2006)}]{Porto:2006bt}%
  \BibitemOpen
  \bibfield  {author} {\bibinfo {author} {\bibfnamefont {R.~A.}\ \bibnamefont
  {Porto}}\ and\ \bibinfo {author} {\bibfnamefont {I.~Z.}\ \bibnamefont
  {Rothstein}},\ }\bibfield  {title} {\bibinfo {title} {{The Hyperfine
  Einstein-Infeld-Hoffmann potential}},\ }\href
  {https://doi.org/10.1103/PhysRevLett.97.021101} {\bibfield  {journal}
  {\bibinfo  {journal} {Phys. Rev. Lett.}\ }\textbf {\bibinfo {volume} {97}},\
  \bibinfo {pages} {021101} (\bibinfo {year} {2006})},\ \Eprint
  {https://arxiv.org/abs/gr-qc/0604099} {arXiv:gr-qc/0604099} \BibitemShut
  {NoStop}%
\bibitem [{\citenamefont {Porto}\ and\ \citenamefont
  {Rothstein}(2008{\natexlab{a}})}]{Porto:2008jj}%
  \BibitemOpen
  \bibfield  {author} {\bibinfo {author} {\bibfnamefont {R.~A.}\ \bibnamefont
  {Porto}}\ and\ \bibinfo {author} {\bibfnamefont {I.~Z.}\ \bibnamefont
  {Rothstein}},\ }\bibfield  {title} {\bibinfo {title} {{Next to Leading Order
  Spin(1)Spin(1) Effects in the Motion of Inspiralling Compact Binaries}},\
  }\href {https://doi.org/10.1103/PhysRevD.78.044013} {\bibfield  {journal}
  {\bibinfo  {journal} {Phys. Rev. D}\ }\textbf {\bibinfo {volume} {78}},\
  \bibinfo {pages} {044013} (\bibinfo {year} {2008}{\natexlab{a}})},\ \bibinfo
  {note} {[Erratum: Phys.Rev.D 81, 029905 (2010)]},\ \Eprint
  {https://arxiv.org/abs/0804.0260} {arXiv:0804.0260 [gr-qc]} \BibitemShut
  {NoStop}%
\bibitem [{\citenamefont {Porto}\ and\ \citenamefont
  {Rothstein}(2008{\natexlab{b}})}]{Porto:2008tb}%
  \BibitemOpen
  \bibfield  {author} {\bibinfo {author} {\bibfnamefont {R.~A.}\ \bibnamefont
  {Porto}}\ and\ \bibinfo {author} {\bibfnamefont {I.~Z.}\ \bibnamefont
  {Rothstein}},\ }\bibfield  {title} {\bibinfo {title} {{Spin(1)Spin(2) Effects
  in the Motion of Inspiralling Compact Binaries at Third Order in the
  Post-Newtonian Expansion}},\ }\href
  {https://doi.org/10.1103/PhysRevD.78.044012} {\bibfield  {journal} {\bibinfo
  {journal} {Phys. Rev. D}\ }\textbf {\bibinfo {volume} {78}},\ \bibinfo
  {pages} {044012} (\bibinfo {year} {2008}{\natexlab{b}})},\ \bibinfo {note}
  {[Erratum: Phys.Rev.D 81, 029904 (2010)]},\ \Eprint
  {https://arxiv.org/abs/0802.0720} {arXiv:0802.0720 [gr-qc]} \BibitemShut
  {NoStop}%
\bibitem [{\citenamefont {Steinhoff}\ \emph
  {et~al.}(2008{\natexlab{b}})\citenamefont {Steinhoff}, \citenamefont
  {Hergt},\ and\ \citenamefont {Schaefer}}]{Steinhoff:2007mb}%
  \BibitemOpen
  \bibfield  {author} {\bibinfo {author} {\bibfnamefont {J.}~\bibnamefont
  {Steinhoff}}, \bibinfo {author} {\bibfnamefont {S.}~\bibnamefont {Hergt}},\
  and\ \bibinfo {author} {\bibfnamefont {G.}~\bibnamefont {Schaefer}},\
  }\bibfield  {title} {\bibinfo {title} {{On the next-to-leading order
  gravitational spin(1)-spin(2) dynamics}},\ }\href
  {https://doi.org/10.1103/PhysRevD.77.081501} {\bibfield  {journal} {\bibinfo
  {journal} {Phys. Rev. D}\ }\textbf {\bibinfo {volume} {77}},\ \bibinfo
  {pages} {081501} (\bibinfo {year} {2008}{\natexlab{b}})},\ \Eprint
  {https://arxiv.org/abs/0712.1716} {arXiv:0712.1716 [gr-qc]} \BibitemShut
  {NoStop}%
\bibitem [{\citenamefont {Steinhoff}\ and\ \citenamefont
  {Schaefer}(2009)}]{Steinhoff:2009hb}%
  \BibitemOpen
  \bibfield  {author} {\bibinfo {author} {\bibfnamefont {J.}~\bibnamefont
  {Steinhoff}}\ and\ \bibinfo {author} {\bibfnamefont {G.}~\bibnamefont
  {Schaefer}},\ }\bibfield  {title} {\bibinfo {title} {{Comment on
  `Spin(1)spin(2) effects in the motion of inspiralling compact binaries at
  third order in the post-Newtonian expansion' and 'Next to leading order
  spin(1)spin(1) effects in the motion of inspiralling compact binaries'}},\
  }\href {https://doi.org/10.1103/PhysRevD.80.088501} {\bibfield  {journal}
  {\bibinfo  {journal} {Phys. Rev. D}\ }\textbf {\bibinfo {volume} {80}},\
  \bibinfo {pages} {088501} (\bibinfo {year} {2009})},\ \Eprint
  {https://arxiv.org/abs/0903.4772} {arXiv:0903.4772 [gr-qc]} \BibitemShut
  {NoStop}%
\bibitem [{\citenamefont {Hergt}\ \emph {et~al.}(2010)\citenamefont {Hergt},
  \citenamefont {Steinhoff},\ and\ \citenamefont {Schaefer}}]{Hergt:2010pa}%
  \BibitemOpen
  \bibfield  {author} {\bibinfo {author} {\bibfnamefont {S.}~\bibnamefont
  {Hergt}}, \bibinfo {author} {\bibfnamefont {J.}~\bibnamefont {Steinhoff}},\
  and\ \bibinfo {author} {\bibfnamefont {G.}~\bibnamefont {Schaefer}},\
  }\bibfield  {title} {\bibinfo {title} {{Reduced Hamiltonian for
  next-to-leading order Spin-Squared Dynamics of General Compact Binaries}},\
  }\href {https://doi.org/10.1088/0264-9381/27/13/135007} {\bibfield  {journal}
  {\bibinfo  {journal} {Class. Quant. Grav.}\ }\textbf {\bibinfo {volume}
  {27}},\ \bibinfo {pages} {135007} (\bibinfo {year} {2010})},\ \Eprint
  {https://arxiv.org/abs/1002.2093} {arXiv:1002.2093 [gr-qc]} \BibitemShut
  {NoStop}%
\bibitem [{\citenamefont {Hartung}\ and\ \citenamefont
  {Steinhoff}(2011{\natexlab{b}})}]{Hartung:2011ea}%
  \BibitemOpen
  \bibfield  {author} {\bibinfo {author} {\bibfnamefont {J.}~\bibnamefont
  {Hartung}}\ and\ \bibinfo {author} {\bibfnamefont {J.}~\bibnamefont
  {Steinhoff}},\ }\bibfield  {title} {\bibinfo {title}
  {{Next-to-next-to-leading order post-Newtonian spin(1)-spin(2) Hamiltonian
  for self-gravitating binaries}},\ }\href
  {https://doi.org/10.1002/andp.201100163} {\bibfield  {journal} {\bibinfo
  {journal} {Annalen Phys.}\ }\textbf {\bibinfo {volume} {523}},\ \bibinfo
  {pages} {919} (\bibinfo {year} {2011}{\natexlab{b}})},\ \Eprint
  {https://arxiv.org/abs/1107.4294} {arXiv:1107.4294 [gr-qc]} \BibitemShut
  {NoStop}%
\bibitem [{\citenamefont {Levi}(2012)}]{Levi:2011eq}%
  \BibitemOpen
  \bibfield  {author} {\bibinfo {author} {\bibfnamefont {M.}~\bibnamefont
  {Levi}},\ }\bibfield  {title} {\bibinfo {title} {{Binary dynamics from
  spin1-spin2 coupling at fourth post-Newtonian order}},\ }\href
  {https://doi.org/10.1103/PhysRevD.85.064043} {\bibfield  {journal} {\bibinfo
  {journal} {Phys. Rev. D}\ }\textbf {\bibinfo {volume} {85}},\ \bibinfo
  {pages} {064043} (\bibinfo {year} {2012})},\ \Eprint
  {https://arxiv.org/abs/1107.4322} {arXiv:1107.4322 [gr-qc]} \BibitemShut
  {NoStop}%
\bibitem [{\citenamefont {Levi}\ and\ \citenamefont
  {Steinhoff}(2016{\natexlab{b}})}]{Levi:2015ixa}%
  \BibitemOpen
  \bibfield  {author} {\bibinfo {author} {\bibfnamefont {M.}~\bibnamefont
  {Levi}}\ and\ \bibinfo {author} {\bibfnamefont {J.}~\bibnamefont
  {Steinhoff}},\ }\bibfield  {title} {\bibinfo {title}
  {{Next-to-next-to-leading order gravitational spin-squared potential via the
  effective field theory for spinning objects in the post-Newtonian scheme}},\
  }\href {https://doi.org/10.1088/1475-7516/2016/01/008} {\bibfield  {journal}
  {\bibinfo  {journal} {JCAP}\ }\textbf {\bibinfo {volume} {01}},\ \bibinfo
  {pages} {008}},\ \Eprint {https://arxiv.org/abs/1506.05794} {arXiv:1506.05794
  [gr-qc]} \BibitemShut {NoStop}%
\bibitem [{\citenamefont {Levi}\ and\ \citenamefont
  {Steinhoff}(2015{\natexlab{b}})}]{Levi:2014gsa}%
  \BibitemOpen
  \bibfield  {author} {\bibinfo {author} {\bibfnamefont {M.}~\bibnamefont
  {Levi}}\ and\ \bibinfo {author} {\bibfnamefont {J.}~\bibnamefont
  {Steinhoff}},\ }\bibfield  {title} {\bibinfo {title} {{Leading order finite
  size effects with spins for inspiralling compact binaries}},\ }\href
  {https://doi.org/10.1007/JHEP06(2015)059} {\bibfield  {journal} {\bibinfo
  {journal} {JHEP}\ }\textbf {\bibinfo {volume} {06}},\ \bibinfo {pages}
  {059}},\ \Eprint {https://arxiv.org/abs/1410.2601} {arXiv:1410.2601 [gr-qc]}
  \BibitemShut {NoStop}%
\bibitem [{\citenamefont {Levi}\ and\ \citenamefont
  {Steinhoff}(2021)}]{Levi:2016ofk}%
  \BibitemOpen
  \bibfield  {author} {\bibinfo {author} {\bibfnamefont {M.}~\bibnamefont
  {Levi}}\ and\ \bibinfo {author} {\bibfnamefont {J.}~\bibnamefont
  {Steinhoff}},\ }\bibfield  {title} {\bibinfo {title} {{Complete conservative
  dynamics for inspiralling compact binaries with spins at the fourth
  post-Newtonian order}},\ }\href
  {https://doi.org/10.1088/1475-7516/2021/09/029} {\bibfield  {journal}
  {\bibinfo  {journal} {JCAP}\ }\textbf {\bibinfo {volume} {09}},\ \bibinfo
  {pages} {029}},\ \Eprint {https://arxiv.org/abs/1607.04252} {arXiv:1607.04252
  [gr-qc]} \BibitemShut {NoStop}%
\bibitem [{\citenamefont {Visser}(2007)}]{Visser:2007fj}%
  \BibitemOpen
  \bibfield  {author} {\bibinfo {author} {\bibfnamefont {M.}~\bibnamefont
  {Visser}},\ }\bibfield  {title} {\bibinfo {title} {{The Kerr spacetime: A
  Brief introduction}},\ }in\ \href@noop {} {\emph {\bibinfo {booktitle} {{Kerr
  Fest: Black Holes in Astrophysics, General Relativity and Quantum
  Gravity}}}}\ (\bibinfo {year} {2007})\ \Eprint
  {https://arxiv.org/abs/0706.0622} {arXiv:0706.0622 [gr-qc]} \BibitemShut
  {NoStop}%
\bibitem [{\citenamefont {Mandal}\ \emph
  {et~al.}(2022{\natexlab{a}})\citenamefont {Mandal}, \citenamefont
  {Mastrolia}, \citenamefont {Patil},\ and\ \citenamefont
  {Steinhoff}}]{Mandal:2022nty}%
  \BibitemOpen
  \bibfield  {author} {\bibinfo {author} {\bibfnamefont {M.~K.}\ \bibnamefont
  {Mandal}}, \bibinfo {author} {\bibfnamefont {P.}~\bibnamefont {Mastrolia}},
  \bibinfo {author} {\bibfnamefont {R.}~\bibnamefont {Patil}},\ and\ \bibinfo
  {author} {\bibfnamefont {J.}~\bibnamefont {Steinhoff}},\ }\bibfield  {title}
  {\bibinfo {title} {{Gravitational Spin-Orbit Hamiltonian at NNNLO in the
  post-Newtonian framework}},\ }\href@noop {} {\  (\bibinfo {year}
  {2022}{\natexlab{a}})},\ \Eprint {https://arxiv.org/abs/2209.00611}
  {arXiv:2209.00611 [hep-th]} \BibitemShut {NoStop}%
\bibitem [{\citenamefont {Kim}\ \emph {et~al.}(2022{\natexlab{a}})\citenamefont
  {Kim}, \citenamefont {Levi},\ and\ \citenamefont {Yin}}]{Kim:2022pou}%
  \BibitemOpen
  \bibfield  {author} {\bibinfo {author} {\bibfnamefont {J.-W.}\ \bibnamefont
  {Kim}}, \bibinfo {author} {\bibfnamefont {M.}~\bibnamefont {Levi}},\ and\
  \bibinfo {author} {\bibfnamefont {Z.}~\bibnamefont {Yin}},\ }\bibfield
  {title} {\bibinfo {title} {{N$^3$LO Spin-Orbit Interaction via the EFT of
  Spinning Gravitating Objects}},\ }\href@noop {} {\  (\bibinfo {year}
  {2022}{\natexlab{a}})},\ \Eprint {https://arxiv.org/abs/2208.14949}
  {arXiv:2208.14949 [hep-th]} \BibitemShut {NoStop}%
\bibitem [{\citenamefont {Damour}\ and\ \citenamefont
  {Nagar}(2007)}]{Damour:2007xr}%
  \BibitemOpen
  \bibfield  {author} {\bibinfo {author} {\bibfnamefont {T.}~\bibnamefont
  {Damour}}\ and\ \bibinfo {author} {\bibfnamefont {A.}~\bibnamefont {Nagar}},\
  }\bibfield  {title} {\bibinfo {title} {{Faithful effective-one-body waveforms
  of small-mass-ratio coalescing black-hole binaries}},\ }\href
  {https://doi.org/10.1103/PhysRevD.76.064028} {\bibfield  {journal} {\bibinfo
  {journal} {Phys. Rev. D}\ }\textbf {\bibinfo {volume} {76}},\ \bibinfo
  {pages} {064028} (\bibinfo {year} {2007})},\ \Eprint
  {https://arxiv.org/abs/0705.2519} {arXiv:0705.2519 [gr-qc]} \BibitemShut
  {NoStop}%
\bibitem [{\citenamefont {Will}(2005)}]{Will:2005sn}%
  \BibitemOpen
  \bibfield  {author} {\bibinfo {author} {\bibfnamefont {C.~M.}\ \bibnamefont
  {Will}},\ }\bibfield  {title} {\bibinfo {title} {{Post-Newtonian
  gravitational radiation and equations of motion via direct integration of the
  relaxed Einstein equations. III. Radiation reaction for binary systems with
  spinning bodies}},\ }\href {https://doi.org/10.1103/PhysRevD.71.084027}
  {\bibfield  {journal} {\bibinfo  {journal} {Phys. Rev. D}\ }\textbf {\bibinfo
  {volume} {71}},\ \bibinfo {pages} {084027} (\bibinfo {year} {2005})},\
  \Eprint {https://arxiv.org/abs/gr-qc/0502039} {arXiv:gr-qc/0502039}
  \BibitemShut {NoStop}%
\bibitem [{\citenamefont {Maia}\ \emph {et~al.}(2017)\citenamefont {Maia},
  \citenamefont {Galley}, \citenamefont {Leibovich},\ and\ \citenamefont
  {Porto}}]{Maia:2017yok}%
  \BibitemOpen
  \bibfield  {author} {\bibinfo {author} {\bibfnamefont {N.~T.}\ \bibnamefont
  {Maia}}, \bibinfo {author} {\bibfnamefont {C.~R.}\ \bibnamefont {Galley}},
  \bibinfo {author} {\bibfnamefont {A.~K.}\ \bibnamefont {Leibovich}},\ and\
  \bibinfo {author} {\bibfnamefont {R.~A.}\ \bibnamefont {Porto}},\ }\bibfield
  {title} {\bibinfo {title} {{Radiation reaction for spinning bodies in
  effective field theory II: Spin-spin effects}},\ }\href
  {https://doi.org/10.1103/PhysRevD.96.084065} {\bibfield  {journal} {\bibinfo
  {journal} {Phys. Rev. D}\ }\textbf {\bibinfo {volume} {96}},\ \bibinfo
  {pages} {084065} (\bibinfo {year} {2017})},\ \Eprint
  {https://arxiv.org/abs/1705.07938} {arXiv:1705.07938 [gr-qc]} \BibitemShut
  {NoStop}%
\bibitem [{\citenamefont {Nagar}\ and\ \citenamefont
  {Rettegno}(2019)}]{Nagar:2018gnk}%
  \BibitemOpen
  \bibfield  {author} {\bibinfo {author} {\bibfnamefont {A.}~\bibnamefont
  {Nagar}}\ and\ \bibinfo {author} {\bibfnamefont {P.}~\bibnamefont
  {Rettegno}},\ }\bibfield  {title} {\bibinfo {title} {{Efficient effective one
  body time-domain gravitational waveforms}},\ }\href
  {https://doi.org/10.1103/PhysRevD.99.021501} {\bibfield  {journal} {\bibinfo
  {journal} {Phys. Rev. D}\ }\textbf {\bibinfo {volume} {99}},\ \bibinfo
  {pages} {021501} (\bibinfo {year} {2019})},\ \Eprint
  {https://arxiv.org/abs/1805.03891} {arXiv:1805.03891 [gr-qc]} \BibitemShut
  {NoStop}%
\bibitem [{\citenamefont {Rettegno}\ \emph {et~al.}(2020)\citenamefont
  {Rettegno}, \citenamefont {Martinetti}, \citenamefont {Nagar}, \citenamefont
  {Bini}, \citenamefont {Riemenschneider},\ and\ \citenamefont
  {Damour}}]{Rettegno:2019tzh}%
  \BibitemOpen
  \bibfield  {author} {\bibinfo {author} {\bibfnamefont {P.}~\bibnamefont
  {Rettegno}}, \bibinfo {author} {\bibfnamefont {F.}~\bibnamefont
  {Martinetti}}, \bibinfo {author} {\bibfnamefont {A.}~\bibnamefont {Nagar}},
  \bibinfo {author} {\bibfnamefont {D.}~\bibnamefont {Bini}}, \bibinfo {author}
  {\bibfnamefont {G.}~\bibnamefont {Riemenschneider}},\ and\ \bibinfo {author}
  {\bibfnamefont {T.}~\bibnamefont {Damour}},\ }\bibfield  {title} {\bibinfo
  {title} {{Comparing Effective One Body Hamiltonians for spin-aligned
  coalescing binaries}},\ }\href {https://doi.org/10.1103/PhysRevD.101.104027}
  {\bibfield  {journal} {\bibinfo  {journal} {Phys. Rev. D}\ }\textbf {\bibinfo
  {volume} {101}},\ \bibinfo {pages} {104027} (\bibinfo {year} {2020})},\
  \Eprint {https://arxiv.org/abs/1911.10818} {arXiv:1911.10818 [gr-qc]}
  \BibitemShut {NoStop}%
\bibitem [{\citenamefont {Mihaylov}\ \emph {et~al.}(2021)\citenamefont
  {Mihaylov}, \citenamefont {Ossokine}, \citenamefont {Buonanno},\ and\
  \citenamefont {Ghosh}}]{Mihaylov:2021bpf}%
  \BibitemOpen
  \bibfield  {author} {\bibinfo {author} {\bibfnamefont {D.~P.}\ \bibnamefont
  {Mihaylov}}, \bibinfo {author} {\bibfnamefont {S.}~\bibnamefont {Ossokine}},
  \bibinfo {author} {\bibfnamefont {A.}~\bibnamefont {Buonanno}},\ and\
  \bibinfo {author} {\bibfnamefont {A.}~\bibnamefont {Ghosh}},\ }\bibfield
  {title} {\bibinfo {title} {{Fast post-adiabatic waveforms in the time domain:
  Applications to compact binary coalescences in LIGO and Virgo}},\ }\href
  {https://doi.org/10.1103/PhysRevD.104.124087} {\bibfield  {journal} {\bibinfo
   {journal} {Phys. Rev. D}\ }\textbf {\bibinfo {volume} {104}},\ \bibinfo
  {pages} {124087} (\bibinfo {year} {2021})},\ \Eprint
  {https://arxiv.org/abs/2105.06983} {arXiv:2105.06983 [gr-qc]} \BibitemShut
  {NoStop}%
\bibitem [{\citenamefont {Cho}\ \emph {et~al.}(2022)\citenamefont {Cho},
  \citenamefont {Porto},\ and\ \citenamefont {Yang}}]{Cho:2022syn}%
  \BibitemOpen
  \bibfield  {author} {\bibinfo {author} {\bibfnamefont {G.}~\bibnamefont
  {Cho}}, \bibinfo {author} {\bibfnamefont {R.~A.}\ \bibnamefont {Porto}},\
  and\ \bibinfo {author} {\bibfnamefont {Z.}~\bibnamefont {Yang}},\ }\bibfield
  {title} {\bibinfo {title} {{Gravitational radiation from inspiralling compact
  objects: Spin effects to fourth Post-Newtonian order}},\ }\href@noop {} {\
  (\bibinfo {year} {2022})},\ \Eprint {https://arxiv.org/abs/2201.05138}
  {arXiv:2201.05138 [gr-qc]} \BibitemShut {NoStop}%
\bibitem [{\citenamefont {Schmidt}\ \emph {et~al.}(2011)\citenamefont
  {Schmidt}, \citenamefont {Hannam}, \citenamefont {Husa},\ and\ \citenamefont
  {Ajith}}]{Schmidt:2010it}%
  \BibitemOpen
  \bibfield  {author} {\bibinfo {author} {\bibfnamefont {P.}~\bibnamefont
  {Schmidt}}, \bibinfo {author} {\bibfnamefont {M.}~\bibnamefont {Hannam}},
  \bibinfo {author} {\bibfnamefont {S.}~\bibnamefont {Husa}},\ and\ \bibinfo
  {author} {\bibfnamefont {P.}~\bibnamefont {Ajith}},\ }\bibfield  {title}
  {\bibinfo {title} {{Tracking the precession of compact binaries from their
  gravitational-wave signal}},\ }\href
  {https://doi.org/10.1103/PhysRevD.84.024046} {\bibfield  {journal} {\bibinfo
  {journal} {Phys. Rev. D}\ }\textbf {\bibinfo {volume} {84}},\ \bibinfo
  {pages} {024046} (\bibinfo {year} {2011})},\ \Eprint
  {https://arxiv.org/abs/1012.2879} {arXiv:1012.2879 [gr-qc]} \BibitemShut
  {NoStop}%
\bibitem [{\citenamefont {Levi}\ and\ \citenamefont
  {Steinhoff}(2014)}]{Levi:2014sba}%
  \BibitemOpen
  \bibfield  {author} {\bibinfo {author} {\bibfnamefont {M.}~\bibnamefont
  {Levi}}\ and\ \bibinfo {author} {\bibfnamefont {J.}~\bibnamefont
  {Steinhoff}},\ }\bibfield  {title} {\bibinfo {title} {{Equivalence of ADM
  Hamiltonian and Effective Field Theory approaches at next-to-next-to-leading
  order spin1-spin2 coupling of binary inspirals}},\ }\href
  {https://doi.org/10.1088/1475-7516/2014/12/003} {\bibfield  {journal}
  {\bibinfo  {journal} {JCAP}\ }\textbf {\bibinfo {volume} {12}},\ \bibinfo
  {pages} {003}},\ \Eprint {https://arxiv.org/abs/1408.5762} {arXiv:1408.5762
  [gr-qc]} \BibitemShut {NoStop}%
\bibitem [{\citenamefont {{Dixon}}(1979)}]{Dixon:1979}%
  \BibitemOpen
  \bibfield  {author} {\bibinfo {author} {\bibfnamefont {W.~G.}\ \bibnamefont
  {{Dixon}}},\ }\bibfield  {title} {\bibinfo {title} {{Extended bodies in
  general relativity: their description and motion}},\ }in\ \href@noop {}
  {\emph {\bibinfo {booktitle} {Isolated Gravitating Systems in General
  Relativity}}},\ \bibinfo {editor} {edited by\ \bibinfo {editor}
  {\bibfnamefont {J.}~\bibnamefont {{Ehlers}}}}\ (\bibinfo  {publisher} {North
  Holland, Amsterdam},\ \bibinfo {year} {1979})\ pp.\ \bibinfo {pages}
  {156--219}\BibitemShut {NoStop}%
\bibitem [{\citenamefont {Pryce}(1948)}]{pryce1948mass}%
  \BibitemOpen
  \bibfield  {author} {\bibinfo {author} {\bibfnamefont {M.~H.~L.}\
  \bibnamefont {Pryce}},\ }\bibfield  {title} {\bibinfo {title} {The
  mass-centre in the restricted theory of relativity and its connexion with the
  quantum theory of elementary particles},\ }\href@noop {} {\bibfield
  {journal} {\bibinfo  {journal} {Proceedings of the Royal Society of London.
  Series A. Mathematical and Physical Sciences}\ }\textbf {\bibinfo {volume}
  {195}},\ \bibinfo {pages} {62} (\bibinfo {year} {1948})}\BibitemShut
  {NoStop}%
\bibitem [{\citenamefont {Newton}\ and\ \citenamefont
  {Wigner}(1949)}]{newton1949localized}%
  \BibitemOpen
  \bibfield  {author} {\bibinfo {author} {\bibfnamefont {T.~D.}\ \bibnamefont
  {Newton}}\ and\ \bibinfo {author} {\bibfnamefont {E.~P.}\ \bibnamefont
  {Wigner}},\ }\bibfield  {title} {\bibinfo {title} {Localized states for
  elementary systems},\ }\href@noop {} {\bibfield  {journal} {\bibinfo
  {journal} {Reviews of Modern Physics}\ }\textbf {\bibinfo {volume} {21}},\
  \bibinfo {pages} {400} (\bibinfo {year} {1949})}\BibitemShut {NoStop}%
\bibitem [{\citenamefont {Boh\'e}\ \emph {et~al.}(2015)\citenamefont {Boh\'e},
  \citenamefont {Faye}, \citenamefont {Marsat},\ and\ \citenamefont
  {Porter}}]{Bohe:2015ana}%
  \BibitemOpen
  \bibfield  {author} {\bibinfo {author} {\bibfnamefont {A.}~\bibnamefont
  {Boh\'e}}, \bibinfo {author} {\bibfnamefont {G.}~\bibnamefont {Faye}},
  \bibinfo {author} {\bibfnamefont {S.}~\bibnamefont {Marsat}},\ and\ \bibinfo
  {author} {\bibfnamefont {E.~K.}\ \bibnamefont {Porter}},\ }\bibfield  {title}
  {\bibinfo {title} {{Quadratic-in-spin effects in the orbital dynamics and
  gravitational-wave energy flux of compact binaries at the 3PN order}},\
  }\href {https://doi.org/10.1088/0264-9381/32/19/195010} {\bibfield  {journal}
  {\bibinfo  {journal} {Class. Quant. Grav.}\ }\textbf {\bibinfo {volume}
  {32}},\ \bibinfo {pages} {195010} (\bibinfo {year} {2015})},\ \Eprint
  {https://arxiv.org/abs/1501.01529} {arXiv:1501.01529 [gr-qc]} \BibitemShut
  {NoStop}%
\bibitem [{\citenamefont {Blanchet}\ \emph {et~al.}(2004)\citenamefont
  {Blanchet}, \citenamefont {Damour}, \citenamefont {Esposito-Farese},\ and\
  \citenamefont {Iyer}}]{Blanchet:2004ek}%
  \BibitemOpen
  \bibfield  {author} {\bibinfo {author} {\bibfnamefont {L.}~\bibnamefont
  {Blanchet}}, \bibinfo {author} {\bibfnamefont {T.}~\bibnamefont {Damour}},
  \bibinfo {author} {\bibfnamefont {G.}~\bibnamefont {Esposito-Farese}},\ and\
  \bibinfo {author} {\bibfnamefont {B.~R.}\ \bibnamefont {Iyer}},\ }\bibfield
  {title} {\bibinfo {title} {{Gravitational radiation from inspiralling compact
  binaries completed at the third post-Newtonian order}},\ }\href
  {https://doi.org/10.1103/PhysRevLett.93.091101} {\bibfield  {journal}
  {\bibinfo  {journal} {Phys. Rev. Lett.}\ }\textbf {\bibinfo {volume} {93}},\
  \bibinfo {pages} {091101} (\bibinfo {year} {2004})},\ \Eprint
  {https://arxiv.org/abs/gr-qc/0406012} {arXiv:gr-qc/0406012} \BibitemShut
  {NoStop}%
\bibitem [{\citenamefont {Mandal}\ \emph
  {et~al.}(2022{\natexlab{b}})\citenamefont {Mandal}, \citenamefont
  {Mastrolia}, \citenamefont {Patil},\ and\ \citenamefont
  {Steinhoff}}]{Mandal:2022ufb}%
  \BibitemOpen
  \bibfield  {author} {\bibinfo {author} {\bibfnamefont {M.~K.}\ \bibnamefont
  {Mandal}}, \bibinfo {author} {\bibfnamefont {P.}~\bibnamefont {Mastrolia}},
  \bibinfo {author} {\bibfnamefont {R.}~\bibnamefont {Patil}},\ and\ \bibinfo
  {author} {\bibfnamefont {J.}~\bibnamefont {Steinhoff}},\ }\bibfield  {title}
  {\bibinfo {title} {{Gravitational Quadratic-in-Spin Hamiltonian at NNNLO in
  the post-Newtonian framework}},\ }\href@noop {} {\  (\bibinfo {year}
  {2022}{\natexlab{b}})},\ \Eprint {https://arxiv.org/abs/2210.09176}
  {arXiv:2210.09176 [hep-th]} \BibitemShut {NoStop}%
\bibitem [{\citenamefont {Kim}\ \emph {et~al.}(2022{\natexlab{b}})\citenamefont
  {Kim}, \citenamefont {Levi},\ and\ \citenamefont {Yin}}]{Kim:2022bwv}%
  \BibitemOpen
  \bibfield  {author} {\bibinfo {author} {\bibfnamefont {J.-W.}\ \bibnamefont
  {Kim}}, \bibinfo {author} {\bibfnamefont {M.}~\bibnamefont {Levi}},\ and\
  \bibinfo {author} {\bibfnamefont {Z.}~\bibnamefont {Yin}},\ }\bibfield
  {title} {\bibinfo {title} {{N$^3$LO Quadratic-in-Spin Interactions for
  Generic Compact Binaries}},\ }\href@noop {} {\  (\bibinfo {year}
  {2022}{\natexlab{b}})},\ \Eprint {https://arxiv.org/abs/2209.09235}
  {arXiv:2209.09235 [hep-th]} \BibitemShut {NoStop}%
\bibitem [{\citenamefont {Levi}\ \emph {et~al.}(2022)\citenamefont {Levi},
  \citenamefont {Morales},\ and\ \citenamefont {Yin}}]{Levi:2022dqm}%
  \BibitemOpen
  \bibfield  {author} {\bibinfo {author} {\bibfnamefont {M.}~\bibnamefont
  {Levi}}, \bibinfo {author} {\bibfnamefont {R.}~\bibnamefont {Morales}},\ and\
  \bibinfo {author} {\bibfnamefont {Z.}~\bibnamefont {Yin}},\ }\bibfield
  {title} {\bibinfo {title} {{From the EFT of Spinning Gravitating Objects to
  Poincar\'e and Gauge Invariance}},\ }\href@noop {} {\  (\bibinfo {year}
  {2022})},\ \Eprint {https://arxiv.org/abs/2210.17538} {arXiv:2210.17538
  [hep-th]} \BibitemShut {NoStop}%
\bibitem [{\citenamefont {Levi}\ and\ \citenamefont
  {Yin}(2022)}]{Levi:2022rrq}%
  \BibitemOpen
  \bibfield  {author} {\bibinfo {author} {\bibfnamefont {M.}~\bibnamefont
  {Levi}}\ and\ \bibinfo {author} {\bibfnamefont {Z.}~\bibnamefont {Yin}},\
  }\bibfield  {title} {\bibinfo {title} {{Completing the Fifth PN Precision
  Frontier via the EFT of Spinning Gravitating Objects}},\ }\href@noop {} {\
  (\bibinfo {year} {2022})},\ \Eprint {https://arxiv.org/abs/2211.14018}
  {arXiv:2211.14018 [hep-th]} \BibitemShut {NoStop}%
\bibitem [{\citenamefont {Levi}\ \emph
  {et~al.}(2021{\natexlab{b}})\citenamefont {Levi}, \citenamefont
  {Mougiakakos},\ and\ \citenamefont {Vieira}}]{Levi:2019kgk}%
  \BibitemOpen
  \bibfield  {author} {\bibinfo {author} {\bibfnamefont {M.}~\bibnamefont
  {Levi}}, \bibinfo {author} {\bibfnamefont {S.}~\bibnamefont {Mougiakakos}},\
  and\ \bibinfo {author} {\bibfnamefont {M.}~\bibnamefont {Vieira}},\
  }\bibfield  {title} {\bibinfo {title} {{Gravitational cubic-in-spin
  interaction at the next-to-leading post-Newtonian order}},\ }\href
  {https://doi.org/10.1007/JHEP01(2021)036} {\bibfield  {journal} {\bibinfo
  {journal} {JHEP}\ }\textbf {\bibinfo {volume} {01}},\ \bibinfo {pages}
  {036}},\ \Eprint {https://arxiv.org/abs/1912.06276} {arXiv:1912.06276
  [hep-th]} \BibitemShut {NoStop}%
\bibitem [{\citenamefont {Levi}\ and\ \citenamefont
  {Teng}(2021)}]{Levi:2020lfn}%
  \BibitemOpen
  \bibfield  {author} {\bibinfo {author} {\bibfnamefont {M.}~\bibnamefont
  {Levi}}\ and\ \bibinfo {author} {\bibfnamefont {F.}~\bibnamefont {Teng}},\
  }\bibfield  {title} {\bibinfo {title} {{NLO gravitational quartic-in-spin
  interaction}},\ }\href {https://doi.org/10.1007/JHEP01(2021)066} {\bibfield
  {journal} {\bibinfo  {journal} {JHEP}\ }\textbf {\bibinfo {volume} {01}},\
  \bibinfo {pages} {066}},\ \Eprint {https://arxiv.org/abs/2008.12280}
  {arXiv:2008.12280 [hep-th]} \BibitemShut {NoStop}%
\bibitem [{\citenamefont {Arun}\ \emph {et~al.}(2009)\citenamefont {Arun},
  \citenamefont {Buonanno}, \citenamefont {Faye},\ and\ \citenamefont
  {Ochsner}}]{Arun:2008kb}%
  \BibitemOpen
  \bibfield  {author} {\bibinfo {author} {\bibfnamefont {K.~G.}\ \bibnamefont
  {Arun}}, \bibinfo {author} {\bibfnamefont {A.}~\bibnamefont {Buonanno}},
  \bibinfo {author} {\bibfnamefont {G.}~\bibnamefont {Faye}},\ and\ \bibinfo
  {author} {\bibfnamefont {E.}~\bibnamefont {Ochsner}},\ }\bibfield  {title}
  {\bibinfo {title} {{Higher-order spin effects in the amplitude and phase of
  gravitational waveforms emitted by inspiraling compact binaries: Ready-to-use
  gravitational waveforms}},\ }\href
  {https://doi.org/10.1103/PhysRevD.79.104023} {\bibfield  {journal} {\bibinfo
  {journal} {Phys. Rev. D}\ }\textbf {\bibinfo {volume} {79}},\ \bibinfo
  {pages} {104023} (\bibinfo {year} {2009})},\ \bibinfo {note} {[Erratum:
  Phys.Rev.D 84, 049901 (2011)]},\ \Eprint {https://arxiv.org/abs/0810.5336}
  {arXiv:0810.5336 [gr-qc]} \BibitemShut {NoStop}%
\bibitem [{\citenamefont {Boyle}\ \emph {et~al.}(2014)\citenamefont {Boyle},
  \citenamefont {Kidder}, \citenamefont {Ossokine},\ and\ \citenamefont
  {Pfeiffer}}]{Boyle:2014ioa}%
  \BibitemOpen
  \bibfield  {author} {\bibinfo {author} {\bibfnamefont {M.}~\bibnamefont
  {Boyle}}, \bibinfo {author} {\bibfnamefont {L.~E.}\ \bibnamefont {Kidder}},
  \bibinfo {author} {\bibfnamefont {S.}~\bibnamefont {Ossokine}},\ and\
  \bibinfo {author} {\bibfnamefont {H.~P.}\ \bibnamefont {Pfeiffer}},\
  }\bibfield  {title} {\bibinfo {title} {{Gravitational-wave modes from
  precessing black-hole binaries}},\ }\href@noop {} {\  (\bibinfo {year}
  {2014})},\ \Eprint {https://arxiv.org/abs/1409.4431} {arXiv:1409.4431
  [gr-qc]} \BibitemShut {NoStop}%
\bibitem [{\citenamefont {Mart{\'\i}n-Garc{\'\i}a}()}]{xAct}%
  \BibitemOpen
  \bibfield  {author} {\bibinfo {author} {\bibfnamefont {J.~M.}\ \bibnamefont
  {Mart{\'\i}n-Garc{\'\i}a}},\ }\href@noop {} {\bibinfo {title} {{xAct}:
  Efficient tensor computer algebra}},\ \bibinfo {howpublished}
  {\url{http://www.xact.es/}}\BibitemShut {NoStop}%
\bibitem [{\citenamefont {Mart{\'\i}n-Garc{\'\i}a}(2008)}]{martin2008xperm}%
  \BibitemOpen
  \bibfield  {author} {\bibinfo {author} {\bibfnamefont {J.~M.}\ \bibnamefont
  {Mart{\'\i}n-Garc{\'\i}a}},\ }\bibfield  {title} {\bibinfo {title} {{xPerm}:
  fast index canonicalization for tensor computer algebra},\ }\href
  {https://doi.org/10.1016/j.cpc.2008.05.009} {\bibfield  {journal} {\bibinfo
  {journal} {Computer physics communications}\ }\textbf {\bibinfo {volume}
  {179}},\ \bibinfo {pages} {597} (\bibinfo {year} {2008})},\ \Eprint
  {https://arxiv.org/abs/0803.0862} {arXiv:0803.0862 [cs.SC]} \BibitemShut
  {NoStop}%
\bibitem [{\citenamefont {Kidder}(1995)}]{Kidder:1995zr}%
  \BibitemOpen
  \bibfield  {author} {\bibinfo {author} {\bibfnamefont {L.~E.}\ \bibnamefont
  {Kidder}},\ }\bibfield  {title} {\bibinfo {title} {{Coalescing binary systems
  of compact objects to postNewtonian 5/2 order. 5. Spin effects}},\ }\href
  {https://doi.org/10.1103/PhysRevD.52.821} {\bibfield  {journal} {\bibinfo
  {journal} {Phys. Rev. D}\ }\textbf {\bibinfo {volume} {52}},\ \bibinfo
  {pages} {821} (\bibinfo {year} {1995})},\ \Eprint
  {https://arxiv.org/abs/gr-qc/9506022} {arXiv:gr-qc/9506022} \BibitemShut
  {NoStop}%
\end{thebibliography}%

\end{document}